\documentclass[twoside,12pt]{article}
\usepackage{epsfig}

\def\ET{\mbox{$E_T$}}
\def\pT{\mbox{$p_T$}}

\def\kT{\mbox{$k_T$}}
\def\jT{\mbox{$j_T$}}
\def\mT{\mbox{$m_T$}}

\def\psat{\mbox{$p_{sat}$}}

\def\KThbt{\mbox{$K_T$}}

\def\sqrtsNN{\mbox{$\sqrt{s_\mathrm{_{NN}}}$}}
\def\sqrts{\mbox{$\sqrt{s}$}}
\def\Npart{\mbox{$N_{part}$}}
\def\NpartMean{\mbox{$\langle\Npart\rangle$}}
\def\Nbinary{\mbox{$N_{bin}$}}
\def\NbinaryMean{\mbox{$\langle\Nbinary\rangle$}}

\def\pizero{\mbox{$\pi^0$}}
\def\kzeros{\mbox{$K^0_s$}}

\def\pbar{\mbox{$\bar{p}$}}
\def\lam{\mbox{$\Lambda$}}
\def\lambar{\mbox{$\bar\Lambda$}}

\def\Ominus{\mbox{$\Omega^-$}}

\def\pbarp{\mbox{$\pbar+p$}}

\def\ePluseMinus{\mbox{$e^++e^-$}}

\def\jpsi{\mbox{$\mathrm{J/}\psi$}}

\def\vtwo{\mbox{$v_2$}}

\def\RAA{\mbox{$R_{AA}(\pT)$}}
\def\IAA{\mbox{$I_{AA}(\pT)$}}
\def\RAB{\mbox{$R_{AB}(\pT)$}}
\def\RCP{\mbox{$R_{CP}(\pT)$}}

\def\TAB{\mbox{$T_{AB}(\vec{b})$}}

\def\Qs{\mbox{$Q_s$}}
\def\Qsmin{\mbox{$Q_{s,min}$}}
\def\Qsmax{\mbox{$Q_{s,max}$}}

\def\LamQCD{\mbox{$\Lambda_{QCD}$}}

\def\xBj{\mbox{$x_{Bj}$}}

\def\epsBj{\mbox{$\epsilon_{Bj}$}}

\def\muB{\mbox{$\mu_B$}}

\def\Tfo{\mbox{$T_{fo}$}}
\def\betaMean{\mbox{$\left<\beta_T\right>$}}

\def\epsx{\mbox{$\epsilon_x$}}
\def\epsp{\mbox{$\epsilon_p$}}

\def\lt{\mbox{$<$}}
\def\gt{\mbox{$>$}}

\def\IntLumi{\mbox{$\int\mathcal{L}dt$}}

\begin{document}

\title{Matter {\it in extremis}: ultrarelativistic nuclear collisions at 
RHIC~\thanks{To be published in Progress in Particle and Nuclear Physics.}}

\author{Peter Jacobs\thanks{pmjacobs@lbl.gov}\ \ and Xin-Nian Wang\thanks{xnwang@lbl.gov}, \\
Nuclear Science Division, \\
Lawrence Berkeley National Laboratory, \\
Berkeley, California 94720}

\maketitle



\begin{abstract}
We review the physics of nuclear matter at high energy density and the
experimental search for the Quark-Gluon Plasma at the Relativistic
Heavy Ion Collider (RHIC). The data obtained in the first three years
of the RHIC physics program provide several lines of evidence that a
novel state of matter has been created in the most violent, head-on
collisions of $Au$ nuclei at $\sqrt{s}=200$ GeV. Jet quenching and
global measurements show that the initial energy density of the
strongly interacting medium generated in the collision is about two
orders of magnitude larger than that of cold nuclear matter, well
above the critical density for the deconfinement phase transition
predicted by lattice QCD. The observed collective flow patterns imply
that the system thermalizes early in its evolution, with the dynamics
of its expansion consistent with ideal hydrodynamic
flow based on a Quark-Gluon Plasma equation of state.

\end{abstract}

\tableofcontents

\section{Introduction}

Quantum Chromodynamics (QCD) is the fundamental theory underlying the
strong interaction between quarks and gluons that dictates the
structure of hadrons and nuclei. The non-Abelian nature of the QCD gauge
symmetry gives rise to confinement of quarks and gluons,
making hadrons the only stable vacuum excitations of the strong interaction. 
Under extreme conditions of high temperature or baryon density the
structure of the vacuum is predicted to change, undergoing a phase
transition that restores the broken symmetries \cite{Lee:1974ma} and
frees quarks and gluons from confinement \cite{Collins:1975ky,Baym:1976yu}. 
Numerical calculation of the equation of state (EOS) in lattice QCD theory
indeed finds a nearly first-order phase transition at a critical
temperature of about $170$ MeV \cite{Karsch:2001vs}.
Such a form of matter, called the Quark-Gluon Plasma (QGP)
\cite{Shuryak:1978ij,Kalashnikov:1979dp,Kapusta:1979fh}, existed in
the early universe a few microseconds after the Big Bang. Quark matter
may still exist today in the core of neutron stars, where the baryon
density exceeds the critical value of the phase transition
\cite{Glendenning:2001pe}.

Soon after the QGP was identified as the excited phase of the QCD
vacuum, it was realized that high energy collisions of heavy nuclei
could create large volumes of matter at high energy density
\cite{Baumgardt:1975qv} and that the energy density achievable at current
or foreseeable nuclear accelerators might be sufficient to create the QGP in
controlled laboratory experiments. Searches for the QGP using
heavy-ion collisions have been carried out over the past three decades
at successively higher energy facilities, beginning at the Berkeley
Bevalac in the early 1980s \cite{Nagamiya:1984kn} and continuing at
the Brookhaven AGS and CERN SPS in the late 1980s
\cite{Harris:1996zx}. The Relativistic Heavy Ion Collider (RHIC), the
subject of this review, was commissioned in 2000.  Future experiments
are planned at even higher energies at the CERN LHC \cite{Giubellino:2003in}.

Experimental results from the AGS and SPS have revealed a rich set of
new phenomena indicating the formation of dense matter
\cite{Heinz:2000bk,Satz:2002ku}, notably the anomalous
suppression of charmonium \cite{Alexopoulos:2002eh} and the broadening or
mass-shift of vector mesons \cite{Agakishiev:1995xb}. However, it has
proven difficult to disentangle hadronic contributions to the observed
signals, and no clear consensus has emerged that a long-lived Quark
Gluon Plasma has been observed in the fixed target heavy ion
experiments at those facilities.

Experiments at RHIC, the world's first heavy ion collider, have
initiated a new era in the study of QCD matter under extreme
conditions. Nuclear collisions at the highest RHIC energy
(nucleon-nucleon CM energy \sqrtsNN=200 GeV) not only produce matter
at the highest energy density ever achieved, but also provide a number
of rare observables that have not been accessible previously.  These
observables are especially clean probes of the hot and dense matter
generated in the collision.  The wealth of data collected and analyzed
in the first three years of RHIC operation indicates that a dense,
equilibrated system is generated in the most violent, head-on
collisions of heavy nuclei. The initial energy density probed by hard
processes and global measurements is estimated to be about two orders
of magnitude larger than that of cold nuclear matter, well above the
critical density for the deconfinement transition predicted by lattice
QCD calculation. The collective behavior of the observed final-state
hadrons provides evidence of early thermalization and hydrodynamic
flow consistent with a Quark-Gluon Plasma equation of state. The matter is
evidently a near-ideal, strongly coupled fluid, and is quite different
from the ideal non-interacting gas expected from QCD at asymptotically
high temperature.

The first set of experimental runs to survey the RHIC physics
landscape has now been completed and it is opportune to pause and
examine where things stand. In this review we will outline the RHIC
heavy ion scientific program and discuss the main experimental results
that support the foregoing conclusions. A large array of data is
available, with over 50 Physical Review Letters published by the four
RHIC experiments thus far. In limited space we cannot review all of
the physics topics under study at RHIC. We therefore concentrate on
those topics for which the data are mature and most clearly address
the issues of QGP formation. There has also been considerable recent
interest in universal properties of high density QCD at low Bjorken
\xBj\
\cite{Iancu:2003xm} and we will discuss RHIC data relevant
to this physics. There are a number of significant topics that we touch
only lightly or omit altogether, among them fluctuations and heavy
quark and quarkonium production. These areas are developing rapidly
and a review at this time would soon become outdated. Detailed
discussions of all aspects of RHIC heavy ion physics can be found in
the proceedings of recent Quark Matter conferences \cite{QM02,QM04}.

The review is organized as follows: section 2 presents a theoretical
overview of the QCD phase transitions, together with theoretical
considerations for analyzing heavy-ion nuclear collisions; section 3
presents the RHIC collider and experiments; section 4 discusses bulk
particle production and the constraints it places on the collision
dynamics; section 5 presents evidence from collective phenomena that
equilibrium is achieved early in the fireball evolution; and section 7
discusses hard probes, concentrating on the theory and measurements of
jet quenching. Section 8 presents a summary and outlook.

\section{QCD and the Quark Gluon Plasma}

Within the Standard Model, the strong interaction between quarks is
mediated through non-Abelian gauge gluon fields, as
described by a theory called Quantum Chromodynamics (QCD), with
\begin{equation}
{ \cal L}_{QCD} =\sum_{i=1}^{n_f}\bar{\psi}_i\gamma_\mu(i\partial^\mu
-gA^\mu_a\frac{\lambda_a}{2}-m_i)\psi_i
-\frac{1}{4} \sum_a F_a^{\mu\nu} F_{a,\mu\nu},
\label{lag-qcd}
\end{equation}
where $\lambda_a$'s are Gell-Mann
matrices in the fundamental representation of $SU(3)$ and
$F_a^{\mu\nu}=\partial^\mu A^\nu_a-\partial^\nu A^\mu_a 
+ igf_{abc}A^\mu_b A^\nu_c$ are gluon field-strength tensors.
The seemingly simple theory is 
very similar in form to Quantum Electrodynamics (QED) but it possesses 
much richer structure because of its many symmetries, including 
the $SU(3)$ gauge symmetry, the approximate chiral symmetry for the 
light quarks that is spontaneously broken in the QCD vacuum. $U_A(1)$ 
symmetry and scale invariance are both broken through quantum 
interactions, leading to $U_A(1)$ (or chiral) and scale (or trace) 
anomalies. These symmetries 
and their breaking dictate the structure of the vacuum and properties 
of strongly interacting matter, including the different phases that
have been the focus of many theoretical and experimental studies
since QCD was established as the theory of strong interactions.

\subsection{Deconfinement}
\label{sec:deconfinement}

Many remarkable features of QCD can be traced to the underlying
$SU(3)$ gauge symmetry of the strong interaction between quarks and
gluons, which both carry color charges. Because of the non-Abelian
self-interaction among gluons, the color charges at short distance are
anti-screened due to color diffusion via gluon radiation, leading to a
weakening of the coupling constant. This asymptotic freedom of the strong
interaction \cite{Gross:1973id,Politzer:1974fr} opens the door for
perturbative studies of the strong interaction within QCD, including
renormalization. The scale dependence of the strong coupling, which at
the leading order is given by
\begin{equation}
\alpha_s(Q^2)=\frac{4\pi}{(11-\frac{2}{3}n_f)\log Q^2/\Lambda_{QCD}^2},
\label{coupling}
\end{equation}
has been successfully tested by many experiments \cite{Bethke:2000ai}.
While the small value of the strong coupling constant implies weak
interaction and thus makes it possible to calculate physical processes
at short distance, the divergence at small energy scale or long
distance signals strong interactions and therefore color
confinement. Indeed, non-perturbative calculations of the heavy quark
potential in lattice QCD show a linear potential at long distance with
a string tension $\kappa \approx 1$ GeV/fm
\cite{Creutz:1980zw,Necco:2001xg}, indicating confinement of quarks to
the domain of hadrons in normal vacuum. Since the average
inter-particle distance becomes smaller at higher temperature or
density, the interaction among quarks and gluons
becomes weaker and confinement will eventually disappear, leading to a new phase
of matter called the Quark Gluon Plasma (QGP) which is distinctly
different from the hadronic phase. The energy density of such
non-interacting QGP is
\begin{equation}
\epsilon_{SB}\equiv
\epsilon_{q+\bar{q}}+\epsilon_g=\left[6n_f\frac{7\pi^2}{120}
+16\frac{\pi^2}{30}\right] T^4
\label{esplionsb}
\end{equation}
with the pressure $P=\epsilon/3$. In contrast, $\epsilon_\pi=3\pi^2T^4/30$
for a massless pion gas.

\begin{figure}
\centering
\includegraphics[width=.48\textwidth]{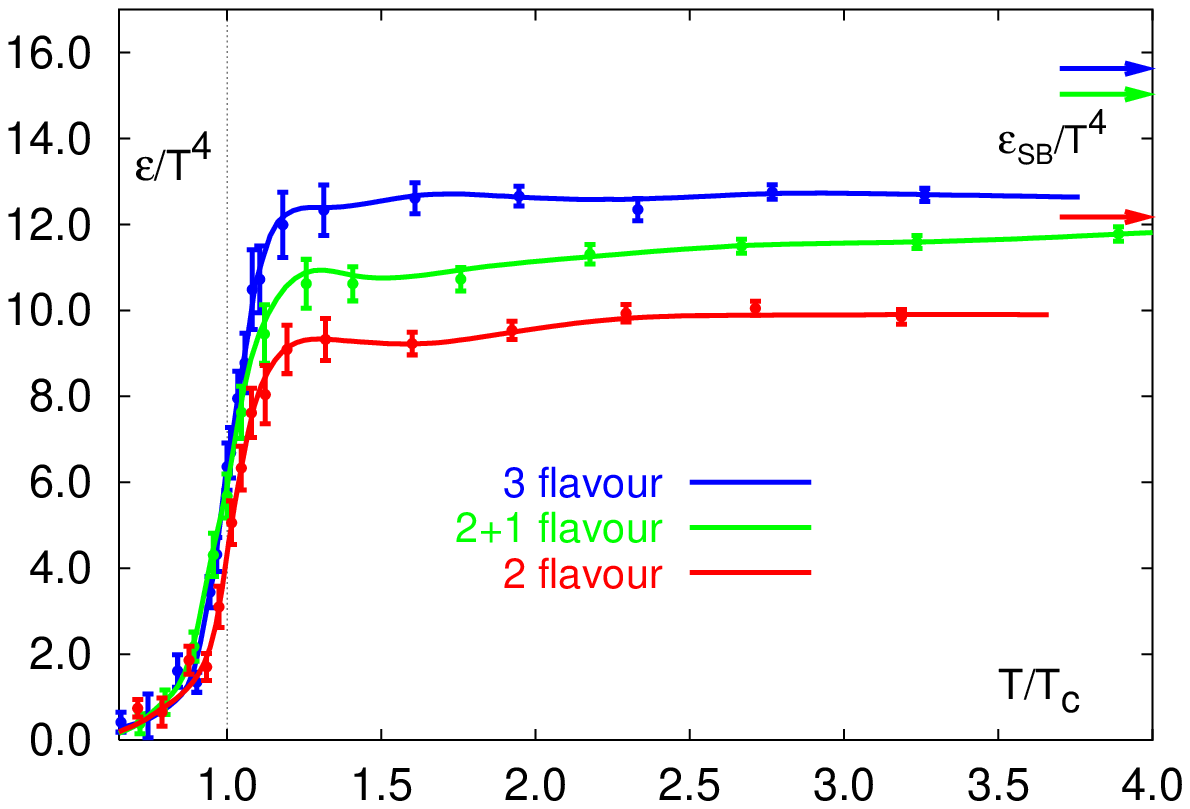}
\includegraphics[width=.48\textwidth]{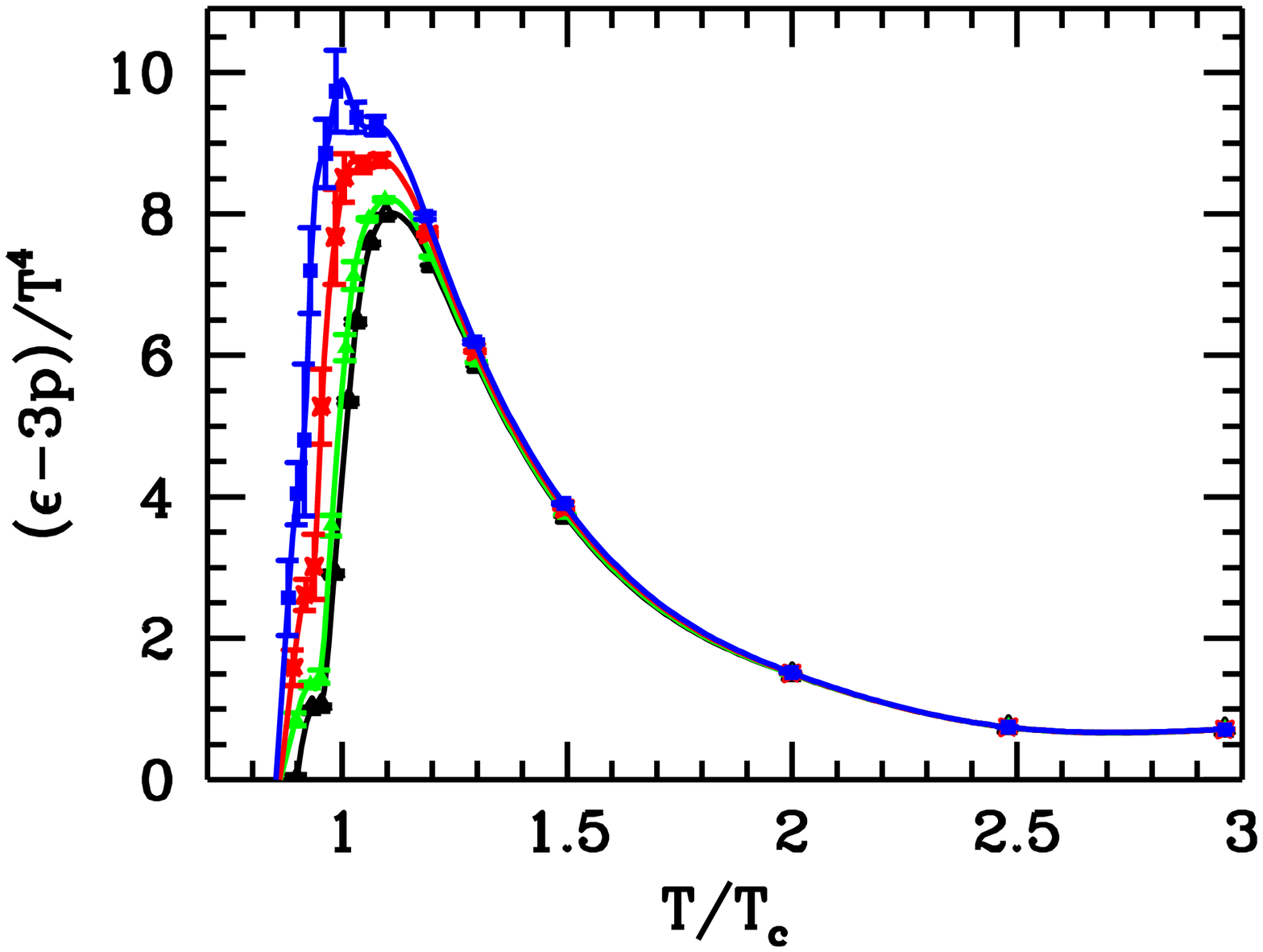}
\caption{Left: The energy density as a function of temperature 
from lattice QCD \protect\cite{Karsch:2001vs}. Arrows show the ideal
gas values $\epsilon_{SB}$ from Eq.~(\protect\ref{esplionsb}).  Right:
Deviation from ideal gas EOS $(\epsilon-3P)/T^4$ at $\mu_B$=0, 210,
410 and 530 MeV (bottom to top) as a function of $T/T_c$
\protect\cite{Fodor:2002sd}.}
\label{lattice}
\end{figure}

Related to confinement, the renormalization of the
coupling constant also breaks the scale invariance of QCD, leading
to the trace anomaly 
$T^\mu_{\;\;\mu}=(\alpha_s/12\pi)F^{\mu\nu}F_{\mu\nu}+m\bar{\psi}\psi$
of the energy-momentum tensor. The non-vanishing value of the
vacuum expectation value of the gluon condensate \cite{Shifman:1979bx},
\begin{equation}
48B\equiv \langle \frac{\alpha_s}{\pi} F^{\mu\nu}F_{\mu\nu}\rangle
\approx 0.02 {\rm GeV}^4
\end{equation}
implies positive pressure and energy density in the vacuum that
confines the weakly interacting quarks and gluons inside a hadron
according to the MIT bag model \cite{Chodos:1974pn}. This concept can
be extended to the equation of state of a non-interacting QGP and
massless pion gas, generating a first-order phase transition at a
temperature $T_c\approx 0.72 B^{1/4}$ for a baryon-free system. Such a
phase transition has indeed been found in numerical simulations of QCD
on the lattice. Fig.~\ref{lattice} shows results from a recent lattice
QCD calculation of the energy density as a function of temperature
\cite{Karsch:2001vs}. The sharp rise of the energy density at
$T_c\approx 170$ MeV for two light quark flavors signals an increase
of the effective number of degrees of freedom. The transition becomes
first order for three light flavors, but the order of the phase
transition is still not clear for a realistic value of the strange
quark mass. However, a sharp cross-over is clearly present.

Shown as horizontal arrows in Fig.~\ref{lattice} (left panel) are the
energy densities of a non-interacting and massless parton gas as given
by Eq.~(\ref{esplionsb}). Lattice QCD results clearly deviate from
these limits, indicating strong interaction among partons even above
the phase transition temperature $T_c$. Higher order corrections to
the perturbative calculation in the weak coupling region do not
improve agreement with the lattice results.  For reasonably small
values of coupling constant $g$ or $\alpha_s$, the perturbative
expansion series up to order $g^4$ is found not to converge
\cite{Arnold:1994ps}. This indicates that massless partons are not a
good basis for perturbative expansion because of the contribution of
soft modes with momentum $k\sim g T$. Resummation of hard thermal
loops (HTL)\cite{Braaten:1990mz} leads to a picture of quasi-particle
modes dominating the interaction in an interacting QGP. Working on the
basis of quasi-particles and solving Dyson's equation in a
self-consistent resummation, an equation of state (EOS) is obtained
that is very close to that of lattice QCD results at temperatures
above but close to the phase transition temperature
\cite{Blaizot:1999ap}.

Since the EOS of an ideal massless QGP is $P=\epsilon/3$, a measure of
deviation from an ideal gas EOS is $\epsilon -3P$. Lattice QCD results
\cite{Fodor:2002sd} (Fig.~\ref{lattice}, right panel)
indicate the presence of strong interactions around the phase
transition temperature. Since the sound velocity is given by
$c_s^2=\partial P/\partial \epsilon$, the softening of the EOS around
the phase transition region would slow down hydrodynamic expansion,
with significant consequences for the evolution of the dense matter
and the relation of observed transverse energy production to the
initial energy density prior to the start of hydrodynamic expansion.

In addition to large energy density, other properties of the quark
gluon plasma will be quite different from those of a hadron gas
or normal nuclear matter. Hadrons as bound states of quarks and
anti-quarks do not exist in the plasma. Lattice studies of the
heavy quark potential find that the confining linear potential
in vacuum disappears once the temperature is above the phase
transition temperature and what is left over is a screened Coulomb
potential \cite{Karsch:2000kv}. This is why heavy quarkonium was
proposed as a good probe of the quark gluon plasma formed in
high-energy heavy-ion collisions \cite{Matsui:1986dk}, since the
quarkonium bound state will be dissolved into an unbound
quark-antiquark pair once the binding potential is screened in the
plasma. Though various theoretical estimates also predict quarkonium
break-up through hadronic interactions
\cite{Matinyan:1998cb,Haglin:1999xs,Lin:1999ad}, thermal gluon
exchange screening of potential in a quark gluon plasma remains the
most efficient way to suppress quarkonium.

Simulations on the lattice are difficult for a system at finite baryon
chemical potential $\mu_B$. However, significant progress in this area
has been made in recent years. The phase transition is found to change
from first-order at high baryon density to a cross over at lower
values of $\mu_B$
\cite{Fodor:2002km,Allton:2003vx}, though the position of the
end-point is still uncertain and depends on the values of quark
masses. Such an end-point in the phase diagram of QCD matter could
have important effects, for instance increased baryon number
fluctuations in heavy-ion collisions at low energy if the end-point is
crossed over in the evolution of the fireball.

At high baryon densities and zero or low temperatures, the attractive
interaction between quarks in the color-anti-triplet and spin-zero
channels could lead to pairing of quarks in iso-singlet spin-zero and
color-anti-triplet states on the Fermi surface, thus forming diquark
condensates in two flavor quark matter
\cite{Alford:1998zt,Rapp:1998zu}. With three flavors of quarks, the
Cooper quark pairs cannot be iso-singlet. The attractive interaction
is found to be favored in the channel where colors and flavors are
locked \cite{Alford:1998mk}. Such color-flavor locking leads to a
plethora of new structures in the phases of dense quark matter which
may generate novel effects in neutron stars where such high densities
could be reached.

\subsection{Chiral Symmetry Restoration}

In addition to the $SU(3)$ local gauge symmetry, the QCD interaction
as given by Eq.~(\ref{lag-qcd}) has the global symmetries of
$SU(3)\times SU_A(3) \times U(1) \times U_A(1)$ in the quark
sector. The global $U(1)$ symmetry is responsible for baryon number
conservation while chiral symmetry, which is approximate for vanishing
quark masses, also leads to many unique properties of QCD.

If the quark fields are decomposed into left and right chirality
components, $\psi_{L,R}=(1\mp \gamma_5)\psi$, the QCD Lagrangian
with three massless quarks is invariant under the transformation,
\begin{equation}
\psi_{L,R} \rightarrow e^{-i\theta_{L,R}^i\lambda^i}\psi_{L,R} \;.
\end{equation}
According to the Noether theorem, this will generate two kinds of
conserved currents or their combination, vector and axial-vector 
currents,
\begin{eqnarray}
V_i^\mu&=&\bar{\psi}\gamma^\mu\frac{\lambda_i}{2}\psi; \nonumber \\
A_i^\mu&=&\bar{\psi}\gamma^\mu\gamma_5\frac{\lambda_i}{2}\psi \;.
\end{eqnarray} 
Identifying each current with a physical state having the
corresponding quantum number results in degeneracy between scalar and
pseudoscalar as well as between vector and psudovector mesons. The
absence of parity doublets led to the the discovery of the spontaneous
breaking of chiral symmetry by the QCD vacuum, manifested by the
non-vanishing of the quark condensate
\begin{equation}
\langle \bar\psi\psi\rangle \equiv \langle \bar\psi_R\psi_L
+\bar\psi_L\psi_R\rangle = -(240{\rm MeV})^3 \;,
\end{equation}
which is directly related to  $f_\pi$, the pion decay constant.
One of the consequences of this spontaneous symmetry breaking
is the existence of massless Goldstone bosons. The absence of the ninth Goldstone
boson is due to the breaking of $U_A(1)$ symmetry by quantum
correction or an anomaly,
\begin{equation}
\partial_\mu A^\mu_0 =\frac{2n_f}{16\pi}\alpha_s 
F_a^{\mu\nu}\tilde{F}_{a,\mu\nu},
\end{equation}
where $\tilde{F}_{a,\mu\nu}=\epsilon_{\mu\nu\alpha\beta}F_a^{\alpha\beta}$.
The non-vanishing vacuum expectation value of this anomaly is
related to what is known as the topological susceptibility of the
vacuum.

All of these order parameters, the quark and gluon condensates and the
topological susceptibility, are believed to be related to the gluonic
structure of the vacuum \cite{Schafer:1998wv}. They dictate all
hadronic properties -- their mass spectra, decay widths and
constants. The medium modification of these condensates at low
temperature and density is calculable within the framework of chiral
perturbation theory \cite{Gasser:1987vb}. At higher temperature and in
the quark gluon plasma phase, they are expected to vanish. The quark
condensate in lattice QCD calculations \cite{Karsch:2001vs} is shown
to disappear above the QCD phase transition temperature characterized
by a sharp cross-over, which coincides with the cross-over of the
expectation value of the Polyakov loop, a measure of
deconfinement. At these temperatures and densities, chiral symmetry is
completely restored and $U_A(1)$ symmetry is also partially
restored. Hadronic properties will be very different from those in
vacuum. There will be mixing of vector and axial vector currents and
all the chiral multiplets will become degenerate.

There are many consequences of the restoration of chiral and $U_A(1)$
symmetry in the dense medium. The vanishing value of the topological
susceptibility gives rise to a reduction of the ninth Goldstone boson,
leading to possible enhancement of $\eta$ mesons
\cite{Kapusta:1996ww,Huang:1996fc}. A rapid cooling of the dense
medium initially in a chirally symmetric state could introduce
instabilities into the evolution of chiral fields leading to the
amplification of soft modes \cite{Rajagopal:1993ah}, though a
realistic treatment of the expansion and cooling of a finite system
can prevent the creation of such disoriented chiral condensates (DCC)
in large domains
\cite{Asakawa:1995wk}. Perhaps the most promising signals of chiral
symmetry restoration are modifications of hadron properties in medium,
{\it i.e.} the masses and widths of vector mesons and axial vector
mesons via mixing, as manifested in the soft dilepton spectra
\cite{Rapp:1999ej}. The enhanced soft dilepton yield below the $\rho$
mass region seen by the CERES experiment at SPS
\cite{Agakishiev:1995xb} may be attributable to the medium modification
of the $\rho$ meson. During the QGP phase, annihilation of thermal
quarks and antiquarks could also contribute to the underlying dilepton
spectra in the intermediate mass region.

\subsection{Perturbative QCD}

The most important practical consequence of asymptotic freedom for the
$SU(3)$ gauge interaction in QCD is the success of perturbative QCD
(pQCD) in describing various processes involving strong interactions at short
distance. Because of the small coupling constant, it is possible to
make a systematic expansion of physical cross sections in the strong
coupling constant when the momentum transfer involved is large. Because
of color confinement in vacuum, both the initial and final observed
particles will be hadrons and therefore will involve strong
interaction at long distance, which is not calculable within the
framework of the perturbative expansion. However, it has been proven to at
least leading power correction ($1/Q^2$) that the cross section can
be factorized into short-distance parts calculable in pQCD
and non-perturbative long distance parts
\cite{Dokshitzer:1980hw,Mueller:1981sg,Collins:1985ue}.

The long distance parts normally involve hadronic wavefunctions, parton
distributions, and hadronization. Though they cannot be calculated
perturbatively, their matrix element definitions are universal,
independent of specific processes. If measured in one process they can
be applied to another process; therein lies the predictive power
of pQCD. The renormalization of these non-perturbative
matrix elements due to initial and final state radiation in pQCD
leads to the Dokshitzer-Gribov-Lipatov-Altarelli-Parisi (DGLAP)
evolution equations
\cite{Gribov:1972ri,Dokshitzer:1977sg,Altarelli:1977zs}.  Given the
experimental measurements of these distributions at one scale, the
DGLAP equations predict their evolution to higher momentum
scales. For example, final-state radiation leads to the DGLAP evolution
equations for parton fragmentation functions $D_{a\rightarrow
h}(z_h,Q^2)$,
\begin{eqnarray}
  \frac{\partial D_{q\rightarrow h}(z_h,Q^2)}{\partial \ln Q^2} & = &
  \frac{\alpha_s(Q^2)}{2\pi} \int^1_{z_h} \frac{dz}{z} 
\left[ \gamma_{q\rightarrow qg}(z)
D_{q\rightarrow h}(z_h/z,Q^2) + \gamma_{q\rightarrow gq}(z) 
D_{g\rightarrow h}(z_h/z,Q^2)\right],  \label{eq:ap1} \nonumber \\
\frac{\partial D_{g\rightarrow h}(z_h,Q^2)}{\partial \ln Q^2} & = &
\frac{\alpha_s(Q^2)}{2\pi} \int^1_{z_h} \frac{dz}{z} \left[
    \sum_{q=1}^{2n_f} \gamma_{g\rightarrow q\bar{q}}(z)
  D_{q\rightarrow h}(z_h/z,Q^2) + \gamma_{g\rightarrow gg}(z)
 D_{g\rightarrow h}(z_h/z,Q^2)\right] \label{eq:ap2}
\end{eqnarray}
where $\gamma_{a\rightarrow bc}(y)$ are the splitting functions of 
the corresponding radiative processes \cite{Altarelli:1977zs}.
These evolution equations have been tested extensively against 
experiment and can now even be used to measure the scale-dependence 
of the running strong coupling constant \cite{Bethke:2000ai}.

The perturbative QCD parton model is based on this factorization
picture of hard processes. In this model the cross section of a
typical hard process can be expressed as the convolution of initial
parton distributions, perturbative parton scattering cross sections and the
parton fragmentation function. Since hard processes happen on a short
time scale in the very earliest stage of high-energy heavy-ion
collisions, they probe of the bulk matter that is formed shortly after
the collision. The pQCD parton model serves as a reliable and tested
framework for the study of these hard probes. Proposed hard probes in
high-energy heavy-ion collisions include Drell-Yan dileptons from
quark-antiquark annihilation, direct photons from quark and gluon
Compton scattering, heavy quark production, and high $p_T$ jets from
hard parton-parton scattering. In this review, we will focus on the
physics of jet propagation in the dense medium.

\section{The Relativistic Heavy Ion Collider}

The Relativistic Heavy Ion Collider (RHIC) \cite{rhic:overview} at
Brookhaven National Laboratory is the world's highest energy
accelerator of heavy nuclei and the world's first polarized proton
collider. In this section we describe the accelerator and the
experiments, together with some theoretical considerations important
for the analysis of ultrarelativistic heavy ion collisions.


\subsection{The Accelerator}
\label{sect:rhic}
 
The collider consists of two independent, concentric acceleration and
storage rings, with a circumference of 3.8 km. All storage ring
magnets are superconducting, cooled to 4.2 K by a 25 kW helium
refrigerator. There are six intersection points, of which four are
currently instrumented with experiments. RHIC can store and collide
beams with masses ranging from protons to the heaviest stable
nuclei. Due to the independence of the rings RHIC has great
flexibility to collide beams of unequal masses, such as protons or
light ions with $Au$ ions. The top collision energy for the heaviest
nuclear beams is
\sqrtsNN=200 GeV per nucleon pair, while the top energy for 
$p+p$ is \sqrts=500 GeV.

\begin{figure}[htbp]
\centering
\includegraphics[height=.3\textheight]{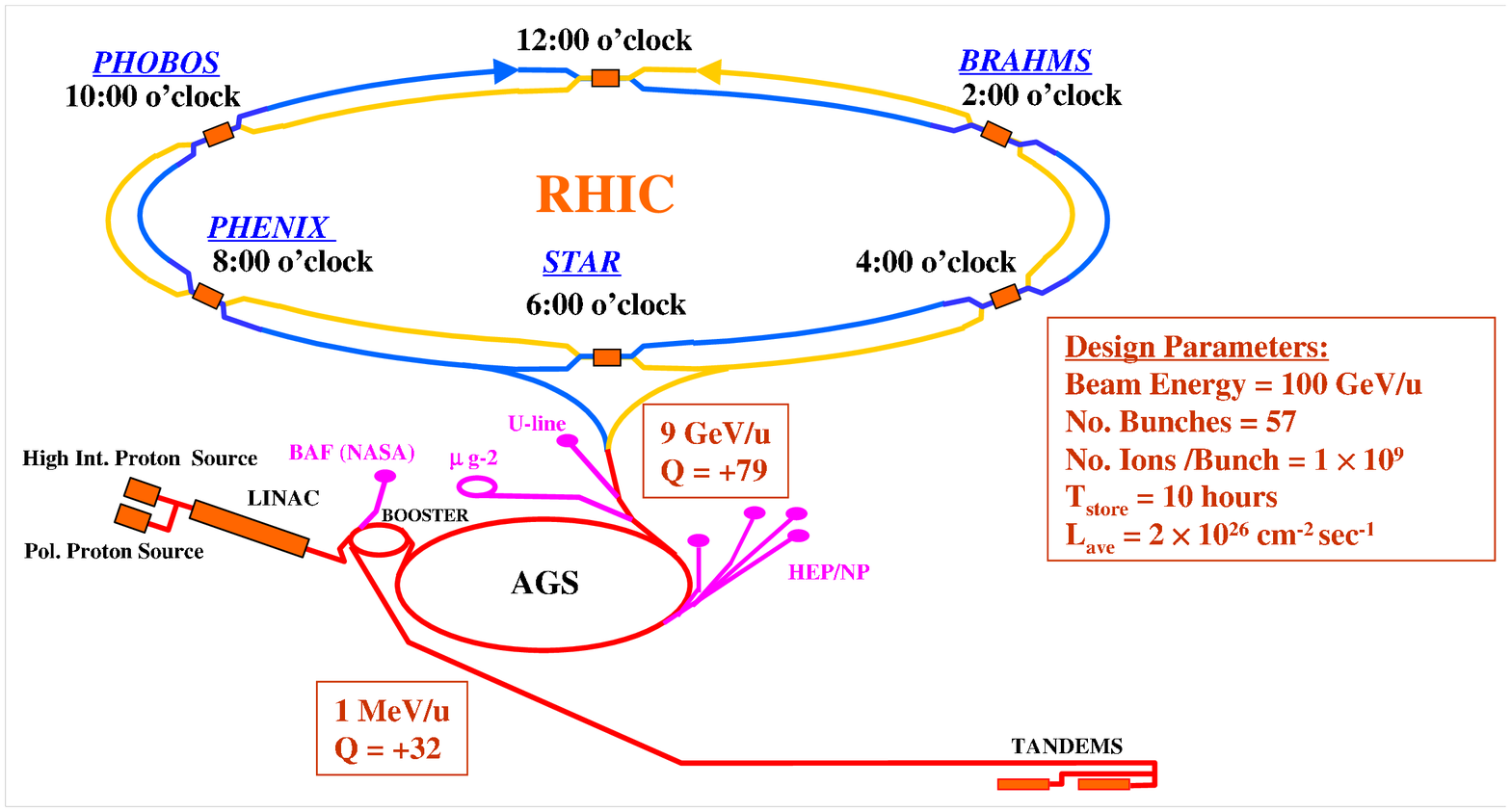}
\caption{The RHIC accelerator complex.}
\label{figRHICLayout}
\end{figure}

The layout of RHIC is shown in Fig. \ref{figRHICLayout}. Heavy ion
beams originate in a pulsed sputter source and are accelerated
successively by a Tandem van der Graaf accelerator, Booster
Synchrotron, and the Alternating Gradient Synchrotron (AGS), where
they are accelerated to 10.8 GeV/nucleon, fully stripped of their
electrons, and injected into RHIC. Polarized protons originate in a
200 MeV Linac and are accelerated by the Booster and the AGS to 24.3
GeV for injection into RHIC. Polarization of protons is maintained by
use of Siberian Snakes \cite{Derbenev:1978hv}. The physics of the RHIC
Spin program is beyond the scope of this review; a recent review and
status report can be found at
\cite{rhic:spin,rhic:spinstatus}.

Acceleration and storage in RHIC utilizes two Radio Frequency (RF)
systems, one at 28 MHz to capture the AGS bunches and accelerate to
top energy, the other at 197 MHz to provide a short collision diamond
($\sigma_L\sim 25$ cm) for efficient utilization of the
luminosity by the experiments. The synchotron phase transition of the
RHIC lattice is at $\gamma_T=24.7$, meaning that all ions except
protons pass through the beam instability at transition.

The main performance specifications of RHIC are given in Table
\ref{table:rhicspec}. For light ions ($A$\lt100), the luminosity 
is limited by beam-beam hadronic interactions, whereas for heavier
ions the luminosity lifetime is limited by intrabeam (intra-bunch)
scattering \cite{rhic:design}. Other processes which significantly
limit the luminosity for heavier ions result from their intense
electromagnetic fields at high energy: Coulomb dissociation and
spontaneous electron-positron production followed by electron capture
\cite{rhic:design}. The luminosity for beams of heavy nuclei scales with 
beam energy as $\gamma^2$.

\begin{table}
\centering
\begin{tabular}{|c||c|c|}
\hline
 & Au+Au & p+p \\
\hline
Beam energy     & $30\rightarrow100$ GeV per nucleon & $30\rightarrow250$ GeV \\
Mean Luminosity at top energy & $2\times10^{26}$ cm$^{-2}$s$^{-1}$ & $1.4\times10^{31}$ cm$^{-2}$s$^{-1}$ \\
Bunches per ring & $60\rightarrow120$ & $60\rightarrow120$ \\
Luminosity lifetime & $\sim10$ hours & \gt10 hours \\
$\beta^*$ & $10\rightarrow1$ m & $10\rightarrow1$ m \\
\hline 
\end{tabular}
\caption{Main design specifications of RHIC for Au+Au and p+p collisions \protect\cite{rhic:overview}.}
\label{table:rhicspec}
\end{table}

Table \ref{table:rhicstatus} lists the beams, energies and integrated
luminosity for the RHIC runs to date. The recently completed runs
labelled ``2004'' have not generated scientific publications as of
this writing. While the luminosities accumulated for heavy ion running
appear at first sight to be small, nuclear geometric effects amplify
hard cross sections approximately as the product of the nuclear masses
$AB$ (Sect. \ref{sect:Glauber}). The rightmost column converts the
integrated luminosity for heavy ions to the equivalent for $p+p$
collisions that would generate the same yield for a process whose
cross section scales as $AB$. In the later runs the store-averaged
luminosities achieved for $Au+Au$ collisions exceed the RHIC design
specifications in Table
\ref{table:rhicspec}.

\begin{table}
\centering
\begin{tabular}{|c|c|c|c|}
\hline
Beams & \sqrtsNN (GeV) & \IntLumi (nb$^{-1}$) & $p+p$ equivalent \IntLumi\ (pb$^{-1}$) \\
\hline
Au+Au & 20       & small & small \\
Au+Au & 130      & 0.02 & 0.8 \\
Au+Au & 200     & 0.24 & 7.9 \\
$\vec{p}+\vec{p}$   & 200       & 1600 & 1.6 \\
d+Au   & 200    & 75 & 3.0 \\
\hline 
Au+Au (2004) & 200 & $\sim$2 &  80 \\
Au+Au (2004) & 62.4 & $\sim$0.05 & 2 \\
\hline
\end{tabular}
\caption{Beam and energy combinations run at RHIC to date. 
\IntLumi\ denotes integrated luminosity
delivered to all four experiments. ``$p+p$ equivalent'' is the
luminosity scaled by the number of binary collisions (note change of
units between columns 3 and 4). Runs labelled ``2004'' were recently
completed and have not generated physics publications as of this
writing.}
\label{table:rhicstatus}
\end{table}

\subsection{The Experiments}
\label{sect:experiments}

The RHIC beams are brought into head-on collision at the intersection
regions. The final dipoles of the lattice are approximately $\pm$10 m
from the collision diamond, with the intervening space free for
detectors. Currently four of the six intersection regions are
instrumented, with two major detectors (PHENIX, STAR) and two smaller
ones (BRAHMS, PHOBOS).

All four experiments contain identical Zero Degree Calorimeters (ZDC)
\cite{ZDC}, which are used for triggering, luminosity monitoring, and event 
characterization. The ZDCs are compact tungsten/fiber hadronic
calorimeters situated immediately downstream of the machine dipole
magnets that define the interaction region, with acceptance 2.5 mrad
centered on the beam direction. The energy flux into this acceptance
is dominated in heavy ion collisions by non-interacting spectator
neutrons. The ZDC energy resolution is sufficient to discriminate
individual beam-velocity neutrons \cite{ZDC}. The luminosity for
$Au+Au$ collisions has been measured with 10\% precision using a
Vernier scan of ZDC coincidence rates
\cite{VanDerMeer,rhic:lumi}.

\paragraph*{PHENIX}

The PHENIX experiment is designed to make precision measurements of a
wide variety of observables, sensitive to multiple time scales in the
evolution of heavy ion collisions. Special emphasis is put on rare
signals (direct photons, lepton pairs, \jpsi\ and $\Upsilon$ families,
jet fragments) that probe the system at the earliest, hot and dense
phase. Measurement of lepton pairs at low \pT\ is a promising tool for
studies of chiral symmetry restoration. Inclusive measurements and
correlations of identified hadrons at high \pT\ are sensitive to
partonic interactions in the medium, and at lower \pT\ probe the late,
hadronic gas stage of the collision.

\begin{figure}[htbp]
\centering
\includegraphics[width=.5\textwidth]{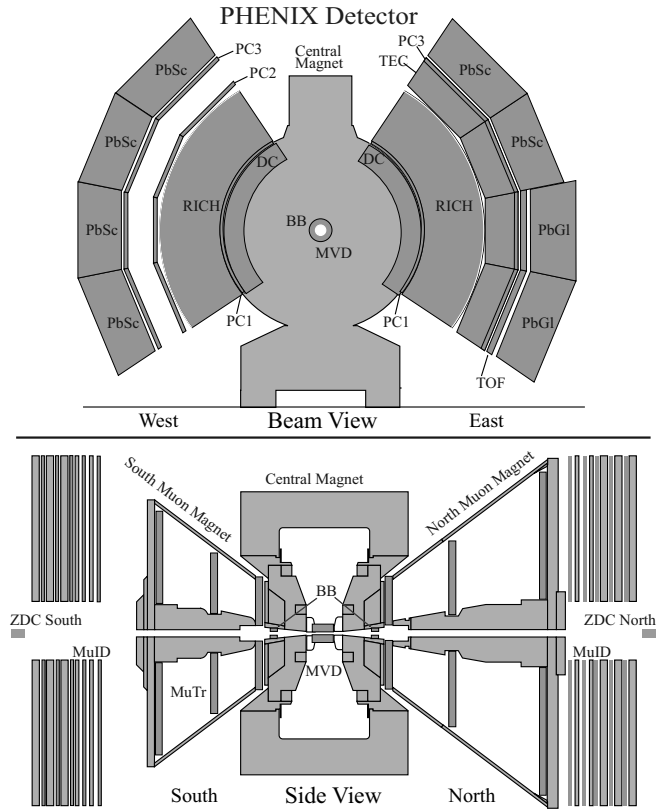}
\caption{The PHENIX detector.} 
\label{figPHENIXLayout} 
\end{figure}

The PHENIX detector \cite{phenix:nim}, shown in
Fig. \ref{figPHENIXLayout}, consists of four independent
spectrometers, two at midrapidity for charged hadrons, electrons, and
photons, and two at forward rapidities for muons. Each has an
acceptance of $\sim1$ steradian. The midrapidity spectrometers have an
axial magnetic field, with tracking for momentum measurements supplied
by drift and pad chambers. Charged particle identification over a
broad momentum range is provided by Time of Flight (TOF), Ring Imaging
Cerenkov (RICH), and Time Expansion Chamber (TEC) detectors, giving
proton identification up to \pT=5 GeV/c. Electrons and photons are
measured in highly granular lead-scintillator and lead glass
calorimeters (EMC). The combination of EMC, RICH and TEC provides a
hadron background rejection factor for electron measurements of $10^4$
over a wide momentum range. The forward muon spectrometers have
acceptance for $\jpsi\rightarrow\mu^+\mu^-$ of $-2.25\lt{y}\lt-1.15$
and $1.15\lt{y}\lt2.44$. The muon arms have a radial magnetic field,
with tracking based on drift chambers followed by a muon identifier
consisting of alternating layers of steel absorber and streamer
tubes. Pion contamination of identified muons is $\sim3\times10^{-3}$.

\paragraph*{STAR}

The STAR experiment has broad physics reach, covering a wide variety
of hadronic and leptonic observables. Large acceptance enables
measurement of a large fraction of the thousands of charged hadrons
produced in a heavy ion collision, to measure correlations and to
search for rare or subtle non-statistical fluctuations indicating new
physics. STAR has robust capabilities to trigger and measure high \pT\
observables (hadron yields and correlations, electrons, photons, and
jets) to investigate partonic interactions in dense
matter. Measurements of \jpsi\ and $\Upsilon$ will probe deconfinement
directly.

\begin{figure}[htbp]
\centering
\includegraphics[height=.35\textheight]{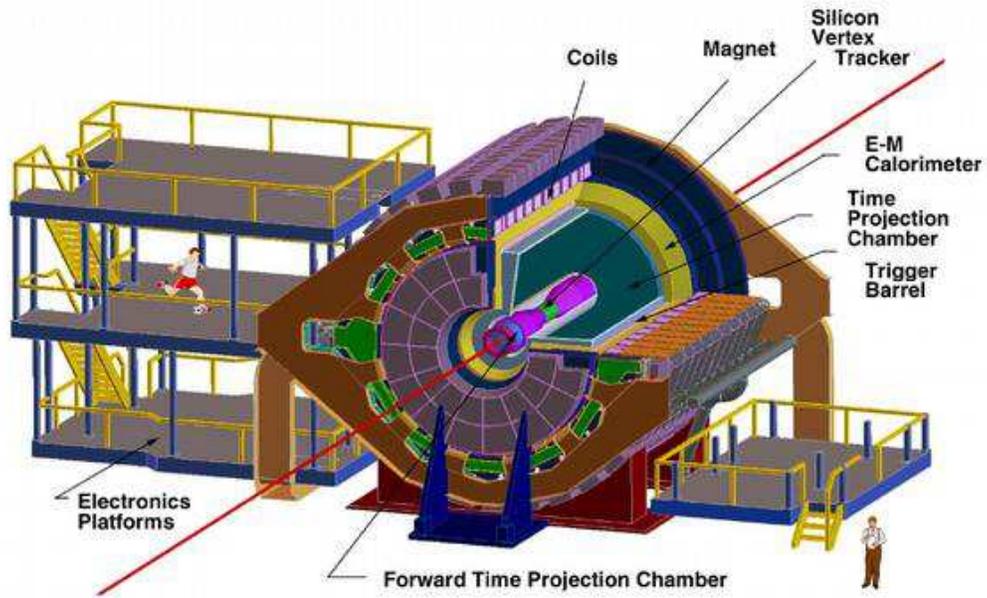}
\caption{The STAR detector.}
\label{STARLayout}
\end{figure}

The STAR detector \cite{star:nim} is shown in fig. \ref{STARLayout}. STAR is based on a warm
solenoidal magnet, with radius 260 cm and maximum field strength 0.5
Tesla, surrounding a variety of detector systems. The main tracking
device is a solenoidal Time Projection Chamber (TPC) with radius 200
cm and full azimuthal acceptance over $|\eta|\lt1.4$. Additional
tracking is provided by inner silicon drift detectors at midrapidity
and forward TPCs at $2.5\lt|\eta|\lt4$. Photons and electrons are
measured by Barrel and Endcap Electromagnetic Calorimeters (EMC), with
full azimuthal acceptance over $-1.0\lt\eta\lt2.0$. Particle
identification is carried out using specific ionization (dE/dx) in the
TPC gas, time of flight, reconstruction of displaced vertices for
weakly decaying particles, and combinatorial invariant mass
methods. The identification of strange baryons and mesons has been
made up to \pT=6 GeV/c and of charmed mesons to 10 GeV/c, with the
measurements currently limited by statistics. Fast triggering utilizes
the ZDCs, forward scintillators (Beam-Beam counters), a barrel of
scintillator slats surrounding the TPC, and the EMC.


\paragraph*{BRAHMS}

\begin{figure}[htbp]
\centering
\includegraphics[width=.45\textwidth]{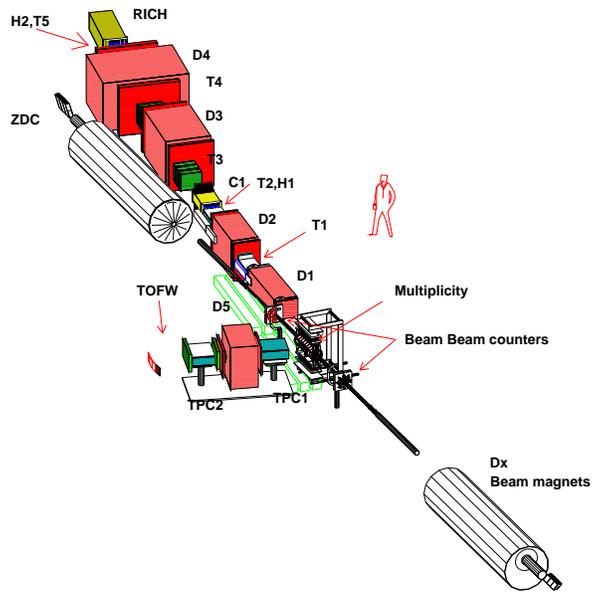}
\caption{The BRAHMS detector.}
\label{figBRAHMSLayout}
\end{figure}

The emphasis of the BRAHMS experiment is on high precision inclusive
measurements and small-angle correlations of primary hadrons over very
broad phase space. BRAHMS indeed has very large coverage, extending
for charged pions to rapidity $y=4$ ($Au$-beam rapidity is 5.37). The
BRAHMS detector \cite{brahms:nim}, shown in
fig. \ref{figBRAHMSLayout}, consists of two independent charged
particle spectrometers: the Forward Spectrometer (FS), with acceptance
0.8 msr measuring momenta $\pT\lt35$ GeV/c for angles relative to the
beam $2.3\lt\theta\lt30$ degrees, and the Mid-Rapidity Spectrometer
(MSR), with acceptance 6.5 msr and angular coverage $30\lt\theta\lt95$
degrees. Momentum measurements and particle identification utilize
Time Projection Chambers (TPCs) in conjunction with Time of Flight
hodoscopes and Threshold and Ring Imaging Cerenkov
detectors. Triggering and global even characterization of heavy ion
events are carried out using a mid-rapidity multiplicity array,
forward scintillator detectors, and the ZDCs.

\paragraph*{PHOBOS}

PHOBOS is designed to carry out a very general search for {\it a
priori} unknown and potentially rare signatures of new physics,
requiring a very large acceptance device that detects almost all
charged particles in each event for a large fraction of all inelastic
collisions. Virtually the entire RHIC phase space is covered by the
PHOBOS multiplicity measurement, with the high trigger and recording
rate allowing offline searches for unusual events or rare
fluctuations. Large-scale phenomena in heavy ion collisions may
generate effects at very low
\pT, requiring the spectrometers to measure down
to $\pT\sim$30 MeV/c. The PHOBOS spectrometers also measure at sufficiently
high \pT\ to be sensitive to jet-related observables.

\begin{figure}[htbp]
\centering
\includegraphics[height=.25\textheight]{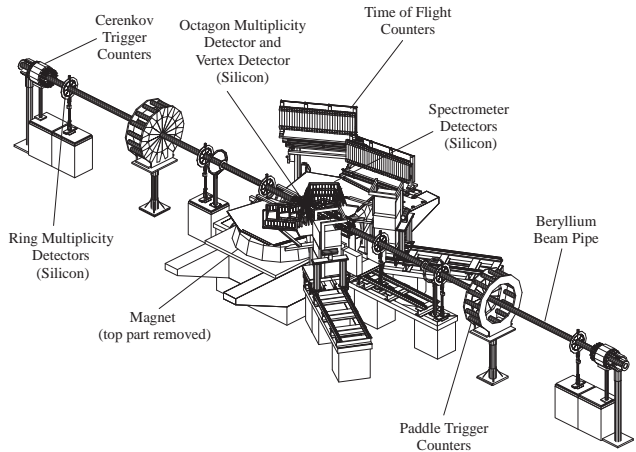}
\caption{The PHOBOS detector.} 
\label{PHOBOSLayout} 
\end{figure}

The PHOBOS experiment \cite{phobos:nim}, shown in
fig. \ref{PHOBOSLayout}, is based almost entirely on silicon pad
detectors. It consists of a multiplicity array covering 11 units of
pseudorapidity, a finely segmented vertex detector, two
small-acceptance midrapidity spectrometers, and trigger detectors. The
spectrometer arms utilize a warm dipole magnet of strength 1.5
Tesla-m. Particle identification is based on measurements of time of
flight and energy loss in the silicon. Special emphasis is put on
measurements at very low transverse momentum, requiring a thin beam
pipe and minimal material in front of the first tracking planes.

\subsection{Geometric Aspects of High Energy Nuclear Collisions}

In proton-nucleus and nucleus-nucleus collisions, multiple scattering
occurs at both the hadronic and the partonic level. Multiple
scattering influences many aspects of the dynamics of high energy
nuclear collisions, in particular the initial parton scattering
responsible for bulk particle production and the spectrum of rare hard
processes. It depends on the geometry of the nucleus-nucleus
collision, which also dictates the geometry of the produced dense
matter. Determination of the collision geometry is a key element in the
study of heavy-ion collisions.

\subsubsection*{Multiple Scattering and the Glauber Model}
\label{sect:Glauber}

Current theoretical treatment of multiple scattering is based on the
Glauber model \cite{Glauber}. In this model, a hadron-nucleus
collision is considered to be a series of multiple hadron-nucleon
scatterings. Neglecting the difference between the hadron and its
excited states between successive scatterings and utilizing the
forward peak of high energy hadron-nucleon elastic scattering, the
total cross section of $h+A$ collisions is \cite{Glauber}
\begin{equation}
\sigma_{hA}=\int d^2b \sum_{n=1}^{A}
\left( \begin{array}{c} A \\ n \end{array} \right)
\left[-\sigma_{hN}t_A(b)\right]^n e^{-\sum_{i=1}^{n}(\Delta q_{zi} R_A/2)^2}
\; ,
\end{equation}
where $t_A(b)=T_A(b)/A$ and $T_A(b)$ is the nuclear thickness function  
defined as a line integral over the nuclear density $\rho_A$,
\begin{equation}
T_A(\vec{b})=\int_{-\infty}^{\infty}dz\rho_A(\vec{b},z).
\label{eq:TA}
\end{equation}
The exponential factor in the above cross section comes from the
interference between different scatterings, assuming a Gaussian form
of the nuclear density distribution. The longitudinal momentum 
transfer is related to the transverse momentum transfer $q_{Ti}$
of each scattering,
\begin{equation}
\Delta q_{zi}\approx (q_{T1}^2+q_{T2}^2 + \cdots + q_{Ti-1}^2)/2E_h,
\end{equation}
and determines the coherence length $\ell_c=1/\Delta q_{zi}$. For high
energy scattering with small transverse momentum transfer, the
coherence length is much longer than the nuclear size $R_A$ and
hadron-nucleus collisions become coherent. The hadron-nucleus cross
section is then given by the familiar Glauber formula,
\begin{equation}
\sigma_{hA}=\int d^2b \left\{ 1-[1-\sigma_{hN}t_A(b)]^A\right\} \; .
\end{equation}
For large nuclei $A \gg 1$, the integrand can be approximated by an
exponential $1-\exp[-\sigma_{hN}T_A(b)]$. For hadronic scattering with
large cross section $\sigma_{hN}$, this cross section is mainly
determined by the nuclear density distribution and is denoted the
geometric cross section. This approach is most relevant for soft
processes.

\begin{figure}[htbp]
\includegraphics[width=.42\textwidth]{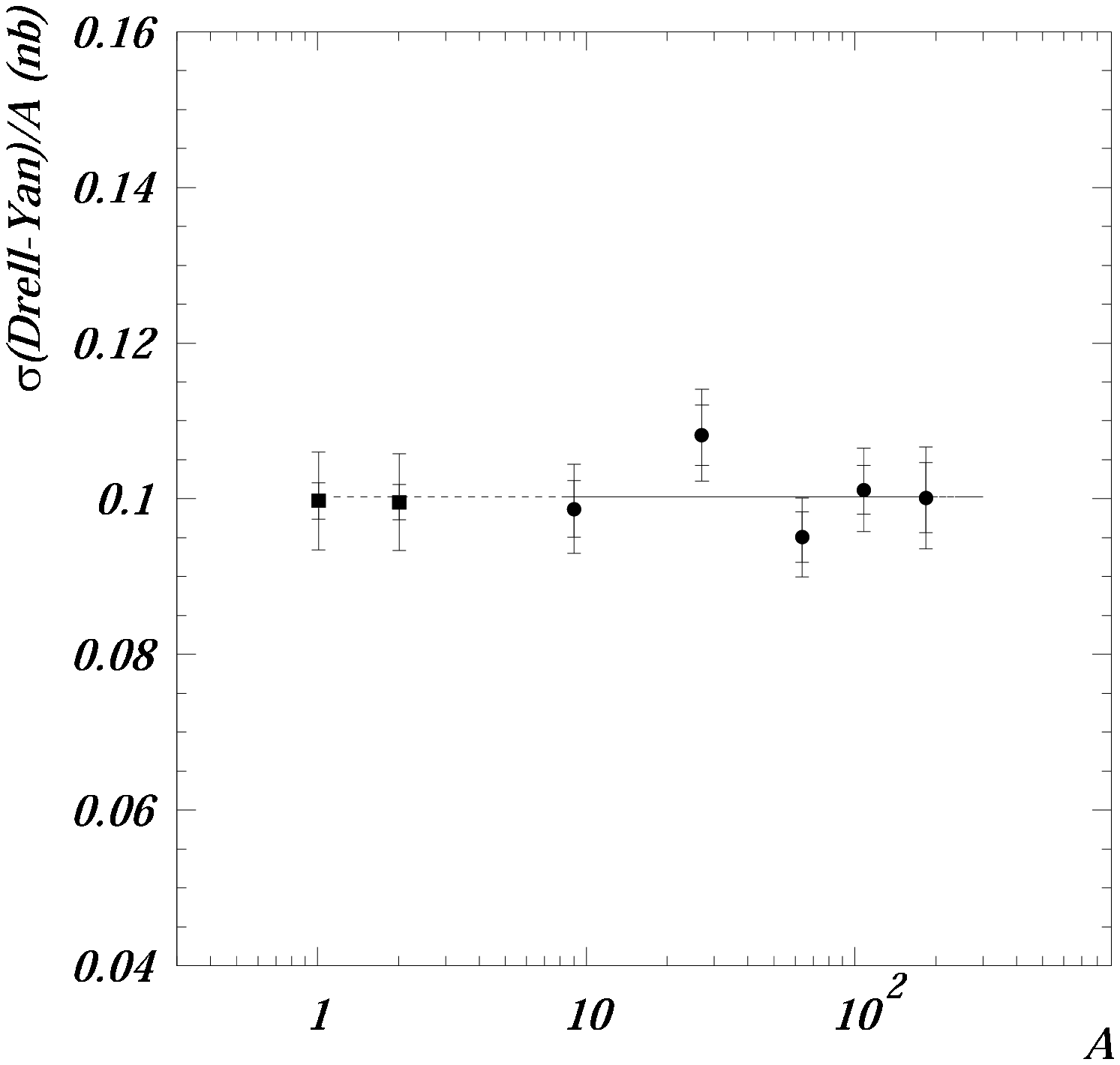}
\includegraphics[width=.57\textwidth]{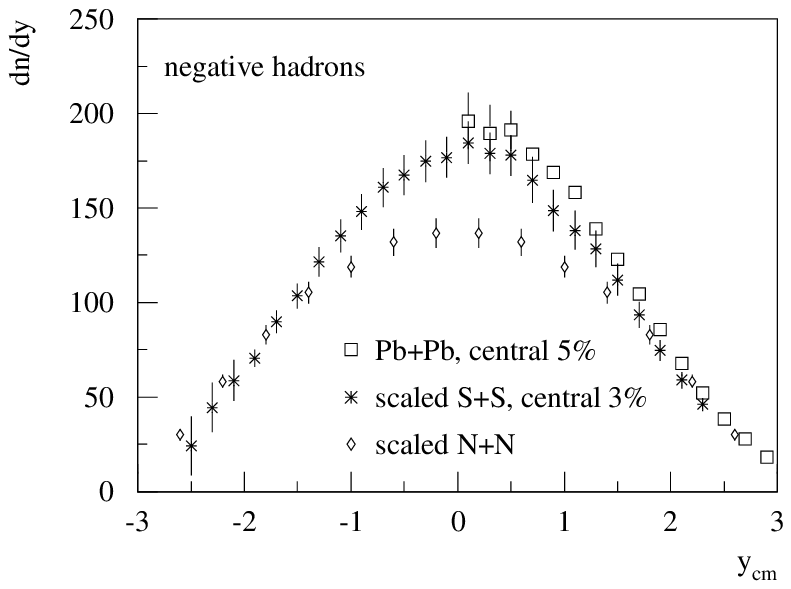}
\caption{Scaling of hard and soft processes in nuclear collisions at the CERN SPS. 
Left: Drell-Yan production cross section in p+A collisions normalized
by atomic mass $A$ (Eq. \ref{eq:GlauberScaling}), from NA50
\protect\cite{NA50DY}.  Right: charged hadron pseudorapidity distribution in central
collisions of equal mass nuclei scaled by the number of nucleon
participants ($\sim2A$), from NA49 \protect\cite{NA49hminus}.}
\label{fig:GlauberScaling}
\end{figure}

For hard processes with large transverse momentum transfer and small
cross section, the coherence length becomes much smaller than the
intra-nucleon distance in the nucleus. In this case all interference
terms drop out and the cross section results from the incoherent
superposition of nucleon-nucleon collisions. The $h+A$ cross section
is then directly proportional to that for $h+N$ collisions:
\begin{equation}
\sigma_{hA}^{\rm hard}=\int d^2b T_A(b)\sigma_{hN}^{\rm hard}
=A \sigma_{hN}^{\rm hard}.
\label{eq:GlauberScaling}
\end{equation}
This expression is proportional to the thickness function, which represents the
number of hadron-nucleon collisions in a $h+A$ collision.
As shown in Fig.~\ref{fig:GlauberScaling} (left panel), experimental
cross sections of Drell-Yan dilepton production with large invariant 
mass at the CERN SPS indeed scale linearly with the atomic 
number to good precision in $p+A$ collisions.

This formula can be extended to the case of $A+B$ collisions by
replacing the thickness function in $h+A$ collisions with the
nuclear overlap function,
\begin{equation}
\TAB=\int{d^2s}{T_A(\vec{s})}{T_B(\vec{b}-\vec{s})},
\label{eq:TAB}
\end{equation}
which represents the number of binary nucleon-nucleon collisions
per unit cross section, 
\begin{equation}
d\Nbinary/d^2b=T_{AB}(b),
\label{binary}
\end{equation}
in $A+B$ collisions at fixed impact parameter $b$.

Since the hard processes are incoherent, their cross sections in $A+B$
collisions should be proportional to the number of binary collisions
\Nbinary. We will see below that certain
hard processes with large momentum transfer $Q^2$ provide sensitive
probes of the matter generated in nuclear collisions through
final-state interaction of their reaction products with the medium.
Deviation from binary scaling of the cross section therefore indicates
novel nuclear effects and provides an experimental observable to
quantify such effects. Isolation of final-state from initial-state
effects requires comparison of systems in which final state effects
are expected to be present or absent. This strategy will play an
important role in our discussion of hard processes.

In contrast to hard processes, soft processes typically have large
cross section and coherence lengths much larger than the nuclear
size. The inclusive cross section and total hadron multiplicity are
then expected to scale as the number of participating (``wounded'')
nucleons (Wounded Nucleon Model \cite{Bialas:1976ed}). In the Glauber
multiple scattering model, the average number of wounded nucleons at a
fixed impact-parameter in $A+B$ collisions is

\begin{eqnarray}
N_{part}(b)=\int d^2s T_A(s)\left[1
  -e^{-\sigma_{NN}T_B(\vec{s}-\vec{b})} \right]
+\int d^2s T_B(\vec{s}-\vec{b}) \left[1
  -e^{-\sigma_{NN}T_A(\vec{s}-\vec{b})} \right]
\label{wounded}
\end{eqnarray}
\noindent
Fig. \ref{fig:GlauberScaling}, right panel, shows the charged particle
pseudorapidity distribution in central collisions of equal mass nuclei
measured at the CERN SPS, which indeed is seen to scale for massive
nuclei as the number of participating nucleons. A particle production
model embodying such coherent processes is the string model
implemented in HIJING Monte Carlo model
\cite{Wang:1991ht,Gyulassy:1994ew}. In this model, a wounded nucleon
becomes an excited string, with the number of produced hadrons
insensitive to the number of scatterings suffered throughout the
multiple scattering process. This leads to hadron multiplicity from
soft interactions proportional to the number of participant nucleons.

We turn now to application of the Glauber model to data analysis. The
impact parameter $b$ is not observable and measurements necessarily
integrate over a finite interval of $b$. Event geometry is tagged by
observables correlated with impact parameter, such as
charged particle multiplicity, whose distributions are binned into
percentiles of the total interaction cross section
(Fig. \ref{fig:CentralityTag}). Using the Glauber model, differential
cross section distributions $d\sigma/d\Nbinary$ and $d\sigma/d\Npart$
are calculated and the weighted mean values \NbinaryMean\ and
\NpartMean\ are found within the same percentile bins of total cross
section. The bins of the measured distribution and model calculation
are equated, under the assumptions that the event-tagging observable
varies on average monotonically (but not necessarily linearly) with
impact parameter and that fluctuations generate negligible mixing of
the bin populations.

The nuclear density in the calculation is taken to be
spherically symmetric, with Woods-Saxon radial dependence:
\begin{equation}
\rho_A(r)=\frac{\rho_0}{1+e^{(r-r_0)/a}}.
\label{eq:WoodsSaxon}
\end{equation}
Typical parameters for $Au$ nuclei are density $\rho_0$=0.169/fm$^3$ and
surface thickness $a=0.535\pm0.027$ fm, the latter derived from
electron scattering data\cite{deJagerEtAl}. The charge radius
$r_0=6.38$ fm is usually increased to $\sim$6.5 fm to account for the neutron
skin thickness.

Two methods are used in the literature to calcuate \Nbinary\ and
\Npart: the {\it Optical} and {\it Monte Carlo} Glauber approaches
(see Appendix A of \cite{star:multlong130} and references therein.)
The Optical Glauber approach is based on a smooth nuclear matter
distribution and numerical evaluation of the analytic Glauber
integrals. The Monte Carlo approach is based on the random
distribution of nucleons according to the Woods-Saxon density, with
nuclear collisions at a given impact parameter modelled by the
incoherent interaction of all nucleon pairs. For central Au+Au
collisions at RHIC energies the practical difference between these
approaches is negligible compared to other uncertainties of the
measurements, but for peripheral collisions the differences can be
significant. Assessment of this uncertainty must be made when
interpreting data that incorporate Glauber calculations.

\begin{figure}[htbp]
\centering
\includegraphics[width=.55\textwidth]{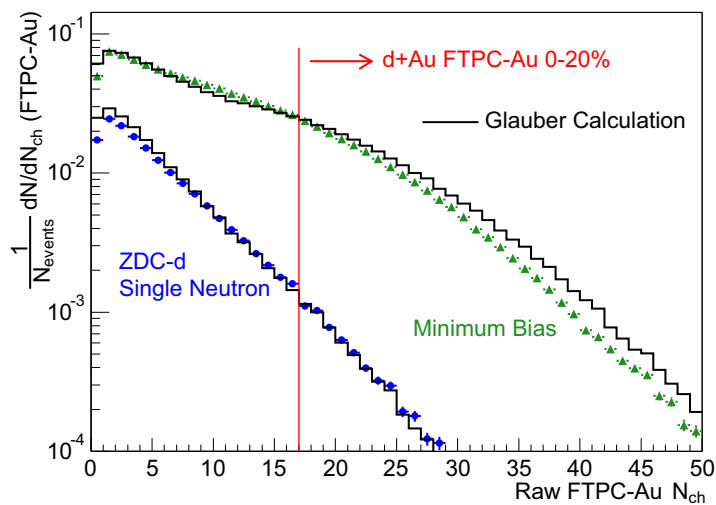}
\caption{Charged particle multiplicity distributions from 200 GeV 
$d+Au$ collisions, from STAR \protect\cite{star:highpTdAu}. Measurement is for
$-3.8\lt\eta\lt-2.8$ (Au-beam direction). Centrality selection and
Glauber calculations (histograms) described in text.}
\label{fig:STARdAuGlauber}
\end{figure}

As a gauge of the accuracy of the Glauber model, Figure
\ref{fig:STARdAuGlauber} compares the results of a Monte Carlo Glauber
calculation to charged particle multiplicity distributions in 200 GeV
$d+Au$ collisions \cite{star:highpTdAu}. The measurements are at
forward rapidity ($-3.8\lt\eta\lt-2.8$, $Au$-beam direction), both for
minimum bias $d+Au$ collisions and for peripheral $d+Au$ collisions
which have a beam-rapidity neutron detected in the deuteron beam
direction (``ZDC-d''). The calculated multiplicity distributions
result from the convolution of the forward charged multiplicity
distribution measured in 200 GeV
\pbarp\ interactions
\cite{UA5Mult} with the \Npart\ distribution from the Glauber model. 
Both the minimum bias and peripheral multiplicity distribution
measurements are well reproduced by the model. In addition, the cross
for such peripheral collisions is calculated to be $(18\pm3)\%$ of
the minimum bias cross section, in good agreement with the measured
fraction $(19.2\pm1.3)\%$\cite{star:highpTdAu}. Figure
\ref{fig:STARdAuGlauber} demonstrates that the Glauber approach provides a
sound basis for modelling geometric effects in nuclear collisions at
RHIC energies.


\subsubsection*{Centrality tagging in $Au+Au$ collisions}

\begin{figure}[htbp]
\includegraphics[width=.45\textwidth]{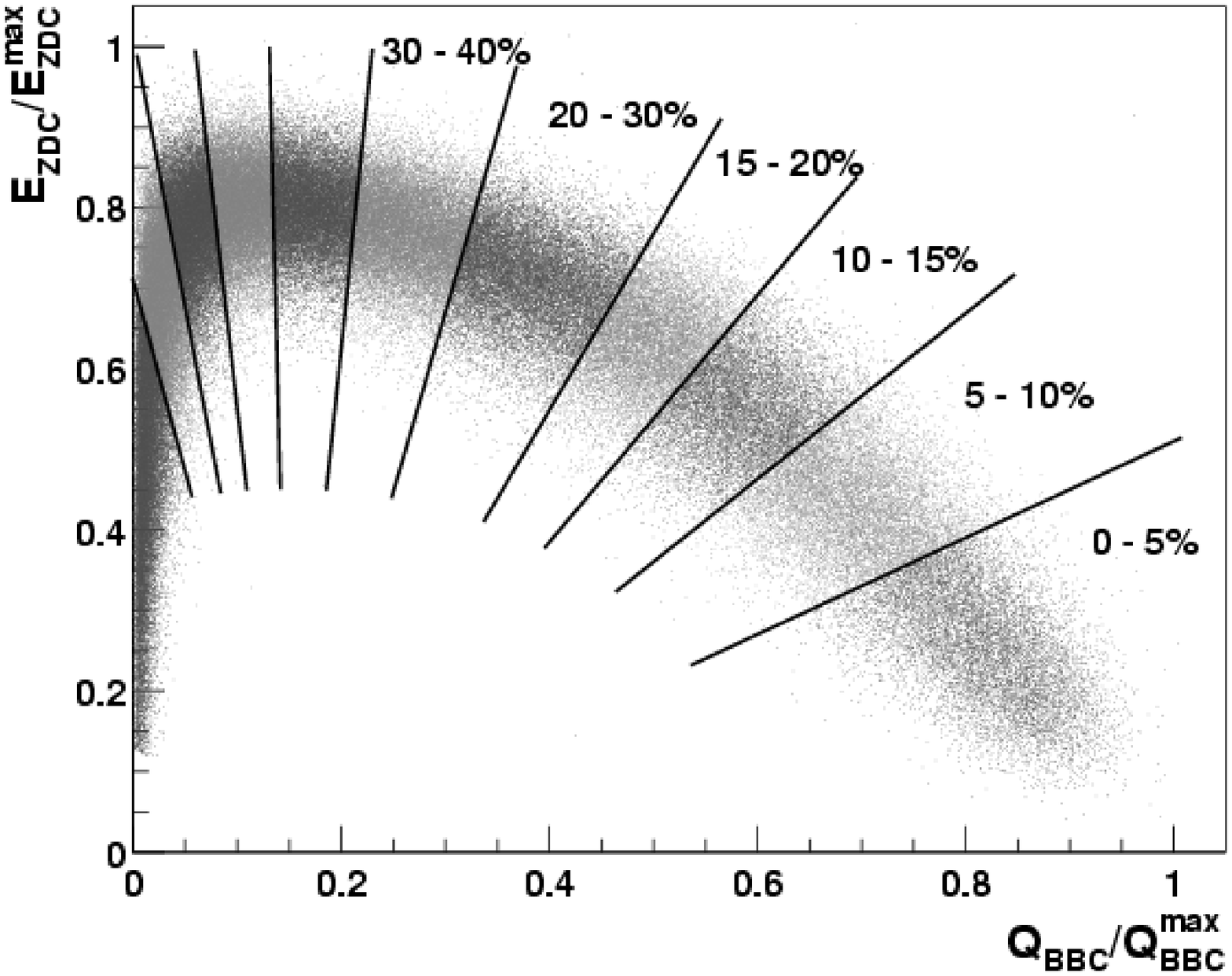}
\includegraphics[width=.55\textwidth]{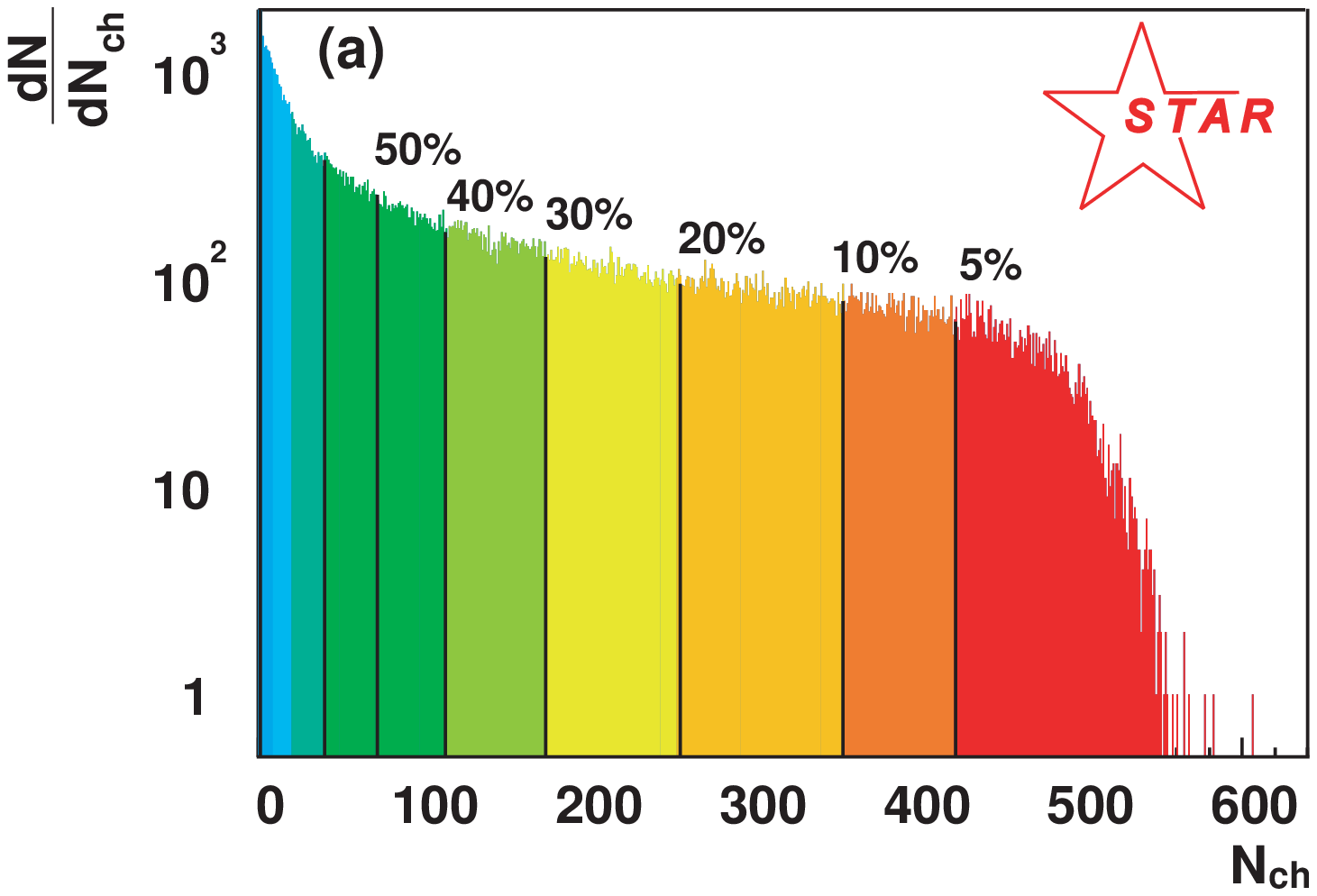}
\caption{Event centrality characterization. Left: scaled ZDC energy
(vertical) vs. forward charged multiplicity (horizontal), from
PHENIX \protect\cite{phenix:pidspectra200}. Right: distribution of charged multiplicity
at mid-rapidity, from STAR \protect\cite{star:flow130}. Both plots are for
minimum bias Au+Au events at \sqrtsNN=200 GeV and show the division
into centrality bins corresponding to percentile intervals of the
total cross section.}
\label{fig:CentralityTag} 
\end{figure}

Fig. \ref{fig:CentralityTag}, left panel, shows the measured
correlation in 200 GeV $Au+Au$ collisions between ZDC energy
(spectator neutrons) and charged particle multiplicity within
$3.1\lt|\eta|\lt3.9$, from PHENIX. In the most peripheral (large
impact parameter) collisions, the number of forward neutrons and
the total multiplicity are both small. Both quantities increase for
decreasing impact parameter, while for the most central collisions the
number of spectator neutrons is again small while the multiplicity is
large. The correlation between these two geometry-sensitive
observables is seen to be strong. The figure also illustrates the
sorting of events into centrality bins corresponding to percentile
intervals of the cross section, with 0-5\% indicating the most central
collisions.

The ZDC energy is small for both the most central and most peripheral
collisions, and this ambiguity limits its utility as a centrality tag. For
many applications the multiplicity distribution alone suffices as a
centrality tag, as shown in Fig. \ref{fig:CentralityTag}, right
panel. The shape of the distribution is dominated by the nuclear
geometry, with the tail at the highest multiplicity governed
by multiplicity fluctuations within the finite measurement aperture
for the most central collisions.

\begin{figure}[htbp]
\includegraphics[width=.44\textwidth]{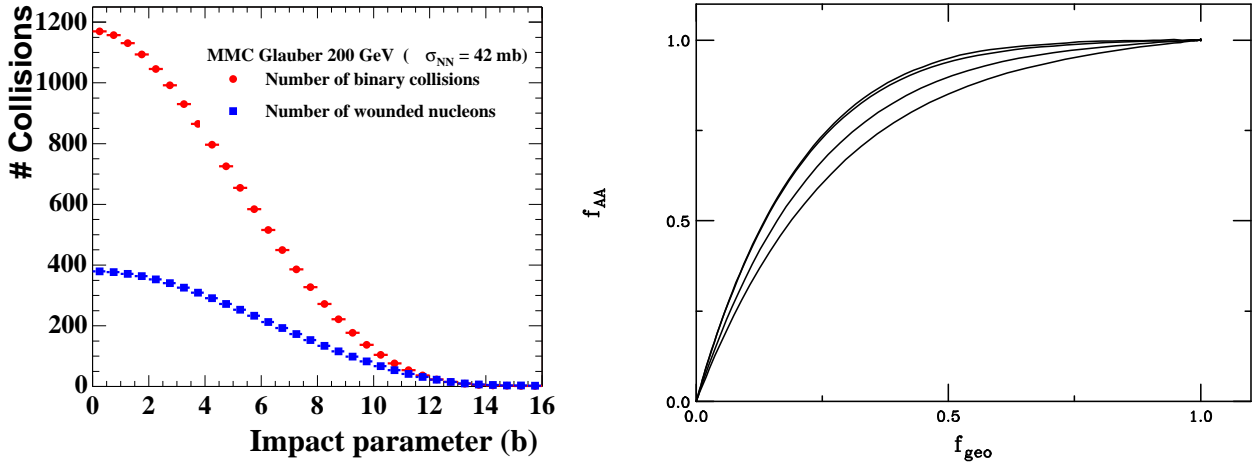}
\includegraphics[width=.52\textwidth]{Vogt_geometry_Fig4.eps}
\caption{Left: \Nbinary\ and \Npart\ vs. impact parameter for Au+Au at RHIC 
energies \protect\cite{snellings_0310019}. 
Right: fraction of binary-scaling cross section vs fraction of
hadronic interaction cross section; figure from \protect\cite{Vogt:geometry}. 
Central events correspond to small
$f_{geo}$. The curves are for (upper to lower) Au+Au, Ag+Ag, Cu+Cu, Al+Al, and O+O collisions. }
\label{fig:Glauber}
\end{figure}

We conclude this section with a discussion of the impact parameter
dependence of \Nbinary\ and \Npart. Figure \ref{fig:Glauber}, left
panel, shows this dependence for $Au+Au$ collisions at RHIC energies,
indicating the strong bias towards central collisions of
binary-scaling processes. For an \Nbinary-scaling process, we define
the fraction of the its total cross section contained in events
with impact parameter $b\lt{b_c}$ as\cite{Vogt:geometry}:

\begin{equation}
f_{AB}=\frac{2\pi}{AB}\int_0^{b_c}bdb\:\TAB.
\label{eq:fAB}
\end{equation}

\noindent
Figure \ref{fig:Glauber}, right panel, shows $f_{AB}$ as a function of
$f_{geo}$, the fraction of the total hadronic interaction cross
section contained in the same impact parameter interval. The
\Nbinary-scaling cross section weights strongly towards
central (small impact parameter) collisions, due purely to nuclear
geometry. As a rough rule of thumb for binary-scaling processes in
symmetric heavy ion collisions, 40\% of the cross section is contained
in the 10\% most central events.

\section{Bulk Particle Production and Initial Conditions}

The formation of a Quark-Gluon Plasma in high-energy heavy-ion
collisions depends critically on the initial parton production at the
earliest stage of the reaction. A number of proposed signals of the
QGP are sensitive to the early conditions and provide direct
measurements of the initial density and other properties of the dense
matter. Alternatively, global observables based on bulk particle
production, such as the rapidity density of hadron multiplicity and
transverse energy, provide a coarse-grained view of the early
dynamics. These observables may be related to the entropy and energy
densities early in the collision evolution, and as such provide
meaningful constraints on the initial conditions that complement the
measurements via direct probes. They can also provide tests of
particle production models and provide a more complete picture of the
dynamics. Our first discussion will focus on properties of bulk
particle production: multiplicity and transverse energy pseudorapidity
($\eta$) densities and net baryon number distributions.

\subsection{Particle Production Models}

Prior to RHIC startup, theoretical model estimates of the initial
conditions and bulk particle production varied over a wide range, due
to uncertainties in modeling soft hadron production and the interplay
between soft and hard processes \cite{Bass:1999zq}. These models
included a pure pQCD parton model \cite{EKRT}, a pQCD parton model in
combination with string model
\cite{Gyulassy:1994ew,Geiger:1992nj}, and a classical Yang-Mills field
model \cite{McLerran:1994ni}. The uncertainties were significantly
reduced with the first publication of RHIC data on charged hadron
multiplicity \cite{Back:2000gw}. New calculations based on the initial
state gluon saturation model \cite{KharzeevLevin} also describe the
energy and centrality dependence well. Here we provide brief
descriptions of three typical models before we present the
experimental data and discuss their implications.

\paragraph*{Two-component model}
\label{sect:hijing}

Mini-jet production in a two-component model was proposed long ago
to explain the energy dependence of the total cross section 
\cite{Gaisser:1985pg} and particle production \cite{Wang:1991qp}
in high-energy hadron collisions. This approach was incorporated into
the HIJING model \cite{Wang:1991ht,Gyulassy:1994ew} to describe
initial parton production in high-energy heavy-ion collisions. In this
two-component model, the nucleon-nucleon cross section at high energy
is divided into collisions with or without hard or semi-hard jet
production processes.  The jet cross section $\sigma_{\rm jet}$ is
assumed to be given by the pQCD parton model. However, the
differential jet cross section has an infrared divergence as the
transverse momentum of the jet goes to zero. An infrared cut-off scale
$p_0$ is introduced to separate hard processes from soft processes
that are not calculable in pQCD. The soft interaction cross section
$\sigma_{\rm soft}$ is a model parameter.  The two parameters,
$\sigma_{\rm soft}$ and $p_0$, are determined phenomenologically by
fitting the experimental data of total $p+p(\bar{p})$ cross sections
within the two-component model \cite{Wang:1991ht,Gyulassy:1994ew}.

The cut-off scale $p_0$ separating non-perturbative and pQCD processes
could in principle depend on both energy and nuclear size in $A+A$
collisions. Using the Duke-Owens parameterization of parton
distributions in the nucleon \cite{Duke:1984gd}, an energy-independent
cut-off scale $p_0=2$ GeV/$c$, and soft cross section $\sigma_{\rm
soft}$, the HIJING model can describe well the experimental data on
total cross sections, hadron multiplicity, and other observables from
$p+p(\bar{p})$ collisions \cite{Wang:1998ww}. The default HIJING
prediction \cite{WangGyulassyMult} agreed with the first data on
$dN_{ch}/d\eta(|\eta|<1)$ of central $Au+Au$ collisions at
$\sqrt{s}=130$ GeV \cite{Back:2000gw}.

In applying the two-component model to nuclear collisions, multiple
mini-jet production is assumed to be incoherent and thus is
proportional to the number of binary collisions \Nbinary. The
soft interaction is however coherent and proportional to the number of
participant nucleons \Npart, according to the Wounded Nucleon
Model \cite{Bialas:1976ed}. Assuming no final state effects on
multiplicity from jet hadronization, the rapidity density of hadron
multiplicity in heavy-ion collisions is then
\begin{equation}
\frac{dN_{ch}}{d\eta}=\frac{1}{2}\NpartMean 
\langle n\rangle_{s} 
+ \langle n\rangle_{h}\NbinaryMean
\frac{\sigma_{\rm jet}^{AA}(s)}{\sigma_{\rm in}},
\label{eq:nch} 
\end{equation}
where $\sigma_{\rm jet}^{AA}(s)$ is the averaged inclusive jet cross
section per nucleon-nucleon interaction in $AA$ collisions. The
average number of participant nucleons and number of binary collisions
for a given impact parameter can be estimated via Glauber model
simulation. Since the parameters $\langle n\rangle_{s}$ and $\langle
n\rangle_{h}$ are determined from $p+p(\bar p)$ collisions, the only
uncertainties are due to nuclear effects on $\sigma_{\rm jet}^{AA}(s)$,
such as parton shadowing.

As an alternative to Eq.~(\ref{eq:nch}), a simpler parameterization 
of the ``two-component'' model is given by \cite{KharzeevNardi}
\begin{equation}
\frac{dN_{ch}}{d\eta}=(1-x)n_{pp}\frac{\NpartMean}{2} + {x}n_{pp}\NbinaryMean,
\label{eq:TwoComponent}
\end{equation}
where $n_{pp}$ is the multiplicity in $p+p$ collisions and $x$ is the
fraction of sources scaling as hard collisions, related to
$\sigma_{\rm jet}^{AA}(s)/\sigma_{\rm in}$.

A final-state parton cascade and hadronic 
rescattering processes can also be introduced into such a pQCD-based model. 
The AMPT model\cite{Lin:2003ah} uses HIJING for the initial
conditions and includes parton and hadronic rescattering 
after-burners. The final-state rescattering is found not to
change the bulk particle production significantly.

\paragraph*{Final State Saturation}
\label{sect:FinalStateSat}

A mini-jet pair with transverse momentum $p_0$ has intrinsic
transverse area of $\pi/p_0^2$. In $A+A$ collisions at high energy,
independent production could result in multiple mini-jet production
within this area, which is quantum
mechanically disallowed. This sets the limit on the number of
independent mini-jet pairs within a total transverse area
$\pi R_A^2$ as $N(p_0)=p_0^2 R_A^2$. The pQCD parton model gives the number of
scattered partons above transverse momentum cutoff $p_0$ as
$N(p_0)=T_{AA}(b)\sigma_{\rm jet}(p_0)$. In the 
final-state saturation model (EKRT \cite{EKRT}), the saturation scale $\psat$
is then defined by the self-consistent solution of
\begin{equation}
T_{AA}(b)\sigma_{\rm jet}(\psat)=\psat^2 R_A^2.
\label{eq:psat}
\end{equation}
\noindent
Below the saturation scale, jet production is correlated and the
divergent mini-jet cross section will be regulated.  The EKRT
final-state saturation model neglects minijets with $p_T<\psat$ and
assumes that the produced parton density (mostly gluons) and the
transverse energy density are dominated by minijets above $\psat$.

Numerical calculation \cite{EKRT} of Eq.~(\ref{eq:psat}) yields
$\psat=0.21A^{0.13}(\sqrts)^{0.19}$ GeV/c, with initially produced gluon
multiplicity $N=1.38A^{0.92}(\sqrts)^{0.38}$. The calculated ratio of
energy density to multiplicity density is found to be very similar to
that of an ideal gas of bosons, $\epsilon/n\simeq 2.70T$, so that for
a wide range of $A$ and \sqrts\ the gluon gas is generated in a
thermalized distribution in this model.

Assuming boost-invariant adiabatic expansion and proportionality
between the produced parton and the observed hadron multiplicities,
the charged hadron density in central collisions is given
by\cite{EKRT}:
\begin{equation}
\frac{dN_{ch}}{d\eta}(b=0)\simeq\frac{2}{3}1.16A^{0.92}(\sqrts)^{0.4}.
\label{eq:EKRT}
\end{equation}
While this expression only applies for symmetric geometry,
it has been extended to non-central collisions \cite{Eskola:2000xq}
and the centrality dependence may be approximated by replacing $A$ with
\NpartMean\ in the above equation.


\paragraph*{Initial State Saturation}
\label{sect:CGC}

More commonly, parton saturation refers to high density effects
in the initial state \cite{Gribov:1983tu,Mueller:1986wy}. 
For an elementary probe (a virtual photon in
deep inelastic scattering, a projectile parton in hadronic collisions)
interacting with a nucleus of mass $A$, the coherence length of the
interaction in the rest frame of the nucleus is $\ell_c
\sim1/(m_N\xBj)$, where $m_N$ is the nucleon mass and the Bjorken
$\xBj$ is the fractional momentum that the struck parton carries. At
sufficiently low $\xBj$, the distribution is dominated by gluons and
the coherence length $\ell_c$ will exceed the nuclear diameter $\sim
2A^{1/3}$. Modification of parton distributions due to the coherence
can be studied within the framework of Glauber multiple scattering in
the rest frame of a nucleus \cite{Huang:1998ii}. More intuitively,
saturation phenomena can be studied as multiple parton interactions in
the infinite-momentum frame.  For a hard process with momentum
transfer $Q$, all gluons within transverse area $1/Q^2$ will
participate coherently in the interaction. Denoting the nuclear gluon
structure function as $xG_A(x,Q^2)$, the density of gluons in the
transverse plane is
\begin{equation}
\rho_A\simeq A \frac{xG(x,Q^2)}{\pi{R}_A^2}\sim{A}^{1/3},
\label{eq:rhoSat}
\end{equation}
\noindent
where $G_A(x,Q^2)\simeq AG(x,Q^2)$ and $G(x,Q^2)$ is the gluon
distribution in a nucleon.

For gluon scattering with cross section $\sigma\sim\pi\alpha_s/Q^2$,
$\sigma\rho_A$ represents the probability of multiple gluon
scattering.  At high $Q^2$ or $\sigma\rho_A\ll1$, the system can be
considered to be dilute, and the perturbative QCD parton model
applies. However, for low $Q^2$ or $\sigma\rho_A\gg 1$, the target
looks black to the probe and the saturation regime is reached. The
boundary where $\sigma\rho_A\simeq1$ defines the {\it saturation
scale} \Qs. The physical process that leads to saturation is the
nonlinear gluon fusion $gg \rightarrow g$, which competes with the
gluon emission process $g\rightarrow gg$.  The emission process
increases the gluon number with rising $Q^2$ according to the normal
DGLAP evolution, while gluon fusion, which is proportional to
$\sigma\rho_A$, reduces the gluon number. Saturation occurs when the
two processes offset each other and the saturation scale is determined
by the self-consistent solution of the equation
\cite{Mueller:1986wy,KharzeevNardi}
\begin{equation}
\Qs^2=\frac{8\pi^2N_c}{N_c^2-1}
\alpha_s(\Qs^2) xG(x,\Qs^2) \frac{A}{\pi{R}_A^2},
\label{eq:QsSat}
\end{equation}
where the gluon distribution $xG(x,\Qs^2)$ is evaluated at
$x=2\Qs/\sqrt{s}$. In the saturation regime,
$xG_A(x,\Qs^2)\propto1/\alpha(\Qs^2)$ and $\Qs^2\propto{A}^{1/3}$.  For large
enough $\Qs$, the strong coupling constant will be small while the
density is high, enabling treatment of the non-linear QCD dynamics by
classical weak coupling methods. This provides the foundation for
semi-classical treatment of the gluon distribution inside large 
nuclei \cite{McLerran:1994ni,McLerran:1994ka}. Often referred to as
Colored Glass Condensate (CGC) model, this approach approximates the gluon
distribution in large nuclei as the Weiszacker-Williams distribution
from the classical Yang-Mills field of randomly distributed color
charges. For a recent review see \cite{Iancu:2003xm}.

The initial saturation phenomenon is generic and is independent 
of the type of hadrons or nuclei being collided. However, the growth 
of $\Qs^2$ as $A^{1/3}$ suggests that saturation phenomena may
occur at higher $x$ (or equivalently, lower \sqrts) in collisions 
of heavy nuclei than in $p+p$ collisions. Deep inelastic scattering 
data from HERA indicate that $\Qs$ scales as\cite{Stasto:2000er}
\begin{equation}
\Qs^2(x)=Q_0^2\left(\frac{x_0}{x}\right)^\lambda,
\label{eq:Qsx}
\end{equation}
\noindent
with $\lambda \sim 0.2-0.3$. The saturation scale in heavy ion collisions 
at RHIC is estimated to be $\Qs^2\simeq2$ GeV$^2$ at $x\sim0.02$
\cite{KharzeevNardi}, roughly the value of $\Qs^2$ at $x\sim10^{-4}$ in
$p+p$ collisions at the same energy.

In contrast to the EKRT final-state saturation model, the
initial-state saturation model assumes that final gluon production is
dominated by gluons below the saturation scale \cite{Mueller:1999fp}
and ignores gluons above the saturation scale, whose yield falls as
$1/\pT^4$. The density of gluons per unit area and unit rapidity
produced in the collision is then given by\cite{Mueller:1999fp}
\begin{equation}
\frac{dN}{d^2bdy}=c\frac{N_c^2-1}{4\pi^2\alpha_s(\Qs^2){N_c}}\Qs^2.
\label{eq:SatGluonDensity}
\end{equation}
Integrating over the transverse area and further assuming the
hadronization coefficient is unity (one produced gluon becomes one
final charged hadron), the observed charged hadron multiplicity in the
most central $Au+Au$ collisions is fitted to obtain the gluon
liberation coefficient $c=1.23\pm 0.20$ \cite{KharzeevNardi}. A
similar value for $c$ results from numerical calculation of
initial parton production within the CGC model \cite{Krasnitz:2000gz}.

With these ingredients, the classical weak coupling treatment of
initial-state saturation gives a prediction for the
centrality dependence of the multiplicity density per participant
pair\cite{KharzeevNardi}:
\begin{equation}
\frac{2}{\Npart}dN/d\eta\simeq0.82\log
\left(\frac{\Qs^2}{\LamQCD^2}\right), 
\label{eq:dNdetaSat}
\end{equation}
where \LamQCD=200 MeV. The centrality dependence results from the
variation of $\Qs^2\sim\rho_{part}$ with the impact parameter 
according to Eq.~(\ref{eq:QsSat}), where $2A/\pi R_A^2$ is replaced 
by $\rho_{part}$ as the transverse density of participant nucleons
at fixed impact-parameter \cite{KharzeevNardi}.

The collision energy and rapidity dependence of the multiplicity is
governed by Eq.~(\ref{eq:Qsx}). Since rapidity $y{\sim}\log(1/x)$, 
the rapidity dependence of $\Qs$ at fixed \sqrts\ 
is $\Qs^2(\pm{y})=\Qs^2(y=0)e^{\pm\lambda{y}}$. In
other words, at forward rapidity one nucleus moves deeper into
the saturation regime (larger \Qs) while the other moves towards the
low density domain. The complete expression for the multiplicity
density in the initial state saturation model is \cite{KharzeevLevin}
\begin{eqnarray}
\frac{dN}{dy}=
 c\Npart\left(\frac{s}{s_0}\right)^{\lambda/2}e^{-\lambda\left|y\right|}
\left[\mathrm{log}\left(\frac{\Qs^2}{\LamQCD^2}\right)-\lambda\left|y\right|\right] \times  \left[1+\lambda\left|y\right|
\left(1-\frac{\Qs}{\sqrts}e^{(1+\lambda/2)|y|}\right)^4\right],
\label{eq:dNdySatFull}
\end{eqnarray}
where $\Qs^2(s)=\Qs^2(s_0)(s/s_0)^{\lambda/2}$. This expression
contains two free parameters $c$ and $\Qs^2(s_0)$, which are fixed at 
one energy, rapidity and centrality.


\subsection{Multiparticle Production}

\begin{figure}
\centering
\includegraphics[height=.25\textheight]{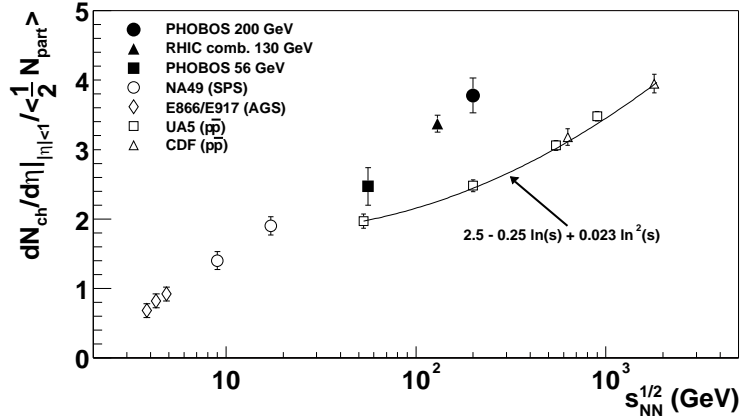}
\caption{\sqrts-dependence of charged particle density per participant 
pair at midrapidity, for central collisions of heavy nuclei
($A\sim200$, open and filled points) and \pbarp\ collisions (points
joined by line). Figure from PHOBOS \protect\cite{phobos:multsqrts}.}
\label{fig:MultPerPartvsSqrts}
\end{figure}

Total charged hadron multiplicities were the first published
experimental data at RHIC \cite{Back:2000gw}.
Fig. \ref{fig:MultPerPartvsSqrts} shows the energy dependence of the
charged particle density at mid-rapidity normalized per participant
pair, for collisions of heavy nuclei ($A\sim200$) and \pbarp.  The
nuclear collision data are from the AGS, SPS and RHIC (the three
highest energy points are from RHIC). The
energy dependence is smooth, with only
logarithmic dependence on \sqrts. The growth for nuclear collisions is
nevertheless faster than that in \pbarp\ collisions, qualitatively in
agreement with the expectation of a larger minijet contribution at
higher energy\cite{WangGyulassyMult}. 

\begin{figure}
\includegraphics[width=.45\textwidth]{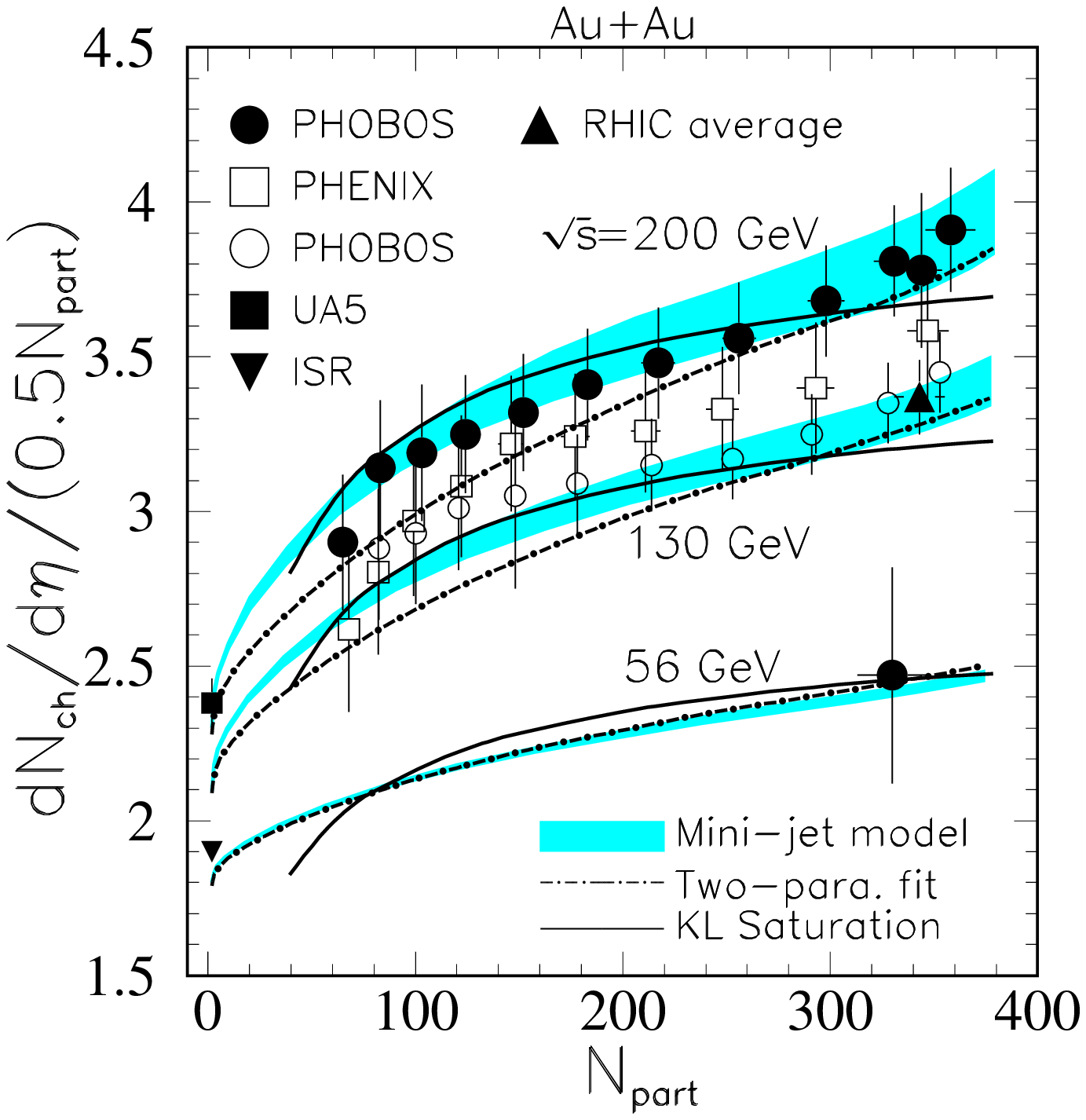}
\includegraphics[width=.5\textwidth]{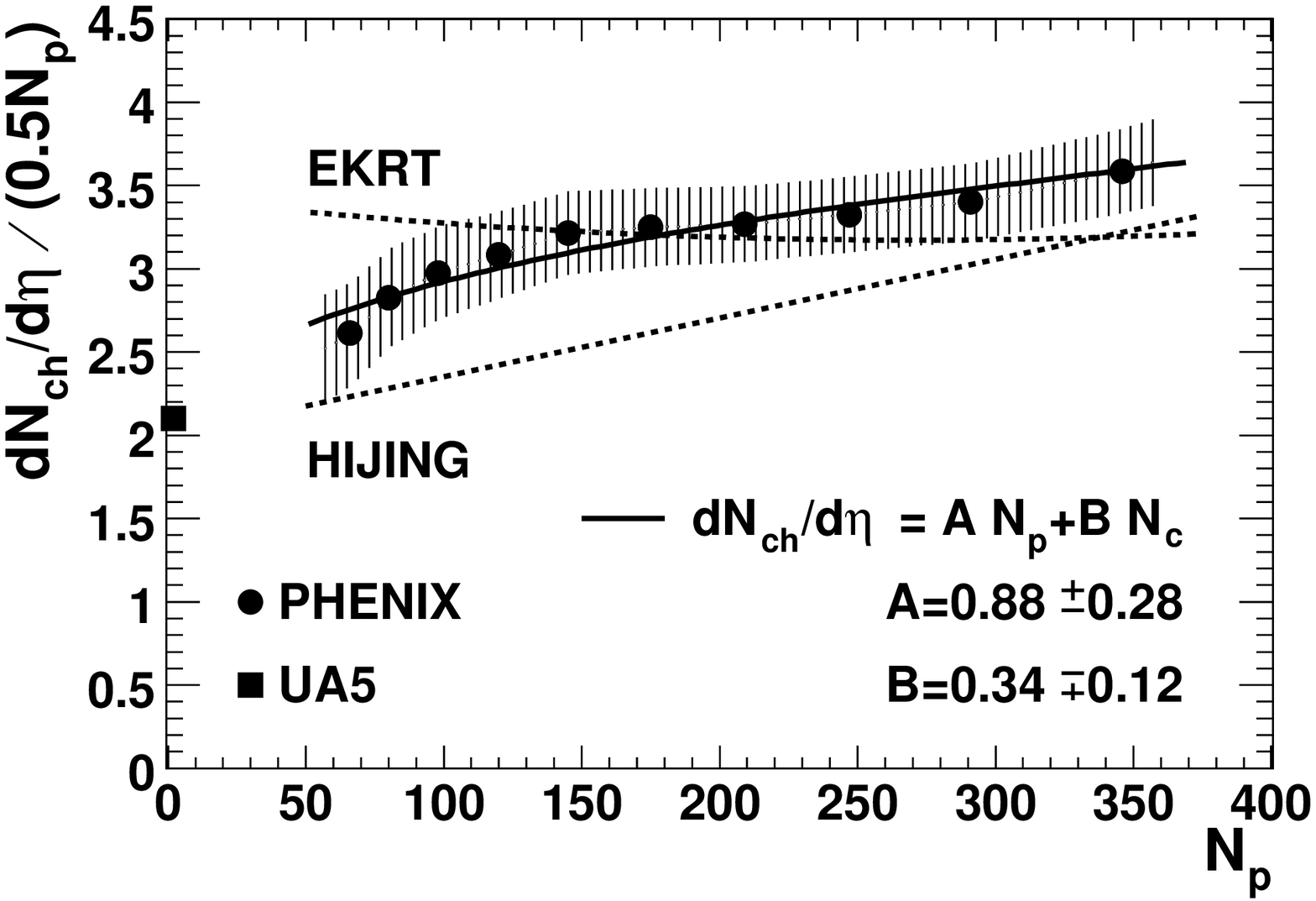}
\caption{Centrality dependence of the charged hadron central rapidity density per participant nucleon pair
for $Au+Au$ collisions from PHOBOS
\protect\cite{Back:2000gw,phobos:mult130200} and PHENIX \protect\cite{phenix:mult130}.
200 GeV \pbarp\ data (leftmost points) from UA5
\protect\cite{Alner:1986xu,Thome:1977ky}. Left: theory calculations from two-component minijet model (shaded bands), 
two-parameter fit (Eq.~(\ref{eq:TwoComponent})) (dot-dashed lines) and
parton saturation model (solid lines) \protect\cite{KharzeevLevin}. Right: data from 130 GeV collisions;
theory calculations from Hijing \protect\cite{Wang:1991ht,Gyulassy:1994ew},
EKRT saturation (Eq. (\ref{eq:EKRT})), and two-parameter
fit(Eq. (\ref{eq:TwoComponent})). }
\label{fig:MultPerPartvsNpart}
\end{figure}

A more differential view of bulk particle production is given by
Fig. \ref{fig:MultPerPartvsNpart}, which shows the centrality
dependence of the multiplicity density per participant pair compared
to several model calculations. In the left panel, the two-component
fit and initial state saturation model are seen to describe within
experimental uncertainties both the centrality dependence and the
growth in multiplicity with energy. In the right panel, the EKRT
prediction apparently fails to describe the centrality dependence,
while the centrality dependence of HIJING is consistent with the data
but the normalization is about 10\% too low except for $p+p$
collisions. The almost linear centrality dependence of the HIJING
result is not characteristic of a two-component model. This effect may
be caused by coherent string fragmentation of the minijet
hadronization, which could modify the binary scaling of the number of
hadrons from jet fragmentation. The two-component minijet model
\cite{Li:2001xa}, which assumes independent fragmentation, describes
the data well (shaded bands, left panel). Both the HIJING model and
the two-component minijet model have parton shadowing that depends on
impact-parameter. Similarly, the saturation models have an impact
parameter-dependent saturation scale.
 
\begin{figure}
\centering
\includegraphics[width=.8\textwidth]{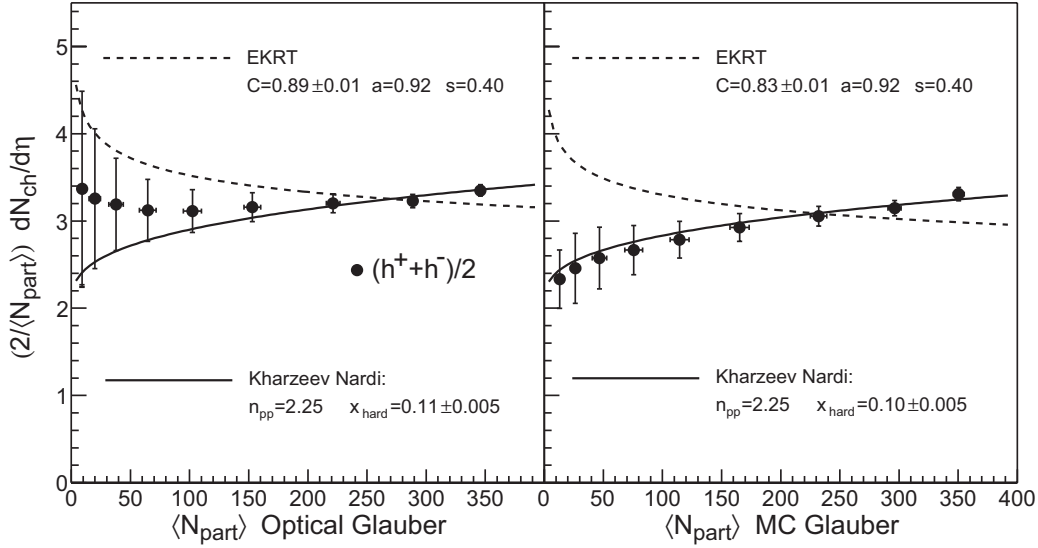}
\caption{$dN_{ch}/d\eta/(\NpartMean/2)$ 
utilizing Optical (left) and MC (right) Glauber calculations with the
same STAR data \protect\cite{star:multlong130}. The error bars for
adjacent data points are highly correlated. Kharzeev-Nardi refers to a
two component soft/hard fit [Eq. \ref{eq:TwoComponent}].}
\label{fig:STARdNchdeta}
\end{figure}

Note, however, that the calculated quantity \NpartMean\ appearing in both
the ordinate and abscissa in Fig. \ref{fig:MultPerPartvsNpart} is
derived using the Monte Carlo Glauber model for the data in both
panels. The uncertainty inherent in this procedure is demonstrated in
Fig. \ref{fig:STARdNchdeta}, which shows the same STAR data in both
panels but with \NpartMean\ calculated via the Optical (left) and
Monte Carlo (right) approach to the Glauber calculation
\cite{star:multlong130}. With the Optical Glauber approach the logarithmic growth at low
\Npart, the hallmark of initial state saturation
(Eq. \ref{eq:dNdetaSat}), is not seen, while the final state
saturation model (EKRT) still disagrees with the data, though less
significantly than with the Monte Carlo approach. Though other
measurements favor the Monte Carlo Glauber
(Fig. \ref{fig:STARdAuGlauber}), the saturation model curves in
Fig. \ref{fig:MultPerPartvsNpart} are calculated using Optical
Glauber, obscuring somewhat the direct comparison to data. The
centrality dependence of the multiplicity density and its comparison
to models therefore requires further clarification, but it is apparent
that multiparticle production in {\it central} collisions can be well
described by a broad range of theoretical approaches, with about 20\%
uncertainty both in theory and in the measurement of the centrality
dependence. All of these models point to initial conditions with high
initial gluon density.

\subsection{Pseudo-rapidity Distributions}

\begin{figure}
\centering
\includegraphics[width=.95\textwidth]{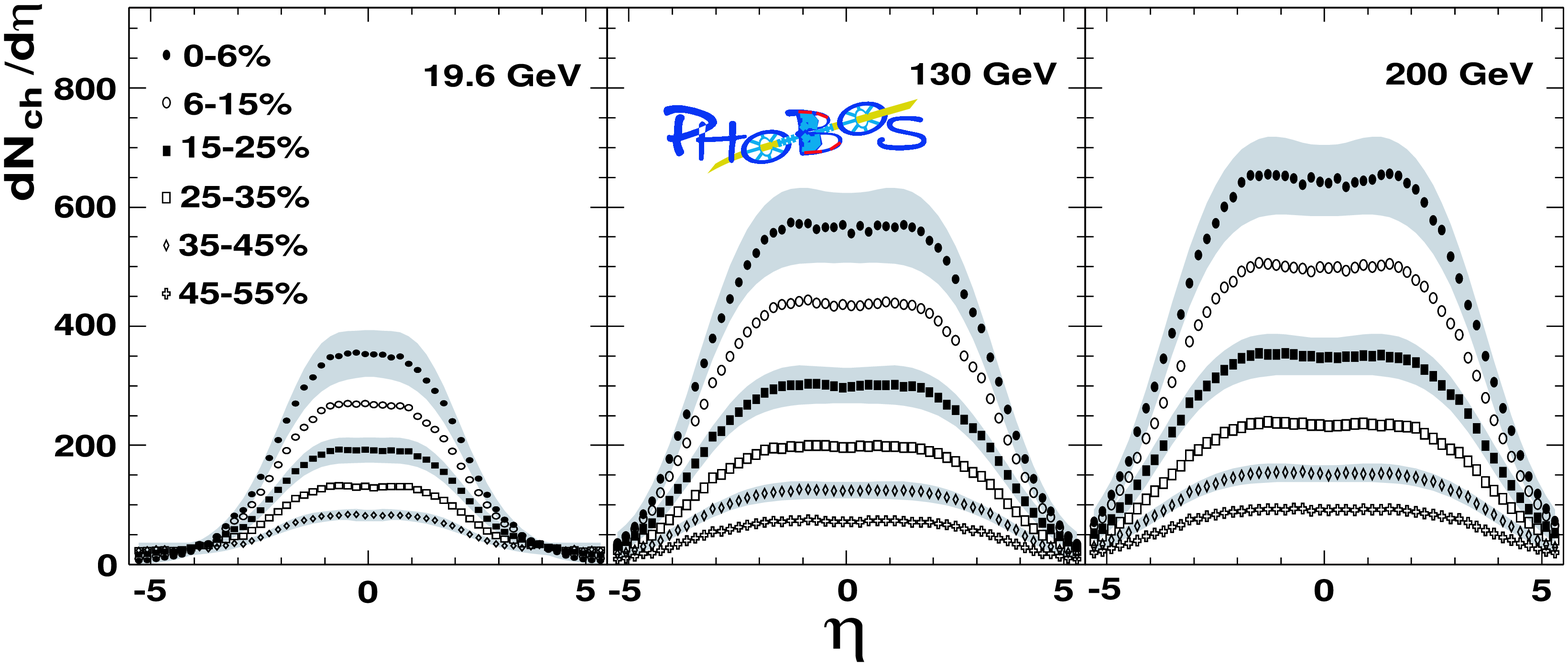}
\includegraphics[width=0.7\textwidth]{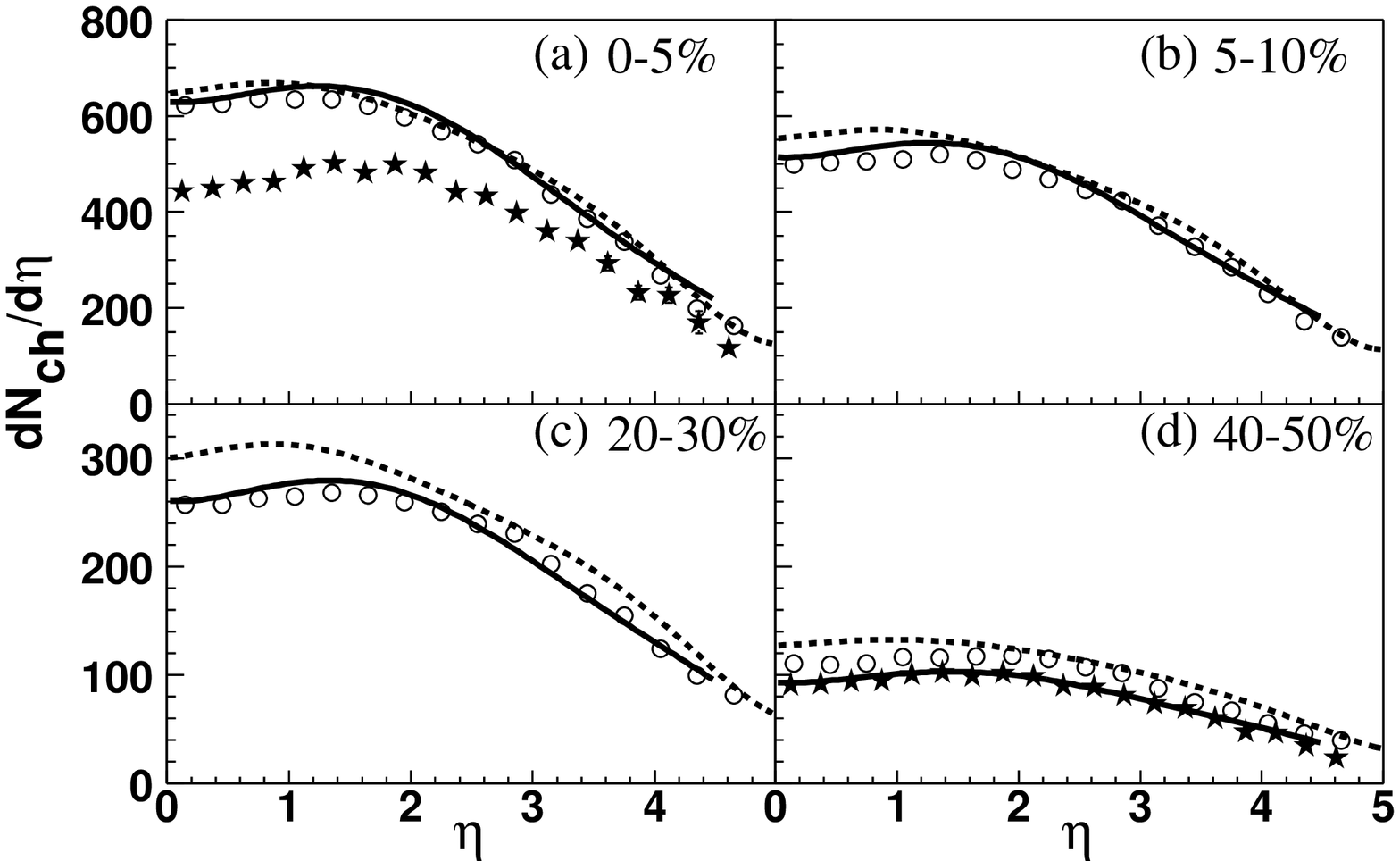}
\caption{Charged hadron distributions over the full RHIC phase space for $Au+Au$ collisions.
Top: Centrality and \sqrtsNN\ dependence of $dN_{ch}/d\eta$ from
PHOBOS \protect\cite{phobos:fragmentation}. Bottom: Centrality dependence of
$dN_{ch}/d\eta$ at $\sqrt{s}=200$ GeV, from
BRAHMS (circles) \protect\cite{brahms:dndeta200}. Stars are \pbarp\
data scaled by \Npart/2 \protect\cite{Alner:1986xu}. Calculations are saturation model (solid)
\protect\cite{KharzeevLevin} and AMPT (dashed) \protect\cite{Lin:2003ah}.  }
\label{fig:PhobosdNchdeta}
\end{figure}

Fig. \ref{fig:PhobosdNchdeta}, upper panels, show the charged particle pseudorapidity
distribution over the full RHIC phase space for Au+Au collisions at
several centralities and energies, measured by PHOBOS
\cite{phobos:fragmentation}. The distributions exhibit two general features: a plateau about
midrapidity which broadens with increasing collision energy, and a
forward region whose width is approximately invariant.  The total
charged multiplicity for central collisions at \sqrtsNN=200 GeV is
$5060\pm250$ \cite{phobos:fragmentation}, indicating a qualitatively
new regime of accelerator-based experimentation in high energy and
nuclear physics.

Figure \ref{fig:PhobosdNchdeta}, lower panels, compare the 200 GeV
data from BRAHMS \cite{brahms:dndeta200} to both saturation
model \cite{KharzeevLevin} and AMPT \cite{Lin:2003ah}
calculations. AMPT combines the HIJING model with partonic and
hadronic final-state rescattering and includes an alternative baryon
pair production mechanism. Good agreement with data is found for both model
calculations.

\begin{figure}
\centering
\includegraphics[width=0.31\textwidth]{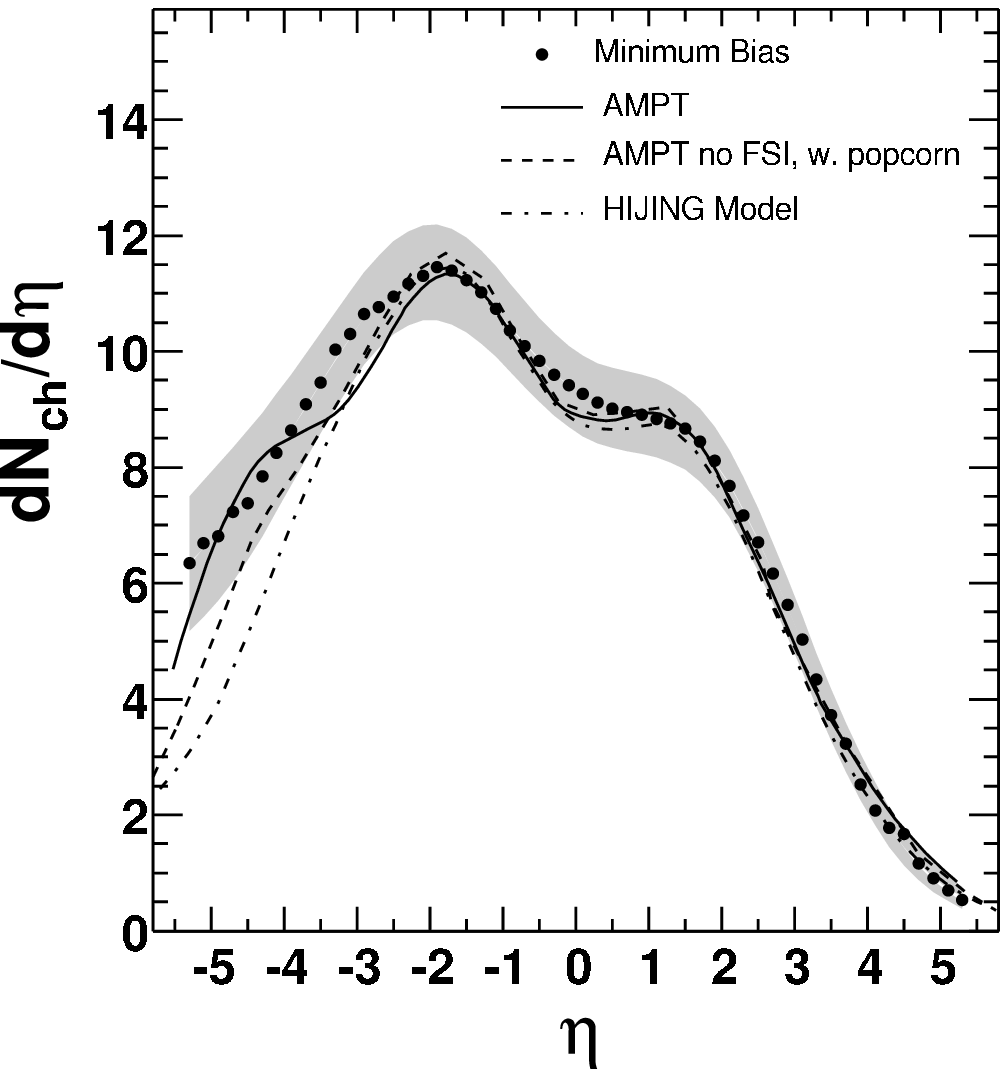}
\includegraphics[width=0.31\textwidth]{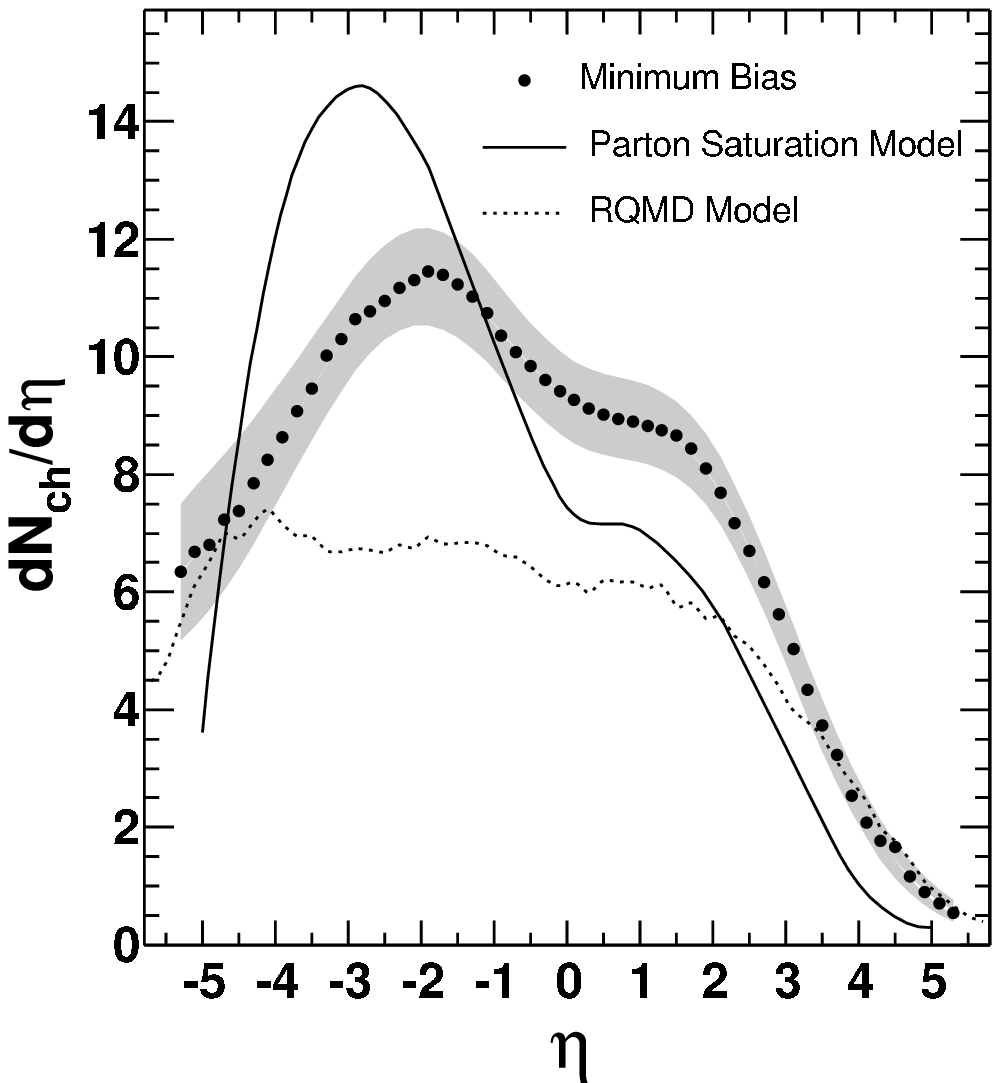}
\hspace{.06\textwidth}
\includegraphics[width=0.3\textwidth]{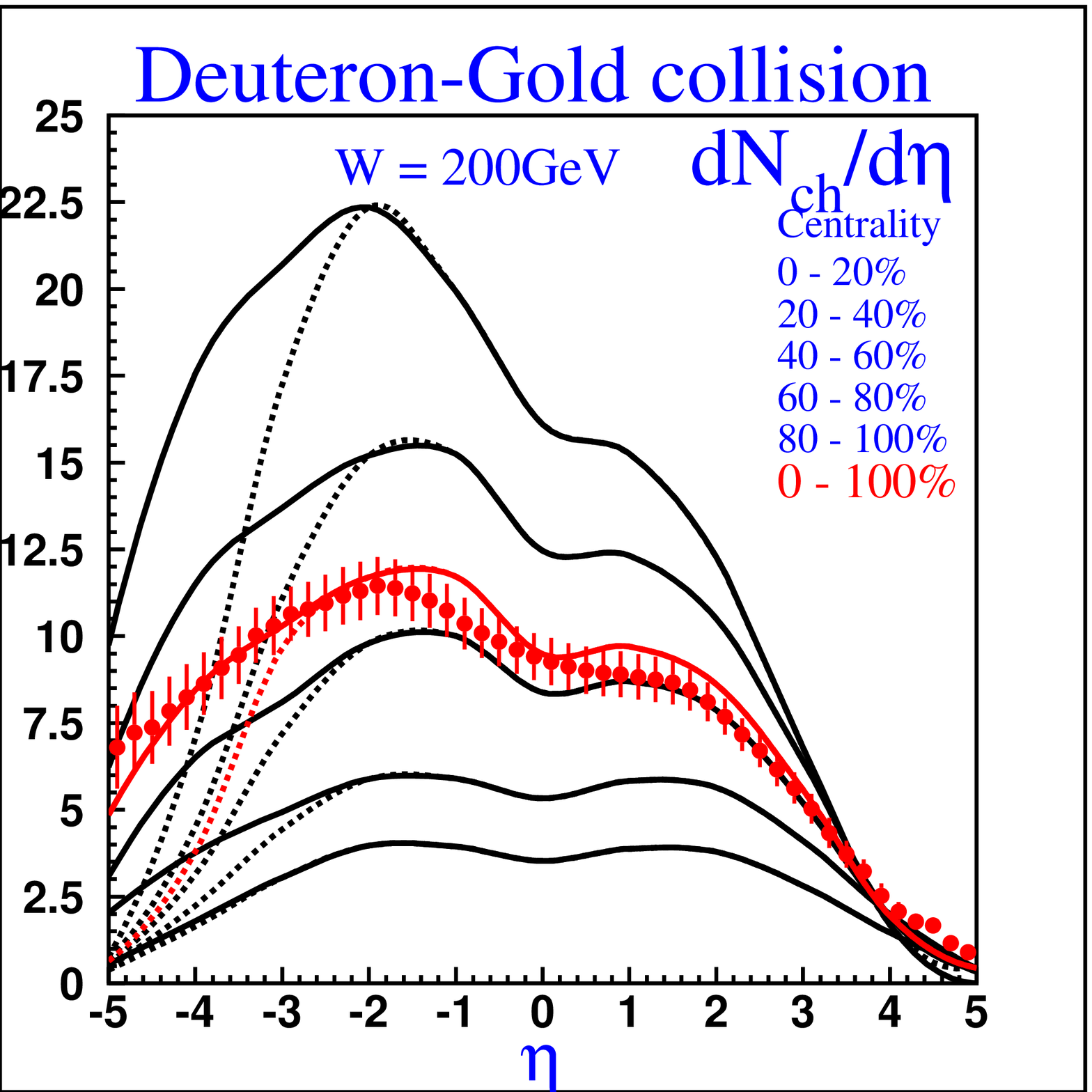}
\caption{$dN_{ch}/d\eta$ for minimum bias d+Au collisions at \sqrts=200 GeV, 
from PHOBOS \protect\cite{phobos:dAuMult}. All panels show the same data. The
left panel compares to model calculations from AMPT \protect\cite{Lin:2003ah}
and Hijing \protect\cite{Wang:1991ht,Gyulassy:1994ew}. The middle and right
panels compare to saturation model calculations \protect\cite{KLM_dAu}
(Eq. [\ref{eq:dNdySatdAu}]) using Optical (middle) and Monte Carlo
(right) Glauber calculations and slightly different proton
saturation parameters.}
\label{fig:dAuMult}
\end{figure}

Study of bulk particle production in $p(d)+A$ collisions may help to
separate initial from final state effects, since a dense medium is not
formed in the central rapidity region in such collisions.  Figure
\ref{fig:dAuMult} shows the charged multiplicity density distribution
for minimum bias $d+Au$ collisions at \sqrts=200 GeV
\cite{phobos:dAuMult} (see also BRAHMS
\cite{Arsene:2004cn}). This measurement
offers a unique probe of particle production, because of the large
projectile asymmetry. All panels show the same data, which are
compared to various model calculations. The data indeed exhibit an
asymmetry in particle production, with larger multiplicity density
towards the direction of the Au beam ($\eta\lt0$).  The left panel of
the figure compares the data to predictions from the HIJING
\cite{Wang:1991ht,Gyulassy:1994ew} and AMPT  models \cite{Lin:2003ah}.  The
predictions agree with the data except at large negative rapidity, the
region of the $Au$-nucleus fragmentation. Evidently the final state
interactions and baryon production mechanisms in AMPT provide a better
description of this region.

Using the weak coupling approach to the saturation regime in the
initial state saturation model, the rapidity 
dependendence of the charged multiplicity in $d+Au$ collisions is given by \cite{KLM_dAu}
\begin{eqnarray}
\frac{dN}{dy}= 
C \frac{S \Qsmin^2(y)}{\alpha_s(\Qsmin^2(y))} 
\left\{\left(1-\frac{\Qsmin(y)}{\sqrts}e^y\right)^4 
+\left[\ln \left(\frac{\Qsmax^2(y)}{\Qsmin^2(y)}\right)+1\right]
\left(1-\frac{\Qsmax(y)}{\sqrts}e^y\right)^4\right\},
\label{eq:dNdySatdAu}
\end{eqnarray}
where $S$ is the the interaction cross section. \Qsmax\ and
\Qsmin\ denote the larger and smaller of the saturation scales in
the deuteron and the $Au$ nucleus, which vary  with
rapidity $y\sim\log(1/x)$ as $e^{\lambda{y}}$ [Eq. \ref{eq:Qsx}]. \Qs\ is assumed to be
the same for the deuteron and proton. Since $S\Qs^2\sim\Npart$
\cite{KharzeevLevin}, $dN/dy\sim\Npart(Au)$ in the $Au$-fragmentation
region and $dN/dy\sim\Npart(d)$ in the deuteron-fragmentation region,
replicating the scaling of the phenomenological Wounded Nucleon Model
\cite{Bialas:1976ed}.  The factor $C$ in Eq. \ref{eq:dNdySatdAu} is
determined from the midrapidity charged hadron density in 130 GeV
Au+Au collisions.

The solid line in the middle panel of Fig. \ref{fig:dAuMult} shows
Eq.(\ref{eq:dNdySatdAu}) utilizing an Optical Glauber calculation
\cite{KLM_dAu}. (We will not discuss the curve labelled ``RQMD''.) The calculation 
overestimates the measured distribution in the $Au$-fragmentation
region and underestimates it in the $d$-fragmentation region. A
revised version of the calculation (erratum to
\cite{KLM_dAu}) utilizing a Monte Carlo rather than Optical Glauber
calculation and a slightly different proton saturation parameter
achieves good agreement with the measurement (right panel).

The summary of this section is similar to that of the previous
section: both pQCD-based models and models incorporating initial state
saturation reproduce the pseudo-rapidity distributions quite well over
very broad phase space. The common feature of these models is again
that the initial energy density is very high, but evidently these
observables are not sufficiently discriminating to distinguish between
the rather different production mechanisms of the models.


\subsection{Rapidity Distributions and Baryon Stopping}
\label{sect:RapidityDistributions}

\begin{figure}
\centering
\includegraphics[width=0.45\textwidth]{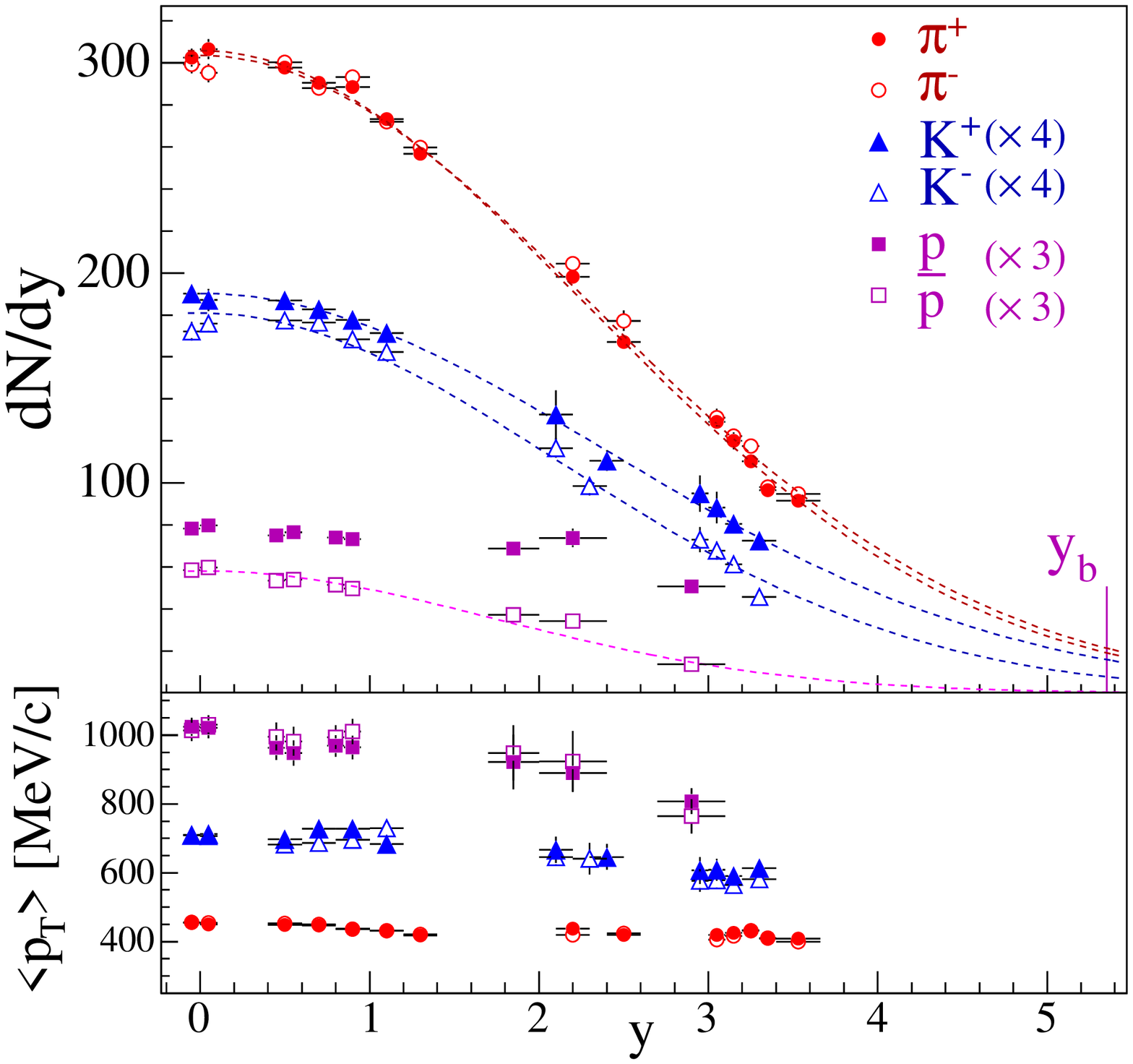}
\includegraphics[width=0.45\textwidth]{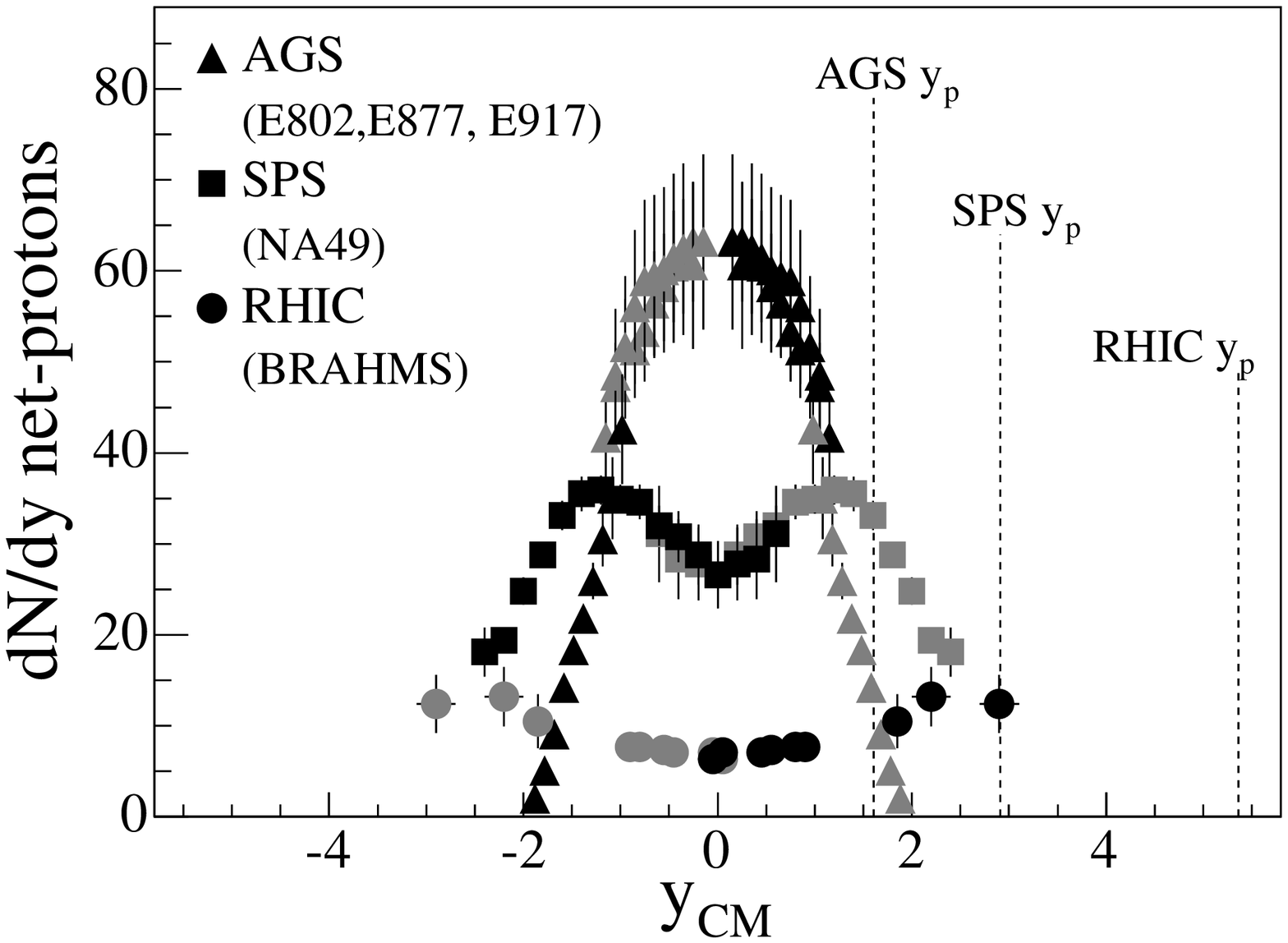}
\caption{Rapidity dependence of \pT-integrated particle yields 
from central $Au+Au$ collisions at 200 GeV, from BRAHMS.  Left: $\pi^\pm$, K$^\pm$, $p$ and \pbar
\protect\cite{Murray:2004gh,brahms:ProtonStopping,Bearden:2004yx}. Dashed lines are Gaussian fits. Lower panel shows rapidity 
dependence of $\langle\pT\rangle$. Right: net protons (difference of
$p$ and \pbar\ yields) compared to lower energy collisions
\protect\cite{brahms:ProtonStopping}. }
\label{fig:BRAHMSrapidity}
\end{figure}

Some time ago, Bjorken postulated that the rapidity distribution at
very high collision energy should develop a plateau in the central rapidity region,
which results from a reaction volume that is invariant under
longitudinal boost \cite{BjorkenHydro}. This assumption leads to
considerable simplification of the hydrodynamic equations and is
common in theoretical treatments of mid-rapidity observables
\cite{KolbHeinzHydroReview} (Sect. \ref{sect:hydro}).
The plateau in pseudo-rapidity density seen in
Fig. \ref{fig:PhobosdNchdeta} suggests that the fireball near
mid-rapidity may indeed be boost-invariant. However, pseudorapidity
$\eta$ only approximates rapidity $y$. Figure
\ref{fig:BRAHMSrapidity}, left panel, shows the {\it rapidity}
dependence of particle production separately for pions, kaons and
protons. While the rapidity distributions are indeed broad, no plateau
is observed and except for protons they are Gaussian in shape. Less
marked but still significant rapidity dependence is also seen for
$\langle\pT\rangle$. These distributions may nevertheless result from
boost-invariant initial conditions of limited extent in rapidity
(e.g. \cite{Morita:2002av,Sollfrank:1997hd}). Note that even at LHC
energies (\sqrtsNN=5500 GeV) the initial energy density computed from
pQCD with saturation scale \psat=2 GeV is not uniform in
rapidity\cite{Eskola:1998hz}, so that a boost-invariant initial
condition may in any case not be the correct high energy limit.

Baryon number is conserved in the collision and its rapidity
distribution should be very different from that of produced
particles. Since rapidity is logarithmic in energy, it is not changed
significantly by rescattering. The net baryon rapidity distribution
observed in the final state is therefore established to a large extent
early in the collision, reflecting the mechanisms of energy transfer
from the colliding nuclei to the fireball. In thermodynamic terms,
finite net baryon number results in finite baryochemical potential
\muB\ (Sect. \ref{sect:StatModel}). While the
distribution of total baryon number is difficult to access
experimentally, it can be deduced from the net proton distribution
$p-\pbar$ (Fig. \ref{fig:BRAHMSrapidity}, right panel), which is
seen to be small but finite at midrapidity. The BRAHMS data indicate a
net baryon density $dN/dy\simeq10$ for central Au+Au collisions at
$y=0$
\cite{brahms:ProtonStopping}, in agreement with other analyses
\cite{phenix:pidspectra200,star:pbar130}.

The mean rapidity loss of leading baryons in lower energy
fixed target interactions of protons with heavy nuclei is
$\left<\Delta{y}\right>\sim2.5$, in contrast to
$\left<\Delta{y}\right>\sim1$ for proton-hydrogen interactions
\cite{Busza:1984rj}. The rapidity distribution of net protons in
Fig. \ref{fig:BRAHMSrapidity}, taken together with conservation of net
baryon number, significantly constrains the full net baryon rapidity
distribution, resulting in
$\left<\Delta{y}\right>\sim2.0\pm0.2$\cite{brahms:ProtonStopping} and
a mean energy loss per participating nucleon of $\Delta{E}=72\pm6$ GeV
in central $Au+Au$ collisions \cite{brahms:ProtonStopping}. In other
words, for interacting nucleons about 70\% of the incoming beam energy
of 100 GeV per nucleon is delivered to the fireball. Central
collisions have about 350 participants, giving $\sim25$ TeV
transferred from the incoming projectiles to final particle
production.  It is notable both that the net baryon density at
midrapidity is small for central collisions relative to the
$2\times197$ nucleons brought into the collision and that it is
finite, indicating transfer of baryon number over 5.5 rapidity
units. 

The conventional mechanism for baryon transport in hadronic and
nuclear collisions is the fragmentation of quark-diquark ($q-qq$)
strings \cite{Andersson:1987gw}. However, such string fragmentation
models underpredict the baryon stopping measured in nuclear collisions
both at SPS \cite{ToporPop:1995cg} and at RHIC energies
\cite{ToporPop:2002gf}. An alternative scenario considers baryon
structure comprising the gauge junction which carries the baryon
number of three quarks in their fundamental representation
\cite{Kharzeev:1996sq,Montanet:1980te}.  An implementation of this
baryon junction mechanism in HIJING/B$\bar{\mathrm{B}}$
\cite{Vance:1998vh} describes well the measured baryon stopping at RHIC
\cite{ToporPop:2002gf}. Other modified string fragmentation models
\cite{Werner:1993uh,Capella:1996th} and diquark rescattering
\cite{Sorge:1995dp,Bass:1998ca} can also provide stronger baryon
stopping power than the conventional string fragmentation. However, it
is not clear at this point which of the baryon transport mechanisms
are dominant in high energy heavy-ion collisions.

\subsection{Transverse Energy and Energy Density}
\label{sect:ET}

Significant transverse energy \ET\ can only be generated during the
collision, through the initial interactions of partons from the
projectiles and the successive interactions among the produced partons
and hadrons. Experimentally, \ET\ is defined as
$\ET=\sum_i{E_i}sin(\theta_i)$, where $i$ sums over all final state
hadrons. The hadron energy $E_i$ is corrected for conserved baryon
number, and $\theta_i$ is the angle relative to the beam
direction. Due to the dynamics of the expansion, \ET\ in a limited
rapidity interval will evolve through the lifetime of the collision.
In the framework of hydrodynamics this dependence is
\cite{Gyulassy:1984ub,Dumitru:2000up}:
\begin{equation}
\frac{\ET(\tau)}{\ET(\tau_0)}=\left(\frac{\tau_0}{\tau}\right)^\delta,
\label{ETtau}
\end{equation}
where $\tau_0$ is the equilibration time. Local thermodynamic
equilibrium is characterized by energy density $\epsilon$, pressure
$p$, and speed of sound $c_0^2=\partial{p}/\partial\epsilon$. If equilibrium is
established at $\tau_0$ and maintained throughout the expansion with
constant $c_0^2$, then $\delta=c_0^2$ and the observed \ET\ is
substantially reduced relative to the initially generated \ET\ due to
the $p\Delta{V}$ work performed during the expansion
\cite{Gyulassy:1984ub}.
Alternatively, if the system falls out of
equilibrium quickly into a free-streaming gas, $\delta=0$ and there will be no
$p\Delta{V}$ work performed during the evolution, so that
\ET\ will remain constant throughout the expansion. 
Final state saturation effects (Sec. \ref{sect:FinalStateSat}) reduce
early pressure, delaying the onset of hydrodynamic behavior and
leading to significant reduction in the observed \ET\
\cite{Dumitru:2000up}. 

It is difficult to disentangle these competing mechanisms based solely
consideration of \ET\ distributions, but the systematic study of
\ET\ together with that of other bulk observables may isolate the
contribution of longitudinal work. If hydrodynamic flow can be shown
to set in early from other considerations, the measured \ET\ will
provide a lower limit to the initially produced transverse energy at
the time of equilibration and thus provide an estimate of the initial
energy density.

\begin{figure}
\centering
\includegraphics[width=0.80\textwidth]{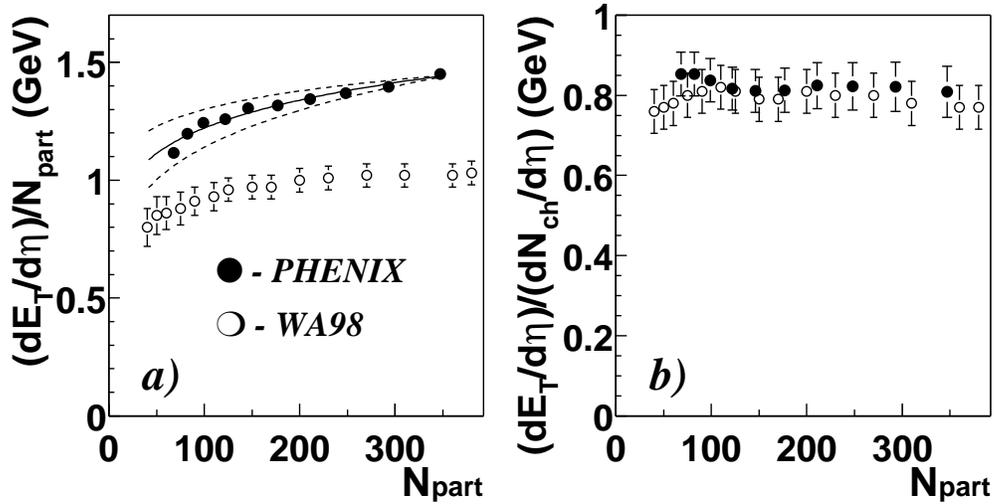}
\caption{Transverse energy \ET\ for $Au+Au$ collisions at \sqrtsNN=130 GeV, 
from PHENIX \protect\cite{phenix:et130} compared to $Pb+Pb$ collisions at
\sqrtsNN=17 GeV \protect\cite{WA98ET}. Left: \ET\ per participant. Right: \ET\ per charged particle. 
The $Pb+Pb$ data in both panels have an overall
normalization uncertainty of $\pm$20\%, not shown.}
\label{fig:PhenixET}
\end{figure}

Fig. \ref{fig:PhenixET} from PHENIX \cite{phenix:et130}
shows the centrality dependence of \ET\ for
\sqrtsNN=130 GeV $Au+Au$ collisions, compared to $Pb+Pb$ collisions at
\sqrtsNN=17 GeV \cite{WA98ET}. In the left panel, \ET\ per 
participant is seen to increase with increasing \sqrts\ for all
centralities. As shown in the right panel, however, \ET\ per charged
particle is largely independent of collision energy, meaning that the
dependence of \ET\ on energy and centrality closely parallels that of
the charged multiplicity in
Figs. \ref{fig:MultPerPartvsNpart}-\ref{fig:PhobosdNchdeta}.  Since
the average $\langle\pT\rangle$ for charged hadrons in $p+p(\bar p)$
collisions increases significantly with $\sqrt{s}$ \cite{UA1spectra},
a constant \ET\ per hadron in heavy-ion collisions implies the
existence of $p\Delta{V}$ work due to hydrodynamic expansion. Indeed,
hydrodynamic calculations assuming onset of equilibration at
$\tau_0\lt 1$ fm/c are able to reproduce approximately the centrality
dependence of \ET\ per charged particle \cite{KolbHeinzHydroReview}.

In the Bjorken picture of boost-invariant free-streaming expansion,
the initial energy density can be expressed in terms of the observed
\ET\ \cite{BjorkenHydro}:
\begin{equation}
\epsBj=\frac{d\ET}{dy}\frac{1}{\tau_0\pi{R}^2}.
\label{eq:epsBj}
\end{equation}
The formation time is taken to be $\tau_0\sim1$ fm/c and initial system size
$R\simeq1.2A^{1/3}$, equal to the nuclear radius.  This expression
relies on the assumption that no $p\Delta{V}$ work is done during the
expansion
\cite{Gyulassy:1984ub} and it therefore represents a lower bound to 
the initial energy density within the hydrodynamic framework. PHENIX
has measured $d\ET/d\eta\simeq540$ GeV\cite{phenix:et130} at
midrapidity for central $Au+Au$ collisions at \sqrtsNN=130 GeV,
resulting in \epsBj=4.6 GeV/fm$^3$. In comparison, NA49 estimates
$\epsBj\sim3$ GeV/fm$^3$ for central $Pb+Pb$ collisions at
\sqrtsNN=17.2 GeV\cite{Margetis:1995tt}.  These values are
provocative: they lie well above the deconfinement energy density
predicted by Lattice QCD calculations
(Sect. \ref{sec:deconfinement}). However, the calculation is based on
the assumption rather than the demonstration that the onset of
hydrodynamic expansion occurs at $\tau_0\simeq1$ fm/c. In the
following sections we will address this question through measurements
sensitive to local equilibration at the early, hot and dense phase of
the collision, in particular elliptic flow.

\section{Collective Phenomena}

A central question at RHIC is the extent to which the quanta produced
in the collision interact and thermalize. Nuclear collisions generate
enormous multiplicity and transverse energy, but in what sense does
the collision generate {\it matter} in local equilibrium which can be
characterized by the thermodynamic parameters temperature, pressure,
and energy density? Only if thermalization has been established can more
detailed questions be asked about the equation of state of the matter.

The initial energy density, whether equilibrated or not, will have
strong spatial gradients due to the geometry of the colliding nuclei
and the dynamics of the collision. Reinteractions among the fireball
constituents will convert these density gradients into pressure
gradients, resulting in collective flow of the matter. Collective flow
is thus a generic consequence of reinteractions, which also lead to
thermalization. It is however not sufficient merely to observe
collective flow, which may be generated both early through partonic
reinteractions and later through interactions in the dense hadronic
gas. It is {\it partonic} thermalization and the {\it partonic}
Equation of State that are of interest, but their signals may be masked
by the hadronically generated flow that must be understood and
unraveled.

In this section we discuss hadronic observables that are sensitive to
collective flow and the degree of thermalization in nuclear collisions
at RHIC. Of particular importance is the azimuthal anisotropy of the
final hadron spectra in non-central collisions (``elliptic flow'')
that results from the conversion of the initial coordinate-space
asymmetry to momentum space via collective expansion. The particle
mass dependence of flow is an especially sensitive observable, since a
common velocity distribution for fluid cells radiating particles of
different mass will result in a characteristic mass dependence of the
momentum spectra.

\subsection{Relativistic Hydrodynamics}
\label{sect:hydro}

Relativistic hydrodynamics provides the theoretical framework to study
collective behavior in high energy collisions, with the first such
attempts dating back to Landau
\cite{Belenkij:1956cd}. We sketch here the basic ideas and compare
hydrodynamic calculations to a wide range of RHIC data. Generally good
agreement is achieved (though with notable exceptions), providing
strong evidence that local equilibrium is established early in the
evolution of the fireball ($\tau\lt1$ fm/c) and that the system
evolves in accordance with ideal hydrodynamics. Some sensitivity to
the equation of state is observed, with preference for a deconfined
phase early in the evolution. Detailed reviews of relativistic
hydrodynamics with applications to RHIC data can be found in
\cite{KolbHeinzHydroReview,Huovinen:2003fa}.

We first present a simple estimate to assess whether hydrodynamics is
a reasonable approach to modeling the dynamics of a deconfined phase
\cite{Huovinen:2003fa}. Consider a two-flavor QGP at temperature
$T\sim200$ MeV, which has partonic density $n\sim4$
fm$^{-3}$. Assuming the Debye screening mass $\mu=gT$ as the typical
momentum transfer in gluon-gluon scattering, the pQCD cross section
$\sigma_{gg{\rightarrow}gg}\sim 3$ mb gives a mean free path
$\lambda=1/\sigma{n}\sim 0.8$ fm. The time between collisions is
therefore an order of magnitude smaller than the expected system
lifetime of a few fm/c, so that the multiple reinteractions necessary
for thermalization may occur.

Hydrodynamic behavior can set in only at a finite time after the
collision, when the produced quanta have interacted and relaxed into
local equilibrium. Hydrodynamics therefore does not address the
earliest moments of the fireball evolution, and its initial conditions
(density distributions and flow velocities) must be imposed on the
basis of other considerations. The initial conditions (entropy,
energy, and net baryon number density) are constrained by comparing to
experimental data such as hadron multiplicities and transverse energy
production.

The hydrodynamic evolution terminates when the system has expanded and
cooled to a degree that the mean free path exceeds system size and
local equilibration can no longer be maintained
(``freezeout''). Generically, two stages of freezeout are expected:
{\it chemical} freezeout occurs when the mean free path for {\it
inelastic} collisions exceeds the system size, whereas {\it kinetic}
freezeout occurs at a later time and a lower temperature, when the {\it
elastic} mean free path also exceeds the system size. The produced
hadrons, dominantly soft, are created continuously at the dilute
periphery of the fireball, according to a specific prescription of
kinetic freezeout. Though these soft hadrons do not directly transmit
signals from the hot and dense early stage of the collision, the
systematic study of the transverse momentum and mass dependence of
soft hadron production can provide substantial evidence for early
pressure build-up and therefore equilibration. Sufficiently detailed
and precise comparison of data and calculations will also be able to
constrain the initial conditions and the EOS at the early stage.

For ideal, non-dissipative hydrodynamics, the energy-momentum tensor
$T^{\mu\nu}(x)$ in the global reference frame for a fluid cell at
space-time coordinate $x$ is given by \cite{KolbHeinzHydroReview}
\begin{equation}
T^{\mu\nu}(x)=[e(x)+p(x)]{u^\mu(x)}{u^\nu(x)}-p(x)g^{\mu\nu},
\label{eq:HydroT}
\end{equation}
\noindent
where $e(x)$ is the energy density, $p(x)$ is the pressure, and
$u^\mu(x)$ is the four-velocity of the cell. Correction for non-ideal
hydrodynamics adds a term that is the product
of the shear viscosity $\eta$ with the thermally averaged gradient of
the velocity field \cite{Teaney:2003pb}. 

The equations of motion result from 
local conservation of energy and momentum,
\begin{equation}
\partial_{\mu}T^{\mu\nu}(x)=0\ (\nu=0,\ldots,3).
\label{eq:HydroConservation}
\end{equation}
\noindent
Additional equations result from the conservation of $M$ different
charges (net baryon number, net strangeness, electric charge),
\begin{equation}
\partial_{\mu}j_i^\mu(x)=0 (i=1\ldots,M),
\label{eq:HydroCurrents}
\end{equation}
\noindent 
where $j_i^\mu(x)=n_i(x)u^\mu(x)$ is the current density in the global
frame and $n_i(x)$ is the local charge density.

Expressions (\ref{eq:HydroConservation}) and (\ref{eq:HydroCurrents})
comprise $4+M$ differential equations for $5+M$ fields: the three
components of the flow velocity, the energy density, the pressure, and
the $M$ charge densities. The system of equations is closed by the
{\it equation of state} (EOS) $p(e,n_i)$, which relates the pressure,
energy density, and conserved charge densities. Most applications of
hydrodynamics to RHIC data use a similar structure for the equation of
state \cite{Huovinen:2003fa}: a plasma phase of massless partons with
a bag constant, a hadronic phase consisting of a gas of free hadrons
and resonances, and a first order phase transition with a large latent
heat connecting the two phases.

The initial conditions at the onset of hydrodynamic expansion must be
specified from external input, usually either the entropy or energy
density, with distribution in the transverse plane according to that
for binary collisions or participants nucleons \cite{Huovinen:2003fa}.
Saturation initial conditions have also been considered
\cite{Hirano:2004rs}. The conventional implementation of freezeout is
via the Cooper-Frye prescription\cite{CooperFrye} which corresponds to an
instantaneous transition from zero to infinite mean free path,
i.e. from ideal hydrodynamics to free streaming. The spectrum of
hadron species $i$ at freezeout is given by

\begin{equation}
E\frac{dN_i}{d^3p}=\frac{g_i}{(2\pi)^3}\int_{\Sigma}
\frac{1}{exp\left(\left(p_{\nu}u^{\nu}-\mu_i\right)/T\right)\pm1}
p^{\mu}d^3\sigma_\mu,
\label{eq:CooperFrye}
\end{equation}

\noindent
where the integral is carried out over the hypersurface $\Sigma(x)$ on
which the freezeout conditions are met. $\mu_i(x)$ is the local
chemical potential for species $i$ and $T(x)$ is the local
temperature.

Having specified the EOS and initial conditions, the differential
equations (\ref{eq:HydroConservation}) and (\ref{eq:HydroCurrents}) are
integrated numerically to freezeout, where the stable hadrons and
resonances are generated according to
Eq.~(\ref{eq:CooperFrye}). Integration of the full three dimensional
hydrodynamic equations is a daunting task \cite{Hirano:2002ds}. The
assumption of longitudinal boost invariance is often made, with the
imposed symmetry reducing the number of coupled equations and
simplifying the numerical problem
considerably\cite{KolbHeinzHydroReview}. This approach is however only
applicable to mid-rapidity observables.

The instantaneous freezeout embodied in the Cooper-Frye prescription
is unphysical. A more realistic though  calculationally  more intensive
transition to on-shell hadrons results from coupling the hydrodynamic
evolution to a kinetic transport
model \cite{Bass:1999tu,Teaney:2001av}. However, at present the
experimentally accessible observables exhibit no significant variation
between this approach and the simpler Cooper-Frye algorithm
\cite{KolbHeinzHydroReview}.

\begin{figure}
\includegraphics[width=.47\textwidth]{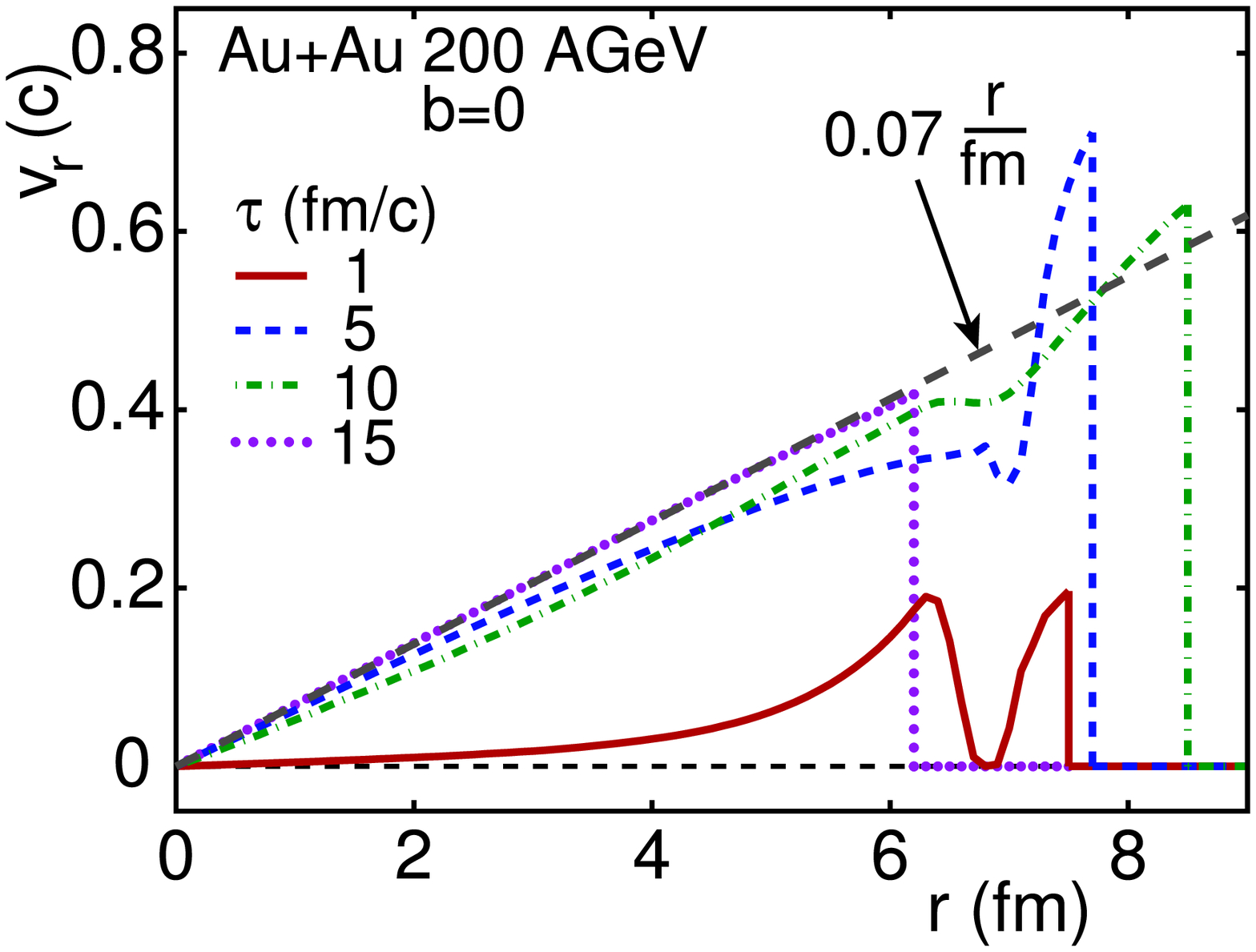}
\includegraphics[width=.47\textwidth]{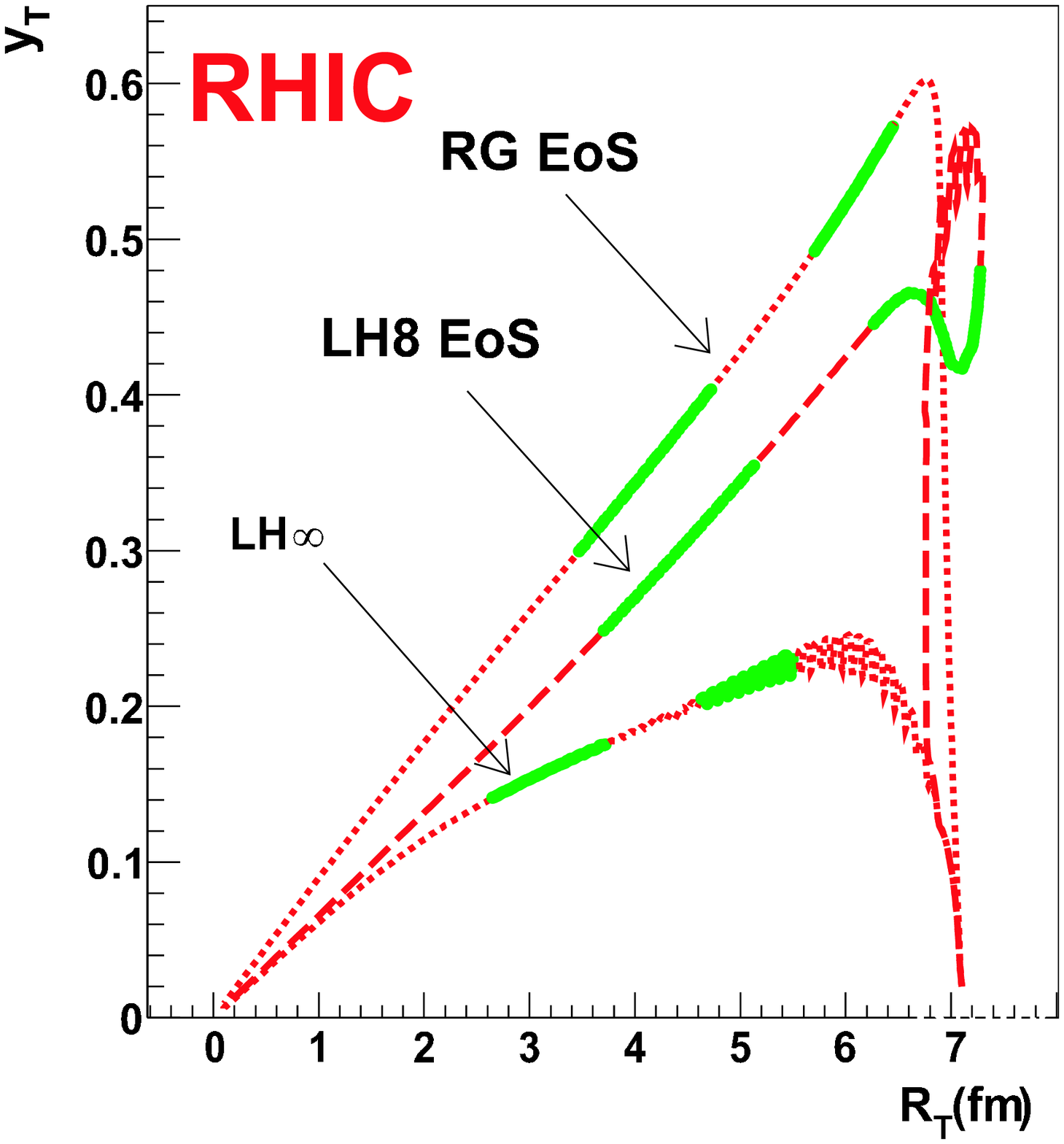}
\caption{Radial dependence of the flow velocity from full hydrodynamic calculations 
of central $Au+Au$ collisions at RHIC. Left: at various proper times
\protect\cite{KolbHeinzHydroReview}. Right: dependence at hadronization on 
equation of state \protect\cite{Teaney:2001av}. The left panel is most
comparable to LH8.}
\label{fig:HydroFlowVelocity}
\end{figure}

Fig. \ref{fig:HydroFlowVelocity} shows the radial profile of the
transverse flow velocity $v_r$ resulting from two different hydrodynamic
calculations for central $Au+Au$ collisions at RHIC
\cite{KolbHeinzHydroReview,Teaney:2001av}. 
The left panel shows the time dependence of $v_r$ with an EOS
incorporating a first order phase transition. In the bulk ($r\lt\sim6$
fm), the buildup of transverse velocity due to pressure is rapid,
achieving a roughly linear gradient $\sim0.07/fm$ that persists for
the lifetime of the fireball. The velocity near the dilute surface
varies strongly with radius at early times due to the initialization
of matter in the mixed phase, which has vanishing pressure gradient,
and in the hadronic phase at the largest radii. This surface feature
is eventually overtaken by the expanding plasma at higher
pressure. The largest system size is achieved at $\tau\sim10$ fm/c; by
15 fm/c the freezeout surface is contracting inwards.

The right panel of Fig. \ref{fig:HydroFlowVelocity} shows the
transverse rapidity $y_T=\tanh^{-1}v_r$ for three different equations
of state \cite{Teaney:2001av}: a hadronic resonance gas (RG), plasma
and hadronic phases linked by a mixed phase with latent heat 0.8
GeV/fm$^3$ (LH8), and mixed and hadronic phases only (i.e. infinite
latent heat, LH$\infty$). The flow profiles are shown for constant
energy density $e=0.45$ GeV/fm$^3$, where in this calculation the
hydrodynamic evolution is terminated and the produced hadrons
propagated further using a kinetic transport model. The resonance gas
EOS is seen to be quite stiff, generating a higher transverse velocity
gradient than those containing a mixed phase. The presence of the
plasma phase also provides significant pressure, generating twice the
flow velocity at large radius than the case where it is absent (LH8 vs
LH$\infty$). Similar to the calculations in the left panel, LH8 also
produces a radial velocity gradient $\sim0.07$/fm late in the
evolution. Note that constant energy density does not correspond to
constant proper time in this calculation. The double-valued loop at large radius for LH8
is also due to matter on the dilute surface initially generated in the
mixed or hadronic phase, freezing out rapidly after modest radial
expansion.

\subsection{Transverse Radial Flow}
\label{sect:RadialFlow}

In the hydrodynamic picture just described, the observed final state
hadrons freeze out from fluid cells that are in local equilibrium but
that have finite transverse velocity relative to the lab frame. Since
the thermal sources for all hadron species are boosted with the same
{\it velocity} distribution, the hydrodynamic expansion should result
in a characteristic mass dependence of the {\it transverse momentum}
spectra for momenta on the order of the particle mass.

\begin{figure}
\includegraphics[width=.55\textwidth]{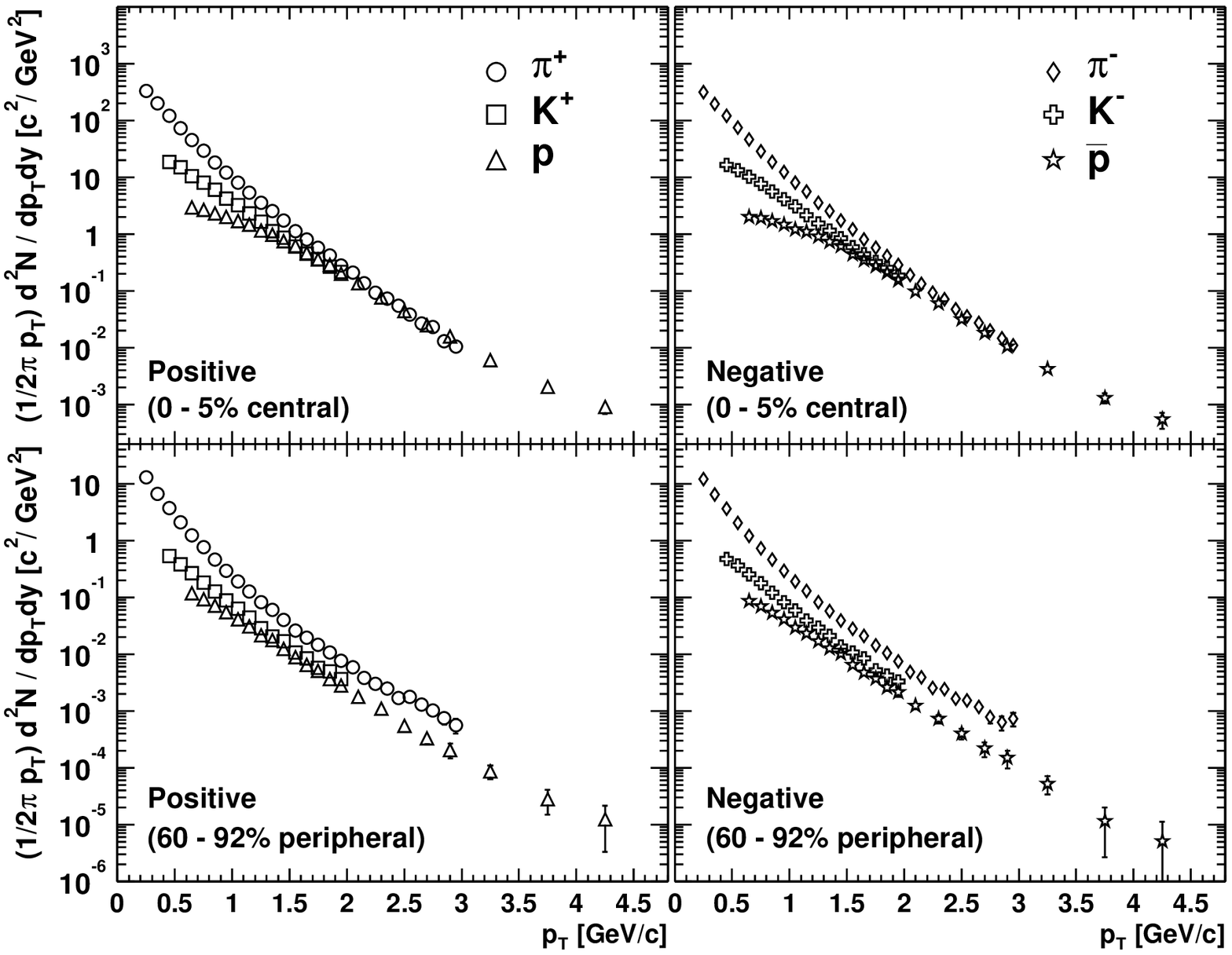}
\includegraphics[width=.45\textwidth]{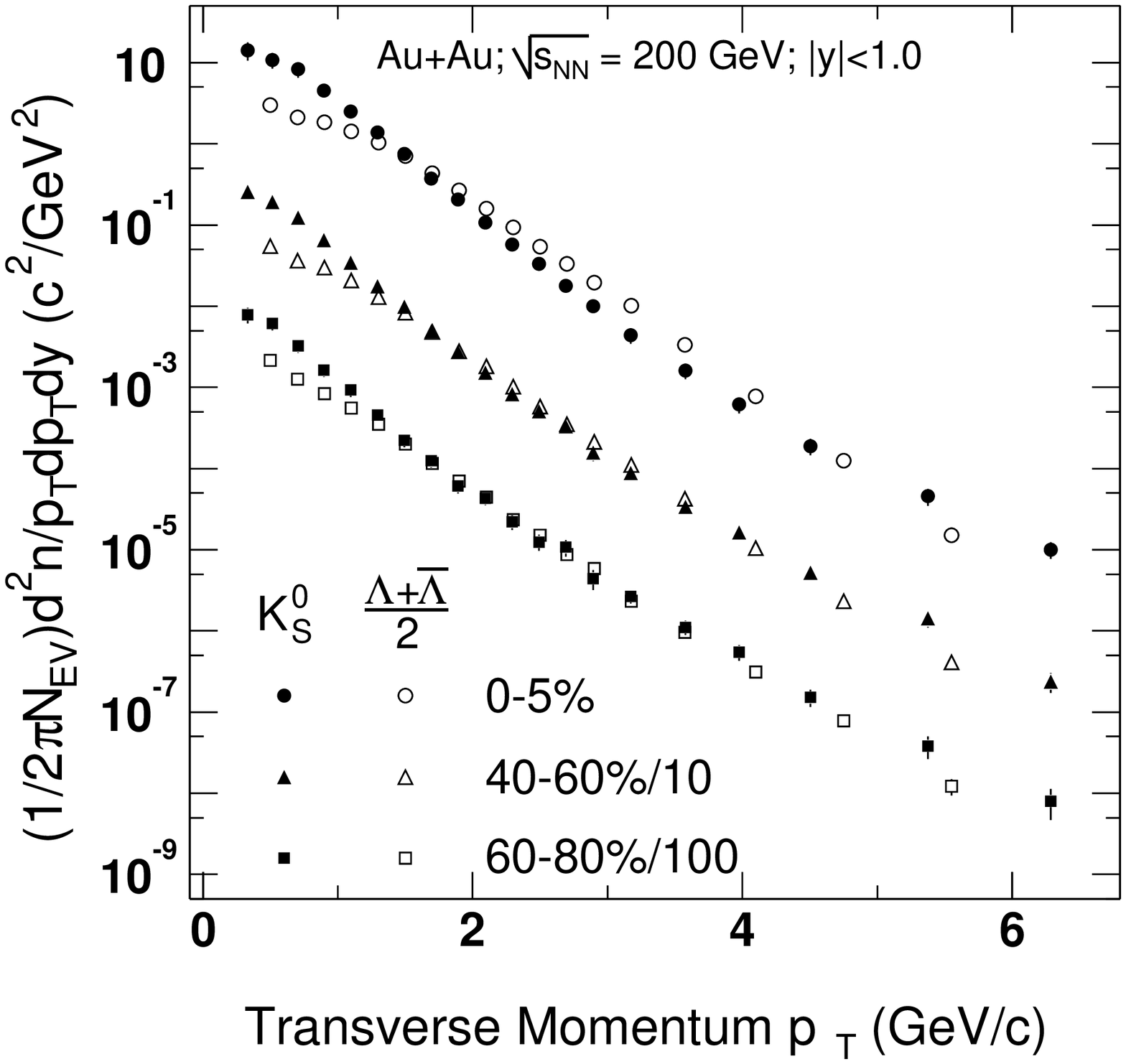}
\caption{Identified particle inclusive spectra for centrality-selected $Au+Au$ 
collisions at \sqrtsNN=200 GeV.  Left: $\pi^\pm$, K$^\pm$, p and
\pbar\ from PHENIX \protect\cite{phenix:pidspectra200}. Right: $\lam+\lambar$
and \kzeros\ from STAR \protect\cite{star:highpTLamK200,Sorensen:2003kp}.}
\label{fig:PIDSpectra}
\end{figure}

We first look at the systematic features of the data. 
Fig. \ref{fig:PIDSpectra} shows inclusive transverse
momentum spectra for $\pi^\pm$, K$^\pm$, p and \pbar\ from
PHENIX \cite{phenix:pidspectra200} (left panel) and $\lam+\lambar$ and
\kzeros\ from STAR \cite{star:highpTLamK200,Sorensen:2003kp} (right panel), 
for centrality-selected $Au+Au$ collisions at \sqrtsNN=200 GeV.
The shape of the baryon spectrum changes qualitatively from
peripheral to central collisions. Relative to the meson yields, the
baryon yields in central collisions are suppressed at low
\pT\ and enhanced at higher \pT, with a marked change of slope at 1-2
GeV/c. For $\pT\gt2$ GeV/c, the baryon yields exceed the meson 
yields (Sect. \ref{sec:recon}).

\begin{figure}
\includegraphics[width=.47\textwidth]{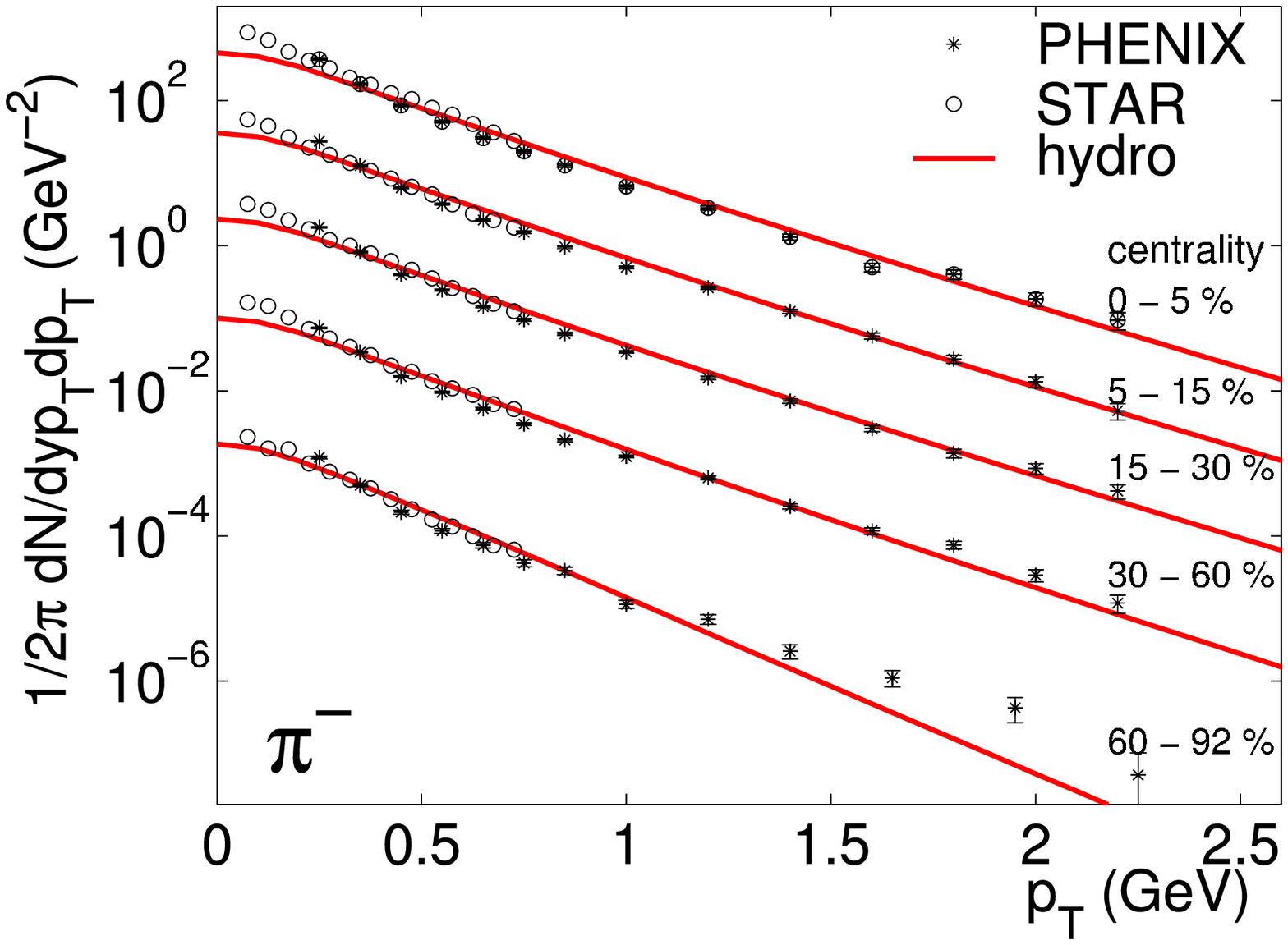}
\includegraphics[width=.47\textwidth]{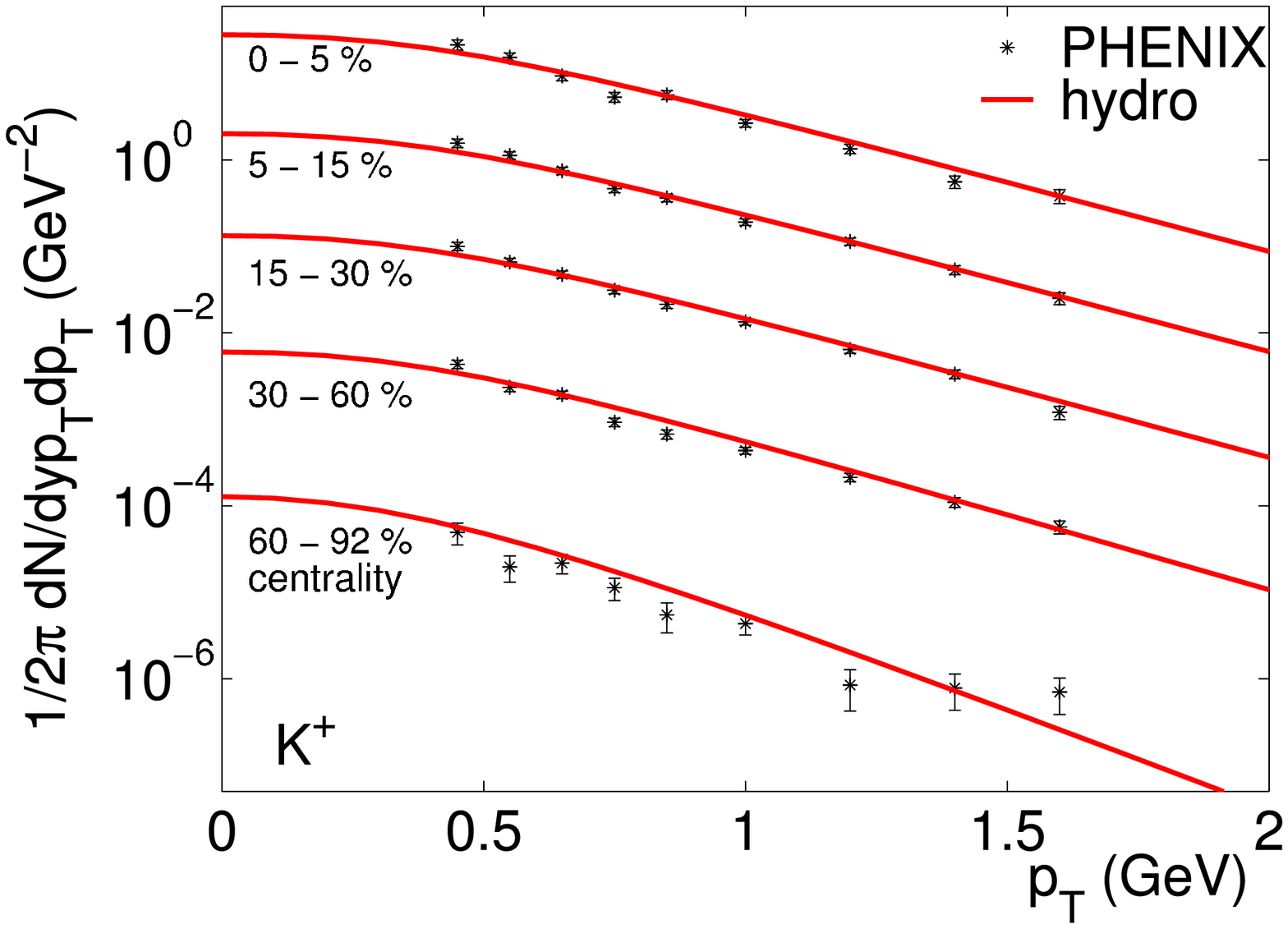}
\includegraphics[width=.47\textwidth]{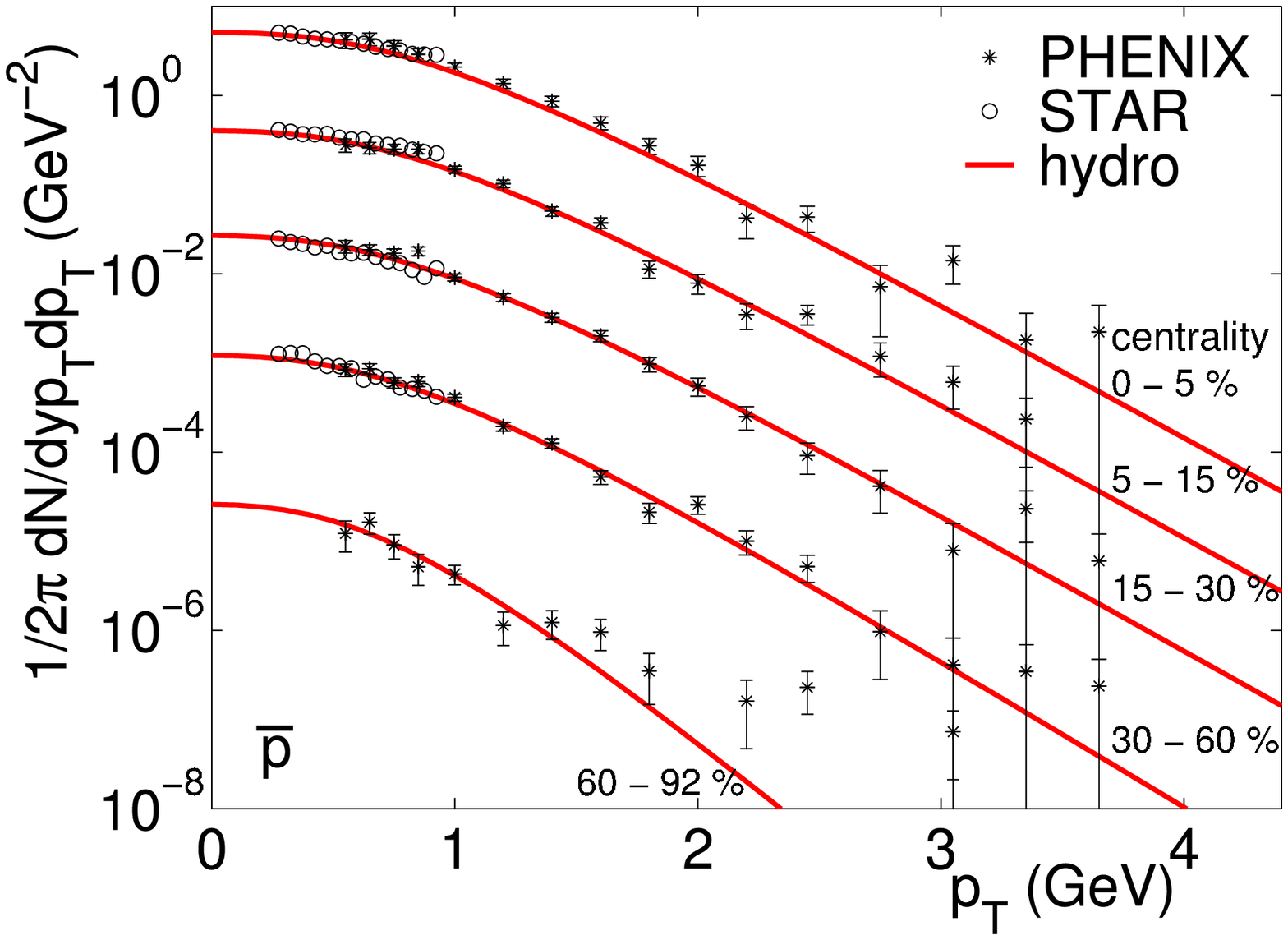}
\includegraphics[width=.53\textwidth]{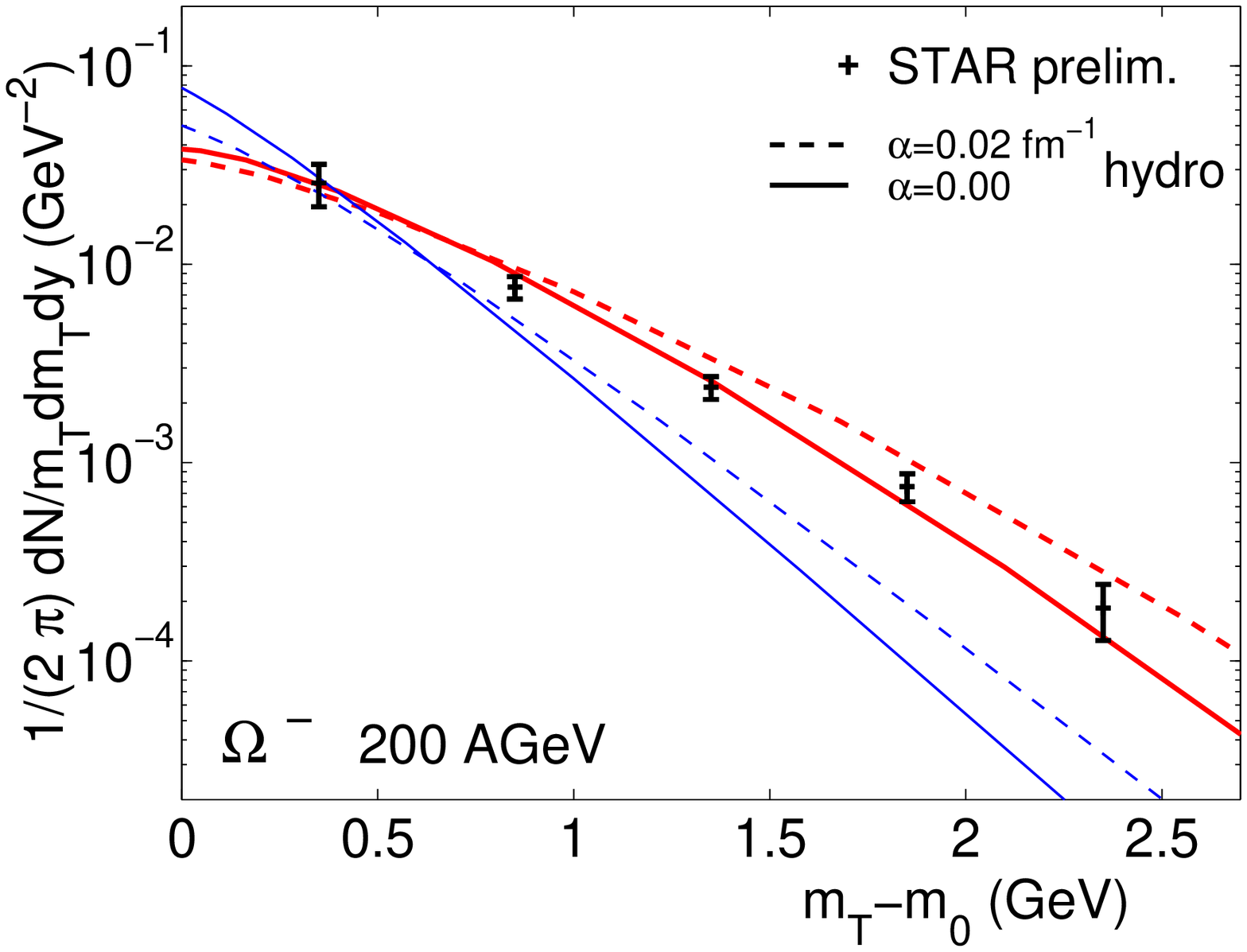}
\caption{Transverse momentum spectra of identified hadrons from $Au+Au$ 
collisions at 
RHIC, compared to hydrodynamic calculations. Upper left, upper right,
and lower left: 130 GeV $Au+Au$ data from PHENIX
\protect\cite{phenix:spectraprl130} and STAR
\protect\cite{star:pbar130,CalderondelaBarcaSanchez:2001np}, calculations from \protect\cite{Heinz:2002un}. 
Lower right: 200 GeV central $Au+Au$ data from STAR \protect\cite{Suire:2002pa}, 
calculations from \protect\cite{KolbHeinzHydroReview}. }
\label{fig:HydroSpectraCompare}
\end{figure}

Fig. \ref{fig:HydroSpectraCompare} compares hydrodynamic calculations
\cite{Heinz:2002un,KolbHeinzHydroReview} to measured \pT\ spectra from
$Au+Au$ collisions of $\pi^-$, K$^+$ and \pbar\
\cite{phenix:spectraprl130,star:pbar130,CalderondelaBarcaSanchez:2001np} 
and $\Omega^-$ \cite{Suire:2002pa}. For the upper left, upper right,
and lower left panels, the parameters of the calculation were
fixed by fitting the $\pi^+$ and \pbar\ distributions in central
collisions, resulting in an equilibration time $\tau=0.6$ fm/c with
temperature $T=340$ MeV and energy density $e=25$ GeV/fm$^3$ at the core
of the fireball. The remaining curves in those panels are then
predictions of the model. Overall agreement with the data is
good. Deviations are seen at low \pT\ for the pions, now understood to
be due to the imposition of chemical equilibrium through to
kinetic freezeout \cite{KolbHeinzHydroReview}. More significant
disagreements are seen for $\pT\gt2$ GeV/c in the most peripheral
collisions. This may delineate the region of applicability of the
hydrodynamic approach, since the fireball is smallest for peripheral collisions
and high \pT\ particles require the greatest number of collisions to
thermalize \cite{KolbHeinzHydroReview}.

The lower right panel of Fig. \ref{fig:HydroSpectraCompare} compares
the \pT\ spectrum of $\Omega^-$ in central 200 GeV $Au+Au$ collisions to
a similar calculation, adjusted for the higher collision energy and
with a chemical non-equilibration EOS in the hadronic phase
\cite{Kolb:2002ve}. The steeper set of
curves results from decoupling at energy density $e=0.45$ GeV/fm$^3$,
corresponding to kinetic freezeout at the onset of the hadronic
phase. The shallower set of curves results from decoupling at the same
time as the pions at $e=0.075$ GeV/fm$^3$, so that the \Ominus\
receives the full boost from the hadronic phase. (Solid lines are for
zero initial flow, dashed lines are for small but finite radial flow
at the onset of hydrodynamic behavior \cite{Kolb:2002ve}). It has been
proposed that multistrange baryons decouple from the flow much earlier
than non-strange hadrons due to absence of strong resonances with
pions \cite{vanHecke:1998yu}. Evidently, purely partonic hydrodynamic
flow (lower curves) does not generate sufficient transverse velocity
to describe the \Ominus\ spectrum and there is significant
contribution from the hadronic phase \cite{KolbHeinzHydroReview}.

The systematic behavior of flow-related observables is extracted by
fitting measured spectra to the phenomenological Blast Wave
parameterization \cite{Siemens:1979pb,SchnedermannEtAl}, which results
from modeling the system at freezeout as an ensemble of transversely
boosted Boltzmann distributions. The transverse velocity distribution
is a parameterization of the radial dependence of the fluid cell
velocity distribution at freezeout from the full hydrodynamic
calculation (Fig. \ref{fig:HydroFlowVelocity}):
\begin{equation}
\beta_T(r)=\beta_s\left(\frac{r}{R}\right)^n,
\label{eq:BWbetar}
\end{equation}
\noindent
where $R$ is the radius at freezeout and $\beta_s$ is the transverse
flow velocity at the surface. $n=1$ reasonably approximates of
the full calculation. Assuming that kinetic freezeout occurs
instantaneously at all radii, the hadronic spectra are given by
\cite{SchnedermannEtAl}:
\begin{equation}
\frac{dN}{\mT d\mT}\propto{\int_0^R}r\ dr\ {\mT} 
I_0\left(\frac{\pT\sinh\rho}{T}\right)
K_1\left(\frac{\mT\cosh\rho}{T}\right),
\label{eq:hydrospectrum}
\end{equation}
\noindent
where $\mT=\sqrt{\pT^2+m^2}$, $\rho(r)=\tanh^{-1}\beta_T(r)$ is the
transverse rapidity and $T$ is the local
temperature. Non-instantaneous freezeout results in additional terms
in the integrand which can change the spectrum shape significantly
\cite{SchnedermannEtAl}.

\begin{figure}
\centering
\includegraphics[width=.65\textwidth]{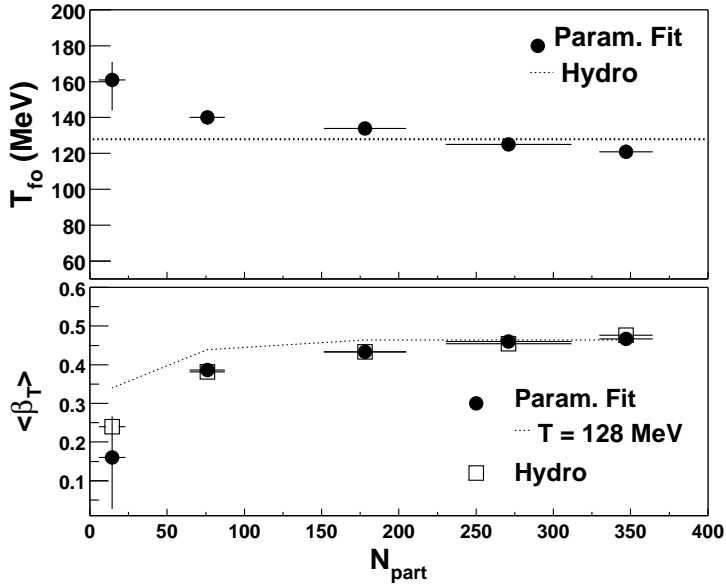}
\caption{Freezeout temperature \Tfo\ and mean flow velocity \betaMean\ for 
centrality-selected 130 GeV $Au+Au$ collisions, from
PHENIX \protect\cite{phenix:pidspectra130}. Filled points are Blast Wave fits to data,
open boxes are from a full hydrodynamic calculation. The dotted line
is for blast wave with centrality-independent \Tfo=128 MeV.
\cite{Kolb:2000fh}. }
\label{fig:Phenix130BetaT}
\end{figure}

Figure \ref{fig:Phenix130BetaT} shows the collision centrality
dependence of the freezeout temperature \Tfo\ and mean flow velocity
\betaMean\ for a Blast Wave fit to the measured $\pi,K$ and $p$ spectra for
130 GeV $Au+Au$ collisions \protect\cite{phenix:pidspectra130}, and
for a full hydrodynamic calculation similar to that in Fig
\ref{fig:HydroSpectraCompare}
\cite{Kolb:2000fh}. The dashed line shows \betaMean\ for fixed
\Tfo=128 MeV, the value used in the full calculation, allowing a
direct comparison of the parameterization with the theory. The
agreement between the Blast Wave parameterization and the full
calculation is good for more central collisions, indicating that the
Blast Wave formulation contains the essential freezeout features of
the hydrodynamic calculation. All approaches show lower
\Tfo\ and higher \betaMean\ for more central collisions, indicating a
longer expansion time until freezeout. Marked deviations are seen only
for the most peripheral collisions, where the region of local
thermalization, if any, may be small and
short-lived. Figs. \ref{fig:HydroSpectraCompare} and
\ref{fig:Phenix130BetaT} suggest that hydrodynamic flow dominates 
the fireball evolution for all but the most peripheral collisions.

\subsection{Anisotropic Flow}
\label{sect:EllipticFlow}

\begin{figure}
\includegraphics[width=.45\textwidth]{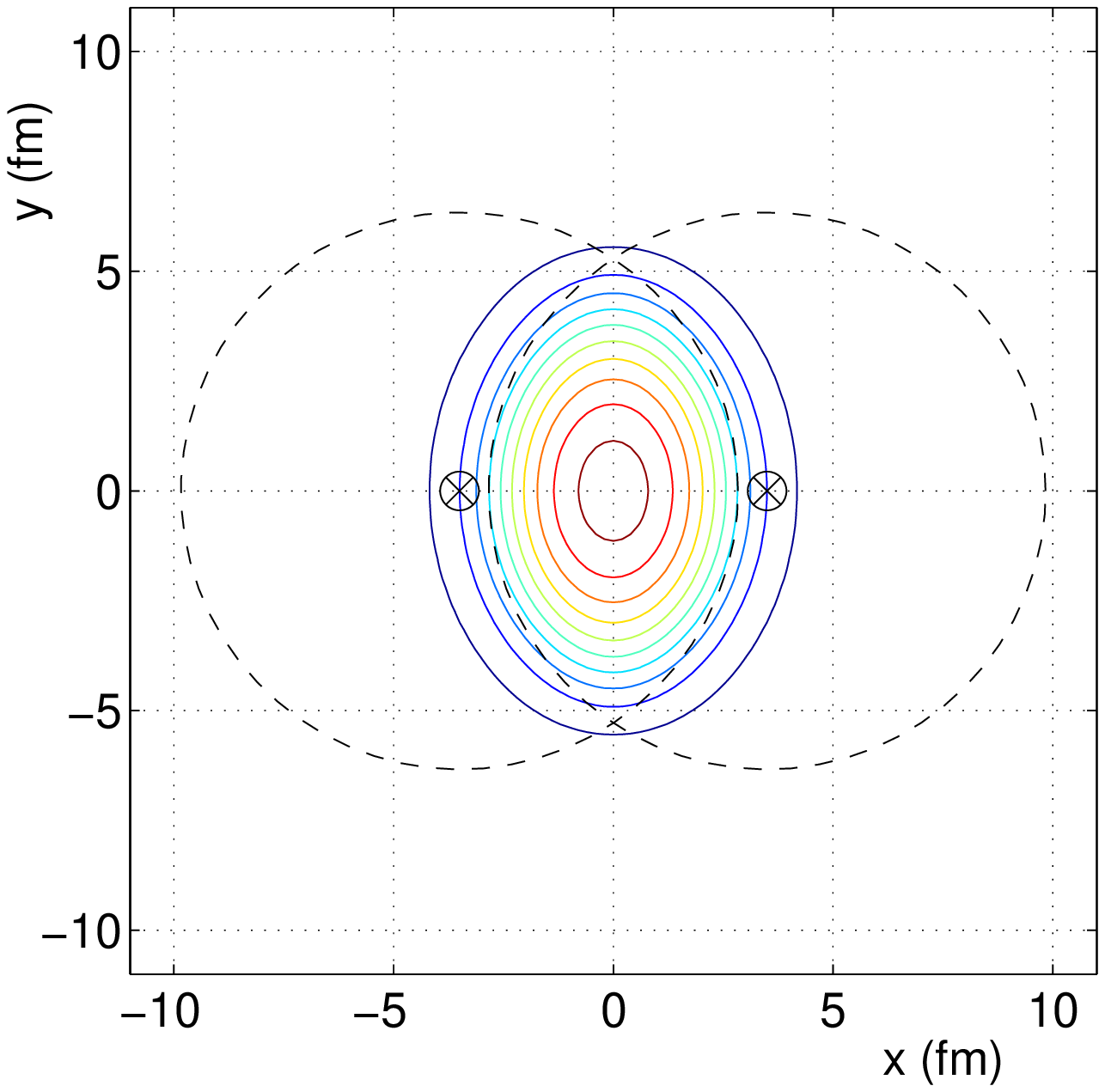}
\includegraphics[width=.54\textwidth]{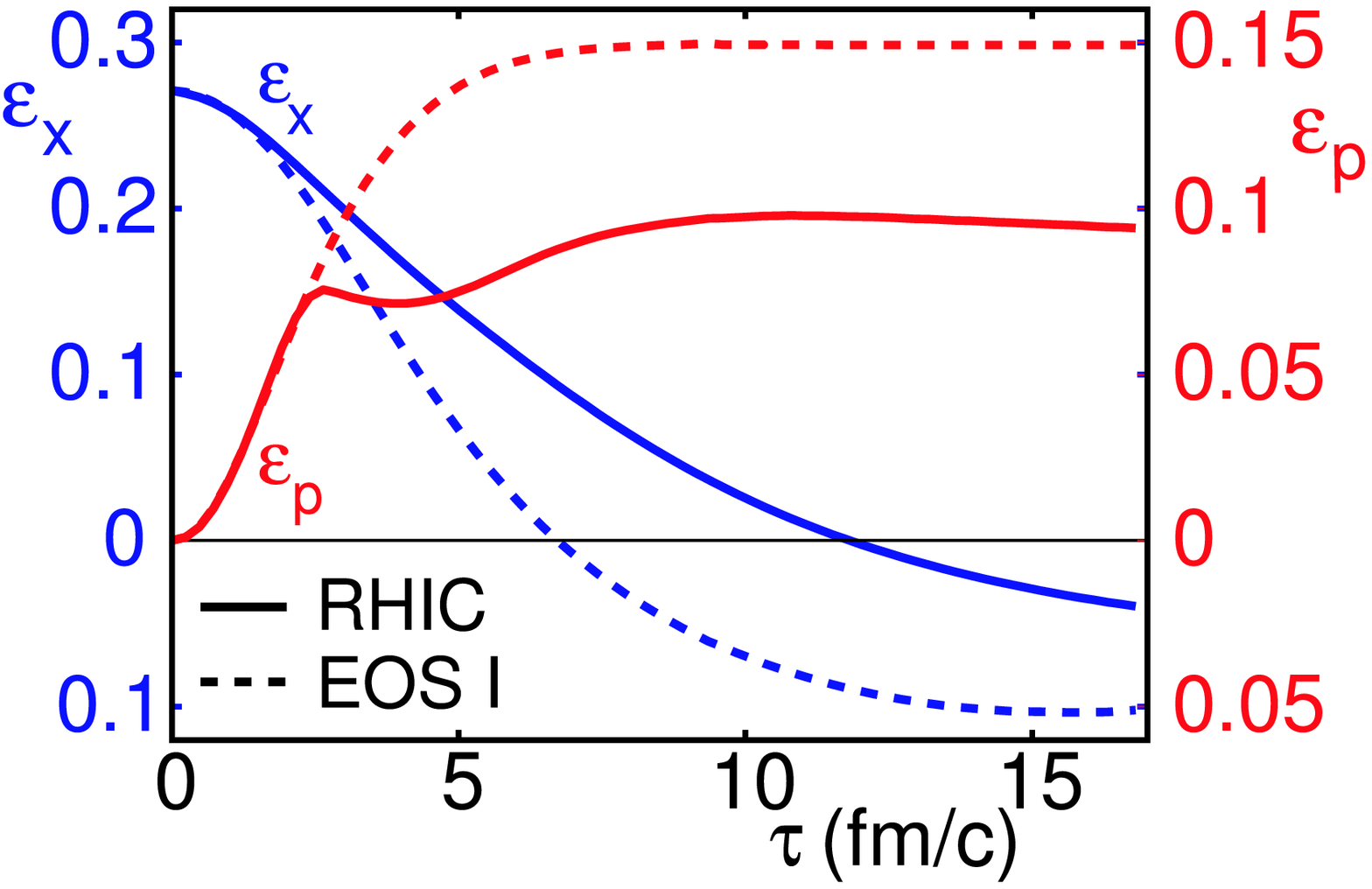}
\caption{
Azimuthal spatial and momentum anisotropy for non-central (b=7 fm) $Au+Au$ 
collisions.
Left: density of binary collisions in the transverse plane
\protect\cite{KolbHeinzHydroReview}. Right: time evolution of spatial
eccentricity $\epsilon_x$ and momentum anisotropy $\epsilon_p$
\protect\cite{Kolb:2002cq}. Solid lines are for EOS with phase transition, dashed
line for massless ideal gas at very high temperature.}
\label{fig:HydroEps}
\end{figure}

In non-central nuclear collisions the overlap region is
azimuthally anisotropic.
Fig. \ref{fig:HydroEps}, left panel, shows the density of
binary collisions in the transverse plane, a common basis for
calculating the initial energy density, for $Au+Au$ interactions with impact parameter b=7 fm
\cite{KolbHeinzHydroReview}. For hydrodynamic evolution,
the initial spatial anisotropy will generate azimuthally anisotropic
pressure gradients that are stronger in the reaction plane
(horizontal in the figure) than perpendicular to it. This
will result in a spatially anisotropic momentum distribution which is
experimentally observable \cite{PoskanzerVoloshin}.

At mid-rapidity the odd harmonics vanish by symmetry and the leading
anisotropy is elliptical. The initial spatial eccentricity is defined
as
\begin{equation}
\epsx=\frac{\left<y^2-x^2\right>}{\left<y^2+x^2\right>} \; ,
\label{eq:HydroEpsx}
\end{equation}
\noindent
where $\left<\ldots\right>$ indicates the average weighted, for
instance, by energy density. The corresponding momentum space
anisotropy in the hydrodynamic framework is
\begin{equation}
\epsp(\tau)=\frac{\int{dx}{dy}(T^{xx}-T^{yy})}{\int{dx}{dy}(T^{xx}+T^{yy})}.
\label{eq:HydroEpsp}
\end{equation}
\noindent
This momentum space anisotropy (``elliptic flow'') results from
interactions within the medium and therefore develops over time as the
fireball evolves. Figure \ref{fig:HydroEps}, right panel, shows the
time evolution of both \epsx\ and \epsp\ for b=7 fm $Au+Au$
collisions. The solid lines are from a hydrodynamic calculation with a
first order phase transition, similar to that in
Fig. \ref{fig:HydroSpectraCompare} \cite{Kolb:2002cq}. The spatial
eccentricity \epsx\ is large at the onset of hydrodynamic flow
($\sim0.27$) but decreases continuously with time. The pressure is
sufficient to drive \epsx\ negative prior to freezeout. In contrast,
\epsp\ grows rapidly but saturates at $\tau\sim6$ fm/c due to the low
pressure in the mixed phase. A small increase in \epsp\ is generated
in the hadronic phase, but almost all of the final momentum asymmetry is
generated in the partonic phase at early time. The early buildup of
momentum anisotropy is seen not only in hydrodynamic calculations but
also in kinetic transport models \cite{Sorge:1997pc,Zhang:1999rs}.
Elliptic flow is thus a key observable of collective hydrodynamic
behavior and thereby thermalization at RHIC. It is predominantly
generated early in the evolution and is potentially sensitive to the
properties of the partonic stage of the collision.

Experimentally, flow is measured by fitting the triple differential
invariant momentum distribution in a Fourier series in azimuthal angle
\cite{PoskanzerVoloshin} (see also \cite{Ollitrault:1992bk}):
\begin{equation}
E\frac{d^3N}{dp^3}=\frac{1}{2\pi}\frac{d^2N}{{\pT}d{\pT}dy}
\left(1+{\sum_{n=1}^\infty}2v_n\mathrm{cos}\left[n\left(\phi-\Psi_r\right)\right]\right),
\label{eq:Flow}
\end{equation}
where $\Psi_r$ represents the orientation of the reaction plane in the
event. The Fourier coefficient $v_n$ measures the asymmetry of order
$n$. By symmetry, $v_1=0$ at midrapidity and the leading term is
elliptic flow \vtwo. $\Psi_r$ is of course not directly measurable,
but it can be estimated based on the azimuthal distribution of all
measured particles in the event. However, since the multiplicity is
finite, the resulting flow coefficients $\tilde{v}_n$ must be
corrected for the finite resolution of the estimated reaction plane
orientation
\cite{PoskanzerVoloshin}.

\begin{figure}
\includegraphics[width=.45\textwidth]{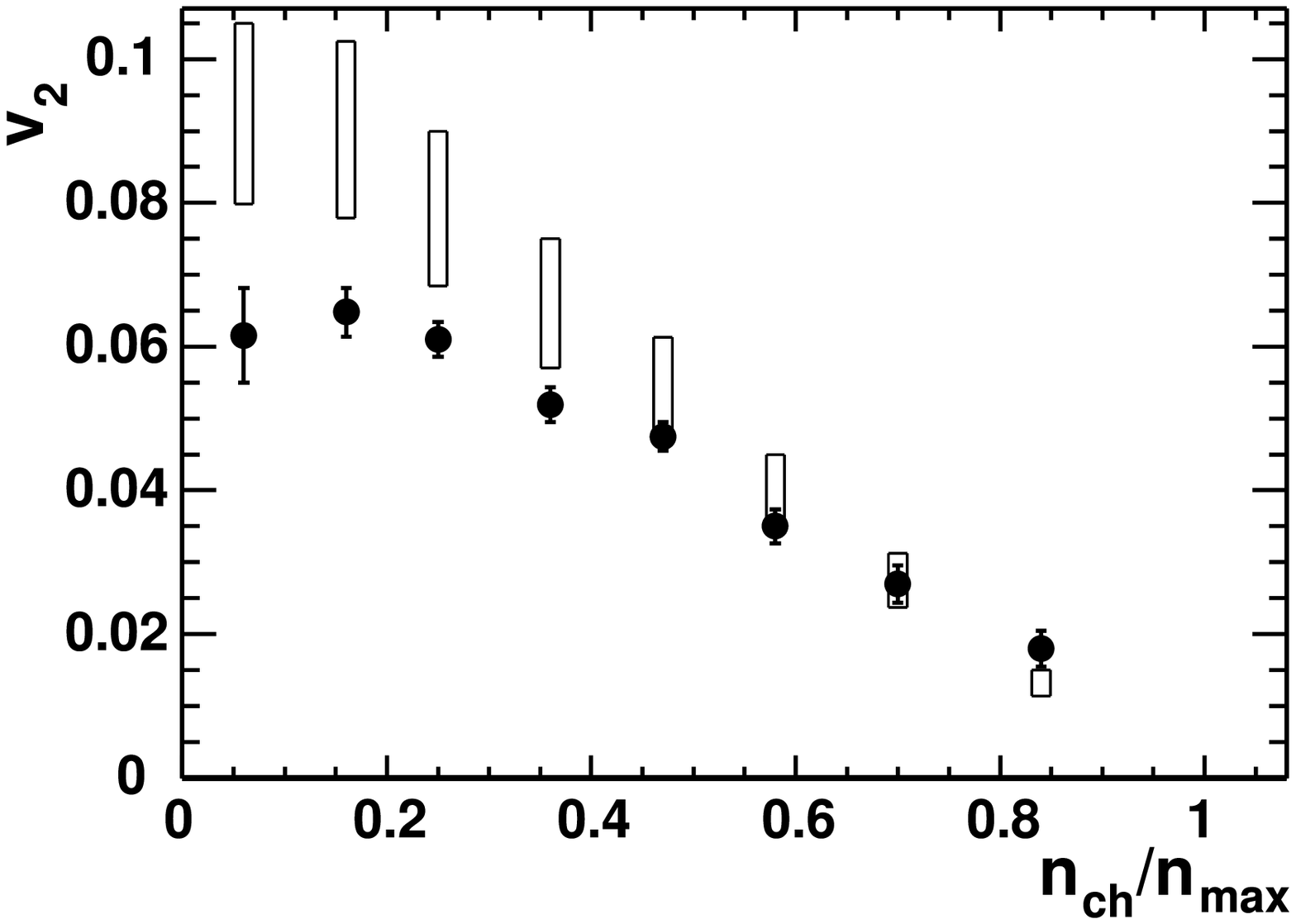}
\includegraphics[width=.45\textwidth]{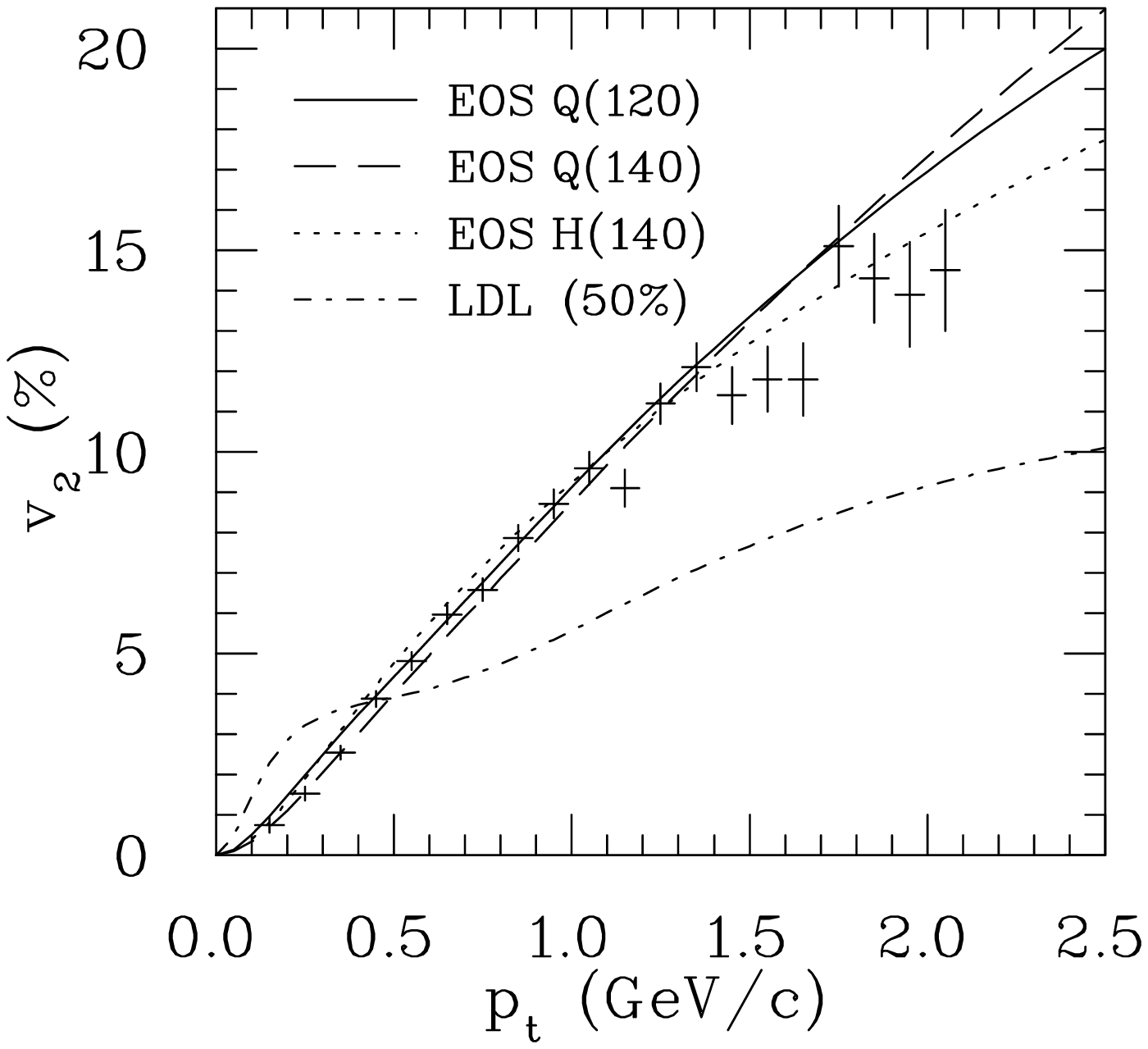}
\caption{
Elliptic flow of charged hadrons from 130 GeV $Au+Au$
collisions, from STAR
\protect\cite{star:flow130}. Left: centrality dependence of \pT-integrated \vtwo. 
Central events correspond to
$n_{ch}/n_{max}\sim1$.  Boxes indicate range of expected values from
hydrodynamic calculations \protect\cite{star:flow130}.  Right: \pT\ dependence
of \vtwo\ for minimum bias collisions, calculations from
\protect\cite{Kolb:2000fh}. }
\label{fig:v2charged}
\end{figure}

Fig. \ref{fig:v2charged} compares \vtwo\ for charged hadrons from 130
GeV $Au+Au$ collisions \cite{star:flow130} with hydrodynamic
calculations. The left panel shows the centrality dependence of \vtwo\
integrated over transverse momentum. Good agreement of the
expectations from hydrodynamics with the data is seen for
$n_{ch}/n_{max}\gt0.5$, corresponding to $b\lt\sim7$ fm. The right
panel shows the \pT\ dependence of \vtwo\ for minimum bias
collisions. Since multiplicity is largest in central collisions and
the anisotropy is largest in peripheral collisions, the greatest
contribution to this measurement comes from intermediate impact
parameters which have both large \vtwo\ and significant
multiplicity. The distribution is compared to hydrodynamic
calculations with various equations of state \cite{Kolb:2000fh}. The
agreement of calculations with data is impressive, though it is
notable that expectations for unidentified charged hadron flow are
very similar for an EOS with a first order phase transition (EOS Q,
$T_{fo}=$120 and 140 MeV) and a purely hadronic resonance gas (EOS
H). The only calculation that is excluded is the low-density limit
(LDL) from a non-hydrodynamic kinetic transport approach
\cite{Kolb:2000fh}.

\begin{figure}
\includegraphics[width=.48\textwidth]{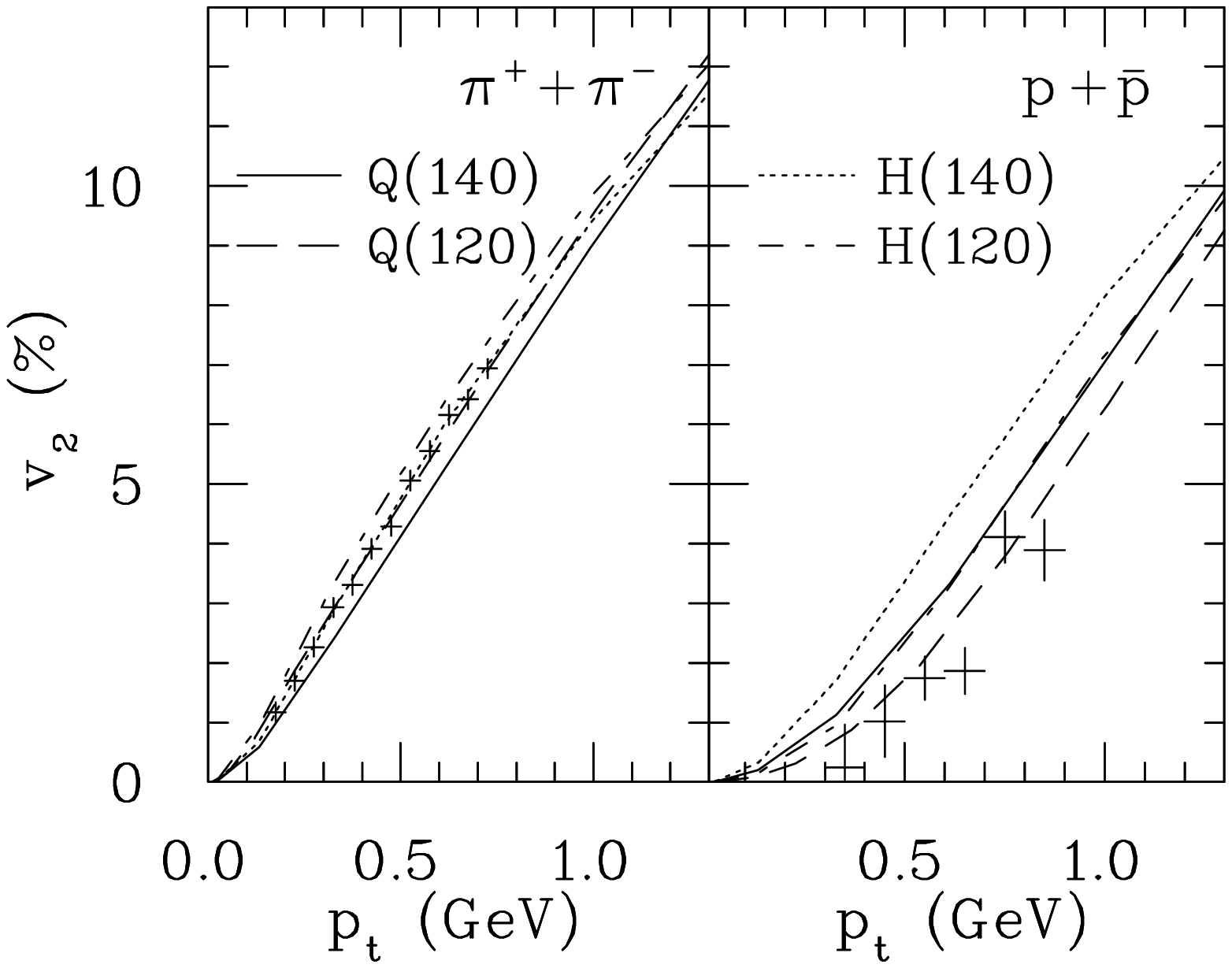}
\hspace{0.1in}
\includegraphics[width=.48\textwidth]{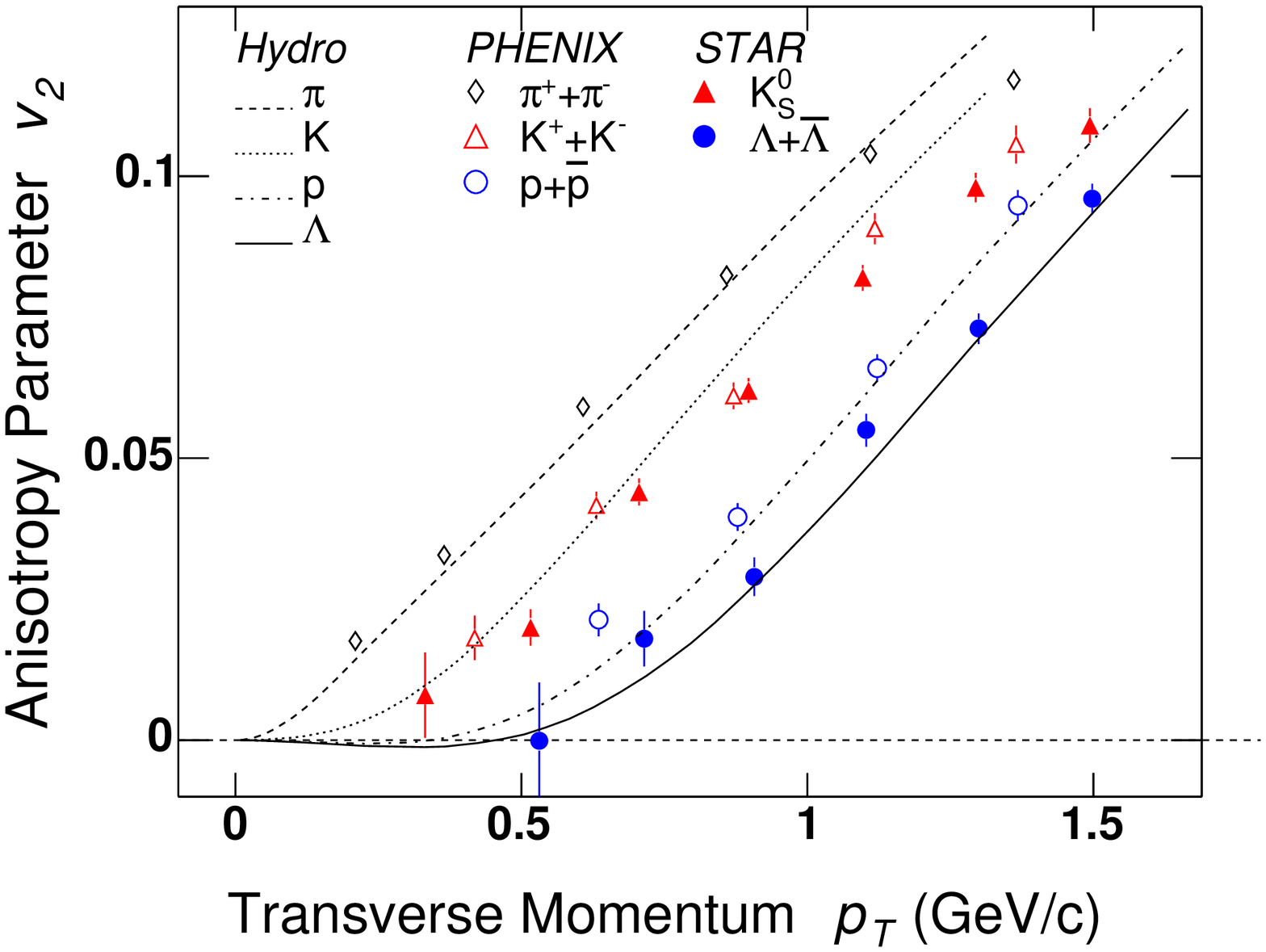}
\caption{\vtwo(\pT) for identified hadrons from STAR \protect\cite{star:pidflow130,star:highpTLamK200} 
compared to hydrodynamic calculations from \protect\cite{Huovinen:2001cy}. 
Left: $\pi^\pm$ and p+\pbar\ for 130 GeV
minimum bias $Au+Au$, figure from \protect\cite{Huovinen:2003fa}. 
Right: $\pi$, K, p and $\Lambda$ from 200 Gev minimum bias 
$Au+Au$ as compiled in \protect\cite{Sorensen:2003kp}.}
\label{fig:v2pid}
\end{figure}

Fig. \ref{fig:v2pid} shows \vtwo(\pT) separately for various
identified particles from STAR
\cite{star:pidflow130}. Similar to the azimuthally averaged
\pT\ distributions in Figs. \ref{fig:PIDSpectra} and
\ref{fig:HydroSpectraCompare},
\vtwo(\pT) also exhibits a mass dependence. Below 2 GeV/$c$, 
proton \vtwo\ is significantly smaller 
than pion \vtwo\ for a given value of \pT. In hydrodynamic terms, the
origin of this effect is the same as the flattening of the inclusive
\pT\ spectrum for higher mass: the velocity boost depletes the 
low \pT\ region in favor of higher \pT\ \cite{Huovinen:2001cy}. This
systematic trend is seen explicitly in the right panel, where
\vtwo\ for different identified hadrons splits according to
the mass, consistent with hydrodynamic predictions
\cite{Huovinen:2001cy}.  At higher $p_T$, the hadronic species
dependence of \vtwo\ changes due to breakdown of the hydrodynamic
model at high $p_T$, a point to which we will return later when
discussing hard probes.

PHOBOS has reported \pT-integrated elliptic flow of charged particles
over very broad phase space ($|\eta|\lt5$)
\cite{phobos:flow130,Back:2004zg}, showing a rapid decrease in integrated \vtwo\ 
away from mid-rapidity. A fully three-dimensional hydrodynamic
calculation \cite{Hirano:2001eu} agrees with the measurements at
mid-rapidity but disagrees significantly at high $\eta$, predicting
only weak variation of \vtwo\ out to $\eta\sim4$. Hydrodynamic
behaviour is thus limited to low \pT\ hadrons ($\lt\sim2$ GeV/c,
depending on particle species) at mid-rapidity for more central
collisions. Outside of these limits the system size and lifetime may
be too small for full thermalization to develop
\cite{KolbHeinzHydroReview}. In contrast, a microscopic calculation based on a 
string model incorporating string excitations and hadonic rescattering
can broadly describe the full phase space distribution of
\pT-integrated elliptic flow \cite{Zabrodin:2001rz}, though detailed
comparison of such an approach to \pT-differential flow and its mass
dependence has not yet been carried out.

The left panels of Fig. \ref{fig:v2pid} also show hydrodynamic
calculations for two EOS (H is a hadronic resonance gas, Q contains a
first order phase transition) and freezeout temperatures 120 and 140
MeV. Pion \vtwo\ exhibits little sensitivity to the EOS or freezeout
temperature, whereas proton \vtwo\ favors EOS Q(120), with a phase
transition and the longest evolution. In
\cite{Teaney:2001av,KolbHeinzHydroReview} it was concluded that a
hadronic resonance gas cannot reproduce the mass dependence of
elliptic flow, which therefore requires a partonic phase. While the
difference between the plasma and purely hadronic scenarios in
Fig. \ref{fig:v2pid} is modest and clear discrimination cannot be made
based on these data, it is nevertheless evident that elliptic flow
does have sensitivity to the EOS in the early stage of the
collision. Detailed model comparisons to the higher precision data in
the right panel and future data, especially multi-strange baryon and
D-meson \vtwo, will sharpen the arguments considerably and may provide
significant constraints on the equation of state.

\subsection{Statistical Distribution of Hadron Yields}
\label{sect:StatModel}

The previous sections discussed the effect of early equilibration on
expansion dynamics and its reflection in the systematic behavior of
transverse momentum spectra and their anisotropies. Equilibration of
the fireball may also be evident in the \pT-integrated
hadronic yields, which should be statistically distributed according
to the thermodynamic conditions at chemical freezeout
\cite{Becattini:1998ii,Becattini:2000jw,Braun-Munzinger:2001ip,Magestro:2001jz,Braun-Munzinger:2003zd}. 
Measured yields at midrapidity from nuclear collisions at RHIC are
available for a wide variety of hadron species. In this section we
compare relative particle abundances to expectations from a
statistical model.

The statistical description of a system with many degrees of freedom
is formulated using the {\it grand canonical} (GC) ensemble, in which
conservation laws are enforced on the average via chemical potentials
$\mu$. The GC partition function for a hadron resonance gas at
temperature $T$ in volume $V$ is\cite{Braun-Munzinger:2003zd}:
\begin{equation}
{\log}Z(T,V,\vec{\mu})=\sum_i{\log}Z_i(T,V,\vec{\mu}),
\label{eq:PartFn}
\end{equation}
\noindent
where $i$ sums over all hadrons with masses less than $\sim$2 GeV/c$^2$ and
$\vec{\mu}=(\mu_B,\mu_S,\mu_Q)$ are the chemical potentials for baryon
number, strangeness and charge. For the hadron species carrying baryon
number $B_i$, strangeness $S_i$ and charge $Q_i$,
\begin{equation}
Z_i(T,V,\vec{\mu})=\frac{Vg_i}{2\pi^2}\int_0^\infty\pm{p^2}dp\log(1\pm\lambda_i{e}^{-\beta\epsilon_i}).
\label{eq:PartFnResonance}
\end{equation}
\noindent
Here $+$ is for fermions and $-$ is for bosons, $g_i$ is the
spin-isospin degeneracy, $\beta=1/T$, $\epsilon_i=\sqrt{p^2+m^2_i}$, and
\begin{equation}
\lambda_i(T,\vec{\mu})=\mathrm{exp}\left(\frac{B_i\mu_B+S_i\mu_S+Q_i\mu_Q}{T}\right).
\label{eq:StatLambda}
\end{equation}
\noindent
Repulsive interactions between hadrons and the effects of resonance
decay are taken into account\cite{Braun-Munzinger:2001ip}. The
imposition of local strangeness and charge neutrality means that the
resulting distributions depend only on the temperature $T$ and
baryochemical potential \muB.

\begin{figure}
\centering
\includegraphics[height=.29\textheight]{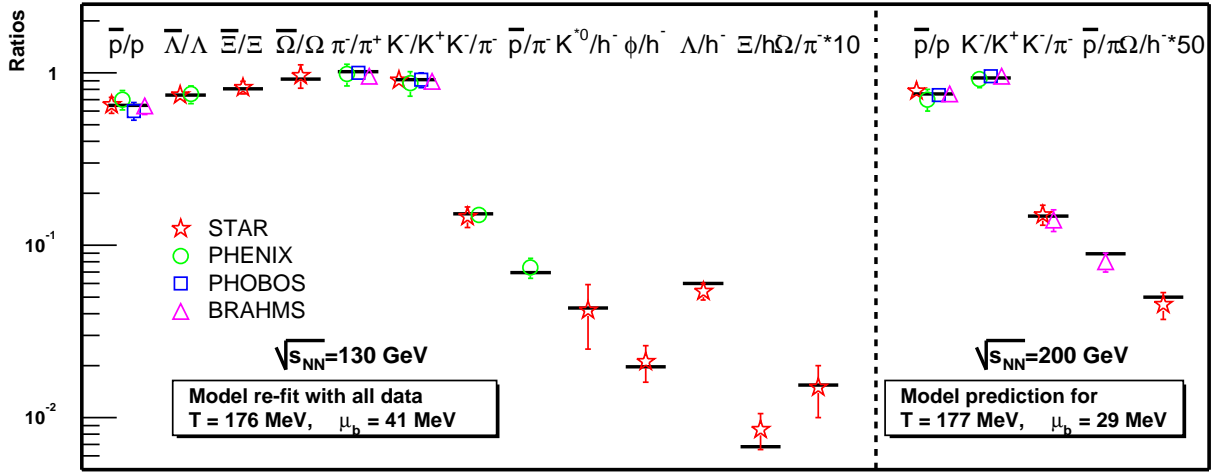}
\caption{Hadronic yield ratios at midrapidity from central 
$Au+Au$ collisions at 130 and 200 GeV. Horizontal bars show the
statistical distributions resulting from fits to the data
\protect\cite{Magestro:2001jz,Braun-Munzinger:2003zd}.}
\label{figYieldRatiosAll}
\end{figure}

Fig. \ref{figYieldRatiosAll} shows the ratios of a wide variety of
hadronic yields measured at midrapidity for central $Au+Au$
collisions. The ratios span three orders of magnitude (note the
scaling of $\Omega/\pi^-$). Also shown are expectations for a
statistically distributed population emitted by an equilibrated medium
with $T=176$ MeV, \muB=41 MeV (\sqrtsNN=130 GeV, left) or $T=177$ MeV,
\muB=29 MeV (200 GeV, right). Agreement between data and model is
good: the hadron population is statistically distributed, consistent
with the model sketched in
Eqs. (\ref{eq:PartFn})-(\ref{eq:StatLambda}). The fitted baryochemical
potential \muB\ is small, indicating low net baryon density in the
medium. The chemical freeze-out temperature $T$ appears to be limited
(i.e. varies little with \sqrts) at a value close to the QCD phase
transition temperature from lattice calculations
(Sec.~\ref{sec:deconfinement}). These phenomena can be accomodated in
a picture in which chemical freeze-out occurs in central nuclear
collisions at the hadronization boundary of a thermalized, deconfined
plasma phase, with the subsequent evolution of the hadronic gas being
moderated by (quasi-)elastic collisions
\cite{Braun-Munzinger:2003zd}. Additional support for this picture is
supplied by the systematics of strangeness production, which may
indicate that strangeness percolates over a much larger volume in
nuclear collisions, as described by a grand canonical ensemble, than in
elementary nucleon-nucleon collisions, which require a canonical
ensemble description with explicit local strangeness conservation
\cite{Braun-Munzinger:2003zd}.

Does the observation of a statistically distributed population of final state
hadrons {\it require} a chemically equilibrated source? A statistically distributed
population may result simply from {\it phase space dominance}
\cite{Koch:2002uq}. Surprisingly, hadron populations are found to be 
statistically distributed in \pbarp\ and even \ePluseMinus\ collisions
\cite{Becattini:1996xt,Becattini:1997rv}. For a final state comprising many particles, a multiplicity
measurement corresponds to integration over a large phase space volume
and details of the matrix element for producing any specific state are
unimportant. If the phase space-averaged matrix elements obey general
scaling rules and do not exhibit strong correlations or dependence on
\sqrts, then the resulting hadron distributions will populate phase
space statistically
\cite{Koch:2002uq}. Discrimination between phase space dominance,
in which $T$ and \muB\ are simply Lagrange multipliers parameterizing
the phase space, and emission from a thermal system, having physical
parameters temperature $T$ and chemical potential \muB, may be
achievable via multiparticle correlation measurements
\cite{Koch:2002uq,Majumder:2003gr}. Such measurements are 
only now maturing  at RHIC (see \cite{Jeon:2003gk} for 
further discussions). The current experimental data are however
consistent with a picture of hadronic freeze-out from a chemically
equilibrated source.

\subsection{Space-time Evolution}

We have so far concentrated on momentum-space observables to infer the
initial conditions, degree of thermalization, and dynamics of the
expansion. Direct measurements of the space-time evolution of the
fireball provide complementary probes of the system dynamics. For
instance, measurement of a long system lifetime may indicate the
presence of a low pressure phase which has stalled the expansion, as
expected from a first order phase transition with a soft equation of
state in the mixed phase \cite{Rischke:1996em}. 

It was recognized long
ago that intensity interferometry of pairs of identical particles is
sensitive to the geometry of the source, both in astrophysical systems
and in elementary particle collisions (Hanbury Brown-Twiss or ``HBT''
interferometry
\cite{HBT,Goldhaber:1960sf}). For bosons, the
wave-function symmetry results in an enhanced coincidence rate for
pairs having small momentum difference, with the momentum range of the
enhancement varying inversely with the space-time dimensions of the
source. The application of intensity interferometry to high energy
nuclear collisions provides a unique probe of the dynamic properties
of the fireball \cite{Heinz:1999rw}. The correlation function of
identical pions encodes the system geometry and expansion dynamics at
kinetic freezeout, late in the fireball evolution. Its projections
relative to the beam direction and to the pair mean momentum reflect
various aspects of the expansion dynamics and space-time extent of the
source. Fortunately, the large multiplicities generated in nuclear
collisions provide the large pair statistics necessary for detailed
investigation of the multi-dimensional correlation function
\cite{TomasikWiedemannHBT}.

The pair correlation function $C(\mathbf{q},\mathbf{K})$ is a function of
the pair's relative 4-momentum $q=p_1-p_2$ and mean momentum
$K=\frac{1}{2}(p_1+p_2)$, and is related to the Wigner
density of the emitting source $S(x,K)$\cite{TomasikWiedemannHBT}:

\begin{equation}
C(\mathbf{q},\mathbf{K})
\equiv
\frac{d^6N}{d\mathbf{p}_1^3d\mathbf{p}_2^3}/
\left(\frac{d^3N}{d\mathbf{p}_1^3}\frac{d^3N}{d\mathbf{p}_2^3}\right)
\approx1+\frac
{\left|\int{d^4x}S(x,K)e^{iq\cdot{x}}\right|^2}
{\left|\int{d^4x}S(x,K)\right|^2}.
\label{eq:CorrFn}
\end{equation}

\noindent
$S(x,K)$ can be understood as the probability that the source emits a
particle with momentum $K$ from space-time point $x$. Experimentally,
$C(\mathbf{q},\mathbf{K})$ is constructed from the ratio of the
measured pair distribution to a distribution of mixed pairs drawn from
different events \cite{Kopylov:1974th}. The measured correlation
function $C(\mathbf{q},\mathbf{K})$ has structure due to Bose-Einstein
statistics, resulting in enhanced probability for small $|\mathbf{q}|$,
and to final state interactions, which mask the BE enhancement and
which must be disentangled to extract geometric quantities.

Within a Gaussian approximation to the spatial distribution of
$S(x,K)$ \cite{Heinz:1999rw}, the correlation function for central
collisions is characterized by its projections onto the orthogonal
{\it longitudinal, outward} and {\it sideward} directions, which are
parallel to the beam, parallel to $\mathbf{\KThbt}$, and perpendicular
to $\mathbf{\KThbt}$ respectively. $\mathbf{\KThbt}$ is the pair
momentum vector perpendicular to the beam. The conjugate radius
parameters are $R_l$, $R_o$ and $R_s$
\cite{Podgoretsky:1983xu,Pratt:1984su,Bertsch:1988db}:

\begin{equation}
C(\mathbf{q},\mathbf{K})\simeq 1+\lambda
\exp\left(-R_l^2q_l^2-R_o^2q_o^2-R_s^2q_s^2\right).
\label{eq:CorrFnRadii}
\end{equation}

In high energy nuclear collisions the source expands longitudinally
and transversely. In a hydrodynamic picture, a pair of identical pions
with small momentum difference are unlikely to be emitted from
different fluid elements with significantly differing velocity. The
parameters $R$ therefore do not reflect the dimensions of the entire
source, but rather the rms widths of the effective source that emits
particles with momentum \KThbt\ (``regions of homogeneity'')
\cite{Akkelin:1995gh}. The systematic dependence of $R$ on pair
momentum is a rich source of information about the expansion dynamics
of the fireball \cite{Heinz:1999rw}. For an infinite Bjorken
(boost-invariant) source at temperature $T$ the longitudinal radius is
\cite{Makhlin:1988gm}

\begin{equation}
R_l^2(\KThbt)\simeq\tau_0^2\frac{T}{\KThbt},
\label{eq:hbtRl}
\end{equation}

\noindent
though corrections for realistic sources are significant
\cite{Wiedemann:1996au}. An extended lifetime of the system 
may be observable via the combination

\begin{equation}
R_o^2-R_s^2\approx\beta_\perp^2\langle\tilde{t}^2\rangle,
\label{eq:HBTRoRs}
\end{equation}

\noindent
which is sensitive to the duration of particle emission. Here,
$\beta_\perp=\KThbt/K_0$ is the transverse velocity of the pair and
$\langle\tilde{t}^2\rangle=\langle{t}^2\rangle-\langle{t}\rangle^2$ is
the variance of the particle emission time. Hydrodynamic calculations
\cite{Rischke:1996em} indicate that a large value of $R_o^2-R_s^2$, or 
specifically the ratio $R_o/R_s\gg1$, is a rather generic indication of
a very soft equation of state stalling the expansion, which can only
arise from the presence of a mixed phase.

\begin{figure}
\includegraphics[width=.45\textwidth]{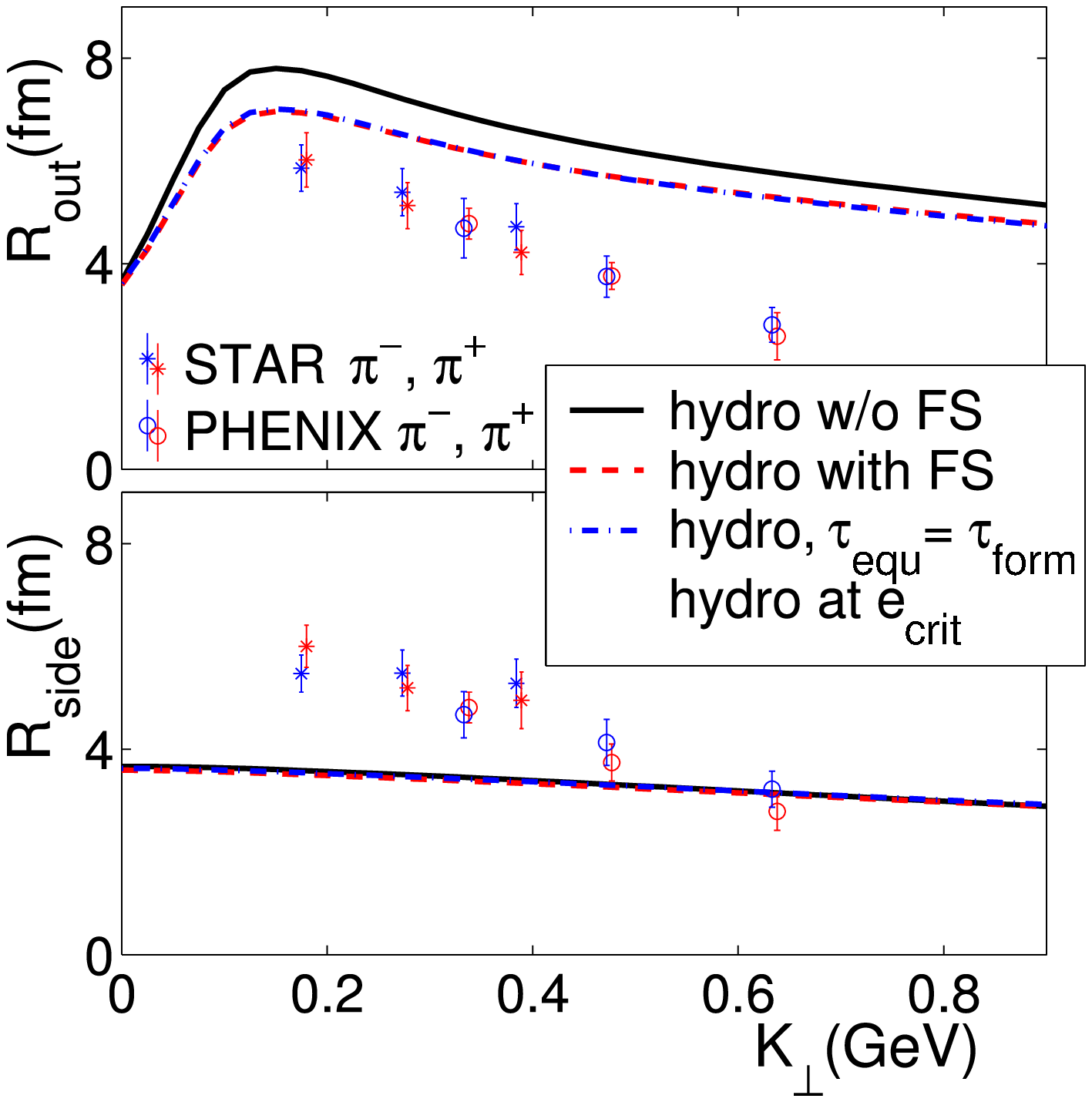}
\includegraphics[width=.52\textwidth]{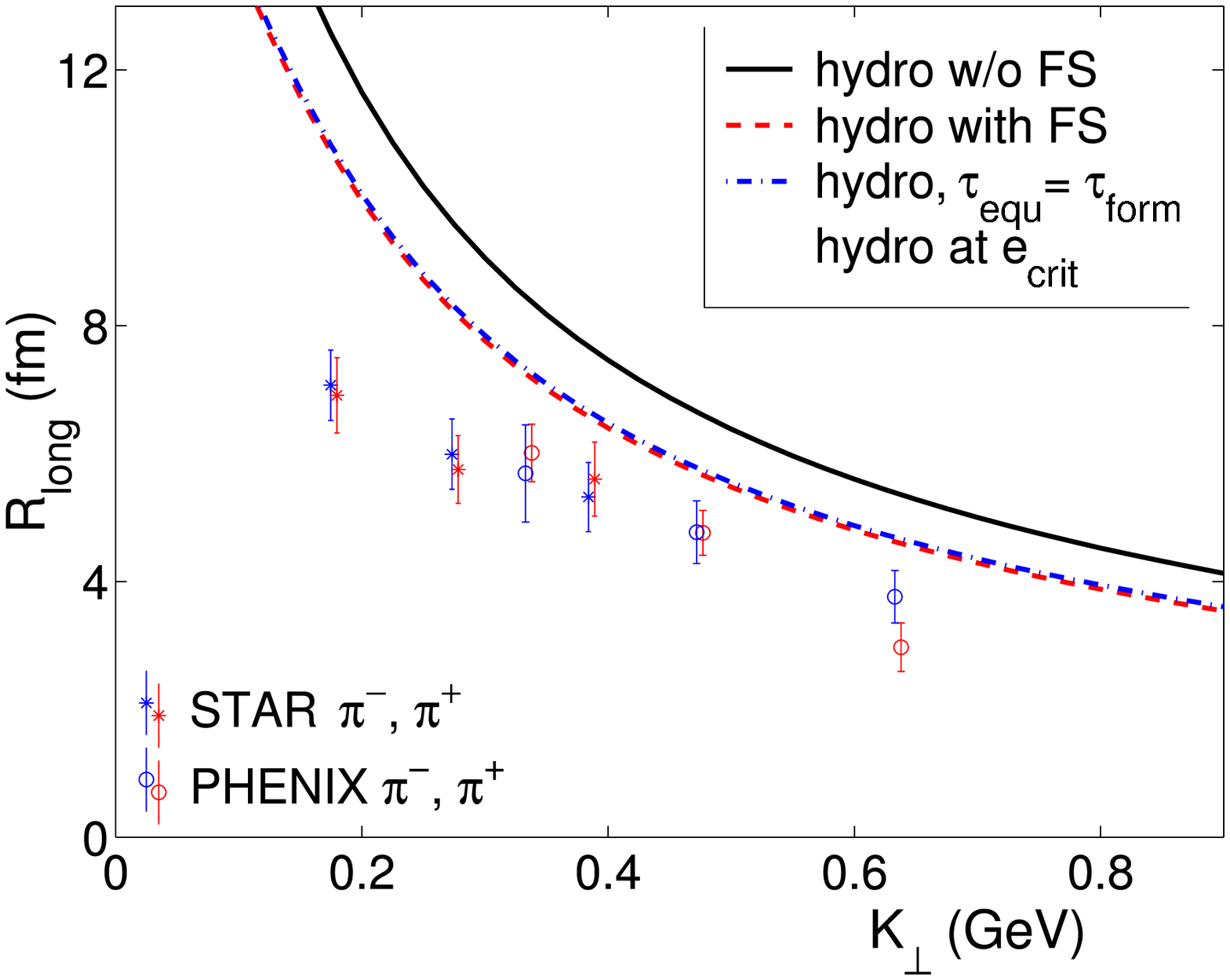}
\caption{Charged pion HBT parameters for 130 GeV central $Au+Au$ collisions from 
STAR \protect\cite{star:HBT130} and PHENIX \protect\cite{phenix:hbt130}, compared to
hydrodynamic calculations \protect\cite{KolbHeinzHydroReview}. Curves are
described in text.}
\label{fig:HBTvsHydro}
\end{figure}

Fig. \ref{fig:HBTvsHydro} shows the measured HBT parameters for
central $Au+Au$ collisions at 130 GeV. The parameters are not large
(\lt8 fm) relative to the radius of a $Au$-nucleus and exhibit
negligible change from measurements with heavy nuclei at much lower
\sqrts\ \cite{star:HBT130}. Most significantly, 
the ratio $R_o/R_s\leq1$ in $0.2\lt\KThbt\lt1.2$ 
GeV/c \cite{star:HBT130,phenix:hbt130,phenix:hbt200},
in contrast to the expectation that $R_o/R_s\sim 1.5$ for a long-lived
source.

The reduction in $R$ with increasing pair \KThbt\ is qualitatively
consistent with expectations from a longitudinally and transversely
expanding source. However, quantitative comparison of the measured
radius parameters to the boost-invariant hydrodynamic calculations in
the figure reveals significant disagreements
\cite{KolbHeinzHydroReview}: neither the magnitude nor the \KThbt\
dependence are well described. Modifications to the calculation such
as earlier freezeout (dotted line), faster buildup of flow at the
partonic stage (short dashed line), or earlier onset of hydrodynamic
behavior (long dashed line) either do not fully rectify the
disagreements or worsen the agreement with inclusive
spectra. Likewise, relieving the boost-invariance condition and
imposing a more realistic treatment of freezeout than the
instantaneous Cooper-Frye algorithm do not fully resolve the
problems. Introduction of finite viscosity generates $R_o/R_s\sim 1$,
but at the expense of significant disagreement with
\vtwo\ measurements \cite{Teaney:2003pb}. A hybrid parton/hadron cascade calculation 
generates $R_o/R_s\sim 1$ \cite{Lin:2002gc}. However, the freezeout in
this calculation occurs earliest at small radii, in contrast to
hydrodynamic calculations where the freezeout surface generically
propagates inward \cite{KolbHeinzHydroReview}.

For non-central collisions, the azimuthal modulation of HBT parameters
relative to the reaction plane orientation provides a more detailed
view of the geometry and dynamics of the source at freezeout
\cite{Wiedemann:1998cr,Heinz:2002au}. Recent measurements of the azimuthal 
dependence of HBT radii \cite{Adams:2003ra} indicate that the source
at freezeout is asymmetric and extends out of the reaction plane
(\epsx\gt0), consistent with a rapid pressure buildup and early
freezeout. This picture should be contrasted with the calculations
shown in Fig. \ref{fig:HydroEps}, right panel, in which hydrodynamic
expansion prior to freezeout lasts long enough to turn the spatial
anisotropy slightly negative (\epsx\lt0, source extended in the
reaction plane).

Reconcilation of HBT radius measurements and hydrodynamic calculations
has not yet been achieved within the currently available theoretical
framework (the ``HBT puzzle'') \cite{KolbHeinzHydroReview}. However,
insofar as the HBT correlations are most sensitive to the system
properties at kinetic freezeout, this may indicate a lack of
understanding of the late expansion stage and ultimate breakup of the
system rather than the dynamics at the earliest, hot and dense
phase. The effect of the source opacity on the correlation function
also remains an open question.

\section{Hard Probes}

The lifetime of the hot and dense matter produced in heavy ion
collisions at RHIC is estimated to be on the order of a few fm/$c$.
Its initial transverse radius is about 6 fm and it undergoes rapid
longitudinal and transverse expansion. Due to the transient nature of
the matter, external probes cannot be used to study its properties.
Fortunately, the dynamical processes that produce the bulk medium also
produce energetic particles through hard processes. These energetic
particles penetrate the bulk matter and reach the detectors as
distinct signals. Study of these energetic particles and their
interaction with the medium, analogous to the method of computed
tomography (CT) in medical science, will yield critical information
about the properties of the matter that is impossible to obtain from
the soft hadrons produced by hadronization of the bulk medium.

Properties of a medium are conventionally studied using the scattering of
particle beams. In deeply inelastic scattering (DIS) experiments, for
example, leptons scatter off a nucleon via photon exchange with
quarks. The response or correlation function of the
electromagnetic currents,
\begin{equation}
W_{\mu\nu}(q)=\frac{1}{4\pi}\int d^4x e^{iq\cdot x}\langle A\mid j^{em}_\mu(0)
j^{em}_\nu(x)\mid A \rangle \; ,
\end{equation}
is a direct measurement of the quark distributions in a nucleon or nucleus,
where 
\linebreak
$j^{em}_\mu(x)=\sum_q e_q\bar{\psi}_q(x)\gamma_\mu\psi_q(x)$ is the
hadronic electromagnetic current.
Such experiments have provided unique information about the partonic
structure of nucleons and nuclei and their 
QCD evolution\cite{Martin:2001es,Pumplin:2002vw}.

The scattering technique is not applicable to the dynamic systems
produced in heavy-ion collisions.  However, it has been shown that the
thermal average of the above correlation function gives the photon
emission rate from the evolving system \cite{McLerran:1985ay}. The
emission rate depends mainly on local temperature or parton density,
while the total yield depends on the entire history of the system
evolution. The properties and dynamics of a strongly interacting
system may therefore be probed via the measurements of photon and
dilepton emission. Additional information is encoded in the resonance
properties and medium modification of the emitted virtual
photons. Screening in a color-deconfined medium leads to dissociation
of bound states, resulting in quarkonium suppression
\cite{Matsui:1986dk}. The color screening arises from the strong
interaction between quarks and gluons at high density and temperature,
which also causes the attenuation of energetic partons propagating
through the medium. Such an effect underlies the phenomenon of jet
quenching and the application of jet tomography to probe the dense matter
generated in high-energy heavy-ion collisions
\cite{Gyulassy:1990ye,Wang:1992xy}.

Charmonium suppression significantly in excess of normal nuclear
absorption has indeed been observed in fixed-target heavy-ion
collisions at the CERN SPS by NA50 \cite{Alexopoulos:2002eh}, leading
to speculation that dense, color deconfined matter has been created in
central $Pb+Pb$ collisions at SPS energy. A direct photon excess over
hadronic sources has been observed at the SPS \cite{Aggarwal:2000th},
and low mass dilepton spectra show signs of medium modifications of
hadron properties \cite{Agakishiev:1995xb}. Reviews of these
results can be found in \cite{Rapp:1999ej,Ko:1997kb,Vogt:1999cu}. At
present, however, the experimental investigation of real and virtual
photon production at RHIC is just beginning. In this section we
concentrate rather on the theory and phenomenology of jet quenching at
RHIC, for which a large body of mature data is available.

Theoretical studies of medium-induced partonic energy loss date back
to an unpublished paper by Bjorken, who calculated the energy loss
due to elastic scattering in a hot medium. A simple estimate is
given by the thermally averaged energy transfer $\nu_{\rm el}\approx
q_\perp^2/2\omega$ of the jet parton to a thermal parton with energy
$\omega$, where $q_\perp$ is the transverse momentum transfer of the
elastic scattering. The resulting elastic energy loss
\cite{Wang:1997yf},
\begin{equation}
\frac{dE_{\rm el}}{dx}=C_2\frac{3\pi\alpha_{\rm s}^2}{2}T^2
\log\left(\frac{3ET}{2\mu^2}\right) \; ,
\end{equation}
is sensitive to the temperature of the medium but is in general small
relative to the radiative energy loss discussed below. Here, $\mu$ is
the Debye screening mass and $C_2$ is the Casimir factor of the
propagating parton in its fundamental representation. Elastic energy
loss can also be calculated within finite temperature QCD field theory
\cite{Thoma:1991fm}, with similar results.

\begin{figure}
\includegraphics[width=.4\textwidth]{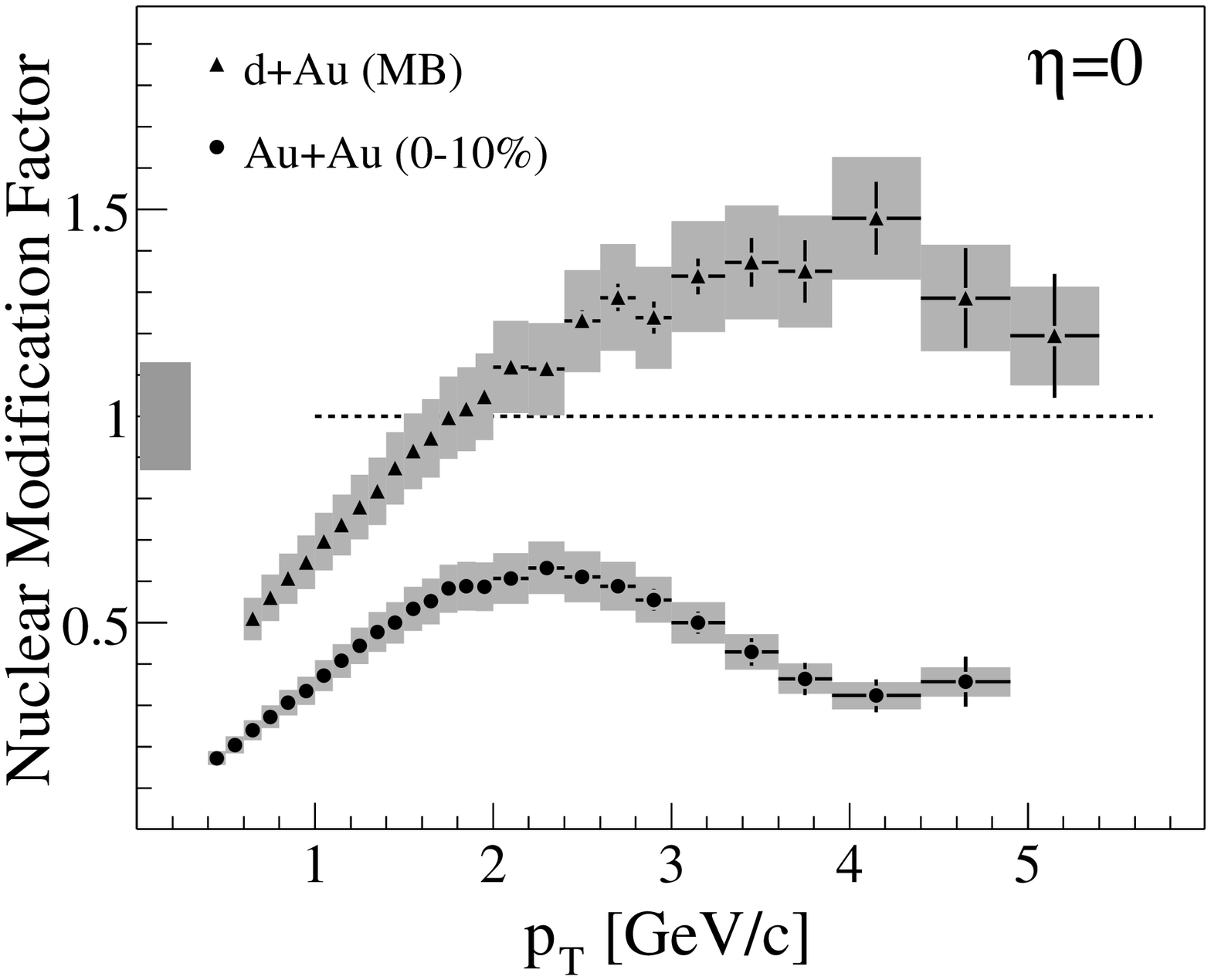}
\includegraphics[width=.48\textwidth]{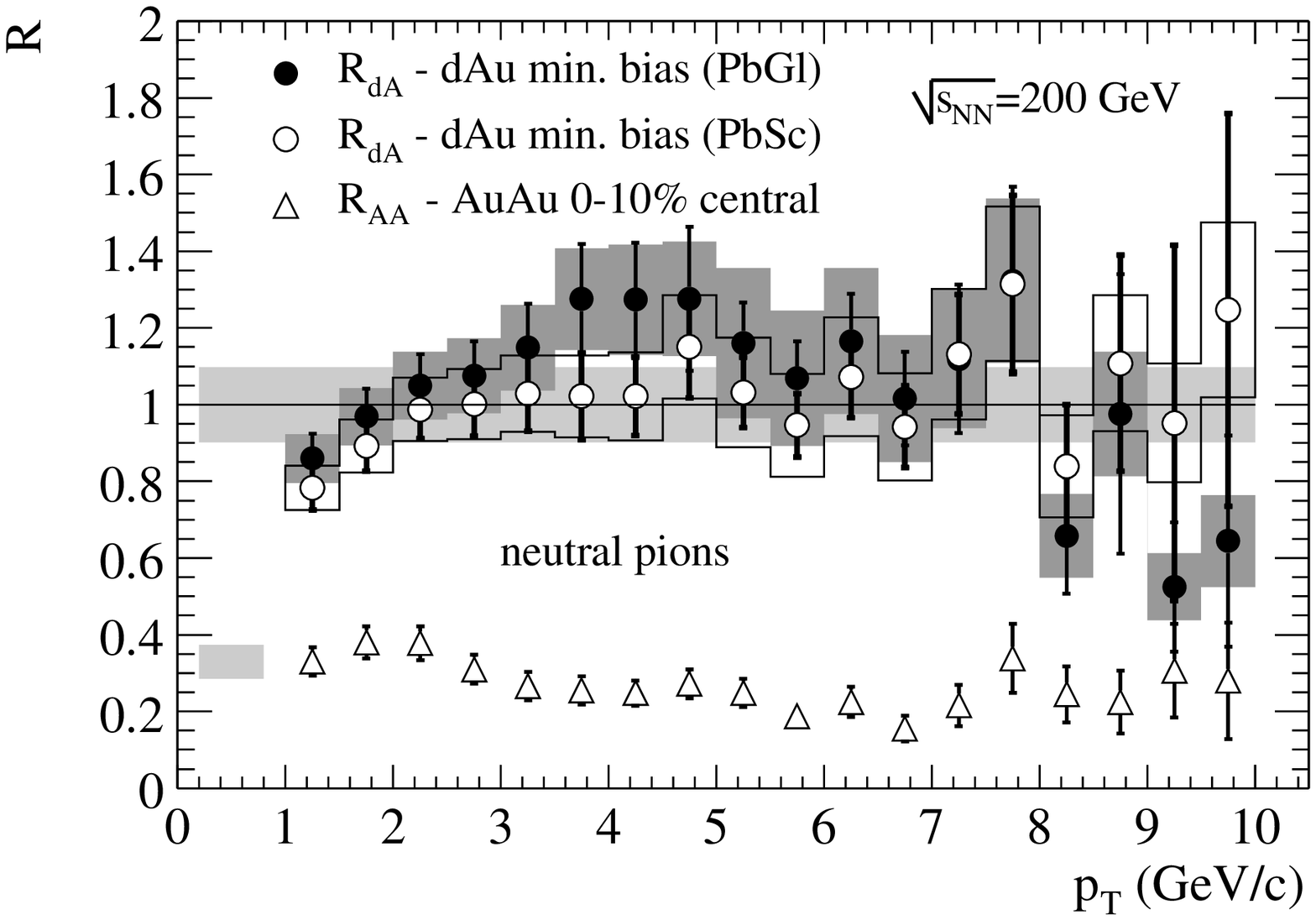}
\includegraphics[width=.4\textwidth]{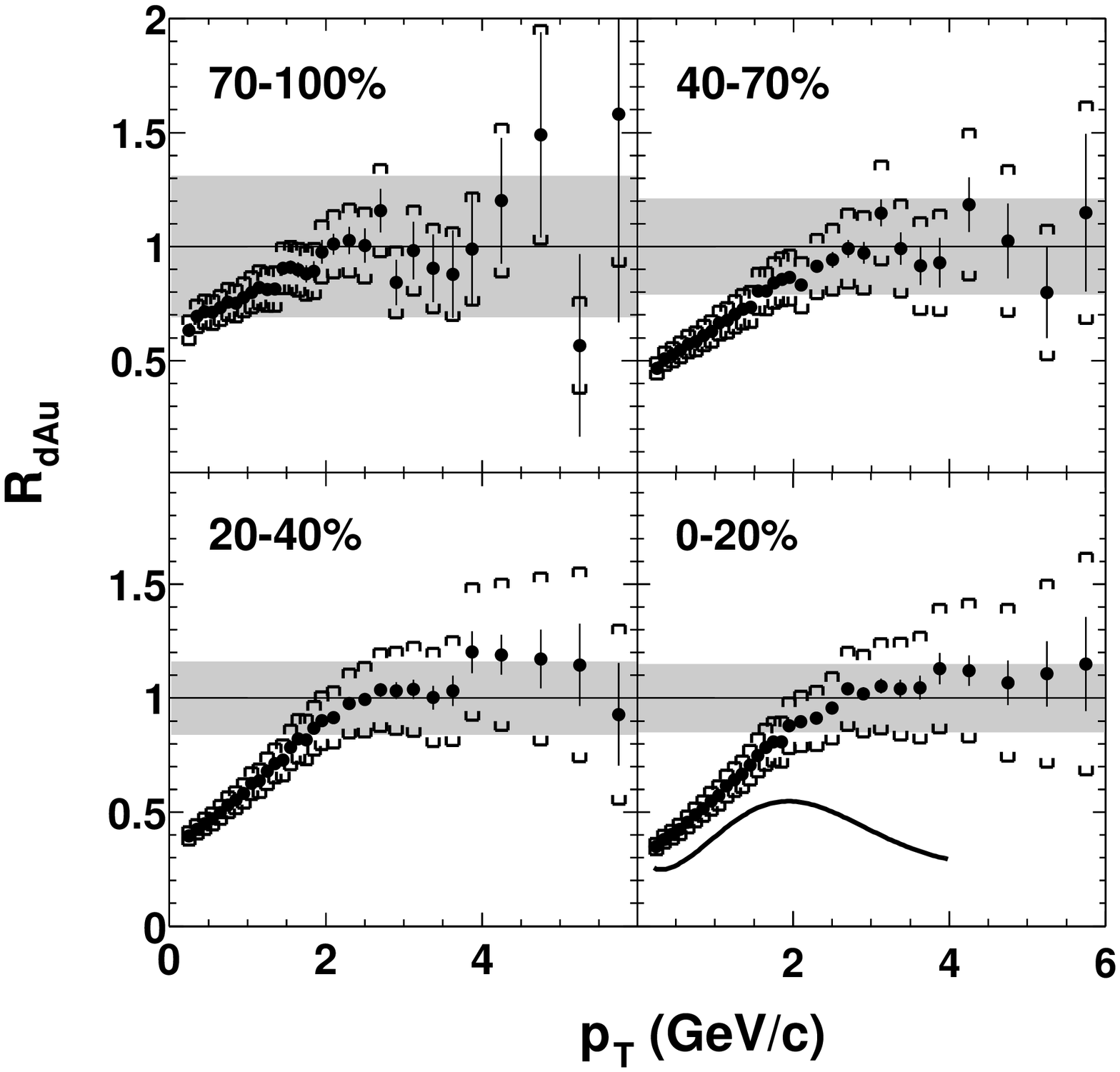}
\includegraphics[width=.48\textwidth]{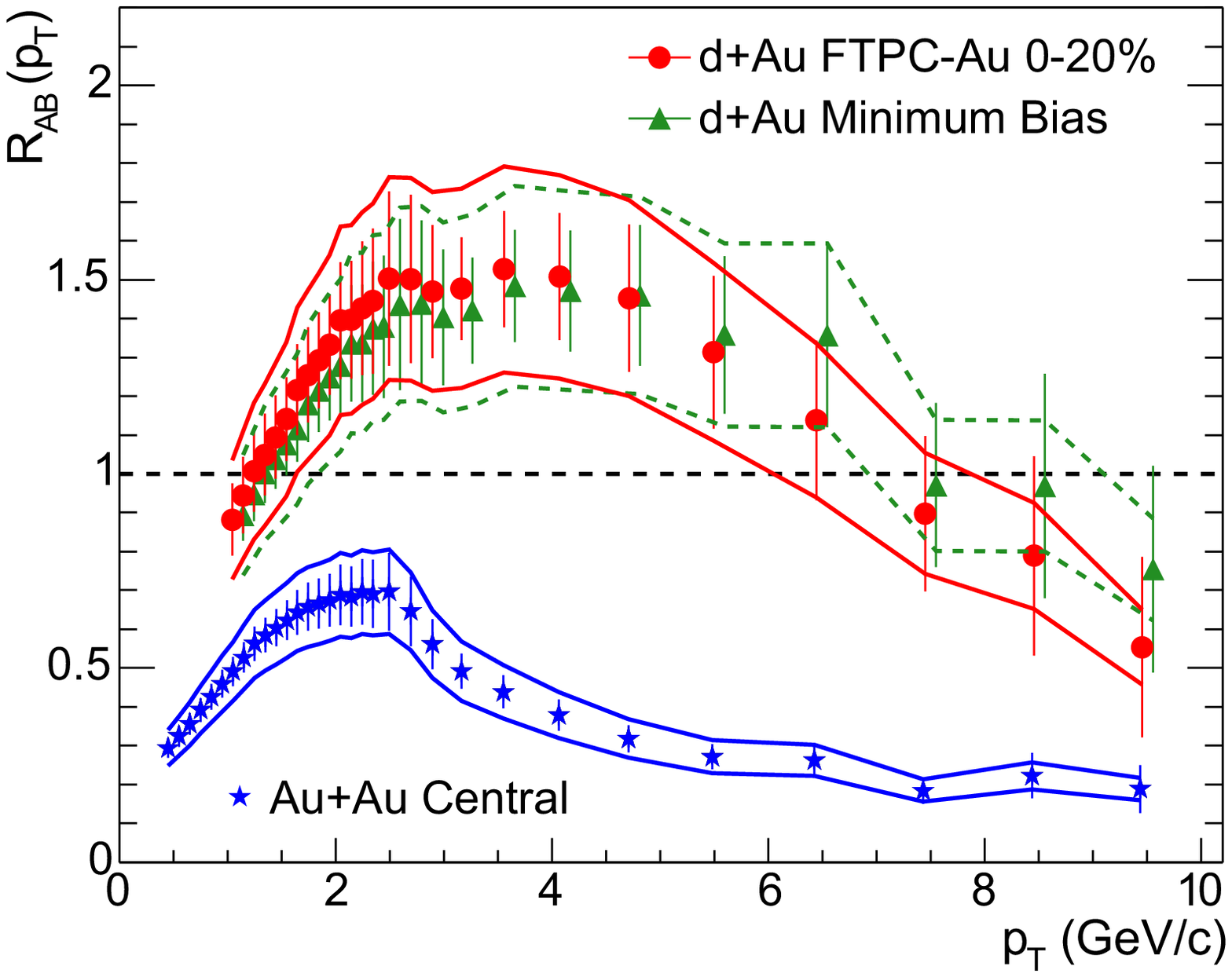}
\caption{Binary collision-scaled ratio of charged hadron and \pizero\ inclusive 
spectra from 
200 GeV $Au+Au$ and $d+Au$ relative to that from $p+p$ collisions, from
BRAHMS\protect\cite{brahms:highpTdAu}(upper left), 
PHENIX\protect\cite{phenix:highpTdAu} (upper right),
PHOBOS\protect\cite{phobos:highpTdAu} (lower left) 
and STAR\protect\cite{star:highpTdAu} (lower right). See Sect.
\ref{sect:InclSupp} for details.}
\label{fig:InclusiveSuppressionPR}
\end{figure}

Radiative partonic energy loss was first estimated using the
uncertainty principle \cite{Brodsky:1993nq}. The first theoretical
study of QCD radiative partonic energy loss, by Gyulassy and Wang
\cite{Gyulassy:1994hr,Wang:1995fx}, modeled multiple parton
scattering using a screened Coulomb potential and found that
Landau-Pomeranchuk-Migdal (LPM) interference effects
\cite{Landau:1953um,Migdal:1956tc} play a crucial role.
Baier {\it et al.} (BDMPS) \cite{Baier:1995bd} later considered gluon
rescattering, which was found to dominate the gluon
radiation induced by multiple scattering in a dense medium.
These initial studies have been followed by
many more recent works on the
subject, including a path integral formulation
\cite{Zakharov:1996fv} and an opacity expansion framework
\cite{Gyulassy:2000fs,Gyulassy:2000er,Wiedemann:2000za} which is
suitable for studying multiple parton scattering in a thin plasma.
The unique feature of radiative energy loss in QCD is its non-linear
dependence on distance, arising from the non-Abelian LPM interference
effect in a QCD medium. In this review, we will take the approach of
twist-expansion \cite{Guo:2000nz,Wang:2001if}, since it connects
naturally the discussions of partonic energy loss in a cold nuclear
medium and hot quark gluon plasma.

\begin{figure}
\includegraphics[width=.48\textwidth]{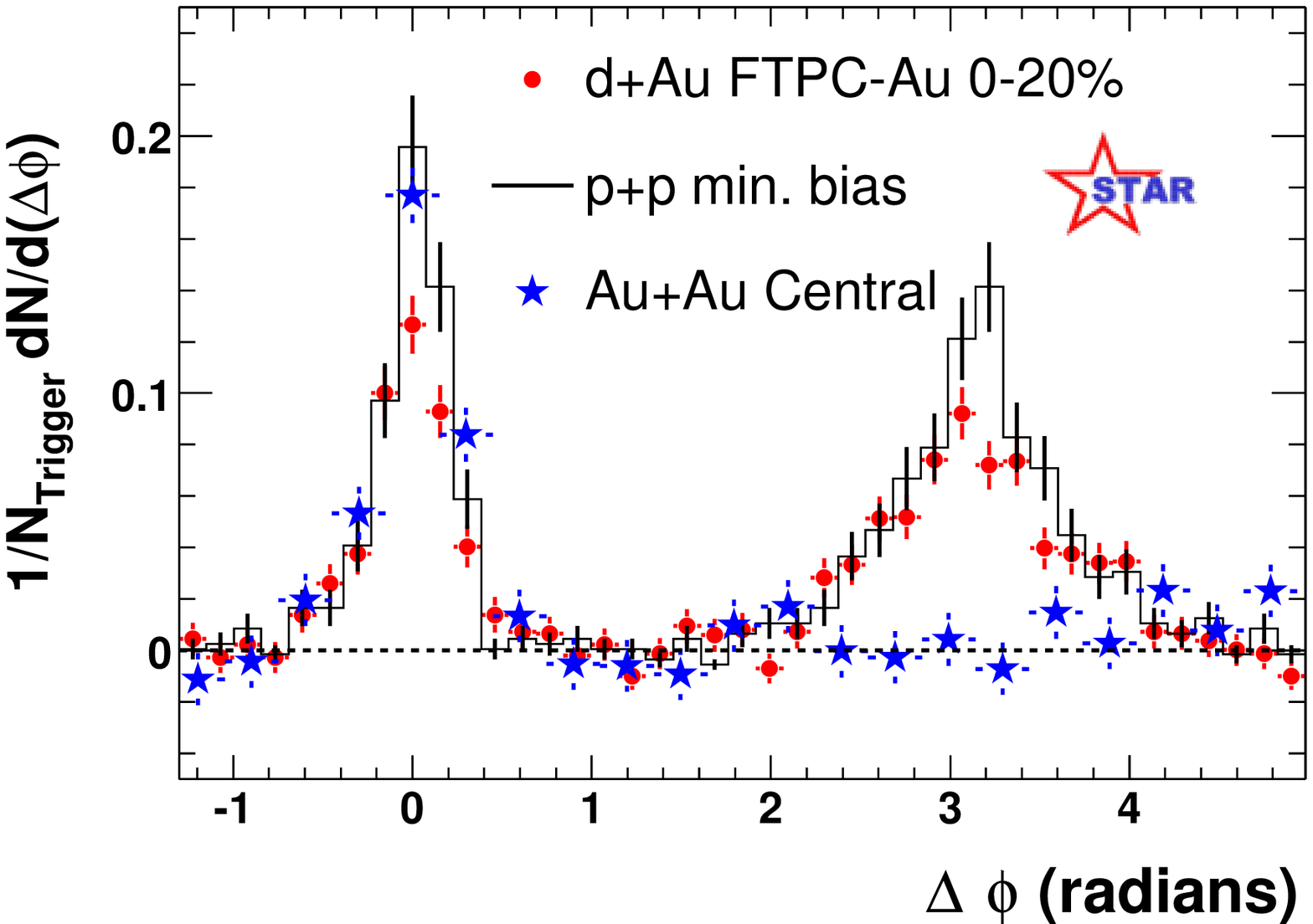}
\includegraphics[width=.48\textwidth]{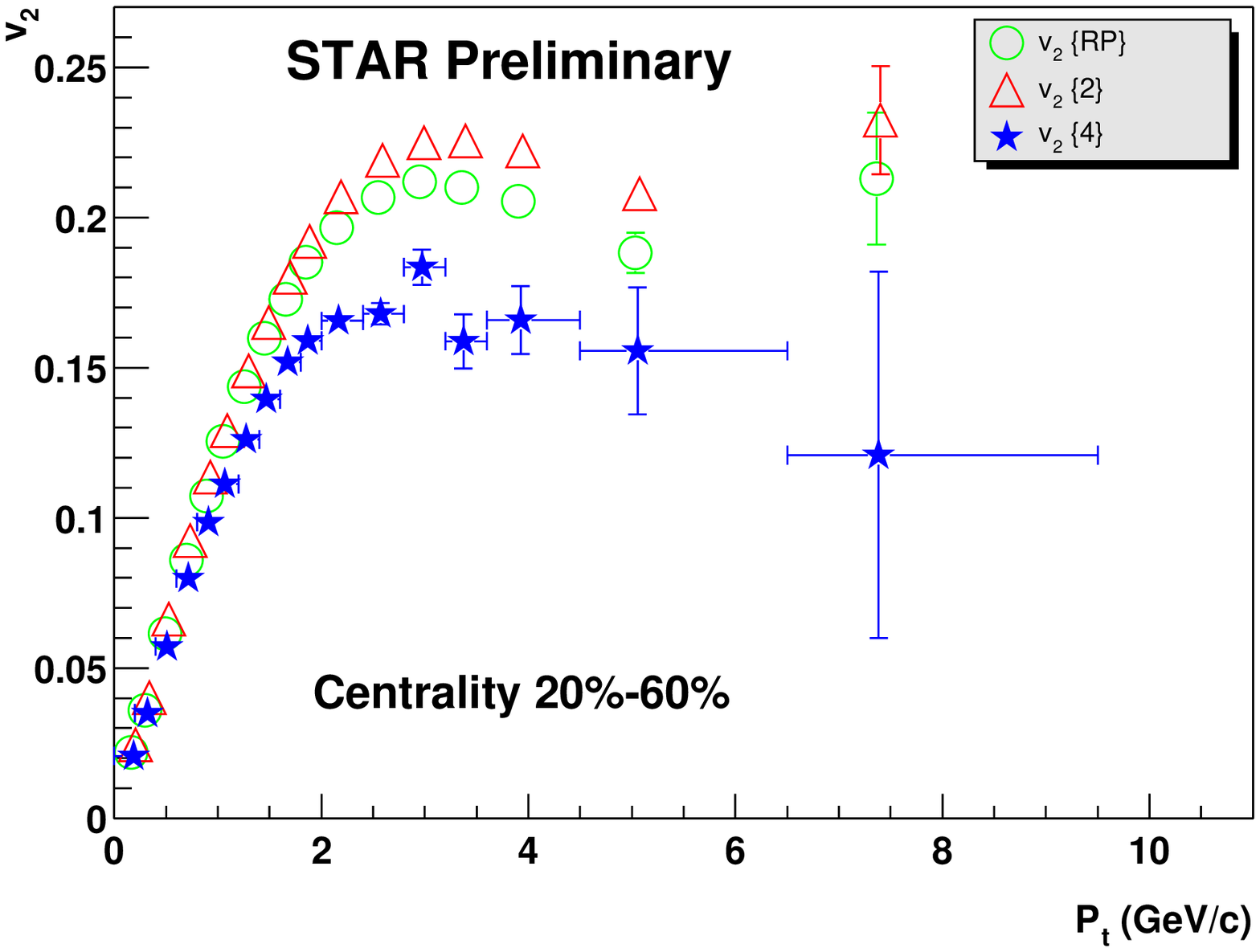}
\caption{Left: Dihadron azimuthal correlations at high \pT\ for $p+p$, 
central $d+Au$ and central $Au+Au$ collisions (background subtracted )
from STAR \protect\cite{star:highpTbtob,star:highpTdAu}. Right:
azimuthal correlation strength with the reaction plane (\vtwo,
Eq. (\ref{eq:Flow})) of high \pT\ hadrons in non-central $Au+Au$
collisions from STAR \protect\cite{Tang:2003kz}. Different symbols
correspond to different methods for calculating \vtwo\
(reaction-plane, 2- and 4-particle cumulant; see
Sect. \ref{sect:HighpTvtwo}).}
\label{fig:CorrelationsPR}
\end{figure}

Before continuing with the theoretical discussion of partonic energy
loss and its effects in heavy ion collisions at RHIC, it is worthwhile
first to gain an impression of the reach of the available data
addressing this physics and the magnitude of the effects under
discussion.  Figures
\ref{fig:InclusiveSuppressionPR} and \ref{fig:CorrelationsPR} show the
most significant high \pT\ measurements made at RHIC thus far.  Both
figures incorporate measurements of \sqrts=200 GeV $p+p$, $d+Au$ and
centrality-selected $Au+Au$ collisions at RHIC, with the simpler $p+p$ and
$d+Au$ systems providing  benchmarks for phenomena seen in
the more complex $Au+Au$ collisions.

Figure \ref{fig:InclusiveSuppressionPR} shows the ratio of inclusive
hadron yields in $Au+Au$ and $d+Au$ to $p+p$, scaled by \NbinaryMean\
to account for trivial geometric effects. A striking phenomenon is
seen: large $p_T$ hadrons in central Au+Au collisions are suppressed
by a factor 5 relative to naive expectations. Conventional nuclear
effects, such as nuclear shadowing of the parton distribution
functions and initial state multiple scattering, cannot account
qualitatively for the suppression. Furthermore, the suppression is not
seen in $d+Au$ but is unique to $Au+Au$ collisions, proving
experimentally that it results not from nuclear effects in the initial
state (in particular, gluon saturation) but from the final state
interaction of hard scattered partons or their fragmentation products
in the dense medium generated in $Au+Au$ collisions
\cite{brahms:highpTdAu,phenix:highpTdAu,phobos:highpTdAu,star:highpTdAu}.

Figure \ref{fig:CorrelationsPR} shows correlations of high \pT\
hadrons. The left panel shows the azimuthal distribution of hadrons
with \pT\gt2 GeV/c relative to a trigger hadron with $\pT^{\rm trig}
\gt4$ GeV/c. A hadron pair drawn from a single jet will generate an
enhanced correlation at $\Delta\phi\sim 0$, as observed for $p+p$,
$d+Au$ and $Au+Au$, with similar correlation strengths and widths. A
hadron pair drawn from back-to-back dijets will generate an enhanced
correlation at $\Delta\phi\sim\pi$, as observed with somewhat broader
width than the near-side correlation peak for $p+p$ and $d+Au$
collisions. However, the back-to-back dihadron correlation is
strikingly absent in central $Au+Au$ collisions, and uniquely in
central $Au+Au$ collisions. If the correlation is indeed the result of
jet fragmentation, this suppression is also due to the final state
interaction of hard scattered partons or their fragmentation products
in the dense medium generated in Au+Au collisions
\cite{star:highpTdAu}. Finally, the right panel shows the finite
azimuthal correlation strength of high \pT\ hadrons with the
orientation of the reaction plane in non-central Au+Au collisions, in
analogy to the elliptic flow seen at low \pT\
(Sect. \ref{sect:EllipticFlow})
\cite{Tang:2003kz}. Since the azimuthal orientation of the initially scattered hard
parton is uncorrelated with that of the reaction plane and the bulk
deformation of the fireball, this correlation can only arise from
final state interactions.

Figures \ref{fig:InclusiveSuppressionPR} and \ref{fig:CorrelationsPR}
present compelling evidence that high \pT\ hadron production in
nuclear collisions at RHIC is profoundly altered by interactions with
the medium created in the collision. We will now discuss in some
detail the theory of partonic energy loss, which provides a unified
description of these phenomena and enables them to be applied as
unique, penetrating probes of the medium.

\subsection{Partonic energy loss and modified fragmentation functions}

In contrast to the QED energy loss of electrons in matter, the QCD
energy loss of partons cannot be measured directly because partons are
not the final, experimentally observed particles. Instead, studies of
partonic energy loss must exploit the particle distributions within a
jet, in particular the modification of the fragmentation functions
$D_{a\rightarrow h}(z,\mu^2)$ which can be directly related to the
energy loss of the leading parton.

The first example we will consider is electron-nucleus deep inelastic scattering (DIS)
\cite{Wang:2001if,Guo:2000nz,Zhang:2003yn}. We study the
semi-inclusive process, $e(L_1) + A(p) \longrightarrow e(L_2) +
h(\ell_h) +X$, where $L_1$ and $L_2$ are the four-momenta of the
incoming and outgoing leptons, and $\ell_h$ is the observed hadron
momentum. The differential cross section for the semi-inclusive
process can be expressed as
\begin{equation}
E_{L_2}E_{\ell_h}\frac{d\sigma_{\rm DIS}^h}{d^3L_2d^3\ell_h}
=\frac{\alpha^2_{\rm EM}}{2\pi s}\frac{1}{Q^4} L_{\mu\nu}
E_{\ell_h}\frac{dW^{\mu\nu}}{d^3\ell_h} \; ,
\label{sigma}
\end{equation}
\noindent
where $p = [p^+,0,{\bf 0}_\perp] \label{eq:frame}$
is the momentum per nucleon in the nucleus,
$q =L_2-L_1 = [-Q^2/2q^-, q^-, {\bf 0}_\perp]$ is the momentum transfer,
$s=(p+L_1)^2$ and $\alpha_{\rm EM}$ is the electromagnetic (EM)
coupling constant. $L_{\mu\nu}$ is the leptonic tensor,
while $W_{\mu\nu}$ is the semi-inclusive hadronic tensor.

In the collinear factorization approximation to the parton model,
the leading-twist contribution to the semi-inclusive cross section can
be factorized into a product of parton distributions, parton
fragmentation functions and the partonic cross section.  Including all
leading log radiative corrections, the lowest order contribution
(${\cal O}(\alpha_s^0)$) from a single hard $\gamma^*+ q$ scattering
can be written as
\begin{eqnarray}
\frac{dW^S_{\mu\nu}}{dz_h}
&=& \sum_q e_q^2 \int dx f_q^A(x,\mu_I^2) H^{(0)}_{\mu\nu}(x,p,q)
D_{q\rightarrow h}(z_h,\mu^2)\, ; \label{Dq} \\
H^{(0)}_{\mu\nu}(x,p,q) &=& \frac{1}{2}\,
{\rm Tr}(\gamma \cdot p \gamma_{\mu} \gamma \cdot(q+xp) \gamma_{\nu})
\, \frac{2\pi}{2p\cdot q} \delta(x-x_B) \, , 
\label{H0}
\end{eqnarray}
where the momentum fraction carried by the hadron is defined as
$z_h=\ell_h^-/q^-$ and $x_B=Q^2/2p^+q^-$ is the Bjorken variable.
$\mu_I^2$ and $\mu^2$ are the factorization scales for the initial
quark distributions $f_q^A(x,\mu_I^2)$ in a nucleus and the fragmentation
functions $D_{q\rightarrow h}(z_h,\mu^2)$, respectively.

The propagating quark in DIS off a nucleus will experience additional
scatterings with other partons from the nucleus. The rescatterings
induce additional gluon radiation and cause the leading quark to lose
energy. This effectively gives rise to additional terms in the
evolution equation, leading to the modification of the fragmentation
functions in a medium. These are called higher-twist
corrections since they involve higher-twist parton matrix elements and
are power-suppressed. We will consider those contributions that
involve two-parton correlations from two different nucleons inside the
nucleus. Generalized factorization is usually applied to these
multiple scattering processes\cite{Luo:1992fz,Luo:1994ui,Luo:1994np}.
In this approximation, the radiative correction to the semi-inclusive
tensor from double quark-gluon scattering is
\begin{eqnarray}
\frac{W_{\mu\nu}^{D,q}}{dz_h}
&=&\sum_q \,\int dx H^{(0)}_{\mu\nu}(xp,q)
\int_{z_h}^1\frac{dz}{z}D_{q\rightarrow h}(z_h/z) 
\frac{\alpha_s}{2\pi} C_A \frac{1+z^2}{1-z}
\int \frac{d\ell_T^2}{\ell_T^4} \frac{2\pi\alpha_s}{N_c}
T^{A}_{qg}(x,x_L)
\;\; , \label{wd1}
\end{eqnarray}
where
\begin{eqnarray}
T^{A}_{qg}(x,x_L)&=& \int \frac{dy^{-}}{2\pi}\, dy_1^-dy_2^-
(1-e^{-ix_Lp^+y_2^-})(1-e^{-ix_Lp^+(y^--y_1^-)}) e^{i(x+x_L)p^+y^-}
\nonumber  \\ &\times & 
\theta(-y_2^-)\theta(y^- -y_1^-)
\frac{1}{2}\langle A | \bar{\psi}_q(0)\,
\gamma^+\, F_{\sigma}^{\ +}(y_{2}^{-})\, F^{+\sigma}(y_1^{-}) \psi_q(y^{-})| A\rangle  \;\;  
\label{Tqg}
\end{eqnarray}
are twist-four parton matrix elements of the nucleus. The fractional
momentum $x_L =\ell_T^2/2p^+q^-z(1-z)$ and $x=x_B=Q^2/2p^+q^-$ is the
Bjorken scaling variable. The dipole-like structure in the effective
twist-four parton matrix results from Landau-Pomeranchuk-Migdal
(LPM) interference in gluon bremsstrahlung
\cite{Landau:1953um,Migdal:1956tc}. After expansion, the first
diagonal term corresponds to the so-called hard-soft process where
gluon radiation is induced by the hard scattering between the virtual
photon and a quark at momentum fraction $x$. The quark is knocked
off-shell by the virtual photon, returning on-shell by radiating a
gluon. The on-shell quark or radiated gluon will then have a secondary
scattering with another soft gluon from the nucleus. The second
diagonal term is due to the double hard process where the quark is
on-shell after the first hard scattering with the virtual photon. The
gluon radiation is then induced by the scattering of the quark with
another gluon that carries finite momentum fraction $x_L$.  The two
off-diagonal terms represent interference between the hard-soft and
double hard processes. In the limit of collinear radiation
($x_L\rightarrow 0$) or when the formation time of the gluon
radiation, $\tau_f\equiv 1/x_Lp^+$, is much larger than the nuclear
size, the two processes interfere destructively, leading to the LPM
interference effect.

Including the virtual corrections and the single scattering
contribution, we rewrite the semi-inclusive tensor in
a factorized form with a nuclear modified fragmentation function,
\begin{eqnarray}
\widetilde{D}_{q\rightarrow h}(z_h,\mu^2)&\equiv&
D_{q\rightarrow h}(z_h,\mu^2)
+\int_0^{\mu^2} \frac{d\ell_T^2}{\ell_T^2}
\frac{\alpha_s}{2\pi} \int_{z_h}^1 \frac{dz}{z} \nonumber \\
&\times &\left[ \Delta\gamma_{q\rightarrow qg}(z,x,x_L,\ell_T^2)
D_{q\rightarrow h}(z_h/z) + 
\Delta\gamma_{q\rightarrow gq}(z,x,x_L,\ell_T^2)
D_{g\rightarrow h}(z_h/z)\right] \, , \label{eq:MDq}
\end{eqnarray}
where $D_{q\rightarrow h}(z_h,\mu^2)$ and
$D_{g\rightarrow h}(z_h,\mu^2)$ are the leading-twist
fragmentation functions. The modified splitting functions are

\begin{eqnarray}
\Delta\gamma_{q\rightarrow qg}(z,x,x_L,\ell_T^2)&=&
\left[\frac{1+z^2}{(1-z)_+}T^{A}_{qg}(x,x_L) +
\delta(1-z)\Delta T^{A}_{qg}(x,\ell_T^2) \right] \frac{2\pi\alpha_s C_A}
{\ell_T^2 N_c\widetilde{f}_q^A(x,\mu_I^2)}\, ,
\label{eq:r1}\\
\Delta\gamma_{q\rightarrow gq}(z,x,x_L,\ell_T^2)
&=& \Delta\gamma_{q\rightarrow qg}(1-z,x,x_L,\ell_T^2). \label{eq:r2}
\end{eqnarray}
This medium correction is very similar in form to that caused by gluon
bremsstrahlung in vacuum that leads to the DGLAP evolution in
Eq.~(\ref{eq:ap2}).

Using the factorization approximation\cite{Luo:1994ui,Osborne:2002st},
we can relate the twist-four parton matrix elements of the nucleus to
the twist-two parton distributions of nucleons and the nucleus,
\begin{eqnarray}
T^A_{qg}(x,x_L)&=&\frac{C}{x_A}
(1-e^{-x_L^2/x_A^2}) \left[f_q^A(x+x_L)\, x_Tf_g^N(x_T)
+f_q^A(x)(x_L+x_T)f_g^N(x_L+x_T)\right] \nonumber \\
&\approx& \frac{\widetilde{C}}{x_A} (1-e^{-x_L^2/x_A^2}) f_q^A(x),
\label{modT2}
\end{eqnarray}
where C is a constant, $x_T=<k_T^2>/2p^+q^-z$ is related to
the intrinsic transverse momentum of gluons inside the nucleus,
$x_A=1/m_NR_A$, $f_q^A(x)$ is the quark
distribution inside a nucleus, and $f_g^N(x)$ is the gluon
distribution inside a nucleon. The coefficient
$\widetilde{C}\equiv 2C x_Tf^N_g(x_T)$ should in principle 
depend on $Q^2$ and $x_T$ but can be approximated as a constant.
A Gaussian distribution in 
light-cone coordinates is assumed for the nuclear distribution,
$\rho(y^-)=n_0 \exp({y^-}^2/2{R^-_A}^2)$, where
$R^-_A=\sqrt{2}R_A m_N/p^+$ and $m_N$ is the nucleon mass. We should
emphasize that the twist-four matrix element is proportional to
$1/x_A=R_Am_N$, {\it i.e.} the nuclear size\cite{Osborne:2002st}.

In the above matrix element, $1/x_Lp^+=2q^-z(1-z)/\mu^2$ is identified
as the formation time of the emitted gluons. For formation time that
is large relative to the nuclear size the above matrix element
vanishes, exhibiting the typical LPM interference effect.  This results
because the emitted gluon (with long formation time) and the leading
quark remain a coherent system while propagating through the
nucleus. Additional scattering will not induce more gluon radiation.

The reduction due to LPM interference of the phase space available for
gluon radiation is critical for applying the LQS formalism (Luo,
Qiu and Sterman \cite{Luo:1992fz,Luo:1994ui,Luo:1994np}) to the
problem under consideration. In the original LQS approach, the
generalized factorization for processes with large final transverse
momentum $\ell_T^2\sim Q^2$ leads to consideration of the leading
contribution in $1/Q^2$, which is enhanced by the nuclear size $R_A
\sim A^{1/3}$. For large $Q^2$ and $A$, the higher-twist contribution
from double parton rescattering that is proportional to $\alpha_s
R_A/Q^2$ will then be the leading nuclear correction. Contributions
from more than two parton rescattering can be neglected. In deriving
the modified fragmentation functions, we however have to take the
leading logarithmic approximation in the limit $\ell_T^2\ll Q^2$,
where $\ell_T$ is the transverse momentum of the radiated gluon.
Since the LPM interference suppresses 
gluon radiation whose formation time ($\tau_f \sim Q^2/\ell_T^2p^+$)
is larger than the nuclear 
size $m_N R_A/p^+$ in our chosen frame, $\ell_T^2$ should then have a 
minimum value of $\ell_T^2\sim Q^2/m_N R_A\sim Q^2/A^{1/3}$,
where $m_N$ is the nucleon mass.
Therefore, the leading higher-twist
contribution proportional to $\alpha_s R_A/\ell_T^2 \sim \alpha_s R_A^2/Q^2$
from double scattering depends quadratically on the nuclear size $R_A$.

\begin{figure}
\includegraphics[width=.48\textwidth]{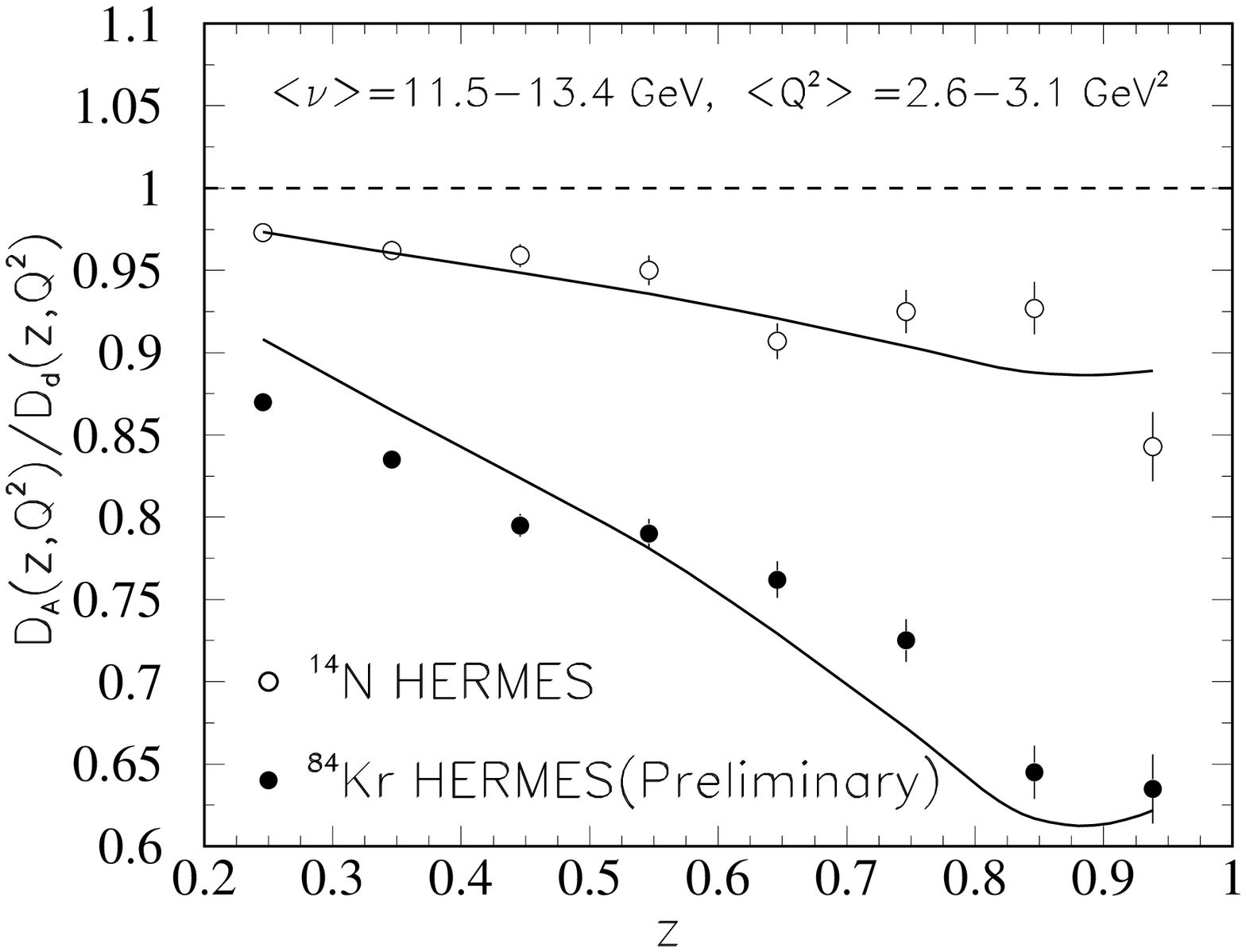}
\includegraphics[width=.48\textwidth]{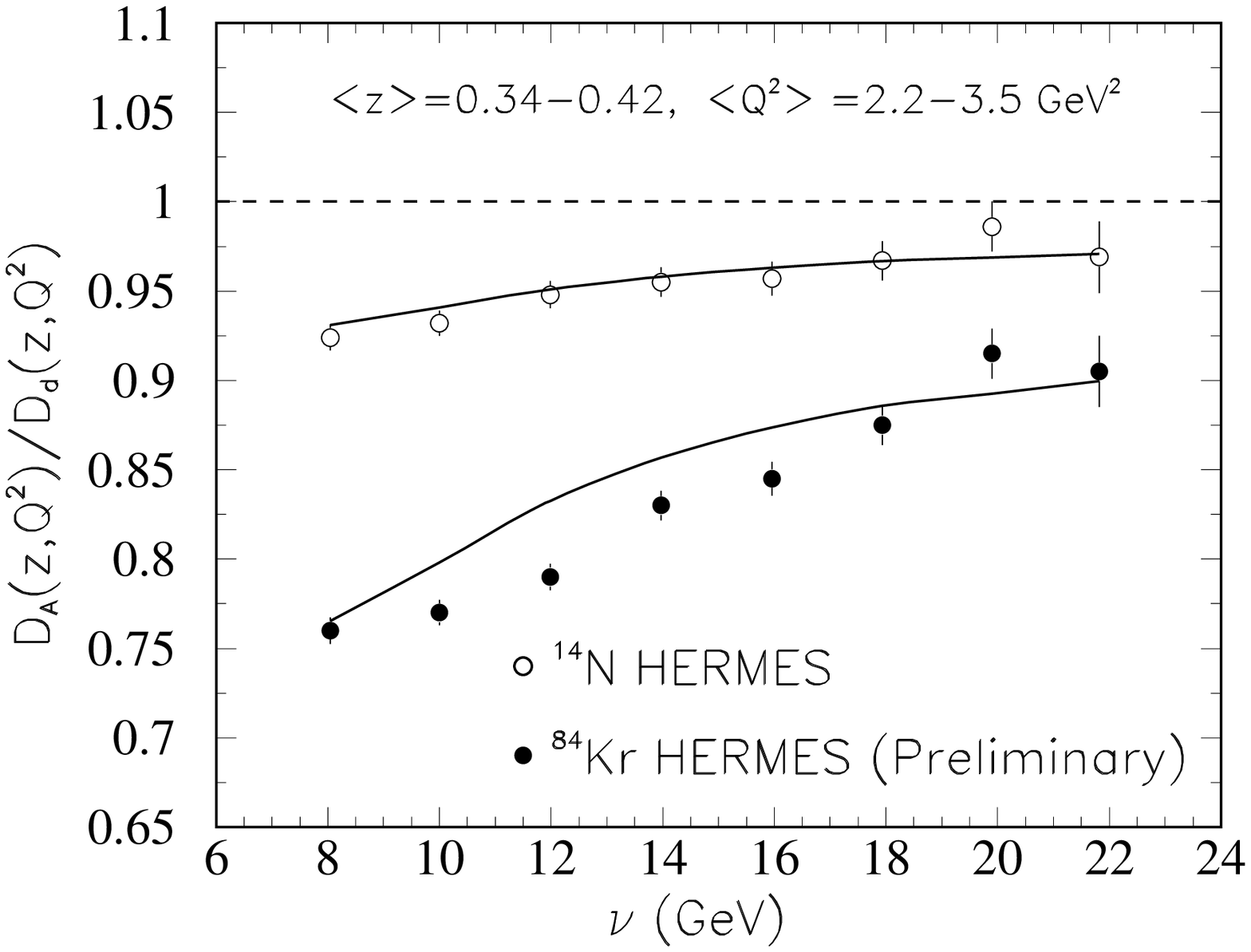}
\caption{Measured ratios of
hadron distributions from DIS off $A$ and $d$ targets, from HERMES
\protect\cite{Airapetian:2000ks,Muccifora:2001zn}. The curves show the predicted 
nuclear modification of the jet fragmentation function, described in text.
Left: vs. fragmentation fraction $z$. Right: vs. energy transfer $\nu$.}
\label{fig:hermes}
\end{figure}

With the assumption of the factorized form of the twist-4 nuclear
parton matrices, there is only one free parameter $\widetilde{C}(Q^2)$
which represents the quark-gluon correlation strength inside nuclei.  Once
it is fixed, the $z$, energy and nuclear dependence of the medium
modification of the fragmentation function can be predicted. Shown in
Fig.~\ref{fig:hermes} are the calculated nuclear modification factors
of the fragmentation functions for $^{14}N$ and $^{84}Kr$ targets,
compared to recent HERMES
data\cite{Agakishiev:1995xb,Airapetian:2000ks}.  The predicted shapes
of the $z$- and $\nu$-dependence agree well with the data. A
remarkable feature of the prediction is the quadratic $A^{2/3}$
nuclear size dependence, which is verified for the first time by an
experiment.  By fitting the overall suppression for one nuclear
target, one fixes the only free parameter in the calculation,
$\widetilde{C}(Q^2)=0.0060$ GeV$^2$ with $\alpha_{\rm s}(Q^2)=0.33$ at
$Q^2\approx 3$ GeV$^2$.

Modification of the fragmentation can be quantified by the quark energy
loss, defined as the momentum fraction carried by the radiated gluon:
\begin{eqnarray}
\langle\Delta z_g\rangle(x_B,\mu^2)
&=& \int_0^{\mu^2}\frac{d\ell_T^2}{\ell_T^2}
\int_0^1 dz \frac{\alpha_s}{2\pi}
 z\,\Delta\gamma_{q\rightarrow gq}(z,x_B,x_L,\ell_T^2) \nonumber \\
&=&\frac{C_A\alpha_s^2}{N_c} \int_0^{\mu^2}\frac{d\ell_T^2}{\ell_T^4}
\int_0^1 \!\!dz [1+(1-z)^2]
\frac{T^{A}_{qg}(x_B,x_L)}{\widetilde{f}_q^A(x_B,\mu_I^2)} \\
\label{eq:loss1}
&=&\widetilde{C}\frac{C_A\alpha_s^2}{N_c}
\frac{x_B}{x_AQ^2} \int_0^1 dz \frac{1+(1-z)^2}{z(1-z)} 
\int_0^{x_\mu} \frac{dx_L}{x_L^2}(1-e^{-x_L^2/x_A^2}),
\label{eq:heli-loss}
\end{eqnarray}
where $x_\mu=\mu^2/2p^+q^-z(1-z)=x_B/z(1-z)$ for
factorization scale $\mu^2=Q^2$.
For $x_A\ll x_B\ll 1$, the leading quark energy loss is roughly
\begin{eqnarray}
\langle \Delta z_g\rangle(x_B,\mu^2)& \approx &
\widetilde{C}\frac{C_A\alpha_s^2}{N_c}\frac{x_B}{Q^2
x_A^2}6\sqrt{\pi}\ln\frac{1}{2x_B}\, .
\label{eq:appr1-loss}
\end{eqnarray}
Since $x_A=1/m_N R_A$, the energy loss $\langle \Delta
z_g\rangle$ thus depends quadratically on the nuclear size.

In the rest frame of the nucleus, $p^+=m_N$, $q^-=\nu$, and
$x_B\equiv Q^2/2p^+q^-=Q^2/2m_N\nu$. The average total energy loss is
$ \Delta E=\nu\langle\Delta z_g\rangle
\approx  \widetilde{C}(Q^2)\alpha_{\rm s}^2(Q^2)
m_NR_A^2(C_A/N_c) 3\sqrt{\pi} \ln (1/2x_B)$.  With the value of $\widetilde{C}$
from the fit, $\langle x_B\rangle \approx 0.124$ in the HERMES
kinematics \cite{Airapetian:2000ks,Muccifora:2001zn} and the average
distance $\langle L_A\rangle=R_A\sqrt{2/\pi}$ for the assumed Gaussian
nuclear distribution, the average quark energy loss $dE/dL\approx 0.5$
GeV/fm in a $Au$ nucleus.

Attenuation of leading hadrons in DIS off nuclear targets has also
been studied in hadronic transport and absorption models
\cite{Kopeliovich:1990sh,Falter:2003di}. In these models, two distinct
types of hadronic absorption are assumed: absorption of fully formed
physical hadrons, and of ``pre-hadrons''. Since the hadronic formation
time is long, attenuation effects on fully formed hadrons are small
and the observed attenuation is attributed to the absorption of
pre-hadrons in the nuclear medium. The pre-hadron can be modeled as a
quark-antiquark dipole, and the interaction of $q\bar{q}$ dipoles with
the nuclear medium should be equivalent to the picture of multiple
parton scattering and induced bremsstrahlung \cite{Wiedemann:2000za}.

\subsection{Energy loss and jet quenching in a hot medium}

To extend the study of modified fragmentation functions to jets in
heavy-ion collisions, we assume a one-dimensional boost invariant
(Bjorken) expansion with transverse gluon density profile
$\rho(r,\tau)=(\tau_0/\tau)\theta(R_A-r)\rho_0$, and $\langle k_T^2\rangle\approx \mu^2$
(the Debye screening mass). The initial jet
production rate is independent of the final gluon density, which is
related to the parton-gluon scattering cross
section $\alpha_s x_TG(x_T)\sim \mu^2\sigma_g$ \cite{Baier:1997sk} so that
\begin{equation}
\frac{\alpha_s T_{qg}^A(x_B,x_L)}{f_q^A(x_B)} \sim
\mu^2\int dr \sigma _g \rho(r,\tau)
[1-\cos(r/\tau_f)],
\end{equation}
where $\tau_f=2Ez(1-z)/\ell_T^2$ is the gluon formation time.
Assuming partons traveling in the transverse direction at the
velocity of light ($r=\tau-\tau_0)$, the fractional energy 
loss from Eq.~(\ref{eq:loss1}) is
\cite{Wang:2002ri}
\begin{eqnarray}
\langle\Delta z_g\rangle &=&\frac{C_A\alpha_s}{\pi}
\int_0^1 dz \int_0^{\frac{Q^2}{\mu^2}}du \frac{1+(1-z)^2}{u(1+u)}
\int_{\tau_0}^{R_A} d\tau\sigma_g\rho(\tau) 
\left[1-\cos\left(\frac{(\tau-\tau_0)\,u\,\mu^2}{2Ez(1-z)}\right)\right].
\end{eqnarray}
Keeping only the dominant contribution and assuming 
$\sigma_g\approx C_a 2\pi\alpha_s^2/\mu^2$ ($C_a$=1 for $qg$, 9/4 for
$gg$), the average energy loss is
\begin{equation}
\langle \frac{dE}{dL}\rangle \approx \frac{\pi C_aC_A\alpha_s^3}{R_A}
\int_{\tau_0}^{R_A} d\tau \rho(\tau) (\tau-\tau_0)\ln\frac{2E}{\tau\mu^2}.
\label{eq:effloss}
\end{equation}
Neglecting the logarithmic dependence on $\tau$, the averaged energy
loss in a one-dimensional expanding system is
$\langle dE/dL \rangle_{1d} \approx (dE_0/dL) (2\tau_0/R_A)$,
where $dE_0/dL\propto \rho_0R_A$ is the energy loss in a static medium
with the initial gluon density $\rho_0$ of the expanding system at
time $\tau_0$. Because of the expansion, the averaged energy loss
$\langle dE/dL\rangle_{1d}$ is suppressed relative to the static case
and does not depend linearly on system size for a fixed value of $\rho_0$.

This form of the energy loss has also been derived in the opacity
expansion framework \cite{Gyulassy:2000gk,Salgado:2002cd}, based on a
model of multiple scattering in a quark-gluon plasma consisting of
randomly distributed scattering centers with screened static potential
\cite{Gyulassy:1994hr}. The magnitude of the momentum transfer is small, limited by
the Debye screening mass $\mu\sim gT$, and amplitudes for multiple
scattering and gluon bremsstrahlung factorize in momentum space. This
leads to an algebraic reaction operator formulation of the radiation
amplitude induced by multiple scattering, which keeps track of the
phase accumulation due to multiple scattering. This approach enables
the iterative evaluation of the radiation spectrum induced by a given
number of scatterings, corresponding to an expansion in opacity
parameter $\xi=\sigma \rho L$. However, this approach cannot be
applied directly to multiple parton scattering in a cold nuclear
medium since the concept of random screened potential is difficult to
justify in that case, and multi-parton correlations are also
important. The extension of the twist expansion to higher twist to
account for multiple parton scattering is also difficult and has yet
to be done. To first order in opacity or leading twist, the two
approaches give the same result for radiative energy loss in a hot
gluon plasma.

In order to calculate the effects of partonic energy loss on the
production of high \pT\ hadrons in nuclear collisions, one can apply a
simpler effective modified fragmentation
function\cite{Wang:1996yh,Wang:1997pe},
\begin{eqnarray}
D_{h/c}^\prime(z_c,Q^2,\Delta E_c) 
=(1-e^{-\langle \frac{\Delta L}{\lambda}\rangle})
\left[ \frac{z_c^\prime}{z_c} D^0_{h/c}(z_c^\prime,Q^2) \right. 
&+&\left. \langle \frac{\Delta L}{\lambda}\rangle
\frac{z_g^\prime}{z_c} D^0_{h/g}(z_g^\prime,Q^2)\right]
\nonumber \\
&+& e^{-\langle\frac{\Delta L}{\lambda}\rangle} D^0_{h/c}(z_c,Q^2),
\label{modfrag} 
\end{eqnarray}
where $z_c^\prime,z_g$ are the rescaled momentum fractions.  The
fragmentation functions in free space $D^0_{h/c}(z_c,Q^2)$ are given
by the BBK parameterization \cite{Binnewies:1995ju}.  The first term
is the fragmentation function of the jet $c$ after losing energy
$\Delta E_c(p_c,\phi)$ due to {\em medium induced} gluon radiation.
The second term is the feedback due to the fragmentation of the
$N_g(p_c,\phi)=\langle \Delta L/\lambda\rangle$ radiated gluons.  This
effective model is found to reproduce the pQCD result from
Eq.(\ref{eq:MDq}) very well, but only when $\Delta z=\Delta E_c/E$ is
set to be $\Delta z\approx 0.6 \langle z_g\rangle$.  Therefore the
actual averaged parton energy loss should be $\Delta E/E=1.6\Delta z$
with $\Delta z$ extracted from the effective model. The factor 1.6 is
mainly due to the effect of unitarity correction in the pQCD calculation.

Since gluons are bosons, there should also be stimulated gluon
emission and absorption by the propagating parton because of the
presence of thermal gluons in the hot medium.  The detailed balance
is crucial for parton thermalization and should also be important for
calculating the energy loss of an energetic parton in a hot
medium. Taking into account such detailed balance in gluon emission,
the asymptotic behavior of the effective energy loss in the opacity
expansion framework is \cite{Wang:2001cs},
\begin{eqnarray}
   {\Delta E\over E}&\approx& {{\alpha_s C_F \mu^2 L^2}\over
   4\lambda_gE} \left[\ln{2E\over \mu^2L} -0.048\right] -
   {{\pi\alpha_s C_F}\over 3} {{LT^2}\over {\lambda_g E^2}} \left[
   \ln{{\mu^2L}\over T} -1+\gamma_{\rm
   E}-{{6\zeta^\prime(2)}\over\pi^2}
\right],
 \end{eqnarray}
where the first term is from the induced bremsstrahlung and the second
term is due to gluon absorption in detailed balance, which effectively
reduces the total partonic energy loss in the medium.

\begin{figure}
\centering
\includegraphics[width=.5\textwidth]{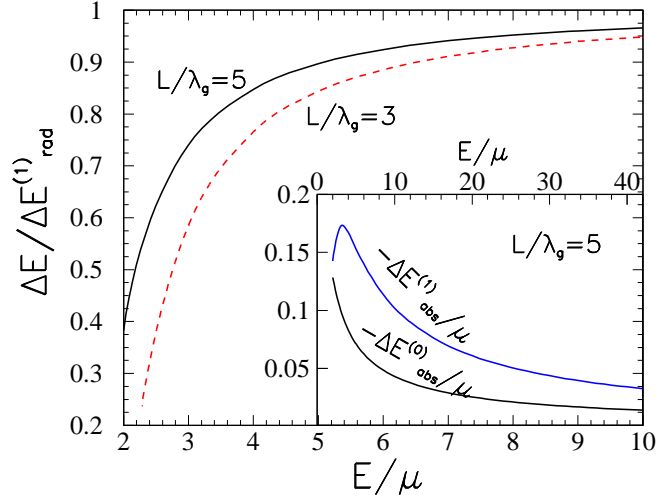}
\caption{The ratio of effective parton energy loss
with absorption ($\Delta E=\Delta E^{(0)}_{abs}+\Delta E^{(1)}_{abs}
+\Delta E^{(1)}_{rad}$) to that without ($\Delta E^{(1)}_{rad}$), as a
function of $E/\mu$. Insert: energy gain via absorption with
rescattering ($\Delta E^{(1)}_{abs}$) and without ($\Delta
E^{(0)}_{abs}$).}
\label{fig:balance}
\end{figure}

Fig.~\ref{fig:balance} shows numerical calculations of the ratio of
radiative energy loss with and without stimulated emission and thermal
absorption, as a function of $E/\mu$ for $L/\lambda_g=3$,5 and
$\alpha_s=0.3$. The insert shows the energy gain via gluon absorption
with and without rescattering.  For very high energy partons the
effect of the gluon absorption is negligible.  However, for
intermediate parton energy the thermal absorption reduces the
effective parton energy loss by about 30-10\%, generating an energy
dependence of the effective parton energy loss in the intermediate
energy region. This energy dependence is parameterized as
\begin{equation}
 \langle\frac{dE}{dL}\rangle_{1d}=\epsilon_0 (E/\mu-1.6)^{1.2}/(7.5+E/\mu),
\label{eq:th-loss}
\end{equation}
The threshold is the consequence of gluon absorption, which competes
with radiation to effectively shut off the energy loss. 
According to Eq.~(\ref{eq:effloss}), the parton energy loss 
and therefore the parameter $\epsilon_0$ in the above equation
depends linearly on the gluon density of the medium. This is the
only property of the medium one can extract from the experimental
measurement of parton energy loss.

\subsection{Cronin Enhancement in $p+A$ Collisions}

Tomography relies upon accurate knowledge of the initial flux of
radiation in order to measure the density of the intervening matter
between the source and the detector. Jet quenching is suitable for
tomography since jet production can be calculated reliably in pQCD, in
good agreement with experimental measurements in high-energy
$p+p(\bar{p})$ collisions. Perturbative calculations and experimental
measurements of jet production cross sections in $p+p$ and $p+A$
collisions effectively calibrate the initial source of the beam of jets.

\begin{figure}
\centering
\includegraphics[height=.35\textheight]{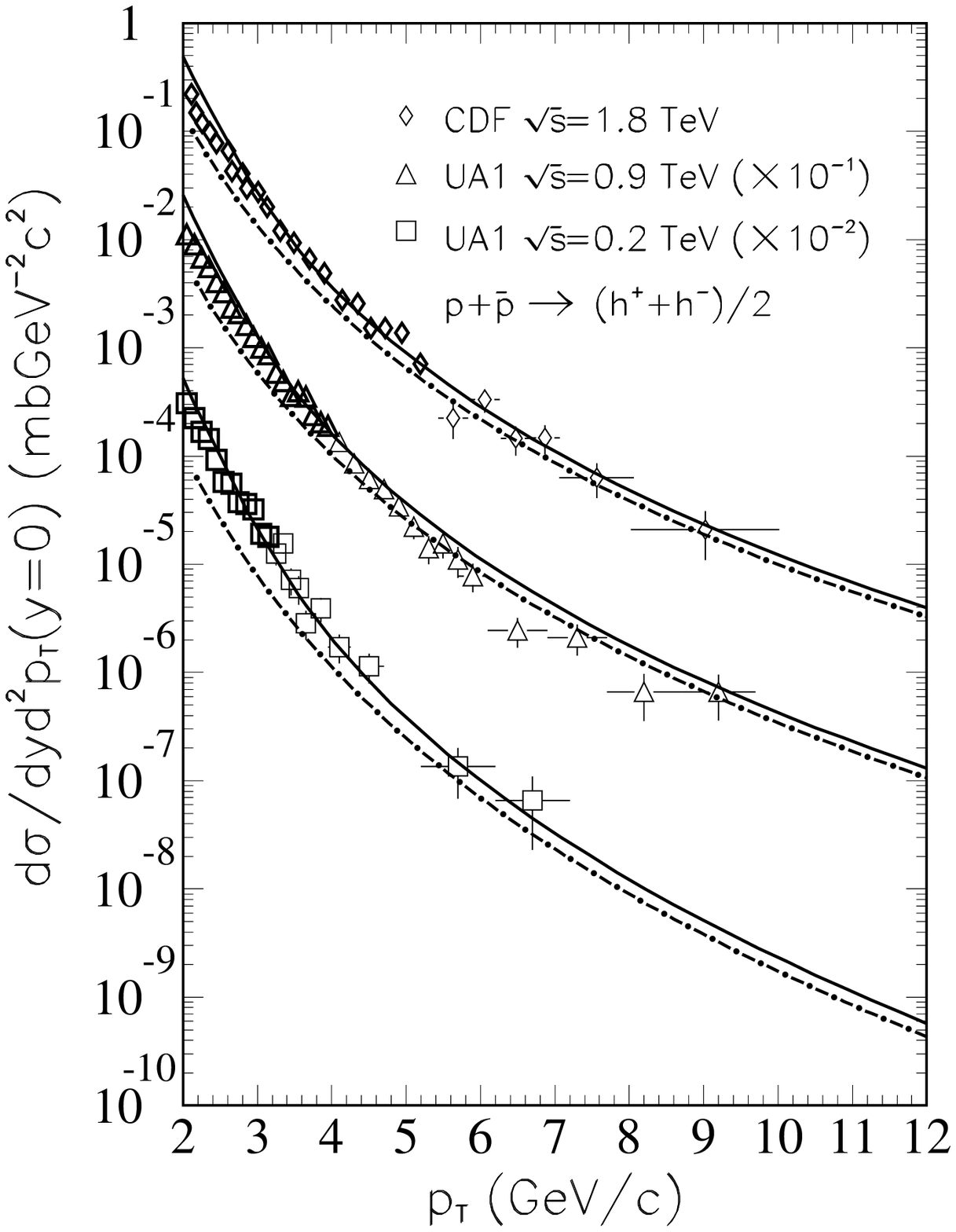}
\includegraphics[height=.32\textheight]{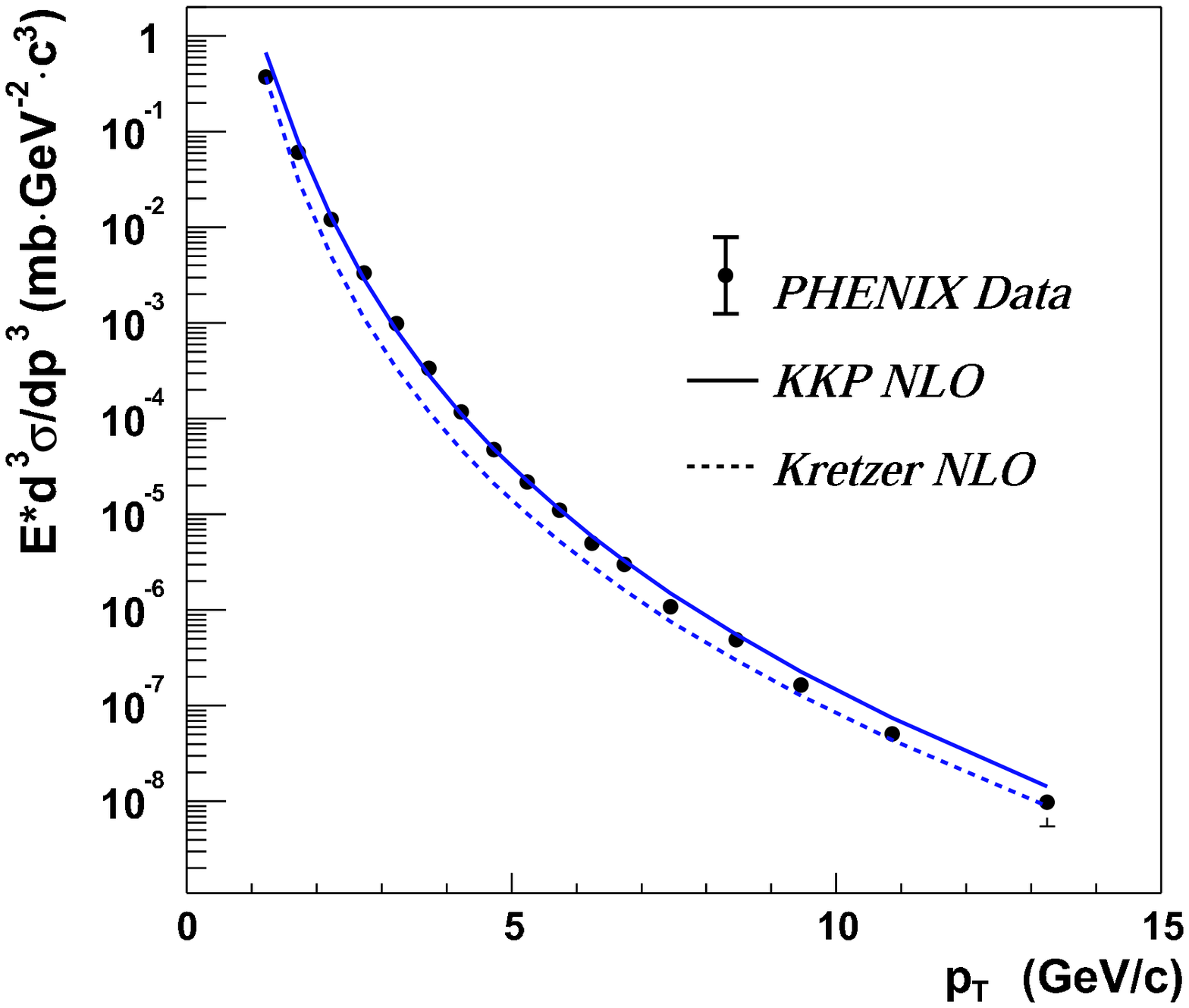}
\caption{Single particle inclusive spectra at collider energies compared to 
parton model and NLO calculations. Left: charged hadrons from $p+\bar{p}$ collisions at $\sqrt{s}=200,
900, 1800$ GeV from UA1 \protect\cite{Albajar:1990an} and CDF
\protect\cite{Abe:1988yu}. Lines indicate parton model calculations with
(solid) and without (dot-dashed) intrinsic $k_T$. Right: \pizero\ from
$p+p$ collisions at \sqrts=200 GeV compared to NLO calculations, from
PHENIX \protect\cite{phenix:pp-pi0}.}
\label{fig-ua1th}
\end{figure}

The simplest observable of jet production in $p+p$ collisions is the 
high \pT\ hadron spectrum resulting from jet fragmentation. In the
collinear factorized parton model, the single
inclusive hadron spectrum is \cite{Owens:1987mp}
\begin{eqnarray}
  \frac{d\sigma^h_{pp}}{dyd^2p_T}&=&K\sum_{abcd}
  \int dx_a dx_b d^2k_{aT} d^2k_{bT} g_p(k_{aT},Q^2)
  g_p(k_{bT},Q^2) \nonumber \\
  &\times& f_{a/p}(x_a,Q^2)f_{b/p}(x_b,Q^2)
   \frac{D^0_{h/c}(z_c,Q^2)}{\pi z_c}
  \frac{d\sigma}{d\hat{t}}(ab\rightarrow cd), 
\label{eq:nch_pp}
\end{eqnarray}
where $f_{a/p}(x)$ is the parton distribution in the proton
(we will use MRSD-$\prime$ parameterization \cite{Martin:1998sq}), 
$D^0_{h/c}(z_c,Q^2)$ is the fragmentation
function of parton $c$ into hadron $h$ derived from $e^+e^-$ data
\cite{Binnewies:1995ju}, and $z_c=p_T/p_{Tc}$ is the momentum fraction of the 
jet carried by a produced hadron. The $K\approx 1-2$ factor accounts
for higher order QCD corrections. At fixed-target energies, NLO
calculations with resummation underestimate the measured hadron
spectra \cite{Catani:1999hs}, indicating the
importance of higher power corrections. This can be remedied
phenomenologically by introducing an intrinsic transverse momentum
smearing $g_p(k_T,Q^2)$, assumed to have a Gaussian form
$g_p(k_T)=e^{-k_T^2/\langle k_T^2\rangle}/\pi\langle k_T^2\rangle$.
This smearing is found to be important for $\sqrt{s}<100$ GeV
\cite{Wang:1998ww}. It is less important at collider energies 
as seen in Fig.~\ref{fig-ua1th}, which shows good agreement between
LO and NLO parton model calculations without $k_T$ and
measured charged hadron and \pizero\ single particle inclusive
spectra. Additional discussion of hadronic spectra and intrinsic
$k_T$ can be found in \cite{Eskola:2002kv,Accardi:2003gp}.

Two nuclear effects must be incorporated into the parton model for an
accurate description of hadron production in $p+A$ collisions, both
arising as a consequence of multiple scattering: the nuclear
modification of the parton distribution functions and nuclear $k_T$
broadening. These effects must also be 
taken into account for quantitative study
of the change in hadron production due to jet quenching in $A+A$
collisions.

A practical approach to multiple scattering
effects calculates the differential cross section for inclusive hadron
production as that for a single hard parton-parton scattering, but
with larger beam parton $k_T$ due to multiple soft partonic scattering
prior to the hard interaction and parton distributions modified
according to DIS measurements off nuclear targets. The single
inclusive hadron production cross section in minimum-bias $p+A$
collisions is then
\begin{eqnarray}
\frac{d\sigma^h_{pA}}{dyd^2p_T}&=&K\sum_{abcd} \int d^2b T_A(b)
\int dx_a dx_b d^2k_{aT} d^2k_{bT} g_A(k_{aT},Q^2,b) g_p(k_{bT},Q^2) 
\nonumber \\ & \times & 
f_{a/p}(x_a,Q^2)f_{b/A}(x_b,Q^2,b) 
\frac{D^0_{h/c}(z_c,Q^2)}{\pi z_c}
\frac{d\sigma}{d\hat{t}}(ab\rightarrow cd), \label{eq:nch_pA}
\end{eqnarray} 
where $T_A(b)$ is the nuclear thickness function normalized to $\int
d^2b T_A(b)=A$. The parton distribution per nucleon inside the 
nucleus of mass $A$ and charge $Z$ at impact parameter $b$,
\begin{equation}
    f_{a/A}(x,Q^2,b)=S_{a/A}(x,b)\left[ \frac{Z}{A}f_{a/p}(x,Q^2)
    +(1-\frac{Z}{A}) f_{a/n}(x,Q^2)\right], \label{eq:shd}
\end{equation}
is assumed to be factorizable into the parton distribution in a
nucleon $f_{a/N}(x,Q^2)$ and the nuclear modification factor
$S_{a/A}(x,b)$, parameterized in various ways
\cite{Eskola:1998df,Hirai:2001np,Li:2001xa} according to the
DIS data.

The initial partonic transverse momentum distribution in a projectile
nucleon striking the target nucleus at impact parameter $b$ is still
assumed to be Gaussian but with a broadened variance
\begin{equation}
\langle k^2_T\rangle_A(Q^2)=\langle k^2_T\rangle_N(Q^2)
    +\delta^2(Q^2)(\nu_A(b) -1).
\end{equation}
The broadening is assumed to be proportional to the mean number of
scatterings $\nu_A(b)$ the projectile suffers inside the nucleus.  The
parameters are fitted to existing fixed-target $p+A$ data at energies up
to $\sqrt{s}=40$ GeV
\cite{Wang:1998ww}

For quantitative comparison of hadronic spectra from $A+B$ and $p+p$
collisions we define the nuclear modification factor
\cite{Wang:2001cy}
\begin{equation}
\RAB=\frac{d\sigma_{AB}/dyd^2\pT}
{\NbinaryMean d\sigma_{NN}/dyd^2\pT}.
\label{eq:RAB}
\end{equation} 
\noindent
\NbinaryMean, the mean number of binary collisions, is discussed in Section 
\ref{sect:Glauber}. In the absence of nuclear effects the cross section for a 
hard process will scale as \Nbinary, making \RAB\
unity. Nuclear-specific effects are measured by the deviation of \RAB\
from unity. In the following we will also utilize a related quantity
\RCP, the binary-scaled ratio of central over peripheral spectra from A+A collisions:
\begin{equation}
\RCP=
\frac{d^2N/d\pT d\eta/\NbinaryMean (central)}
{d^2N/d\pT d\eta/\NbinaryMean (peripheral)}
\label{eq:RCP}
\end{equation} 
\noindent
\RAB\ and \RCP\ contain similar physics, though
\RCP\ may at times be preferable from the standpoint of 
experimental uncertainties.

\begin{figure}
\includegraphics[width=.48\textwidth]{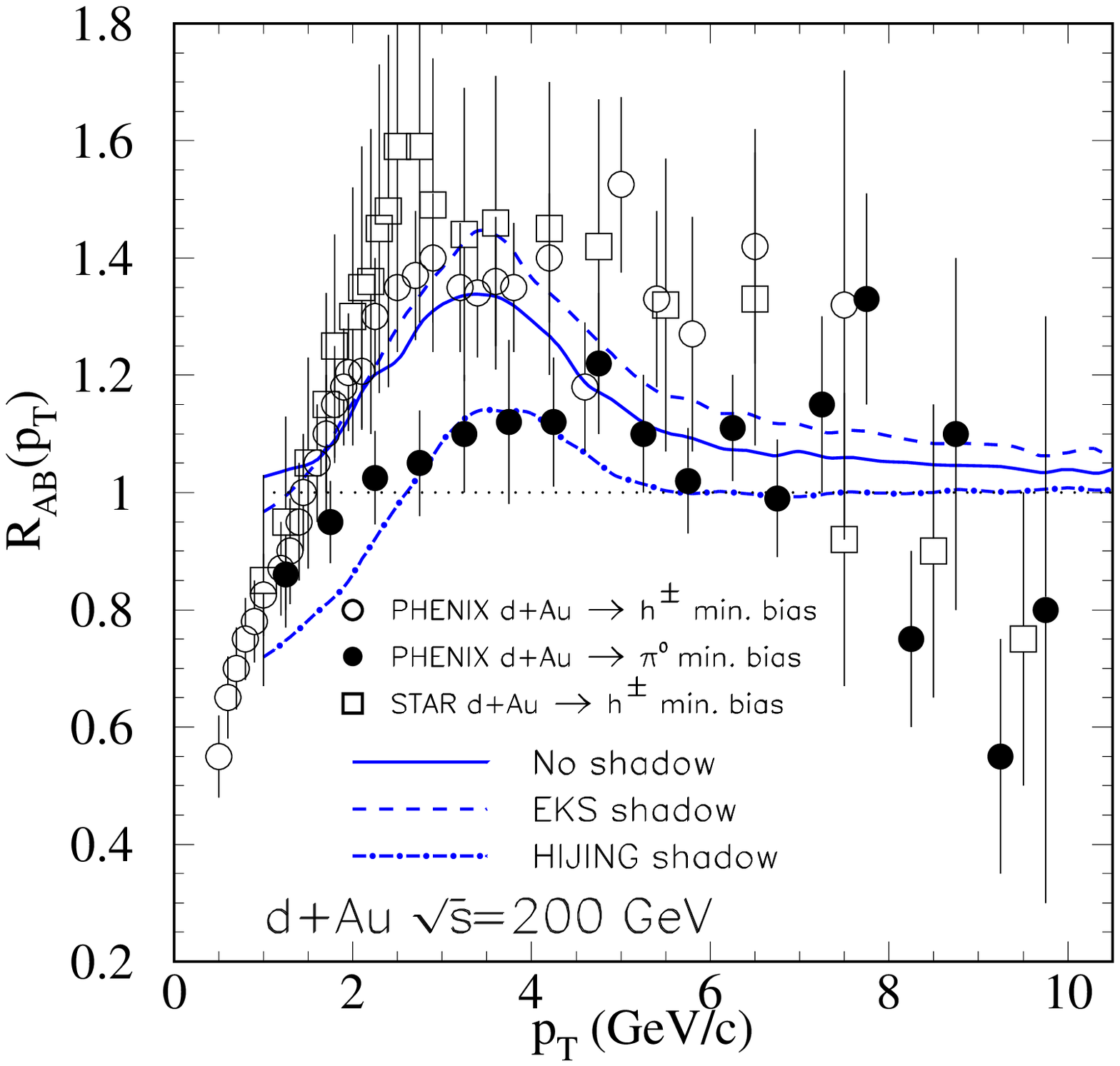}
\includegraphics[width=.48\textwidth]{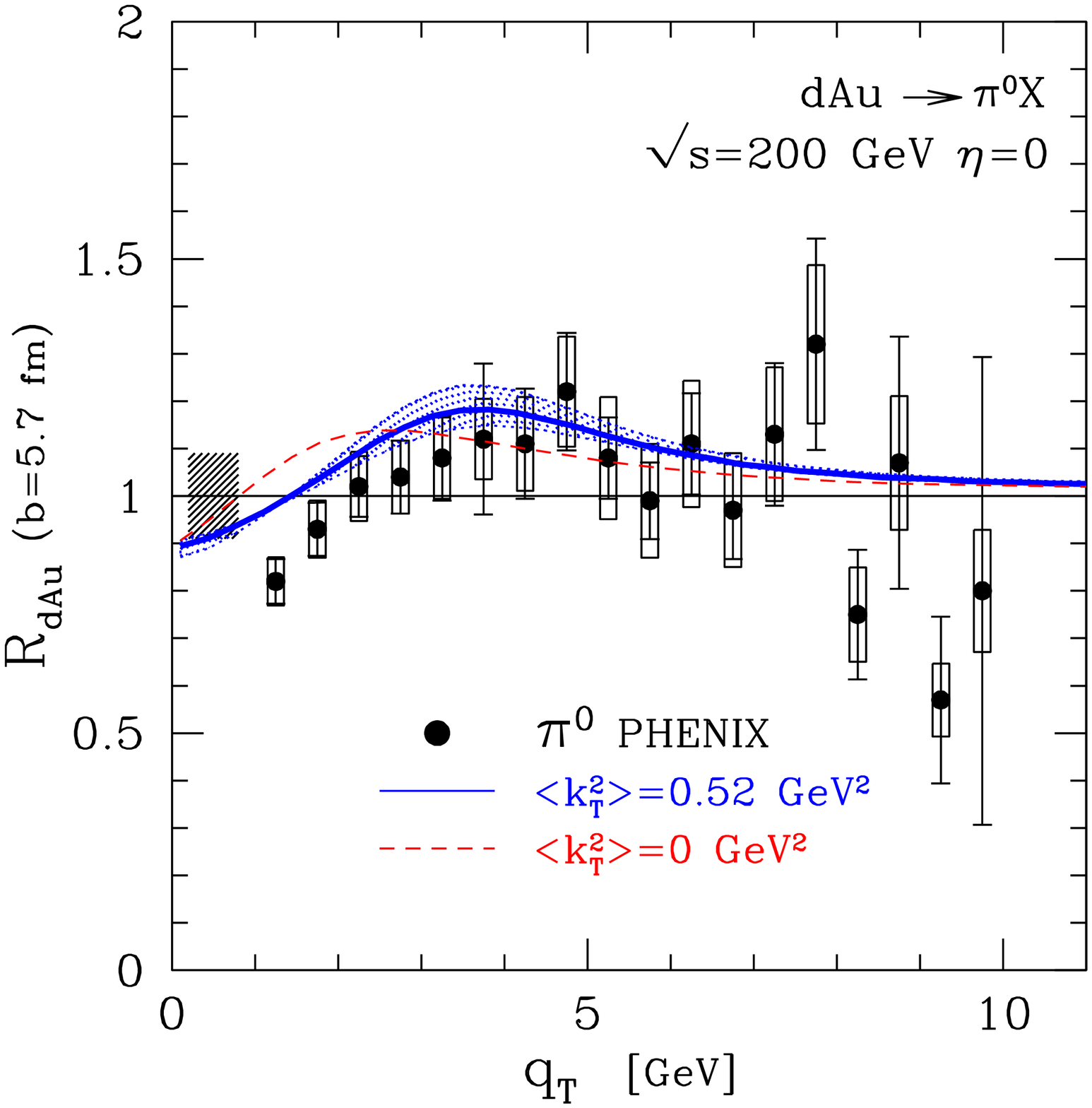}
\caption{\RAB\ (eq. \ref{eq:RAB}) for midrapidity \pizero\ and charged hadrons 
from d+Au relative to p+p collisions at 200 GeV, from PHENIX
\protect\cite{phenix:highpTdAu} and STAR \protect\cite{star:highpTdAu}. Left: data compared to 
first predictions of the Cronin effect in $p+Au$ collisions at
$\sqrt{s}=200$ GeV \protect\cite{Wang:1998ww}. Right: data compared to calculated
Cronin effect for \pizero\ from a Glauber-eikonal model of multiple
scattering \protect\cite{Accardi:2003jh}. The solid and dashed lines show
variation with intrinsic \kT, dotted band shows uncertainty due to
infrared regulator $p_0=1.0\pm0.1$.}
\label{fig:Cronin}
\end{figure}

Fig.~\ref{fig:Cronin} shows \RAB\ for \pizero\ and charged hadrons
from d+Au collisions at 200 GeV, measured by PHENIX
\cite{phenix:highpTdAu} and STAR \cite{star:highpTdAu} 
(see also PHOBOS \cite{phobos:highpTdAu} and BRAHMS
\cite{brahms:highpTdAu}). The curves in the left panel show the first
prediction of the nuclear modification of hadron spectra in $p+A$
collisions within the parton model \cite{Wang:1998ww}. The enhancement
at intermediate \pT, known as the Cronin effect, is due in this
calculation to broadening of \kT\ from multiple scattering. The
enhancement disappears at large \pT, as do all higher-twist
effects. The modest Cronin enhancement predicted for $p+A$ collisions
at this energy is confirmed by the measurements. The Cronin effect in
$A+A$ collisions at this energy should be of similar magnitude.

The Cronin effect has also been studied in multiple parton scattering
models which utilize the eikonal Glauber framework
\cite{Wang:1997yf,Wang:2001cy,Accardi:2003jh}. Interference effects
play an important role, for instance the absorption arising from
single scattering cancels part of the double scattering contribution,
leading to a $1/\pT^2$ dependence of the yield enhancement due to
double scattering \cite{Wang:1997yf}. The dominant double scattering
contribution occurs when one scattering is soft and most of the jet's
\pT\ comes from one hard scattering. At lower \pT\ the absorptive 
correction dominates double scattering and the coherence of the
interaction suppresses the spectrum relative to binary scaling. The
transition between the coherent suppression and Cronin enhancement
occurs at a scale $p_0$, where parton-nucleon scattering is no longer
a power-law like process. Fig. \ref{fig:Cronin}, right panel, shows
such a calculation \cite{Accardi:2003jh} compared to data.

\subsection{High \pT\ hadron suppression in $A+A$ collisions}
\label{sect:InclSupp}

\begin{figure}
\centering
\includegraphics[height=.3\textheight]{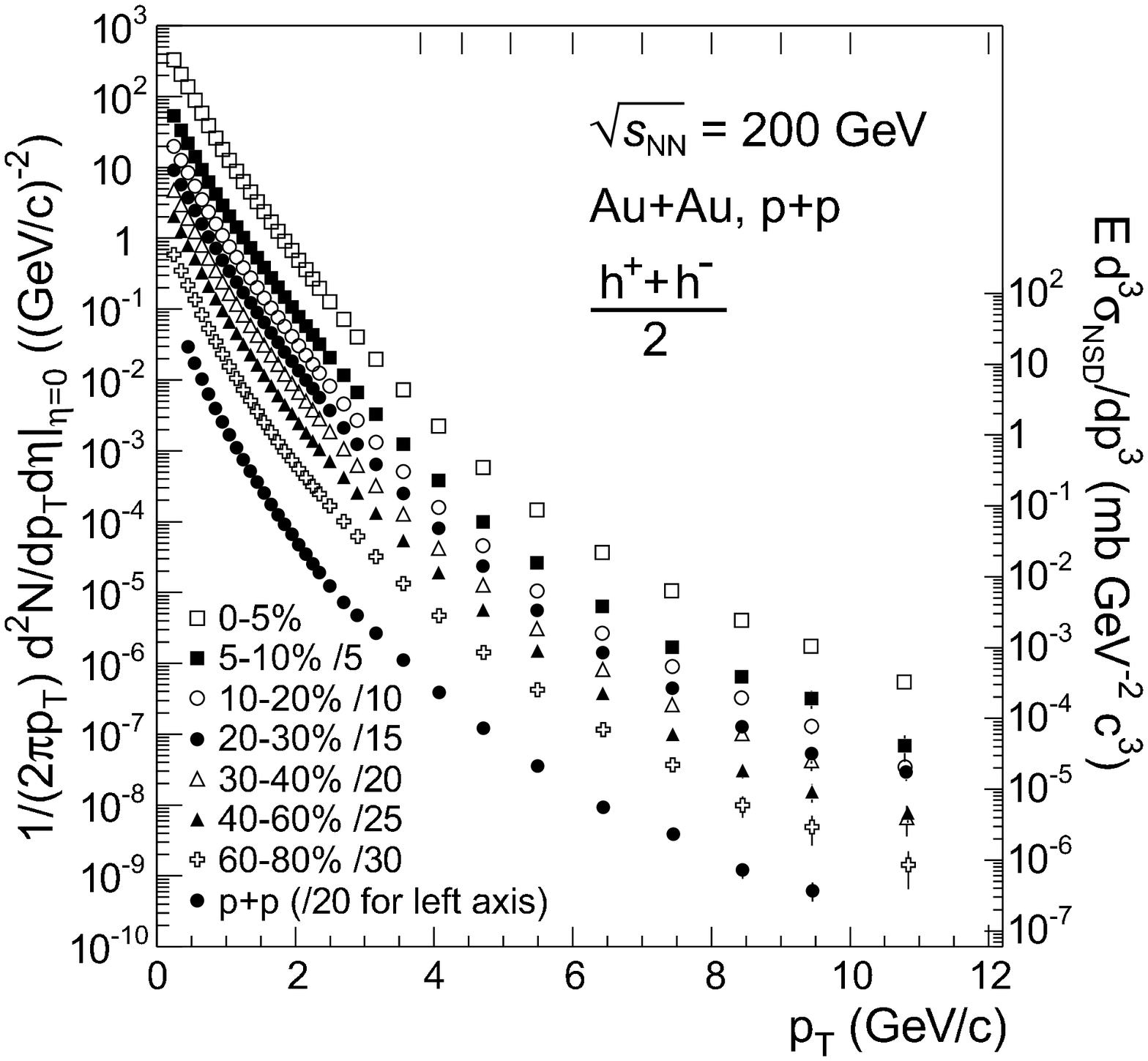}
\includegraphics[height=.3\textheight]{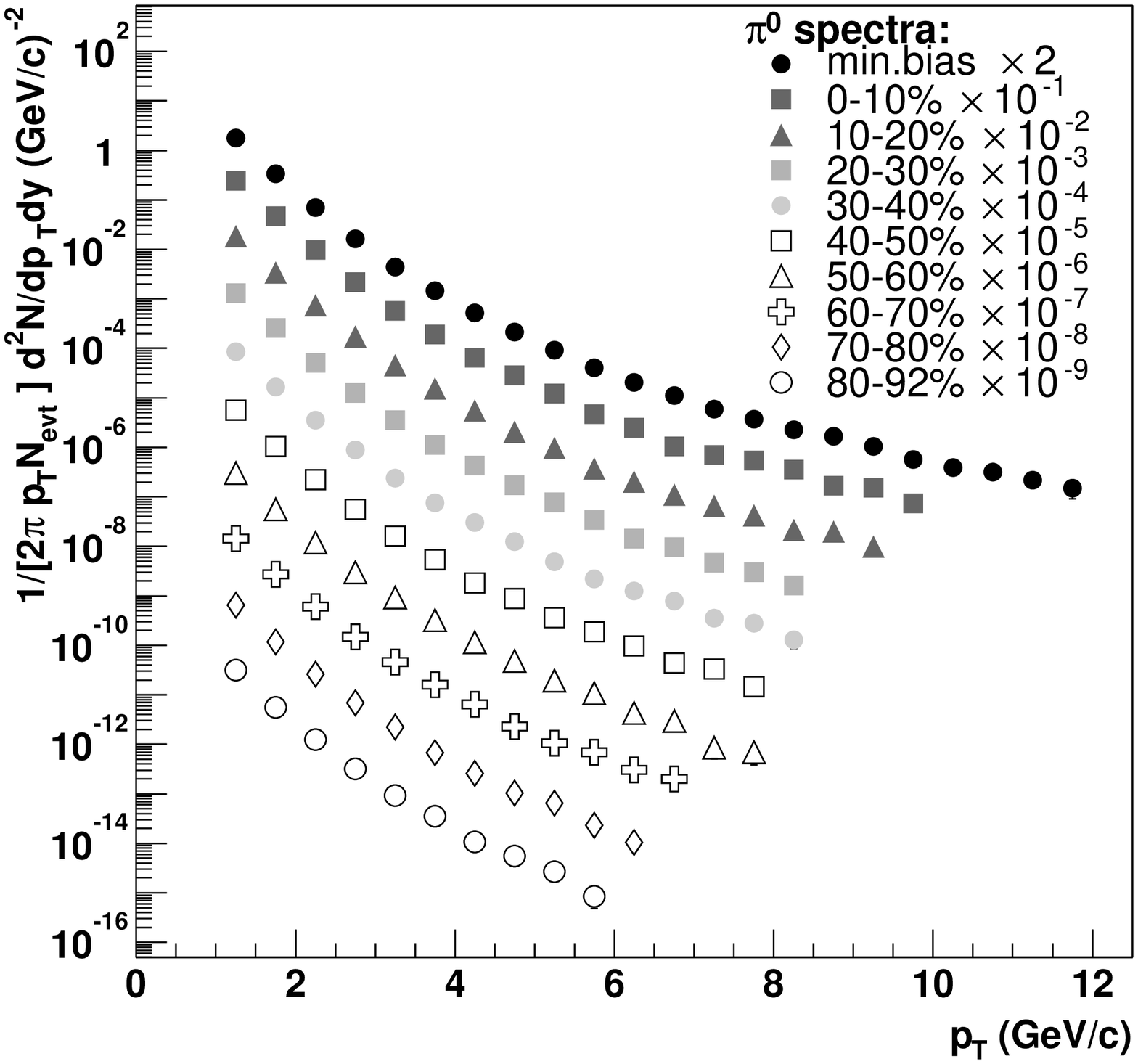}
\caption{High \pT\ single inclusive hadron spectra from 
centrality-selected Au+Au collisions at 
\sqrtsNN=200 GeV. Left: charged hadrons from STAR
\protect\cite{star:highpTAuAu200}. Right: \pizero\ from PHENIX
\protect\cite{phenix:highpT-pi0-200}.}
\label{fig:HighptSpectra}
\end{figure}

The foregoing analysis of hard scattering and high \pT\ hadron
production at midrapidity in $p+p$ and $p(d)+A$ collisions
demonstrates that the elementary production cross sections and nuclear
multiple scattering effects are well understood, providing a
foundation for jet tomographic studies in $A+A$ collisions. Within the
energy loss picture of multiple parton scattering and induced gluon
bremsstrahlung, high
\pT\ jet fragmentation will be softened and the final hadron spectra
suppressed in $A+A$ collisions \cite{Wang:1992xy}. The effective
partonic energy loss can therefore be deduced from measurements of
high \pT\ hadron suppression, providing a direct measurement of the
initial gluon density. The experimental input for this study is shown
in Fig.~\ref{fig:HighptSpectra}. Charged hadron and \pizero\ single
particle inclusive spectra from centrality-selected $Au+Au$ collisions
have been measured with high precision over a very broad \pT\ range
\cite{star:highpTAuAu200,phenix:highpT-pi0-200}.

The inclusive hadron spectrum in $A+A$ collisions can be calculated
using a LO pQCD model \cite{Wang:1998ww,Wang:2003mm},
\begin{eqnarray}
  \frac{d\sigma^h_{AA}}{dyd^2p_T}&=&K\sum_{abcd} 
  \int d^2b d^2r dx_a dx_b d^2k_{aT} d^2k_{bT}  t_A(r)t_A(|{\bf b}-{\bf r}|) 
  g_A(k_{aT},r)  g_A(k_{bT},|{\bf b}-{\bf r}|) 
  \nonumber \\ &\times&
  f_{a/A}(x_a,Q^2,r)f_{b/A}(x_b,Q^2,|{\bf b}-{\bf r}|)
  \frac{D_{h/c}^\prime (z_c,Q^2,\Delta E_c)}{\pi z_c}  
  \frac{d\sigma}{d\hat{t}}(ab\rightarrow cd), \label{eq:nch_AA}
\end{eqnarray}
with medium modified fragmentation functions $D_{h/c}^\prime$
given by Eq.~(\ref{modfrag}).

Assuming one-dimensional longitudinal expansion and gluon density
$\rho_g(\tau,r)$ proportional to the transverse density of participant
nucleons, the impact parameter and path length dependence of the
energy loss is [Eq.~(\ref{eq:effloss})]
\begin{equation}
\Delta E(b,r,\phi)\approx \langle\frac{dE}{dL}\rangle_{1d}
\int_{\tau_0}^{\Delta L} d\tau\frac{\tau-\tau_0}{\tau_0\rho_0}
\rho_g(\tau,b,\vec{r}+\vec{n}\tau),
\label{total-loss}
\end{equation}
\noindent
where $\Delta L(b,\vec{r},\phi)$ is the path length in matter for a
jet produced at $\vec{r}$ and traveling in direction $\vec{n}$ with
azimuthal angle $\phi$ relative to the reaction plane, in a collision
with impact-parameter $b$. $\rho_0=\langle \rho_g\rangle(\tau_0)$ 
is the average 
initial gluon density in central collisions at $\tau_0$ 
 and $\langle dE/dL\rangle_{1d}$ 
is the average parton energy loss
in a one-dimensionally expanding medium with an initial uniform gluon 
density $\rho_0$. The corresponding energy loss 
in a static medium with a uniform gluon density 
$\rho_0$ within radius $R_A$ is \cite{Wang:2002ri}
$dE_0/dL=(R_A/2\tau_0)\langle dE/dL\rangle_{1d}$.
We will use the parameterization in Eq.~(\ref{eq:th-loss})
for the effective energy dependence.

\begin{figure}
\includegraphics[width=.5\textwidth]{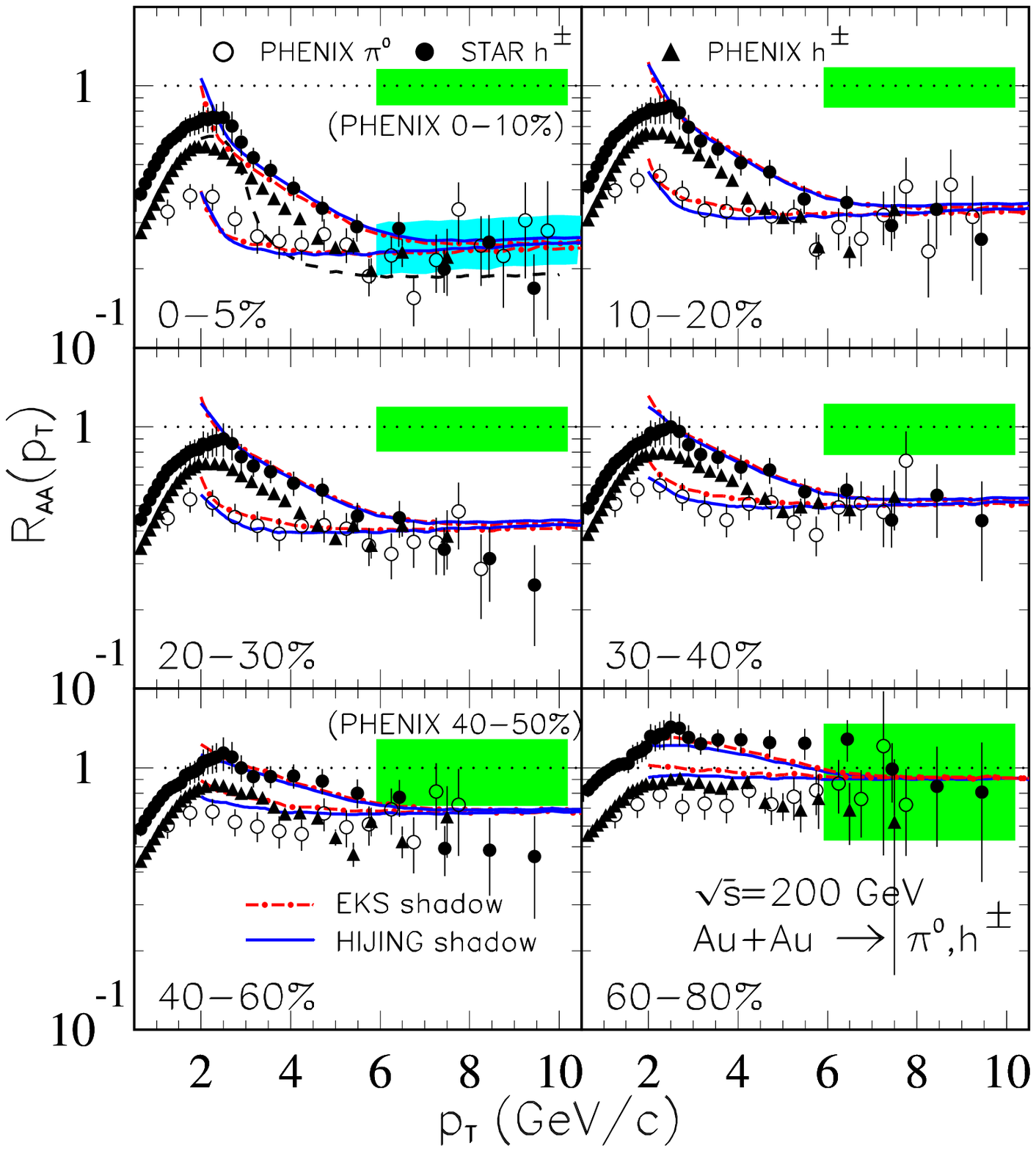}
\includegraphics[width=.5\textwidth]{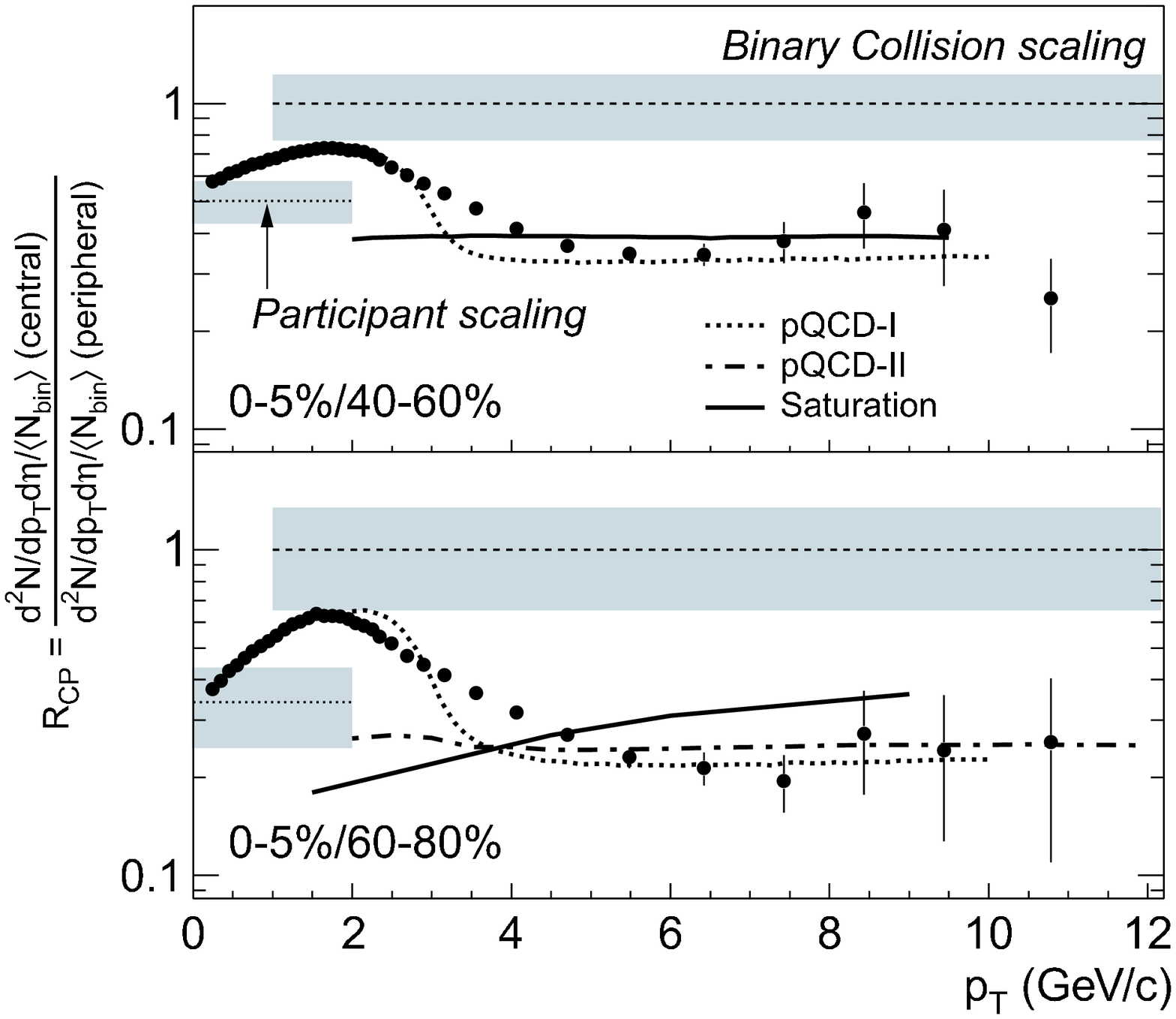}
\caption{Binary collision-scaled ratios of spectra shown in Fig. \ref{fig:HighptSpectra}
\protect\cite{star:highpTAuAu200,phenix:highpT-pi0-200}(see also PHOBOS
\protect\cite{phobos:highpT200,phobos:highpTdAu} and BRAHMS
\protect\cite{brahms:highpTdAu}).
Left: Centrality dependence of \RAA\ vs. \pT. Theory curves are from
calculations described in text. Boxes at \RAA=1 indicate the
(multiplicative) uncertainty due to binary collision scaling.  The
hatched region on the central collision theory curve indicates
variation in $\epsilon_0$ of $\pm0.3$ GeV/fm (Eq. \ref{eq:th-loss}).
Right:
\RCP\ [Eq.~(\ref{eq:RCP})] 
for central (0-5\%) over peripheral (40-60\%, 60-80\%) spectra, 
for charged hadrons from STAR \protect\cite{star:highpTAuAu200}.}
\label{fig:HadronSuppression}
\end{figure}

Fig. \ref{fig:HadronSuppression}, left panel, shows \RAA\ derived from
the data in Fig. \ref{fig:HighptSpectra}, with additional charged
hadron measurements from PHENIX
\cite{phenix:highpTAuAuCharged200}. The figure shows comparable
charged hadron \RAA\ from STAR (filled circles) and PHENIX
(triangles), which are seen to differ in overall normalization by
$\sim$20\%. This difference can be attributed to the $\sim$20\%
difference in normalization of the $p+p$ reference spectrum used in
the two analyses (denominator of Eq. \ref{eq:RAB})
\cite{star:highpTAuAu200,phenix:highpTAuAuCharged200}. The measurements 
are however consistent within systematic uncertainties 
and for the purpose of this discussion the difference can be taken as
an indication of the precision of the measurements. The figure shows
a marked effect: high \pT\ hadron production in central $Au+Au$
collisions is suppressed by a factor five relative to binary collision
scaling. The suppression has strong centrality dependence, with hadron
yield in the most peripheral collisions consistent with binary
collision scaling.  Similar results have been obtained by PHOBOS
\cite{phobos:highpT200,phobos:highpTdAu} and BRAHMS
\cite{brahms:highpTdAu}. This
suppression stands in strong contrast to the {\it enhancement} in high
\pT\ hadron production observed for $d+Au$ collisions in
Fig. \ref{fig:InclusiveSuppressionPR} and \ref{fig:Cronin}. The
strong yield suppression in central Au+Au contrasted with the
enhancement in d+Au collisions clearly demonstrates that the central
Au+Au suppression at high \pT\ is due to final-state interactions in
the fireball generated by the collision.

The solid lines in Fig. \ref{fig:HadronSuppression}, left panel, are
parton model calculations according to Eq.~({\ref{eq:nch_AA}). The
fit to the observed factor of five suppression at large \pT\ in the more
central collisions results in the parameters $\epsilon_0=1.07$
GeV/fm, $\mu=1.5$ GeV for the effective quark energy loss in
Eq.~(\ref{eq:th-loss}).  The difference between charged hadron and
\pizero\ spectra at intermediate \pT\ is attributed to a
non-perturbative component of hadron production, which we will return to in
the next subsection.  To demonstrate the sensitivity to the
parameterized partonic energy loss in the intermediate \pT\ region,
the dashed curves show
\RAA\ for charged hadrons excluding the soft component \cite{Wang:2003mm}.
Once these parameters are fixed by fitting to central collisions, the
centrality dependence of the suppression is a prediction of the model.

The suppression factor at
the highest measured \pT\ is shown for charged hadrons in
Fig. \ref{fig:HadronSuppression}, right panel
\cite{star:highpTAuAu200} (\RCP\ 
extends the \pT\ reach and reduces the systematic uncertainties
relative to \RAA\ \cite{star:highpTAuAu200}).
The suppression again is seen to be a factor five, independent of
\pT. This \pT-(in)dependence is well reproduced by the pQCD calculations
incorporating partonic energy loss (pQCD-I \cite{Wang:2003mm},
pQCD-II \cite{Vitev:2002pf}). The \pT-independence of the suppression 
is not a trivial effect: it results from the subtle
interplay between Cronin enhancement, shadowing of the structure
functions, and jet quenching in a parton model with energy loss \cite{Vitev:2002pf}.

The suppression for \pizero\ (left panel) varies only slightly over a
very broad \pT\ range, down to $\pT\sim$1-2 GeV/c. A constant or
logarithmic jet energy dependence of the energy loss as in cold
nuclear matter generates a suppression factor that slowly rises with
\pT\ \cite{Wang:2002ri}. Thus, the energy dependence of the partonic
energy loss seen in A+A collisions is quite different from that in DIS
off cold nuclei. This is an indication of detailed balance at play in
a thermal system. Detailed balance also leads to a slight rise in the
calculation of \RAA\ at $\pT\lt4$ GeV/$c$, where the fragmentation
picture gradually loses its validity and hadron production may be
influenced by non-perturbative effects, especially for kaons and
baryons. For $\pT\gt5$ GeV/$c$, these non-perturbative effects are
expected to diminish and indeed the \pizero\ and charged hadron
suppression factors converge in that region.

Partonic energy loss calculations reproduce the centrality dependence
of the suppression well. The solid curves in the left panel of
Fig.~\ref{fig:HadronSuppression} are calculated using partonic energy
loss from Eq.~(\ref{eq:th-loss}) with parameters fit to the central
collision data for \pizero\ and charged hadrons. The centrality
dependence is determined by the averaged total energy loss
[Eq.~(\ref{total-loss})], which results in an effective surface
emission of the surviving jets. Jets produced at the core of the
overlap region are strongly suppressed, since they lose the largest
amount of energy. The centrality dependence of the suppression is
found to be dominated by the geometry of the produced dense matter
rather than the length dependence of the parton energy loss. The
non-Abelian nature of the path length dependence of the energy loss
must be tested in other ways, in particular via central $A+A$
collisions with varying nuclear size.

Transverse expansion of the bulk medium can also be considered in the
parton model calculation. Since the total energy loss in
Eq.~(\ref{eq:effloss}) depends on a path integral of the parton
density profile $\rho(r,\tau)$, the transverse expansion will increase
the duration of parton propagation in matter. However, the expansion will
also accelerate the reduction of the parton density along the
path. These effects compete and the final total parton energy loss
remains approximately the same as in the case without transverse
expansion \cite{Gyulassy:2001kr}.

It was proposed recently that gluon saturation effects
(Sect. \ref{sect:CGC}) can extend well beyond the saturation momentum
scale \Qs, resulting in hadron suppression relative to binary scaling
(\RAB\lt1) for hadron $\pT\sim5-10$ GeV/c at RHIC energies
\cite{Kharzeev:2002pc} as shown by the solid lines in the right panel
of Fig.~\ref{fig:HadronSuppression}.  Since this suppression
originates in the properties of the incoming nuclear wave-function,
hadron production in $d+Au$ collisions should also be suppressed by
this mechanism \cite{Kharzeev:2002pc}.  Experimentally, however, an
enhancement in mid-rapidity hadron production in $d+Au$ is seen
instead (Figs. \ref{fig:InclusiveSuppressionPR} and \ref{fig:Cronin}
\cite{phenix:highpTdAu,star:highpTdAu,phobos:highpTdAu,brahms:highpTdAu}),
even in central $d+Au$ collisions \cite{star:highpTdAu} where
saturation effects should be most pronounced. The observed enhancement
is at variance with saturation model expectations
\cite{Kharzeev:2002pc}.

The saturation model calculation has since been developed to include
the Cronin effect due to classical elastic scattering
\cite{Kharzeev:2003wz}. Quantum evolution
cancels the Cronin enhancement, however, leading to suppression of
high \pT\ hadrons at high energies or large rapidities. Suppression at
large rapidity might however not be a unique signature of gluon saturation.
For instance, limiting fragmentation of valence quark jets could
also result in hadron suppression at forward rapidity \cite{Accardi:2004ut}.
Discrimination of these mechanisms may be possible 
by utilizing di-hadron correlations, which we discuss below.

\subsection{Parton Recombination}
\label{sec:recon}

The hadron suppression factor \RAA\ in
Fig.~\ref{fig:HadronSuppression} is different for charged hadrons and
\pizero\ in the intermediate $\pT\sim2-5$ GeV region. 
More detail of this effect is seen in 
Fig.~\ref{fig:RCPpid}. The left panel shows that \pizero\ are
strongly suppressed in central collisions whereas
protons are not suppressed in this region. The right panel shows
similar systematic behavior  for strange meson and
baryon yields. It will also be seen that the azimuthal
anisotropy (\vtwo) depends on hadron flavor at intermediate
\pT, with the anisotropy of baryons larger
than that of mesons \cite{star:highpTLamK200}.

\begin{figure}
\includegraphics[width=.45\textwidth]{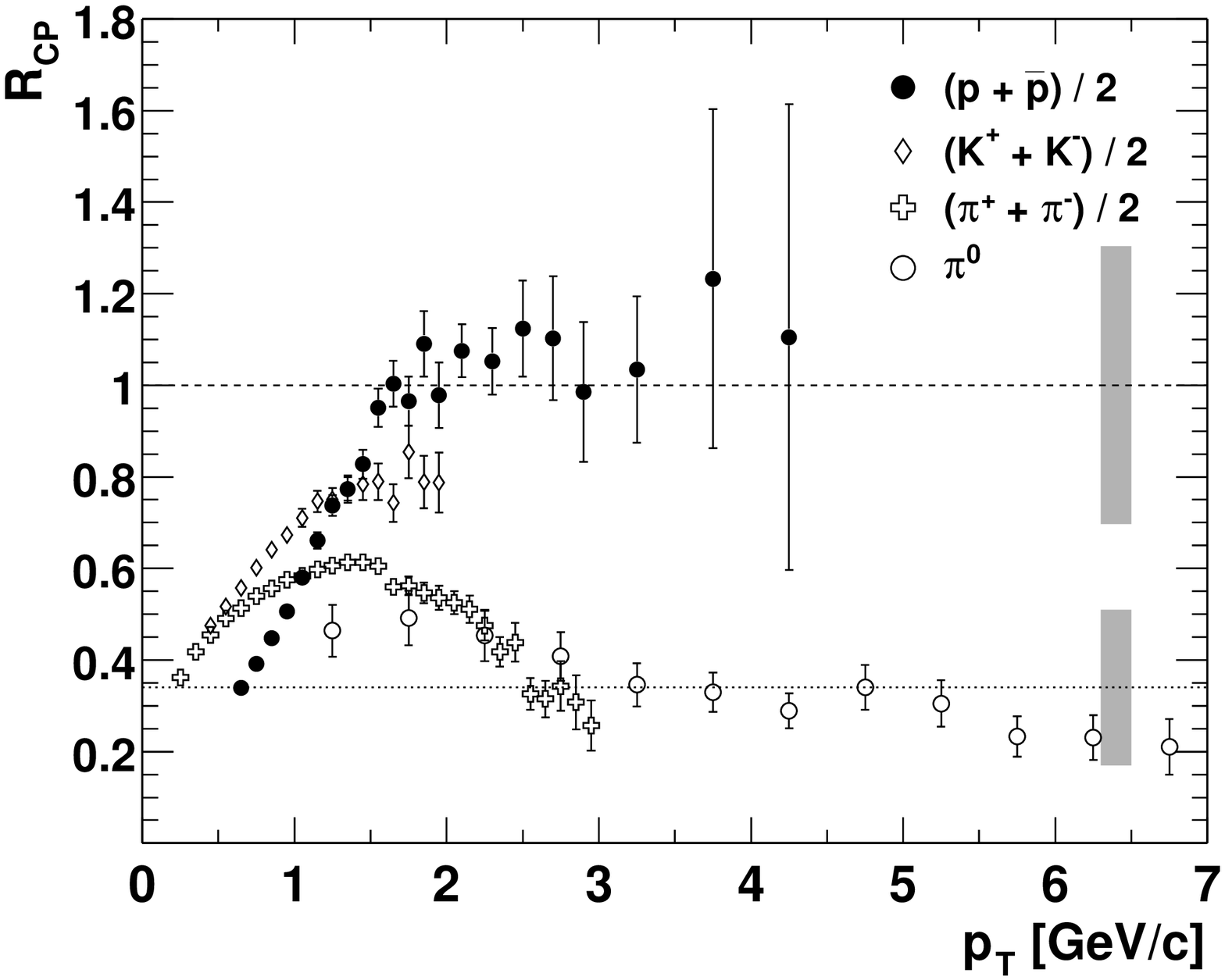}
\includegraphics[width=.54\textwidth]{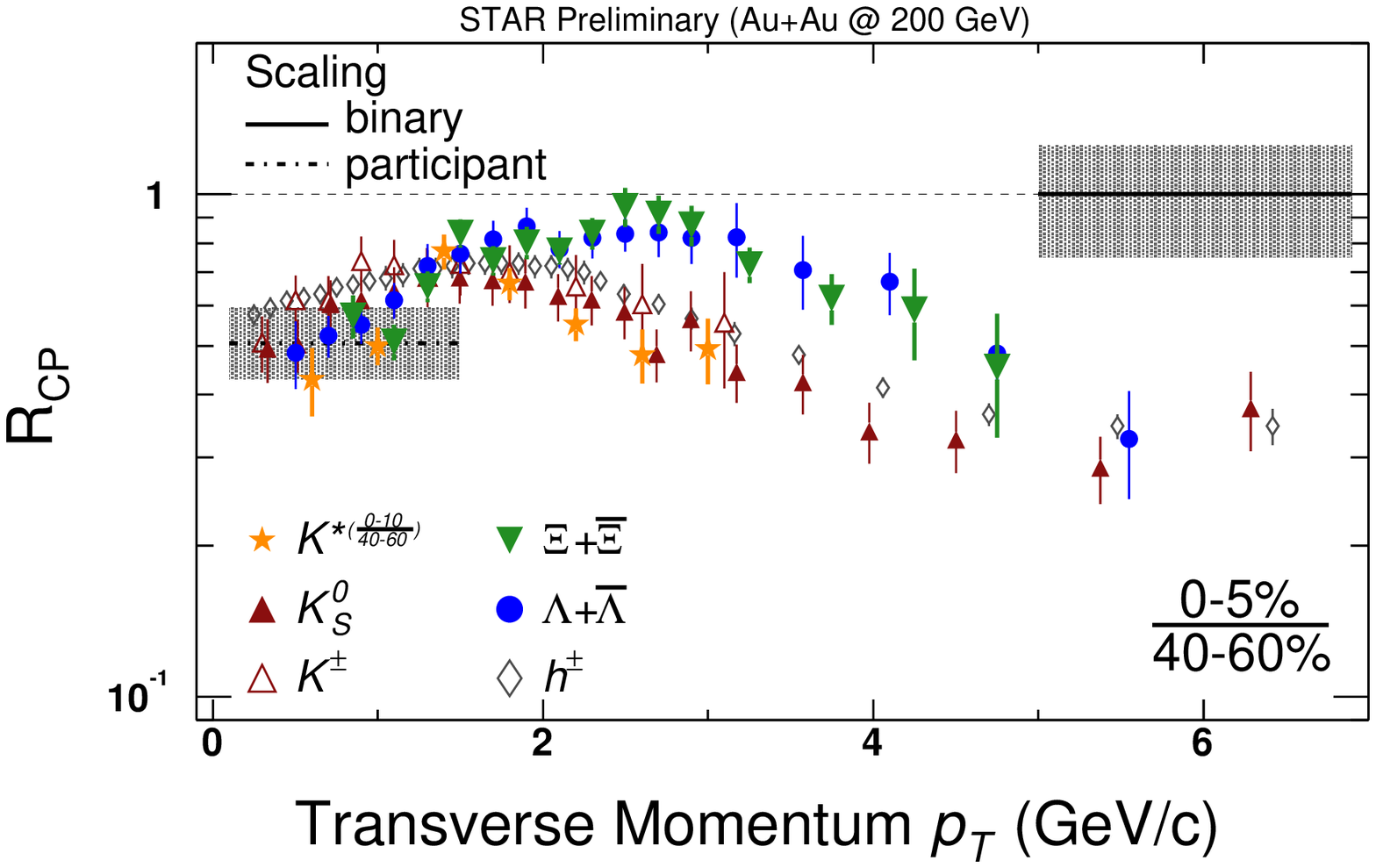}
\caption{Binary collision-scaled ratio of yields \RCP\ (eq. \ref{eq:RCP}) from
central relative to peripheral collisions  for
identified particles. Left: protons and pions from PHENIX
\protect\cite{phenix:pidspectra200}. Right: strange mesons and baryons from
STAR \protect\cite{star:highpTLamK200,Lamont:2004qy} compared to 
charged hadrons from Fig. \ref{fig:HadronSuppression}.
}
\label{fig:RCPpid}
\end{figure}

In the intermediate \pT\ region, where the parton energy loss is
strongly influenced by the medium through absorption of thermal
partons, fragmentation or hadronization should also be modified by the
presence of other partons. These nonperturbative effects invalidate
the picture of independent parton fragmentation in the vacuum. For
instance, the leading parton may pick up another parton from the
medium before hadronization, a scenario referred to as recombination
or coalescence. Such recombination processes were long ago proposed as
the dominant mechanism for hadron production in the beam fragmentation
region of hadron collisions \cite{Das:1977cp,Hwa:1980pn}. Since
recombination is a non-perturbative process, hadron wavefunctions
should be taken into account along with the initial parton
distributions. Alternatively, the recombination processes can be
described as higher-twist corrections to the fragmentation functions,
since they involve higher-twist matrix elements of the overlap between
two- or three-parton operators and the final hadronic state. Such an
approach has been applied to $D$ meson production to
explain the $D^+ - D^-$ asymmetry in the forward rapidity region of
$h+p$ collisions \cite{Braaten:2002yt}.

To take this effect into account in the simple parton model a nuclear
dependent soft component, assumed to be proportional to $\langle
N_{\rm binary}\rangle$, is added to kaon and baryon fragmentation
functions so that in central $Au+Au$ collisions the ratio
$(K+p)/\pi\approx 2$ at $p_T\sim 3$ GeV/$c$ and the ratio approaches
its value for $p+p$ collisions at $p_T>5$ GeV/$c$. The resulting
suppression for charged hadrons and its centrality dependence agree
well with the STAR data (Fig. \ref{fig:HadronSuppression}). The
$h^{\pm}$ and $\pi^0$ suppression are related via the $(K+p)/\pi$
ratio:
$R_{AA}^{h^{\pm}}=R_{AA}^{\pi^0}[1+(K+p)/\pi]_{AA}/[1+(K+p)/\pi]_{pp}$.
It is apparent from the data that $(K+p)/\pi$ converges for $Au+Au$
and $p+p$ collisions at $p_T>5$ GeV/$c$. Since such nonperturbative
effects are caused by presence of other produced partons, thermalized
or not, a qualitatively similar dependence of the Cronin enhancement
on hadron species should be seen in $p(d)+A$ collisions. Such an
effect is indeed observed \cite{star:dAupiKp}, though the enhancement
relative to $p+p$ collisions in the ratio of proton to pion yields at
intermediate \pT\ is markedly smaller in $d+Au$ than in $Au+Au$
collisions.

The observation of the flavor dependence of the suppression factor
and, in particular, the flavor dependence of the azimuthal anisotropy
(Sect. \ref{sect:HighpTvtwo}) have spurred new developments in the
recombination approach to hadron production in heavy-ion collisions
\cite{Fries:2004ej}. Parton recombination in heavy-ion collisions
was first investigated by Hwa and Yang \cite{Hwa:2002zu,Hwa:2003bn} 
and was later studied in detail in
terms of coalescence or recombination models
\cite{Voloshin:2002wa,Molnar:2003ff,Fries:2003vb,Greco:2003xt}. These
models generally assume two components of high \pT\ hadron production
in heavy-ion collisions. The recombination of partons from the bulk
medium dominates the production of low and intermediate \pT\ hadrons,
while high \pT\ hadrons result mainly from the fragmentation of parton
jets after propagating through the bulk medium and losing energy. In
these models, the number of mesons formed from parton recombination is
\cite{Greco:2003xt}
\begin{eqnarray}
N_M=g_M\int p_1\cdot d\sigma_1 p_2\cdot d\sigma_2 \frac{d^3p_1}{(2\pi)^32E_1}
\frac{d^3p_2}{(2\pi)^32E_2} 
f_q(x_1;p_1)f_{\bar{q}}(x_2;p_2) f_M(x_1,x_2;p_1,p_2),
\label{eq:Reco}
\end{eqnarray}
where $d\sigma$ denotes the differential element of the space-like
hypersurface of hadronization and $f_M(x_1,x_2;p_1,p_2)$ is the
coalescence probability given by the Wigner distribution function of
the meson in terms of constituent quarks. The statistical factor $g_M$
takes into account the number of internal quantum states in forming a
colorless meson from a colored quark and antiquark. The formula for
baryon production from parton recombination is similar, except that it
involves the baryon coalescence probability from three constituent
quarks. The quark (or antiquark) distributions should contain both
soft and hard components. The soft component arises from a thermalized
quark-gluon plasma with collective radial and elliptic flow though the
connection between the consituent quarks appearing in
Eq. (\ref{eq:Reco}) and the massles partons of a chirally restored
plasma is not specified at present in the model. The hard component is given by
the minijet distribution from a pQCD calculation, with transverse
momentum reduced in accordance with the observed partonic energy loss.

In the low and intermediate \pT\ regions, the hadron spectra are
dominated by recombination of thermal partons. Such a description may
be considered as a model for hadronization of the quark gluon
plasma. In this region the hadron spectra are determined by the
underlying parton spectra and the statistical factors $g_{M,B}$. If an
exponential form is assumed for the parton spectra, the proton to pion
ratio in a simple recombination model will be independent of
transverse momentum and is determined only by the ratio of statistical
factors $g_B/g_M$
\cite{Fries:2003vb}. If the contribution of $\Delta$-decay to the
proton yield is included and the contribution of higher resonance
decay to pions is excluded,
$p/\pi^-\sim1$. Fig.~\ref{fig:BaryonMesonRatio}, left panel, shows the
measured $p/\pi^-$ ratio for central $Au+Au$ collisions at
\sqrtsNN=200 GeV from PHENIX \cite{phenix:pidspectra200} compared to
the calculation by Greco, Ko and Levai \cite{Greco:2003mm}, one of
several recombination models \cite{Hwa:2002zu,Fries:2003vb}.  The
calculation describes the data well. The $p/\pi$ ratio indeed reaches
a value of about unity, much larger than the value achieved in $p+p$
collisions \cite{star:dAupiKp} and in gluon jets from $e^+e^-$
annihilation \cite{Abreu:2000nw}. The \pT\ dependence at lower \pT\
can be attributed to the contribution to the pion spectra from
resonance decays, whose importance diminishes at higher
\pT. Recombination models also successfully describe azimuthal
asymmetry measurements, as discussed in Sect. \ref{sect:HighpTvtwo}.

\begin{figure}
\includegraphics[width=.5\textwidth]{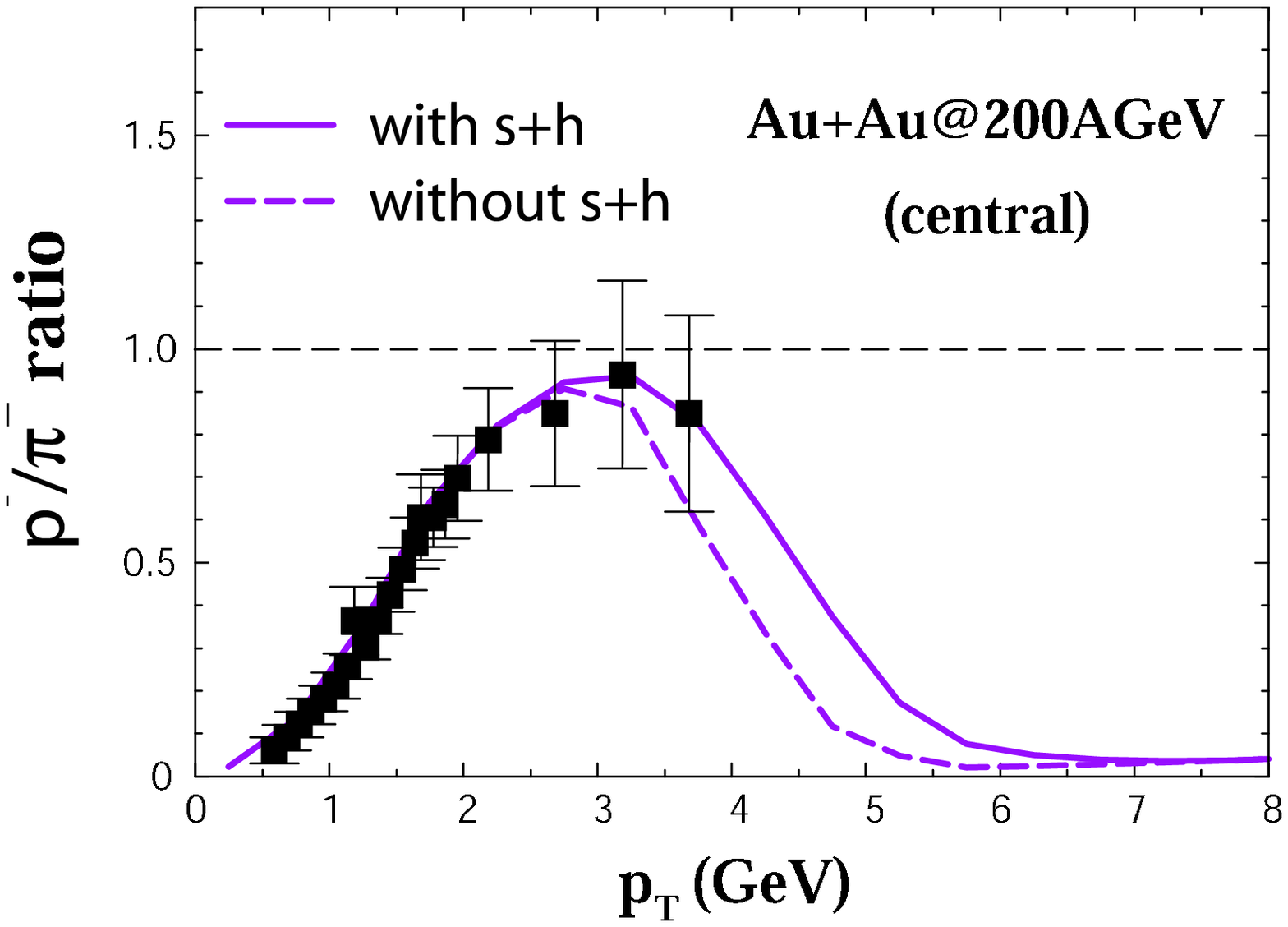}
\includegraphics[width=.45\textwidth]{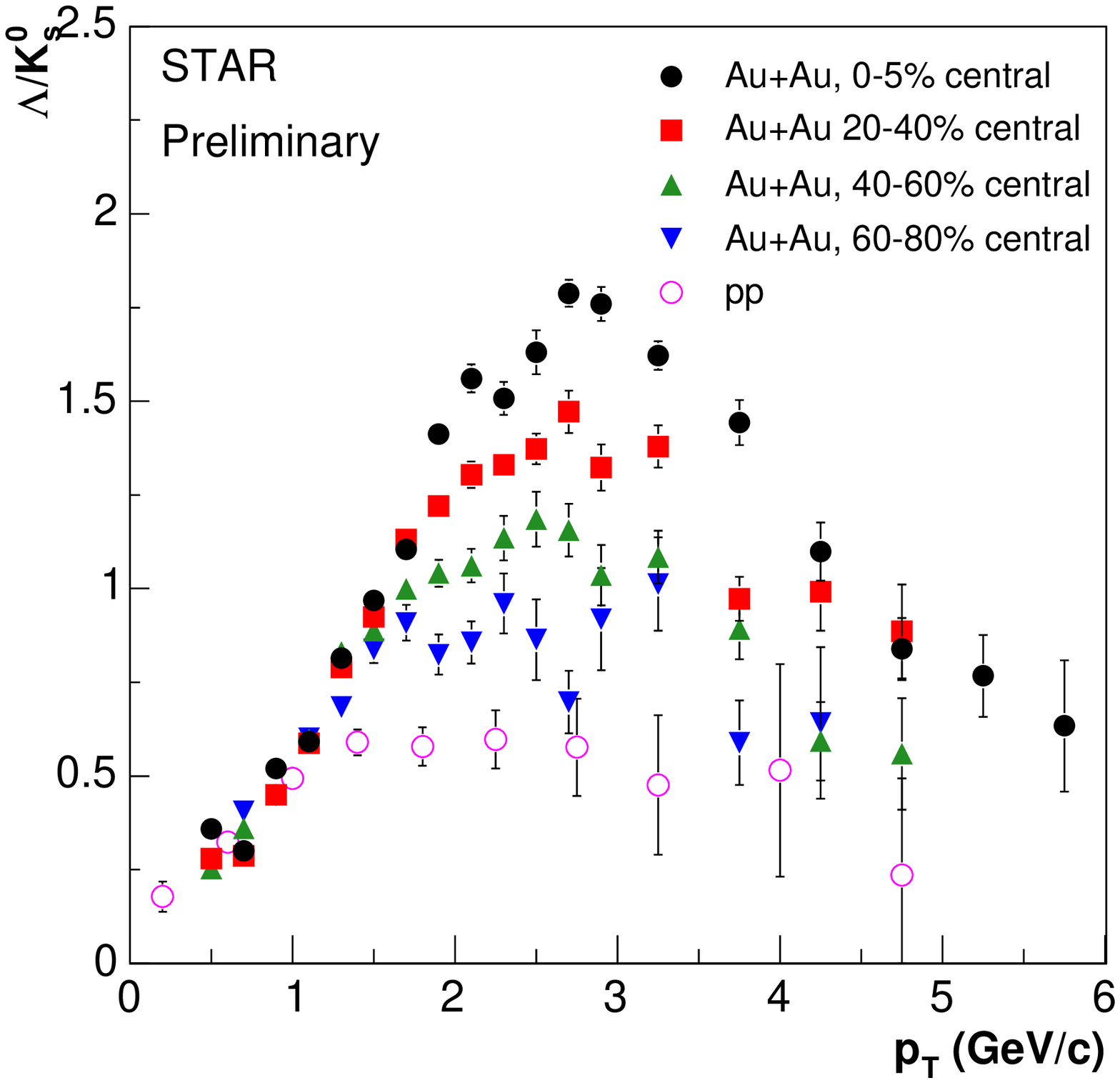}
\caption{\pT\ dependence of the ratio of baryon to meson yields at 
\sqrtsNN=200 GeV. Left: $\pbar/\pi^-$ for central $Au+Au$ collisions, from PHENIX
\protect\cite{phenix:pidspectra200}. Lines are recombination model
calculations \protect\cite{Greco:2003mm} with (solid) and without
(dashed) soft+hard contributions due to coalescence of minijet partons
with thermalized partons. Right: \lam/\kzeros\ for $p+p$ and
centrality-selected $Au+Au$ collisions, from STAR
\protect\cite{Lamont:2004qy}.}
\label{fig:BaryonMesonRatio}
\end{figure}

At large transverse momentum the underlying parton spectrum is
dominated by pQCD minijet production, which  has power-law
\pT\ dependence. The recombination mechanism in this case will 
always generate higher meson than baryon yields. For increasing
\pT, fragmentation will eventually dominate
recombination and the baryon/meson ratio should revert to that of
parton fragmentation in vacuum, consistent with the
expectation that recombination is a higher-twist process that is
suppressed at high \pT. This \pT-dependence of the relative yield
of mesons and baryons is common to a variety of recombination models
\cite{Hwa:2002zu,Fries:2003vb,Greco:2003xt}. It is in broad agreement 
with the data in Fig. \ref{fig:BaryonMesonRatio}, right panel, which
shows a large enhancement of \lam/\kzeros\ for more central $Au+Au$
collisions at intermediate \pT\ but convergence of the \lam/\kzeros\
ratio for all Au+Au centralities and p+p collisions at higher \pT.

The model of Greco, Ko and Levai includes an additional pick-up
process by leading hard partons, corresponding to the recombination of
a parton from the hard component and a parton from the soft component
\cite{Greco:2003xt}. Such soft+hard
recombination can be significant at intermediate \pT\ as shown by the
dashed line in Fig.~\ref{fig:BaryonMesonRatio}, left panel. This
process will be critical for addressing the observed two-particle
correlation of the same-side jet that will be discussed in the next
subsection.

Recently, a jet fragmentation model was proposed \cite{Hwa:2003ic} in
which parton recombination is the hadronization mechanism even in
vacuum. The medium effect at intermediate $p_T$ is a natural
consequence of the hadronization in medium in this model, which also
leads to flavor dependence of the Cronin enhancement in $d+Au$
collisions \cite{Hwa:2004zd,Hwa:2004yi}. Quantitative tests of recombination
models and the determination of their parameters are best done using
dihadron correlations. These studies are now underway.

As discussed in previous sections, measurements of low \pT\ inclusive
spectra and elliptic flow can be reproduced well by calculations
incorporating ideal hydrodynamics. It is to be expected that such a
hydrodynamic picture will break down at high \pT\ where partonic cross
sections are small and the mean free path is long, meaning that hard
partons cannot be brought to thermal equilibrium as readily as soft
partons. In this region, pQCD dynamics should dominate. To study the
transition between these regimes at intermediate \pT, hybrid
calculations have been carried out that combine hydrodynamics with a
pQCD parton model
\cite{Hirano:2003pw}. Since collective flow gives higher 
mass hadrons a larger \pT\ boost, it is expected that the effects of
collective flow will extend to larger \pT\ for baryons than for light
mesons. Such a two-component approach can indeed describe the observed
$p/\pi$ and \lam/\kzeros\ enhancements in $Au+Au$ collisions. However,
since the effects are purely kinematic it also predicts the same
enhancement for heavy mesons. Preliminary data
(\cite{Lamont:2004qy,star:QM04summary,Frawley:2004gj}, see also
Fig. \ref{fig:RCPpid} right panel) indicate that the enhancement is
rather more dependent on whether the hadron is a meson or baryon than
on its mass. However, heavier mesons such as the $\phi$ may have
smaller hadronic cross sections than baryons and not flow effectively
with the bulk medium during the hadronic stage. An interesting test
will be the measurement of azimuthal anisotropy of $\phi$ meson
spectra, which may be relatively insensitive to the hadronic dynamics
at the late stages of the expansion.

\subsection{Dihadron Spectra and Jet Quenching}

Jets are produced in pairs in leading order pQCD, and high \pT\
dihadron correlations in $p+p$ collisions should exhibit the
back-to-back jet structure of the underlying hard parton-parton
scattering. Jet quenching due to partonic energy loss in nuclear
collisions is expected to modify such correlations.

As we have shown in Fig. \ref{fig:CorrelationsPR}, left panel, the
relative azimuthal angle distributions of high \pT\ hadron pairs in
$p+p$ and $d+Au$ collisions exhibit the two-peak feature
characteristic of back-to-back jet pairs but the away-side correlation
vanishes in central $Au+Au$ collisions, consistent with the predicted
phenomenon of mono-jet production due to jet quenching
\cite{Pluemer:1995dp}. In these measurements the trigger 
hadron has $\pT^{\rm trig} \gt4$ GeV/c, with the relative azimuthal
angle plotted for all other hadrons having $2\lt\pT\lt\pT^{\rm trig}$
\cite{star:highpTbtob,star:highpTdAu}. All particles lie within
$|\eta|\lt0.7$. For $Au+Au$ collisions the effects of elliptic flow on
the dihadron distribution must be taken into account
\cite{star:highpTbtob}. The distributions are normalized to the number
of trigger hadrons, thereby measuring the probability to find an
associated hadron. The figure shows background-subtracted
distributions, where the background is defined as the yield in an
azimuthal interval orthogonal to the trigger where the correlated
yield in $p+p$ collisions is seen to be small. 

The strength and width of the small-angle peak ($\Delta\phi\sim0$) are
similar in all three distributions. The relative probability that the
small-angle hadron pair has the same vs. opposite charge sign is similar to that
measured in jets from \ePluseMinus\ annihilation (``charge ordering''
\cite{Abreu:1997wx,star:highpTbtob}). The contribution of resonances to 
this peak is estimated to be negligible
\cite{star:highpTbtob}. This series of evidence leads to the conclusion that the small-angle
correlation in all systems from $p+p$ to central $Au+Au$ results
dominantly from the fragmentation of jets. The large-angle peak
($\Delta\phi\sim\pi$) has similar strength and width in $p+p$ and
$d+Au$ collisions, indicating a hadron pair produced by the
fragmentation of a back-to-back jet pair. The large-angle high \pT\
dihadron correlation is however markedly suppressed in central $Au+Au$
collisions, suggesting significant suppression of the back-to-back jet
yield.

The strength of the correlation in $Au+Au$ can be quantified by
integrating the yield in the correlation peak and comparing to the
$p+p$ correlation via \cite{star:highpTbtob}
\begin{equation}
\IAA=\frac
{\int_{\Delta\phi_1}^{\Delta\phi_2}d\left(\Delta\phi\right)
\left[C^{AuAu}-B\left(1+2v_2^2\cos\left(2\Delta\phi\right)\right)\right]}
{\int_{\Delta\phi_1}^{\Delta\phi_2}d\left(\Delta\phi\right)C^{pp}},
\label{eq:IAA}
\end{equation}
where $C^{AuAu}$ and $C^{pp}$ are the measured dihadron azimuthal
angle distributions in $Au+Au$ and $p+p$ prior to background
subtraction. The subtracted background
$B\left(1+2v_2^2\cos\left(2\Delta\phi\right)\right)$ results from a
fit to the $Au+Au$ distribution in the background interval orthogonal
to the trigger direction. The elliptic flow coefficient \vtwo\ is
determined by independent measurements \cite{star:highpTflow}.  The
integration interval $\left[\Delta\phi_1,\Delta\phi_2\right]$ spans
either the small-angle or back-to-back peaks in the $p+p$
distribution. Figure
\ref{fig:STARcorrelations}, left panel, shows the centrality
dependence of \IAA\ for both same-side (upper) and back-to-back
(lower) dihadron pairs. As is also apparent in
Fig. \ref{fig:CorrelationsPR}, the same-side correlation strength for
all $Au+Au$ centralities is similar to $p+p$. In contrast, the
back-to-back correlation strength in $Au+Au$ is seen to vary markedly from
peripheral collisions, with strong correlation, to central collisions
where it is negligible \cite{star:highpTbtob}.
It is compelling to attribute the back-to-back suppression in central
collisions to jet quenching.

The population of jets contributing to the near-side and back-to-back
correlations may differ for two reasons. First, the requirement of a
pair of hadrons above threshold biases the near-side contribution to
jets with larger initial energy. Second, if the partonic energy loss
in the core of the fireball is large, the trigger will bias towards
those jets that are produced near the surface of the reaction volume,
heading outward. The jets that generate the same-side correlation
would therefore punch through the medium after losing a limited amount
of energy and fragment essentially in vacuum, leading to similar
near-angle dihadron correlations from $p+p$ to central $Au+Au$, as
observed. The jets recoiling against the trigger will however be
biased towards the population heading into the core of the fireball
and will be strongly suppressed, as observed. The total energy of the
jets, suppressed or not, is of course conserved but it may be
redistributed to softer hadrons with much broader angular
distributions. In the picture of partonic energy loss due to induced
radiation, the emitted gluons could further interact with other
partons and dissipate their energy to the medium. The hadronic
structure of the jet will thereby be lost, leading to strong
suppression of the back-to-back high \pT\ dihadron correlation. This
picture of trigger bias is also consistent with the suppression of
inclusive production at high \pT\ relative to binary collision scaling
(Fig. \ref{fig:HadronSuppression}), since jets will be emitted
dominantly from the surface and not the volume of the fireball.

\begin{figure}
\includegraphics[width=.48\textwidth]{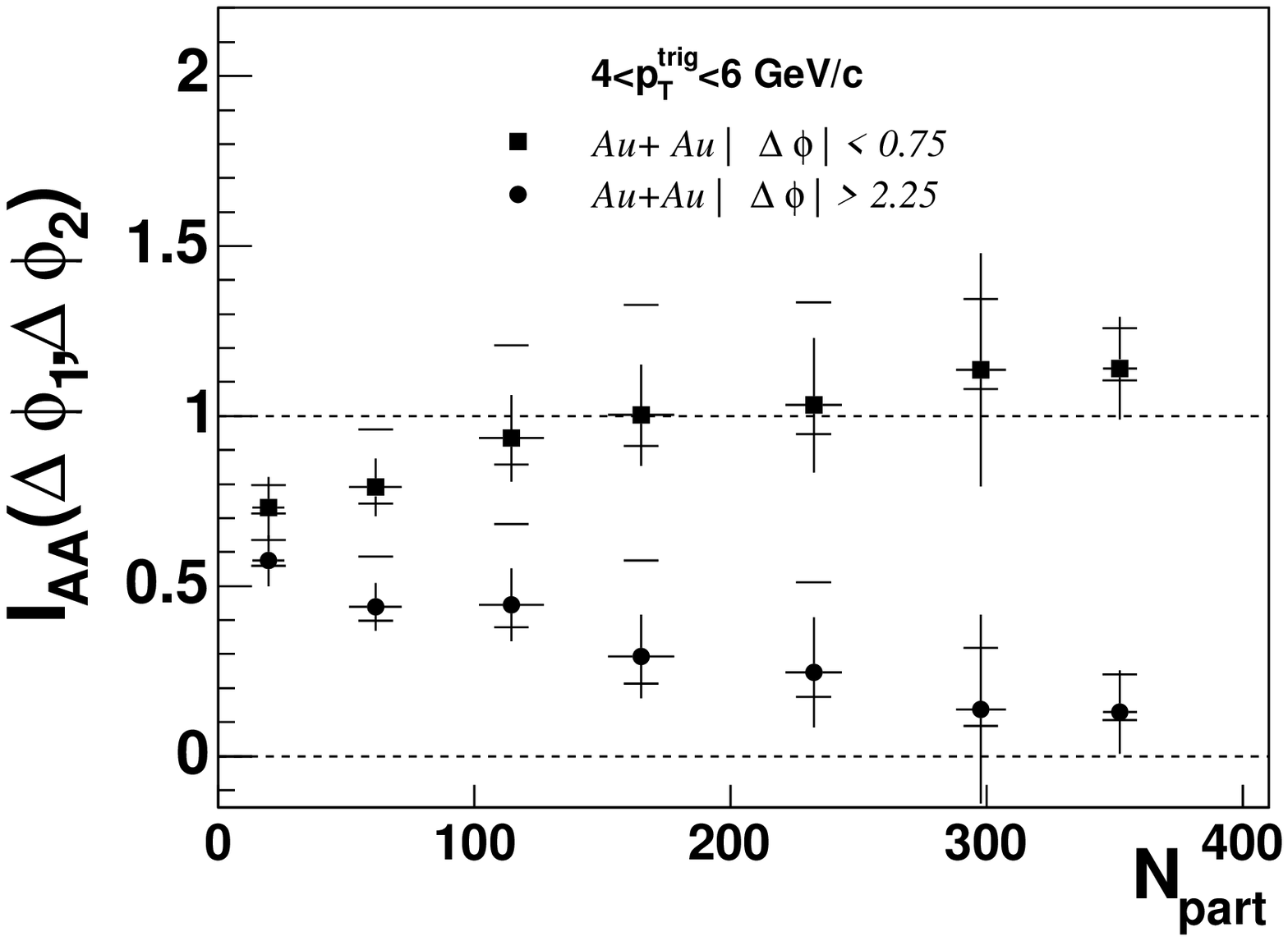}
\includegraphics[width=.48\textwidth]{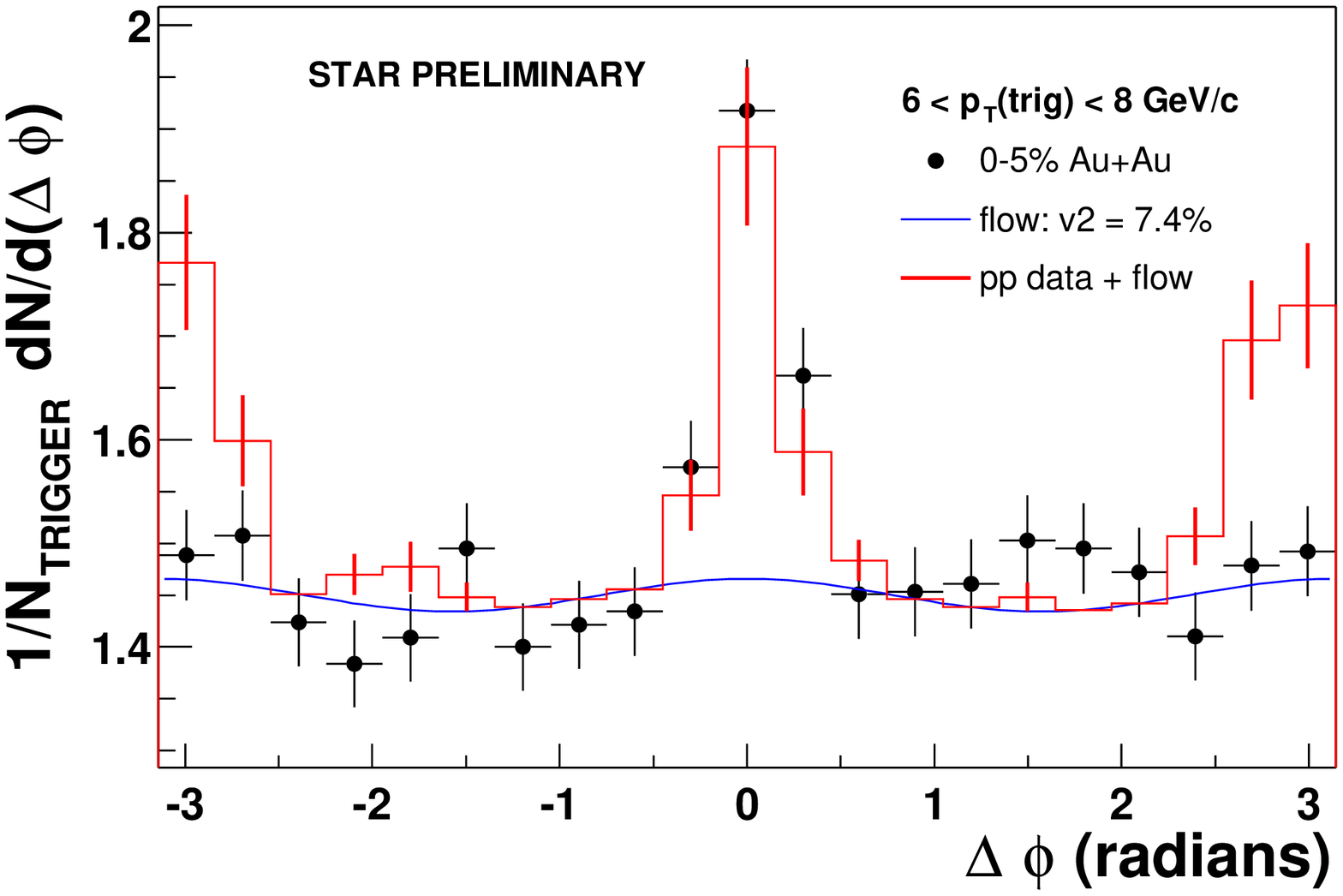}
\caption{High \pT\ dihadron distributions from 200 GeV $Au+Au$ and $p+p$ collisions, 
from STAR. Left: centrality dependence of near-angle and back-to-back
dihadron correlation strength in $Au+Au$ relative to $p+p$ (\IAA,
Eq. [\ref{eq:IAA}]) for $\pT(trig)\gt4$ GeV/c
\protect\cite{star:highpTbtob}. Central collisions correspond to large \Npart. 
Right: dihadron azimuthal distributions for $\pT(trig)\gt6$ GeV for
central $Au+Au$ (data points) and for $p+p$ plus a model of the
$Au+Au$ background (histogram) \protect\cite{Hardtke:2002ph}. Note the
rotation of the horizontal axis relative to
Fig. \ref{fig:CorrelationsPR}, left panel.}
\label{fig:STARcorrelations}
\end{figure}

It should be noted, however, that the correlations discussed thus far
have $\pT^{\rm trig} \gt4$ GeV/c. 
While the systematic features of near-angle
correlations in central $Au+Au$ are similar to those from jet fragmentation in
\ePluseMinus\ annihilation, Fig. \ref{fig:BaryonMesonRatio} shows that 
the inclusive hadron population in the range of $\pT^{\rm trig}\sim 4$
GeV/$c$ is significantly different in central $Au+Au$ collisions than
in vacuum jet fragmentation. It therefore cannot be excluded that
non-perturbative mechanisms contribute to the dihadron phenomenology
in Figs. \ref{fig:CorrelationsPR} and
\ref{fig:STARcorrelations}, left panel. 
Fig. \ref{fig:STARcorrelations}, right panel, shows the same dihadron
correlation analysis described above but for $\pT^{\rm trig} \gt6$
GeV/c \cite{Hardtke:2002ph}. The points are the measured distribution
for central $Au+Au$, in this case not background subtracted, while the
histogram is the sum of the measured $p+p$ correlation plus a model of
the central $Au+Au$ background including elliptic flow. The near-angle
peaks are again seen to be similar, while the back-to-back correlation
strength for central $Au+Au$ is again seen to be strongly suppressed
relative to the $p+p$ excess above background. The persistence of the
back-to-back suppression to higher \pT\ argues that it is indeed due
to jet quenching.

In the direction opposite the triggered hadron, the azimuthal angle
distribution measures the single inclusive hadron distribution of the
away-side jet. The width of the peak in $p+p$ collisions is
characteristic of the jet profile and is determined by both the
distribution of the intrinsic \pT\ perpendicular to the jet axis and
the relative transverse momentum between the two back-to-back jets.
In $d+Au$ collisions, initial multiple parton scattering may broaden
the dijet relative \pT\ and thus the back-to-back dihadron
correlation, but the integrated correlation strength should remain
approximately the same as in $p+p$ collisions since the integrated
area under the away-side peak is essentially determined by
the fragmentation function of the jet. In central $Au+Au$ collisions,
however, partonic energy loss will suppress the leading hadron
distribution in a jet, thus leading to the reduction of the away-side
jet peak as indeed shown by the experimental data.

For the small-angle pairs, the shape of the correlation in $p+p$
collisions is entirely determined by the intrinsic \pT\ perpendicular
to the jet axis. The integrated value of the correlation strength is
related to the ratio of dihadron and single hadron fragmentation
functions
\cite{Majumder:2004wh}. This correlation 
remains approximately the same in $p+p$, $d+Au$ and $Au+Au$
collisions, indicating jet fragmentation in all cases. According to
the picture of parton energy loss, a parton with reduced energy
fragments outside the dense medium, giving rise to a leading dihadron
correlation similar to that in the absence of energy loss. The lost
energy will be carried by the radiated gluons which in turn will only
contribute to soft hadrons along the direction of the triggered
hadron.

To study the back-to-back dihadron correlation, one can again apply a LO
pQCD parton model. The spectrum of back-to-back dihadrons
from the independent fragmentation of back-to-back jets can be
calculated as
\begin{eqnarray}
  E_1E_2\frac{d\sigma^{h_1h_2}_{AA}}{d^3p_1d^3p_2}&=&\frac{K}{2}\sum_{abcd} 
  \int d^2b d^2r dx_a dx_b d^2k_{aT} d^2k_{bT} 
 t_A(r)t_A(|{\bf b}-{\bf r}|) 
  g_A(k_{aT},r)  g_A(k_{bT},|{\bf b}-{\bf r}|)  \nonumber \\
  & & \hspace{-1.in}\times f_{a/A}(x_a,Q^2,r) 
  f_{b/A}(x_b,Q^2,|{\bf b}-{\bf r}|) D_{h/c}(z_c,Q^2,\Delta E_c)
  D_{h/d}(z_d,Q^2,\Delta E_d) \nonumber \\
 & & \hspace{-1.in}\times
 \frac{\hat{s}}{2\pi z_c^2 z_d^2} \frac{d\sigma}{d\hat{t}}(ab\rightarrow cd)
 \delta^4(p_a+p_b-p_c-p_d).
 \label{eq:dih}
\end{eqnarray}
Let hadron $h_1$ be the trigger hadron with $p_{T1}=p_T^{\rm
trig}$. We define the hadron-triggered fragmentation function (FF) as
the back-to-back correlation with respect to the triggered hadron:
\begin{equation}
  D^{h_1h_2}(z_T,\phi,p^{\rm trig}_T)=
  \frac{d\sigma^{h_1h_2}_{AA}/d^2p^{\rm trig}_T dp_Td\phi}
  {d\sigma^{h_1}_{AA}/d^2p^{\rm trig}_T},
\label{frg-htrig}
\end{equation}
similar to the direct-photon triggered FF in $\gamma$-jet events
\cite{Wang:1996yh,Wang:1997pe}. Here, $z_T=p_T/p^{\rm trig}_T$ and
integration over $|y_{1,2}|<\Delta y$ is implied. In a simple parton
model the dijets will be precisely back-to-back, but the initial parton
\pT\ distribution will give rise to a Gaussian
angular distribution.  In addition, \pT\ smearing within a jet must be
taken into account using a Gaussian distribution with a width of
$\sim0.6$ GeV/$c$.

Fig.~\ref{fig:BackToBackTheory}, left panel, shows the calculated
back-to-back correlations for charged hadrons in $Au+Au$ collisions
compared to the STAR data \cite{star:highpTbtob}. The same energy loss
that is used to calculate single hadron suppression 
also describes well the observed back-to-back suppression
and its centrality dependence. In central $Au+Au$ collisions, multiple
parton scatterings that induce partonic energy loss can also generate
smearing in the transverse momentum of the final parton before
fragmentation. This can further suppress the back-side correlation at
its peak \cite{Hirano:2003hq}. However, after
integration over the azimuthal angle, the smearing due to final state
$p_T$ broadening does not influence significantly the 
integrated suppression factor of the hadron-triggered FF.

\begin{figure}
\includegraphics[width=.46\textwidth]{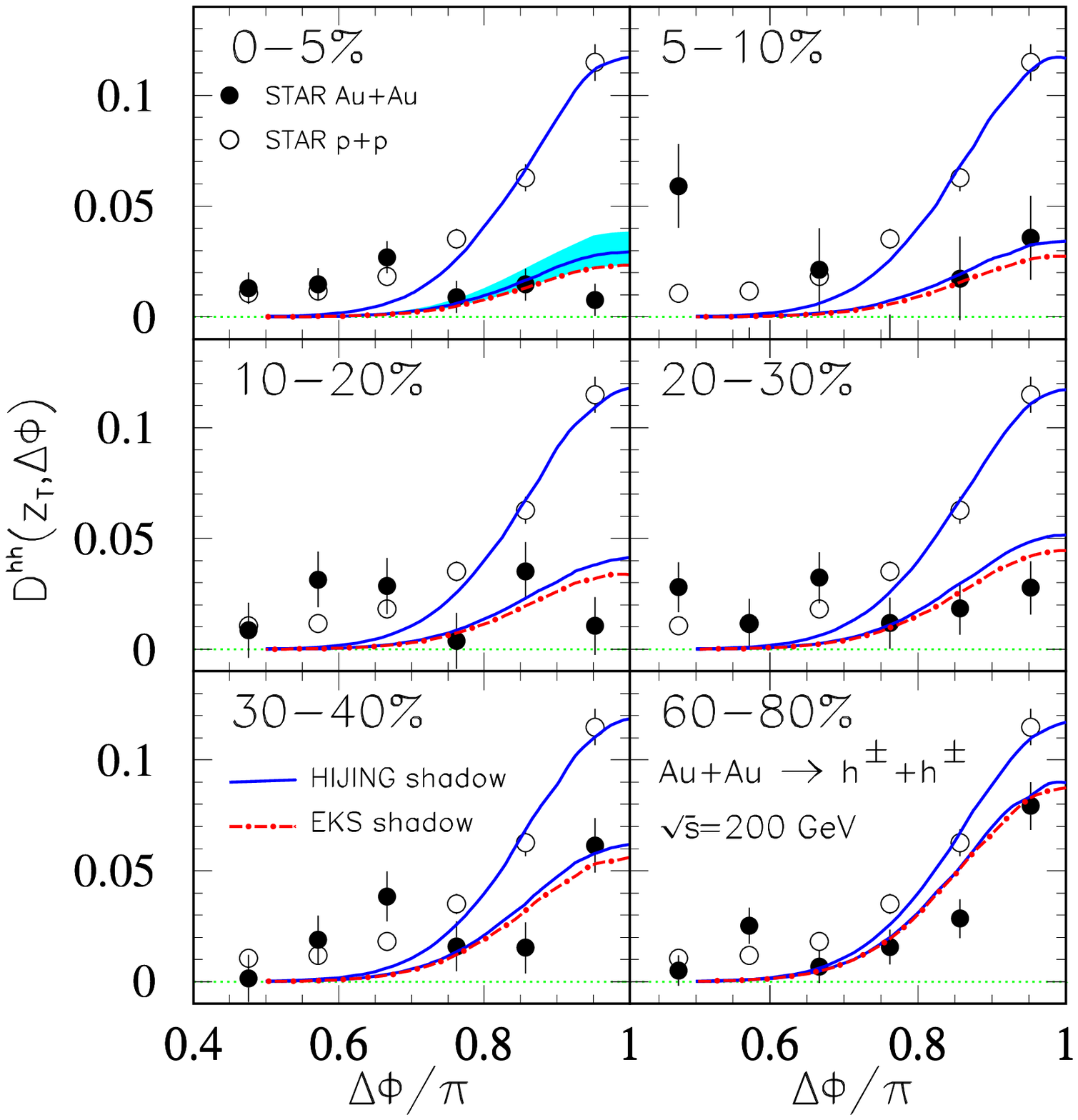}
\includegraphics[width=.52\textwidth]{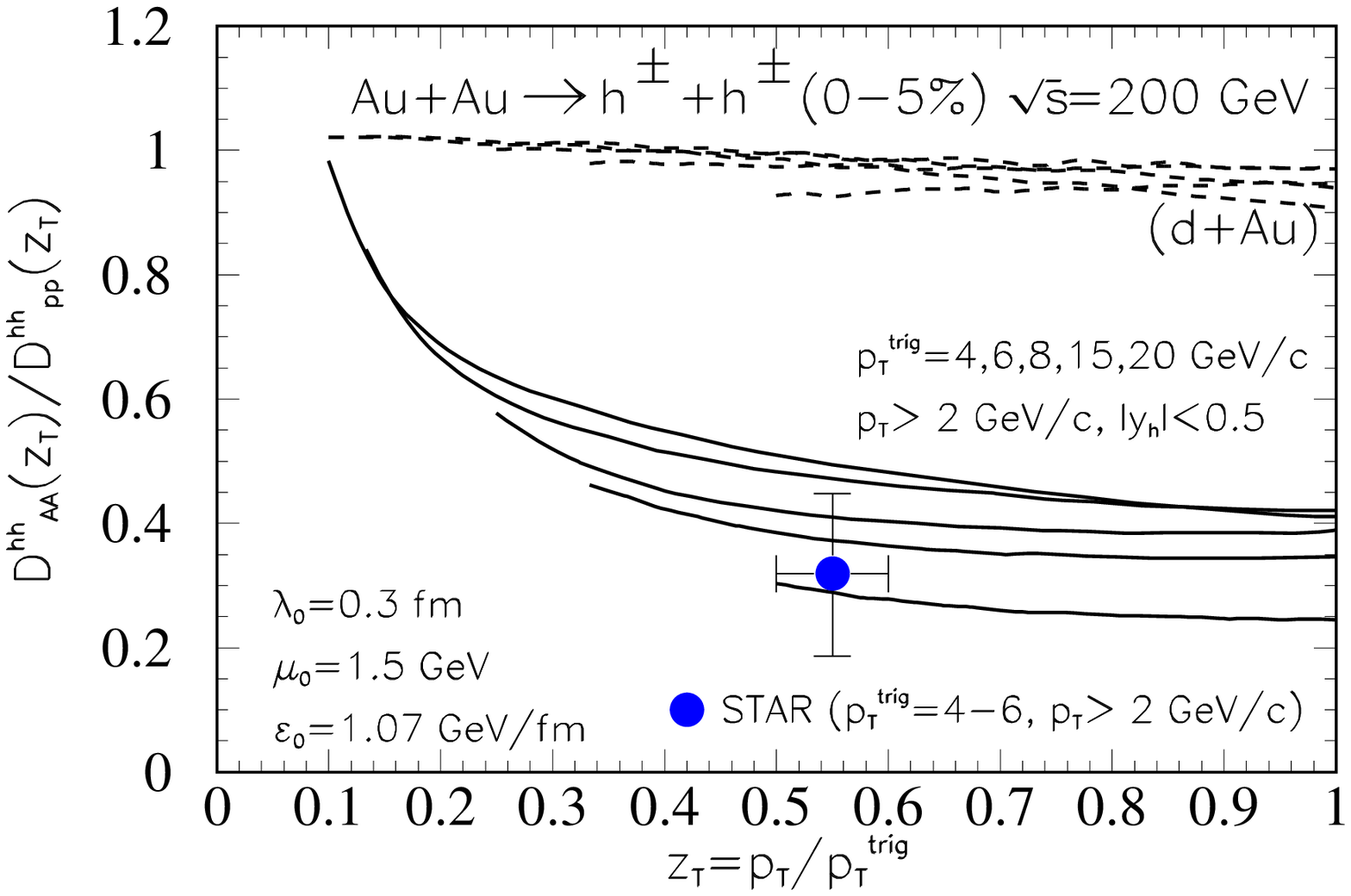}
\caption{Left: Back-to-back dihadron correlations 
for $Au+Au$ (lower curves) and $p+p$ (upper curves), with
$4{\lt}p^{\rm trig}_T\lt6$ GeV/$c$, $2{\lt}p_T{\lt}p^{\rm trig}_T$
GeV/$c$, and $|y|<0.7$, compared to background-subtracted STAR data \protect\cite{star:highpTbtob}.
Right: The suppression factor for hadron-triggered fragmentation
functions in central (0-5\%) $Au+Au$ (d+Au) collisions, compared to
STAR data \protect\cite{star:highpTbtob}.  }
\label{fig:BackToBackTheory}
\end{figure}

The hadron-triggered fragmentation function is obtained by integrating
over $\phi$, $D^{h_1h_2}(z_T,p_T^{\rm trig})=\int_{\pi/2}^{\pi} d\phi
D^{h_1h_2}(z_T,\phi,p^{\rm trig}_T)$. The dihadron suppression factor
defined by STAR \cite{star:highpTbtob} (Eq. [\ref{eq:IAA}]) is 
\begin{equation}
I_{AA}(z_T, p_T^{\rm trig})\equiv 
\frac{D^{h_1h_2}_{AA}(z_T,p_T^{\rm trig})}
{D^{h_1h_2}_{pp}(z_T,p_T^{\rm trig})},
\end{equation}
which is the modification factor of the hadron-triggered FF.
Fig.~\ref{fig:BackToBackTheory}, right panel, shows the suppression
factors of the hadron-triggered FF's for different values of $p^{\rm
trig}_T$ in central $Au+Au$ collisions, compared to a STAR data point
obtained by integrating the observed correlation over
$\pi/2<|\Delta\phi|<\pi$. The dashed lines illustrate the small
suppression of back-to-back correlations due to the initial nuclear
$k_T$ broadening in $d+A$ collisions.  The strong QCD scale dependence
of the fragmentation functions on $p^{\rm trig}_T$ is to a large
extent canceled in the suppression factor. The approximately universal
shape reflects the weak \pT\ dependence of the hadron spectra
suppression factor in Figs. \ref{fig:HadronSuppression}, due to the
unique energy dependence of parton energy loss.

Preliminary studies of back-to-back dihadron distributions with low
associated \pT\ have revealed a very broad azimuthal distribution at
low $z_T$ for central $Au+Au$ collisions, with no evidence of jet-like
azimuthal correlations \cite{star:mombalance1,star:mombalance2}. The
measured distributions are consistent with statistical momentum
balance by a large ensemble of recoiling hadrons
\cite{Borghini:2000cm} even for $p_T^{\rm trig}\gt6.5$ GeV/c 
\cite{star:mombalance1}. These preliminary results
suggest that the soft hadrons in the recoiling jet may be
significantly broadened, resulting from thermalization of the emitted
gluons and \pT\ broadening of the jet in the medium. More extensive
studies of the modification of the FF at low $z_T$ with higher
$p_T^{\rm trig}$ are needed to elucidate the picture.

The dihadron correlation on the same side also places important
constraints on coalescence models. The same-side correlations
characteristic of jet structure exclude the recombination of
uncorrelated partons from the thermal medium as the dominant mechanism
for hadron production at intermediate $p_T$. Such a jet structure on
the near side rather favors coalescence of a fast parton from the hard
scattering with a slower parton from the medium. In the case of
hard-soft coalescence, both the shape and strength of the correlation
may be different from vacuum fragmentation. In particular, if the
thermal medium develops longitudinal flow, coalescence of a leading
parton with a thermal parton will lead to broadening of the
correlation in rapidity.

\subsection{High $p_T$ Azimuthal Anisotropy}
\label{sect:HighpTvtwo}

Non-central collisions generate an initially anisotropic reaction zone
(Fig. \ref{fig:HydroEps}, left panel), with the long axis perpendicular to
the reaction plane. In pQCD, the final state partons following a hard
scattering have {\it a priori} no correlation with the azimuthal
orientation of the reaction plane. However, the average path length of
parton propagation in the medium will vary with the azimuthal angle
relative to the reaction plane, leading to an azimuthal dependence of
the total partonic energy loss. Thus, azimuthal anisotropy or ``elliptic
flow'' of high \pT\ hadrons relative to the reaction plane is a
consequence of partonic energy loss and is insensitive to effects of
initial state interactions \cite{Wang:2000fq,Gyulassy:2000gk}.

The experimental techniques for studying azimuthal anisotropy at high
\pT\ are the same as at low \pT\ (Sect. \ref{sect:EllipticFlow}). The strength 
of the correlation is measured by the second Fourier coefficient
\vtwo\ of the azimuthal distribution [Eq.~(\ref{eq:Flow})]. 
Special care must be 
taken to account for so-called ``non-flow'' effects contributing to
\vtwo\ which arise from multiparticle correlations that are unrelated 
to correlations with the reaction plane \cite{PoskanzerVoloshin}. 
Non-flow effects may result, for instance, 
from resonance decays, momentum conservation, and particularly at high
\pT, intra- and inter-jet hadron correlations. The standard method of 
correlating each high \pT\ hadron with the reaction plane
is equivalent to a
dihadron correlation analysis \cite{PoskanzerVoloshin}. Higher order
cumulants have been shown to be markedly less sensitive to non-flow
effects than the two-particle correlation methods
\cite{Borghini:2000sa,Borghini:2001vi}. 
It has recently been argued, however, that higher order cumulants are
susceptible to fluctuation effects and may thereby underestimate the
true correlation strength \cite{Miller:2003kd}. The two-particle and
higher order correlation measurements thus may bracket the true flow
\vtwo\ and are usually reported together. 

\begin{figure}
\includegraphics[width=.54\textwidth]{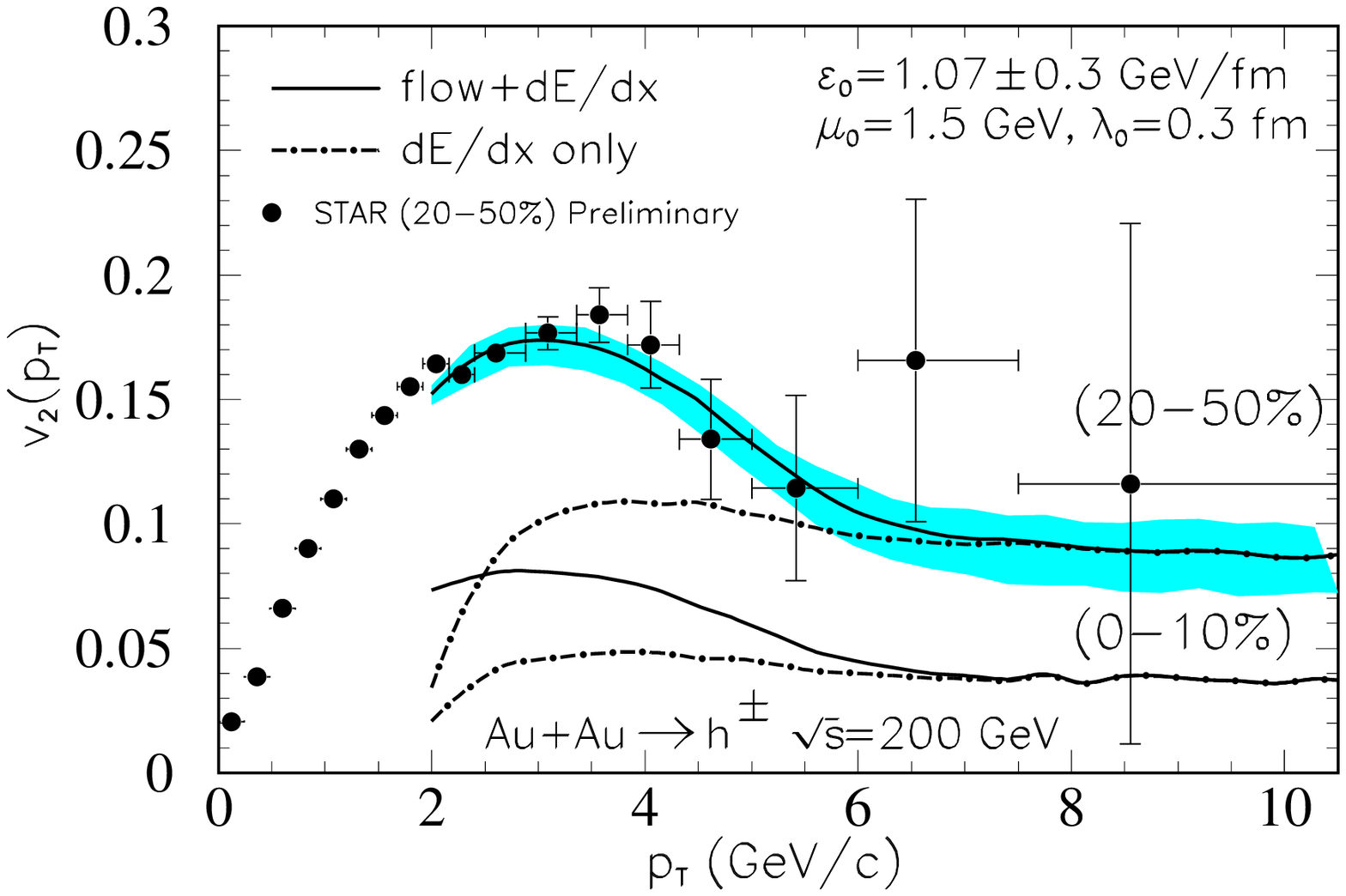}
\includegraphics[width=.45\textwidth]{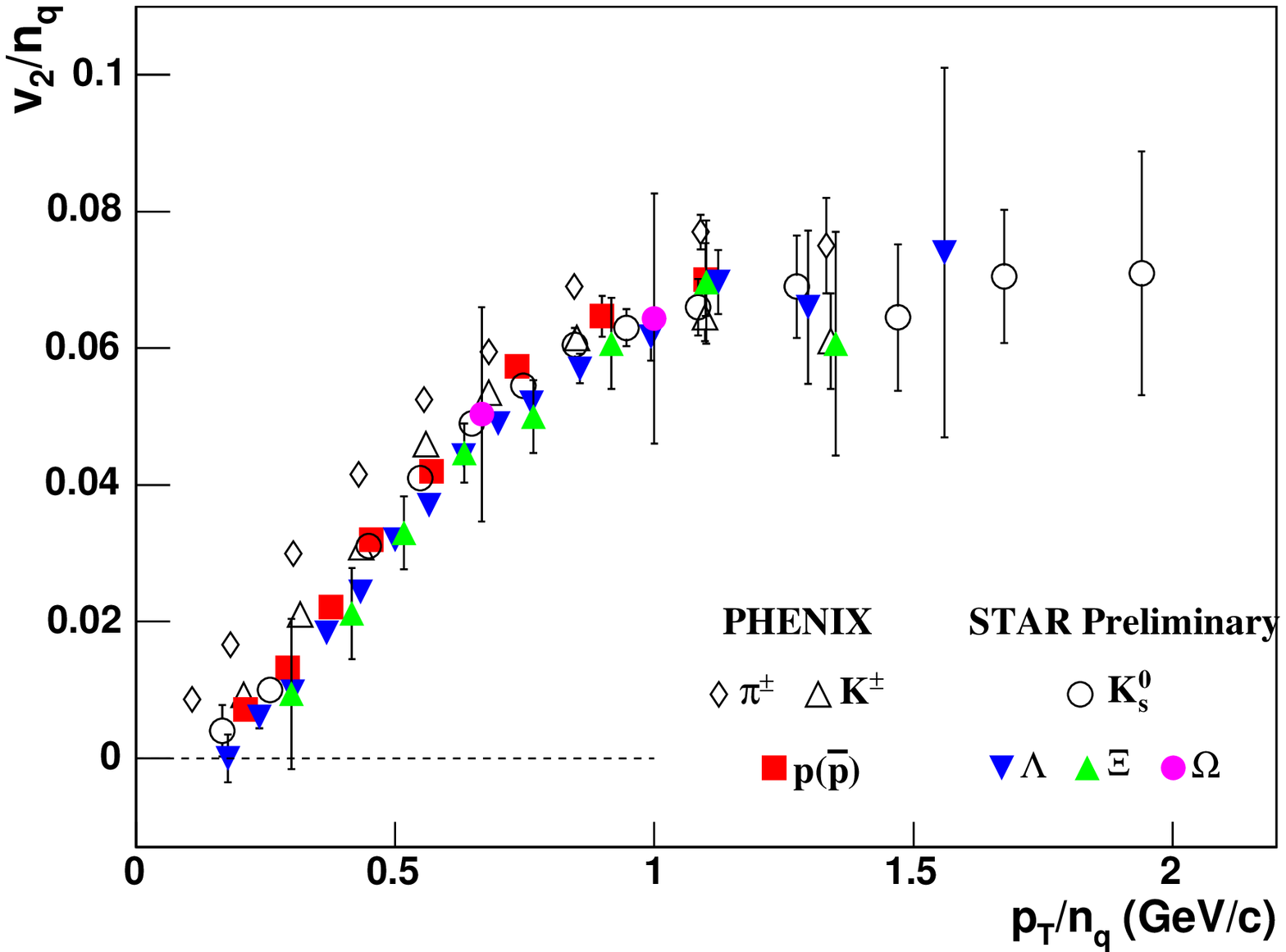}
\caption{Left: azimuthal anisotropy \vtwo\ from 4-particle cumulant analysis of
200 GeV $Au+Au$ collisions from STAR \protect\cite{Snellings:2003mh} compared
to parton model calculations incorporating partonic energy
loss. Right: \pT\ dependence of \vtwo\ for identified particles, with
both axes scaled by the number of constituent quarks; data from PHENIX
and STAR, figure from \protect\cite{Xu:2004pc}.}
\label{fig:Highptv2}
\end{figure}

Fig. \ref{fig:CorrelationsPR}, right panel, shows the azimuthal
anisotropy \vtwo\ in non-central $Au+Au$ collisions at 200 GeV for the
reaction plane (circles) and 2- (triangles) and 4-particle (stars) cumulant methods
\cite{Tang:2003kz}.
The measured \vtwo\ is large, in the sense that its magnitude is
similar to the eccentricity due to the initial spatial anisotropy of
the collision [Eq.~(\ref{eq:HydroEpsx})]
\cite{Shuryak:2001me,Voloshin:2002wa}.  Since the spatial eccentricity
of the fireball is diluted by expansion
\cite{KolbHeinzHydroReview,Huovinen:2003fa}, 
such a large \vtwo\ at high \pT\ may
result from strong partonic energy loss at the earliest, hot and dense
phase of the evolution. 

In the parton model approach, the azimuthal dependence of the partonic
energy loss for non-central heavy-ion collisions is given by
Eq.~(\ref{total-loss}). The anisotropic energy loss in the effective
modified fragmentation functions of the pQCD parton model can be used
to obtain the azimuthally anisotropic hadron spectra at high \pT.
Fig.~\ref{fig:Highptv2}, left panel, shows \vtwo\ for charged hadrons
generated from partonic energy loss (dot-dashed) compared to the
measured
\vtwo\ from a 4-particle cumulant analysis
\cite{Tang:2003kz,Snellings:2003mh}. The energy loss
extracted from high \pT\ inclusive suppression can account for the
observed azimuthal anisotropy at large \pT\gt6 GeV/$c$ (dot-dashed line). 

The azimuthal anisotropy at high \pT\ can also depend on the
transverse expansion, which depletes the gluon density more rapidly
than one-dimensional expansion and dilutes the initial geometric
anisotropy.  The depletion in gluon density is however compensated by
longer propagation time in the medium, giving rise to roughly the same
total energy loss. In realistic calculations the azimuthal anisotropy
of partonic energy loss is found to decrease slightly relative to the
case of no transverse expansion
\cite{Gyulassy:2001kr} and quantitative analyses
should take this effect into account.  Note also that in the current
parton model calculations a hard-sphere nuclear distribution is
used. A more accurate calculation would employ the more realistic
Wood-Saxon distribution which would reduce the high $p_T$ $v_2$
\cite{Drees:2003zh}. However, partonic energy loss occurs mainly at
early time due to rapid expansion, making the final result relatively
insensitive to the shape of edge of the reaction zone.

The observed \vtwo\ at intermediate \pT\ is larger than the simple
parton model calculation. This discrepancy may be attributable to
effects of parton coalescence. If the difference is due to the flow of
kaons and baryons generated by coalescence, they must have
$\vtwo\approx0.23$ for 20-50\% collisions and $\vtwo\approx0.11$ for
0-10\% collisions. The total \vtwo\ incorporating this effect is shown
by the solid lines in Fig. \ref{fig:Highptv2}, left panel.

A remarkable phenomenological scaling for particle-identified \vtwo\
is seen in Fig. \ref{fig:Highptv2}, right panel, where both the
horizontal (\pT) and vertical (\vtwo) axes have been scaled by the
number of constituent quarks in the hadron ($n_q=2$ for mesons,
$n_q=3$ for baryons). The scaled distributions for all hadrons except
pions collapse within experimental uncertainties to a universal curve,
indicating that constituent quark dynamics may be driving the
\vtwo\ of hadrons at intermediate \pT. Such scaling emerges from a recombination model based 
an exponential partonic spectrum \cite{Fries:2003kq,Fries:2004ej},
with the requirement that all partons contribute equally to the
hadron's momentum ({\it i.e.} the Wigner function of the hadron is a
$\delta$-function in momentum space). Relaxation of the
$\delta$-function condition is found however to generate negligible
change in \vtwo\ at intermediate
\pT
\cite{Fries:2003kq,Fries:2004ej}, so that the observed scaling provides robust support for 
recombination as the mechanism underlying hadronization at
intermediate \pT. Pions are excused from following the
scaling because they arise dominantly from $\rho$-decay
\cite{Greco:2004ex}.


\begin{figure}
\includegraphics[width=.45\textwidth]{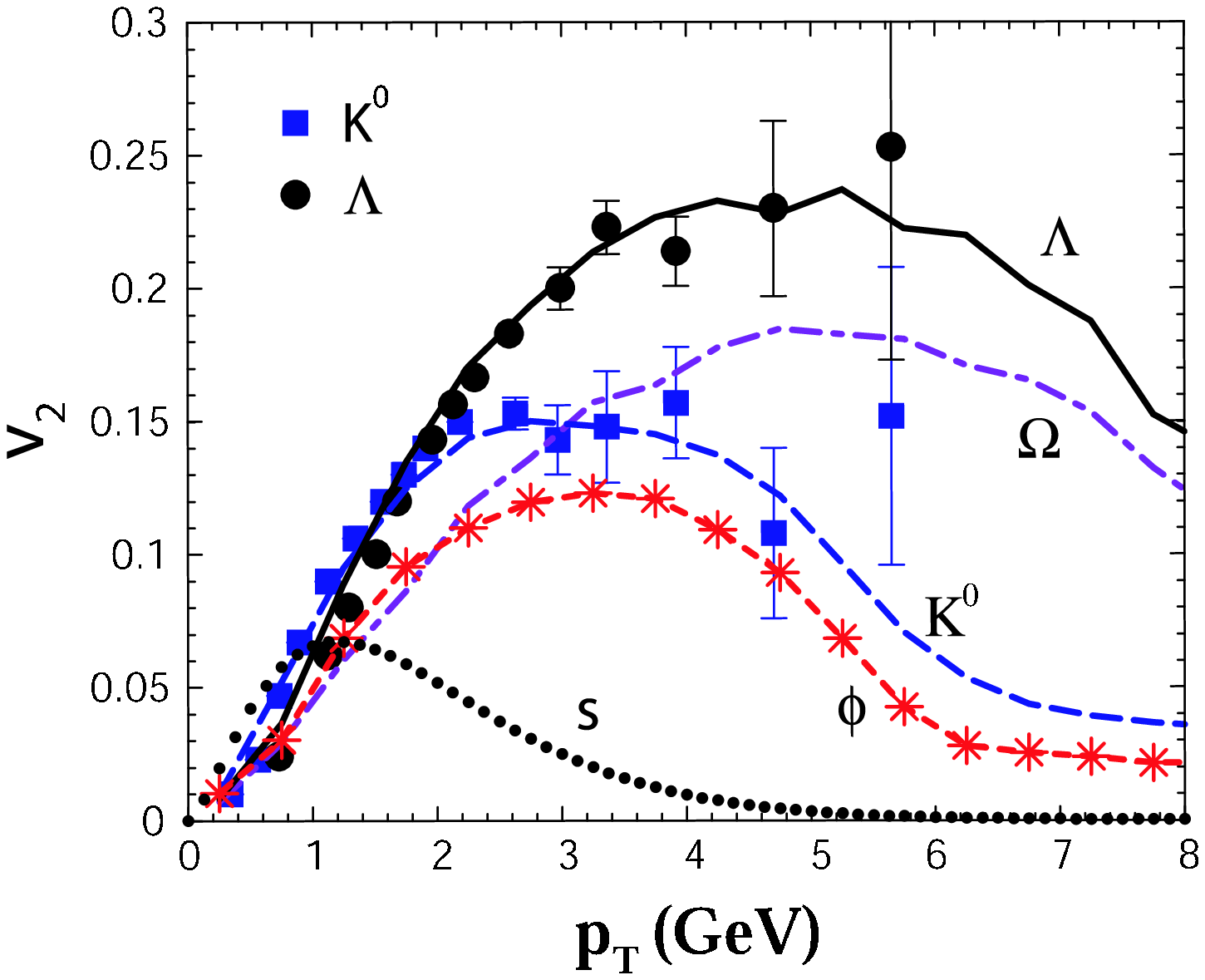}
\includegraphics[width=.45\textwidth]{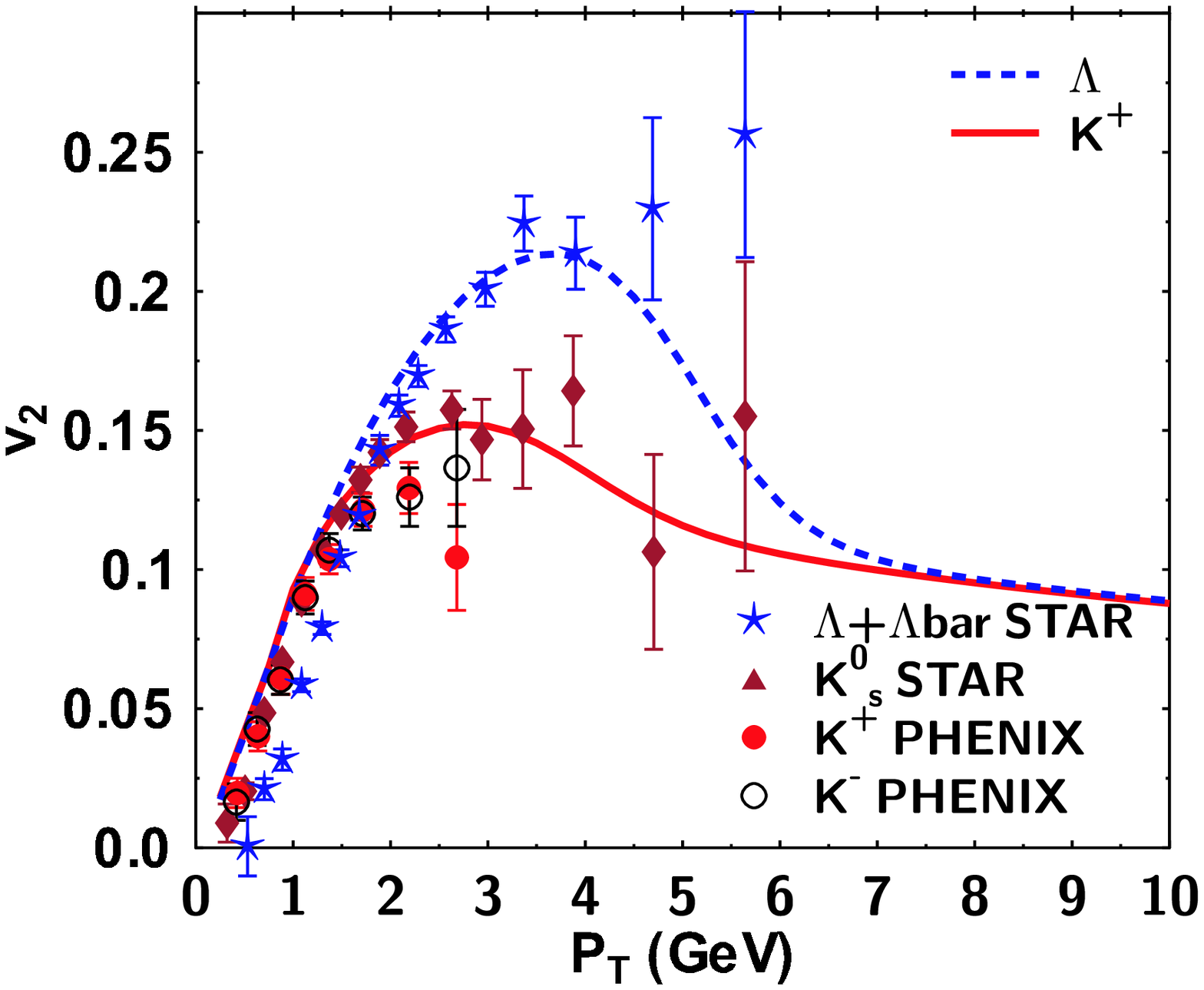}
\caption{
\vtwo\ of $K$ and \lam\ \protect\cite{star:highpTLamK200} compared to coalescence
model calculations.  Left: calculations from
\protect\cite{Greco:2003mm}. Model predictions for $\phi$ and $\Omega$ also
shown.  Dotted curve is strange quark anisotropy.  Right: calculations
from \protect\cite{Fries:2003kq}.  }
\label{fig:Highptv2a}
\end{figure}


If hadrons at intermediate \pT\ are produced through recombination of
a fast parton from a jet and with soft partons from the thermal
medium, their azimuthal anisotropy relative to the reaction plane
should depend both on the azimuthal dependence of the partonic energy
loss and on the elliptic flow of the soft partons. In such models the
functional dependence of $v_2$ on
\pT\ for soft partons must be specified, with parameters fixed by
fitting to the data. The predictive power of the model lies in
the flavor dependence. Since baryons receive a larger relative
contribution from coalescence, with the effect extending to larger
\pT\ than for mesons, splitting of \vtwo(\pT) for baryons and mesons
should be observed. Fig.~\ref{fig:Highptv2a} shows measured
$\vtwo(\pT)$ for $K$ and $\Lambda$ compared to two
coalescence model calculations
\cite{Fries:2003kq,Greco:2003mm}.  The models reproduce \vtwo\ of all
species well up to
\pT$\sim$4-5 GeV/c. As for inclusive spectra, coalescence effects on
\vtwo\ diminish for $p_T>5$ GeV where $\vtwo(\pT)$ for all
hadron species should become the same, driven by partonic energy loss.

An additional test of partonic energy loss results from the
differential study of high \pT\ dihadron correlations relative to the
reaction plane orientation.  Fig.~\ref{fig:StarJetplane} shows a
preliminary study of the high \pT\ dihadron correlation for
non-central (20-60\%) $Au+Au$ collisions, with the trigger hadron
situated alternatively in the azimuthal quadrants centered on the
reaction plane (``in-plane'') or those orthogonal to it
(``out-of-plane'')
\cite{Tang:2004vc,Filimonov:2004qz}. The same-side dihadron correlation in 
both cases is similar to that in $p+p$ collisions. In contrast, the
suppression of the back-to-back correlation depends strongly on the
relative angle between the trigger hadron and the reaction plane. This
systematic dependence is also consistent with the picture of partonic
energy loss: the path length in medium for a dijet oriented out of the
reaction plane is longer than in the reaction plane
(Fig. \ref{fig:HydroEps}, left panel), leading to correspondingly
larger energy loss.

\begin{figure}
\centering
\includegraphics[width=.57\textwidth]{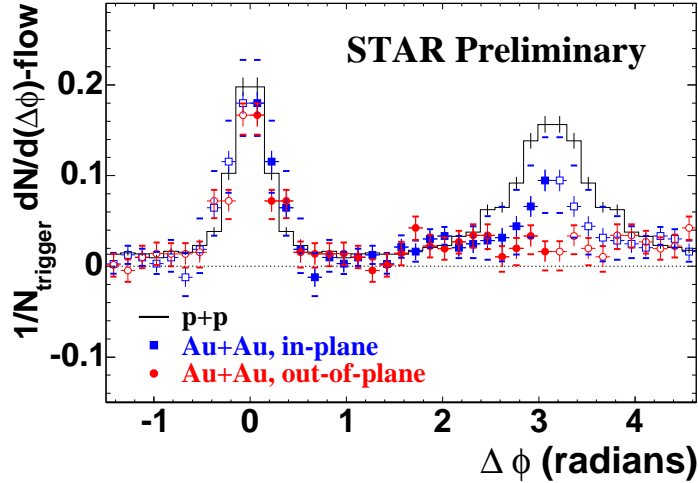}
\caption{Background-subtracted high \pT\ dihadron correlation for different 
orientations of the trigger hadron relative to the reaction
plane \protect\cite{Tang:2004vc,Filimonov:2004qz}.}
\label{fig:StarJetplane}
\end{figure}

\subsection{Partonic Energy Loss vs. Hadronic Absorption}

From the analyses within a parton model of single inclusive and
dihadron spectra and the azimuthal anisotropy $v_2(p_T)$, the average
energy loss for a 10 GeV quark propagating through the expanding
medium created at initial time $\tau_0=0.2$ fm/$c$ in 200 GeV central
Au+Au collisions is $\langle dE/dL\rangle_{1d}\approx 0.85 \pm 0.24$
GeV/fm.  This is equivalent to $dE_0/dL\approx 13.8 \pm 3.9$ GeV/fm in
a static and uniform medium over a distance $R_A=6.5$ fm at initial
time $\tau_0=0.2$ fm.  The value is about 30 times higher than the
quark energy loss in cold nuclei, as extracted from HERMES DIS data.
Since the partonic energy loss in the thin plasma limit is
proportional to the gluon number density, this indicates that the
initial gluon density at $\tau_0=0.2$ fm/$c$ reached in central
$Au+Au$ collisions at 200 GeV is about 30 times higher than the gluon
density in a cold $Au$ nucleus.  This number is consistent with the
estimate from the measured rapidity density of charged hadrons
\cite{phobos:multsqrts} using the Bjorken scenario \cite{BjorkenHydro},
assuming duality between the number of initial gluons and final
charged hadrons.  Since total parton energy loss is only sensitive to
the gluon density of the medium [Eq.~(\ref{eq:effloss})], this is the
only property that can be extracted from the suppression of single and
dihadron spectra. To extract the initial {\it energy} density, additional
measurements such as jet broadening are required. Alternatively, the
energy density can be estimated from global measurements. Given
the measured total transverse energy $dE_T/d\eta\approx 540$ GeV or
about 0.8 GeV per charged hadron in central $Au+Au$ collisions at
$\sqrt{s}=130$ GeV \cite{phenix:et130}, the initial energy density is
50-100 times that in cold nuclear matter (Sec. \ref{sect:ET}).

In the above analyses of RHIC data, the mechanism dominantly responsible
for jet quenching is partonic energy loss prior to hadronization of
the jet. While this picture is in good accord with the observed high
\pT\ phenomena, it is essential to ask whether {\it hadronic} 
interactions, specifically the interaction of hadronic 
jet fragments with the medium, can at least in part generate the 
observed high \pT\ phenomena and contribute substantially to the 
jet quenching \cite{Falter:2002jc,Gallmeister:2002us,Cassing:2003sb}. 
Some simple considerations already argue against this scenario. 
The formation time of hadrons with energy $E_h$ and 
mass $m_h$ is $t_f=(E_h/m_h)\tau_f$, where the rest
frame formation time $\tau_f\sim0.5-0.8$ fm/c. Thus, a 10 GeV/c pion
has formation time $\sim50$ fm/c and is unlikely to interact as a
fully formed pion in the medium.
Since the formation time depends on the boost, the suppression 
due to hadronic absorption with constant or slowly varying
cross section should turn off with rising \pT, at variance with 
observations (Fig. \ref{fig:HadronSuppression}, right panel).
A detailed hadronic transport calculation
\cite{Cassing:2003sb} leads to a similar conclusion: the
absorption of formed hadrons in the medium cannot account by a large
factor for the observed suppression, and the suppression is attributed
to medium interactions of ``pre-hadrons'' which have short formation
time and constant cross section, in other words properties similar to
those of colored partons \cite{Cassing:2003sb}. Further consideration
of the available high \pT\ data \cite{Wang:2003aw} also supports the
conclusion that jet quenching in heavy ion collisions at RHIC is the
consequence of partonic energy loss. In particular, large \vtwo\ at
high \pT\ and the systematics of the small-angle dihadron correlations
are difficult to reconcile with the hadronic absorption scenario.

Azimuthal anisotropy of the spectra in non-central collisions
arises from the initial spatial eccentricity of
the dense medium. The eccentricity decreases rapidly with time due to
transverse expansion \cite{KolbHeinzHydroReview,Huovinen:2003fa}, so that momentum
anisotropy must be generated early, $\sim$few fm/$c$. The finite
anisotropy observed at high \pT\ (Fig. \ref{fig:CorrelationsPR}, right
panel) is therefore unlikely to have significant contribution from the
absorption of formed hadrons.

\begin{figure}
\centering
\includegraphics[width=.48\textwidth]{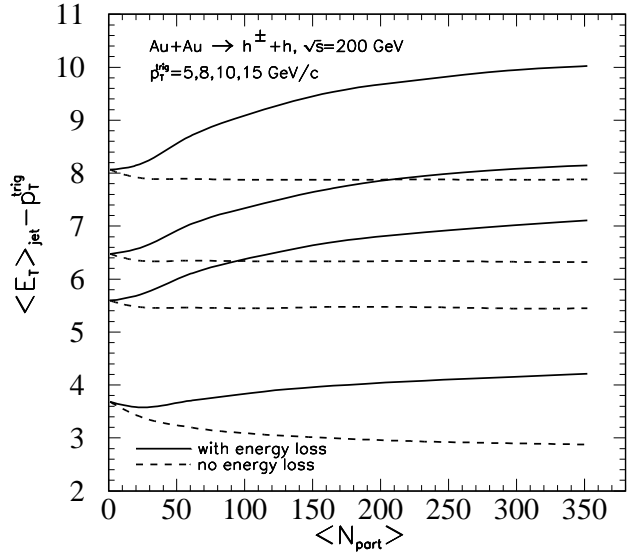}
\caption{
Average transverse energy $\langle E_T\rangle^{jet}$ of the
initial partons that produce a final hadron with $p_T^{\rm trig}$ as a
function of $\langle N_{\rm part}\rangle$ for different values of
$p_T^{\rm trig}$ (lowest curves for smallest \pT).}
\label{fig:avet-jet}
\end{figure}

As we have shown in Fig.~\ref{fig:CorrelationsPR}, left panel, 
the back-to-back dihadron correlation is suppressed in 
central $Au+Au$ collisions
relative to $p+p$ and $d+Au$ collisions, while the same-side
correlations are similar in all three systems. 
In the framework of jet fragmentation, the same-side correlation
measures the conditional distribution of the second leading hadron
within a jet in coincidence with the trigger hadron, essentially the ratio
of dihadron to single hadron fragmentation functions. Insofar as both
hadrons originate from the same fragmenting parton, this ratio will
largely be independent of the parton energy.  Gluons radiated via
energy loss will generate soft hadrons and will enhance the
conditional distribution at small $z$. A recent study of the dihadron
fragmentation functions shows that the conditional (or triggered)
dihadron distribution at large $z$ within a jet is quite stable
against radiative evolution (or energy loss), while soft secondary
hadrons are enhanced \cite{Majumder:2004wh}. 

Hadronic absorption, on the other hand, will generally suppress the
leading and secondary hadron independently, thereby suppressing the
conditional distribution to the same degree as the single particle
inclusive yield. This may not be the case if the hadronic absorption
is very strong and only jets originating in a thin surface shell
generate trigger hadrons which are not suppressed, but we consider
this scenario to be unrealistic. Calculations in a parton model with
energy loss show that even the trigger-biased jet population generated
near the surface loses on average about 2 GeV of energy
\cite{Wang:2003aw}, as seen in Fig.~\ref{fig:avet-jet}. This energy
will be carried by soft hadrons correlated with the trigger, and the
total jet energy for fixed trigger hadron $p_T^{\rm trig}$ should be
larger in $Au+Au$ collisions than in $p+p$ collisions by about 2
GeV. We will discuss the prospects for observing the radiated energy
carried by soft hadrons using dihadron measurements in section
\ref{sect:HardProbesOutlook}.

If hadronic absorption suppresses high \pT\ hadrons and jets, it
should also do so in heavy-ion collisions at SPS energy. Hadronic
spectra at this energy vary strongly with \pT\ and are very sensitive
to initial transverse momentum broadening and parton energy loss
\cite{Wang:1998ww}. The measured $\pi^0$ spectrum in central $Pb+Pb$
collisions appears to exhibit only the expected Cronin enhancement
relative to a $p+p$ reference, with no suppression observed
\cite{Aggarwal:1998vh,Aggarwal:2001gn,Wang:1998hs}. However, the
\pizero\ yield for central collisions is seen to be suppressed relative to that
for peripheral collisions \cite{Aggarwal:2001gn}. It has been been
pointed out recently \cite{d'Enterria:2004ig} that the $p+p$ reference
used in this study contains significant uncertainties, and a
reassessment also reveals a possible high \pT\ \pizero\ suppression
for central $Pb+Pb$ relative to $p+p$ collisions. Recent analysis of
dihadron correlations shows that both same-side and back-to-back
jet-like correlations are not suppressed in central collisions at the
SPS, though the back-side distribution is broadened
\cite{Agakichiev:2003gg}. The question of high \pT\ hadron suppression
at the SPS and its connection to jet quenching therefore remains
open. More generally, study of the \sqrts\ dependence of the
suppression will provide an essential cross-check of our
understanding of these phenomena. High \pT\ hadron production results
from the recently completed 62 GeV $Au+Au$ run at RHIC are now
becoming available \cite{Back:2004ra}.

We conclude that the data provide no clear support for hadronic absorption
as the dominant mechanism underlying the observed high \pT\
suppression phenomena. We consider large {\it partonic} energy loss in
the dense medium formed in central $Au+Au$ collisions at RHIC to be
well established. We now discuss future measurements that will exploit
this new phenomenon as a precision probe of the medium.


\subsection{Hard probes: an outlook}
\label{sect:HardProbesOutlook}

The discovery of jet quenching opens a new era in the study of dense
QCD matter. Higher precision data extending to larger \pT\ will map in
detail the modification of the fragmentation functions and the energy
dependence of the energy loss. The ultimate measurement of modified
FF's in heavy ion collisions will be the direct photon triggered FF.
Measurements of charmed mesons will also provide key tests of the
parton energy loss scenario of jet quenching; recent theoretical
studies reveal features of heavy quark energy loss that are measurably
different from those of light quarks and gluons
\cite{Dokshitzer:2001zm,Djordjevic:2003qk,Zhang:2003wk,Armesto:2003jh}.
Dihadron fragmentation and correlations within the same jet will also
be important, for instance the broadening of the dihadron angular
correlation which probes the average momentum transfer of the
interaction with the medium that may be related directly to the energy
density \cite{Majumder:2004wh}.

\subsubsection*{Direct Photon Tagged Jet Quenching}

The definition of direct photon-triggered fragmentation functions is
similar to the hadron-triggered fragmentation function in
Eq.~(\ref{frg-htrig}), replacing the triggered hadron with a direct
photon \cite{Wang:1996yh,Wang:1997pe}. This corresponds to the
measurement of the hadron $p_T$ distribution in the opposite azimuthal
direction to the triggered photon. 
Since a direct photon in the central rapidity
region ($y=0$) is always accompanied by a jet in the opposite azimuth
with roughly equal transverse energy, the $p_T$ distribution of
particles is directly related to the fragmentation function of a jet 
with known initial energy, $E_T^{\rm jet}\approx E_T^{\gamma}$. By
comparing the extracted jet fragmentation function in $A+A$ to that in
$p+p$ collisions, the modification of the fragmentation function
and thereby the partonic energy loss can be measured directly.

Fig.~\ref{fig:DirectGamma} shows examples of the the suppression
factors of the direct photon-tagged jet fragmentation function for
average partonic energy loss $dE/dx=1$ GeV/fm.  They are similar in
shape to the hadron-triggered fragmentation function.  The presently
measured single particle inclusive and dihadron suppression generate
an average energy loss of about $dE/dx=0.85$ GeV/fm for a 10 GeV quark
in central $Au+Au$ collisions at RHIC. According to the parameterized
energy dependence in Eq.~(\ref{eq:th-loss}), the average energy loss
for a 15 GeV quark should be about 1.26 GeV, slightly larger than that
used to generate the curves in the figure. The initial mean free path
that fits the inclusive and dihadron data, $\lambda_0=0.3$ fm, is
consistent with 1-2 fm used in the calculation after correction for
the one-dimensional expansion.

\begin{figure}
\includegraphics[width=.47\textwidth]{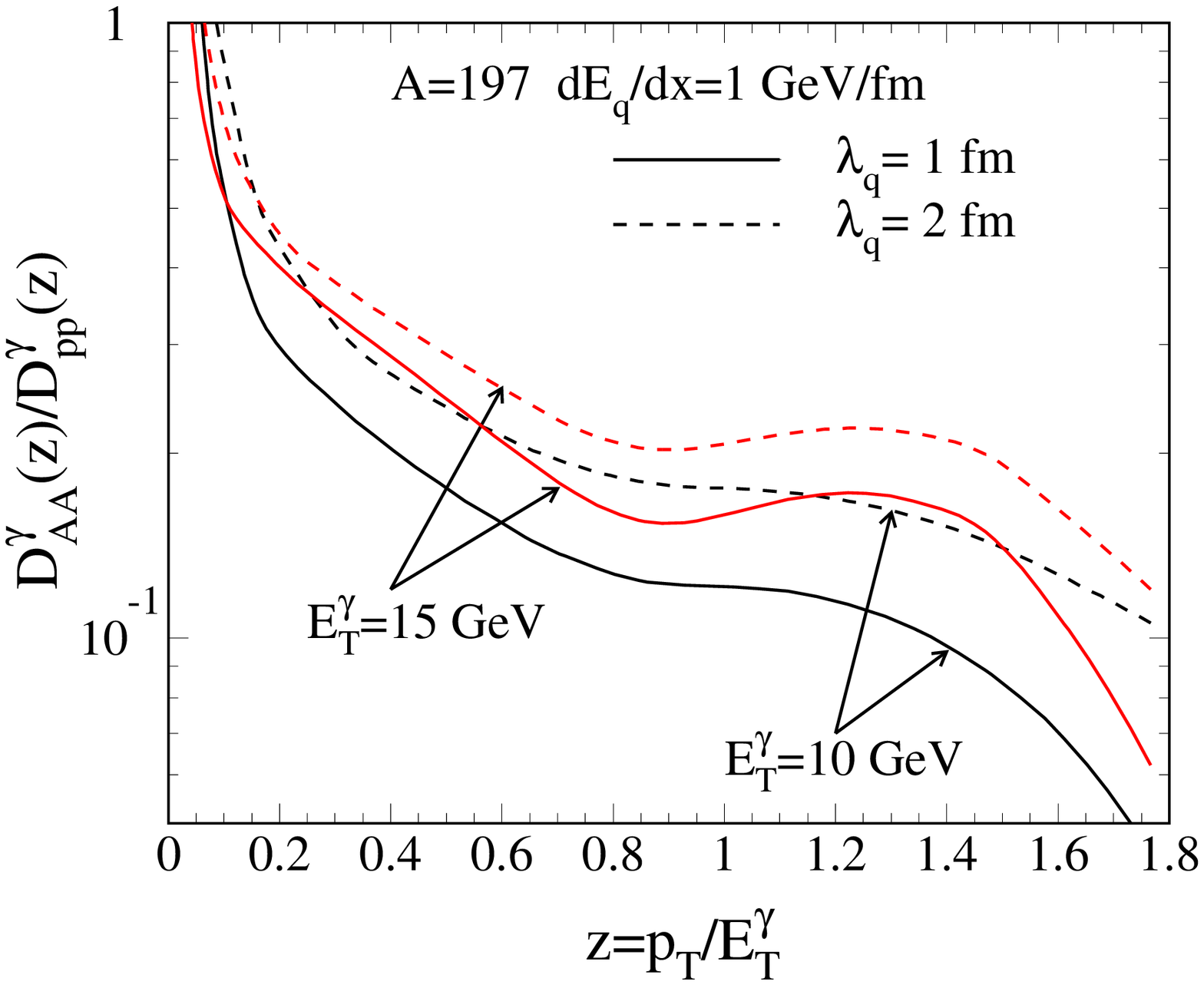}
\includegraphics[width=.53\textwidth]{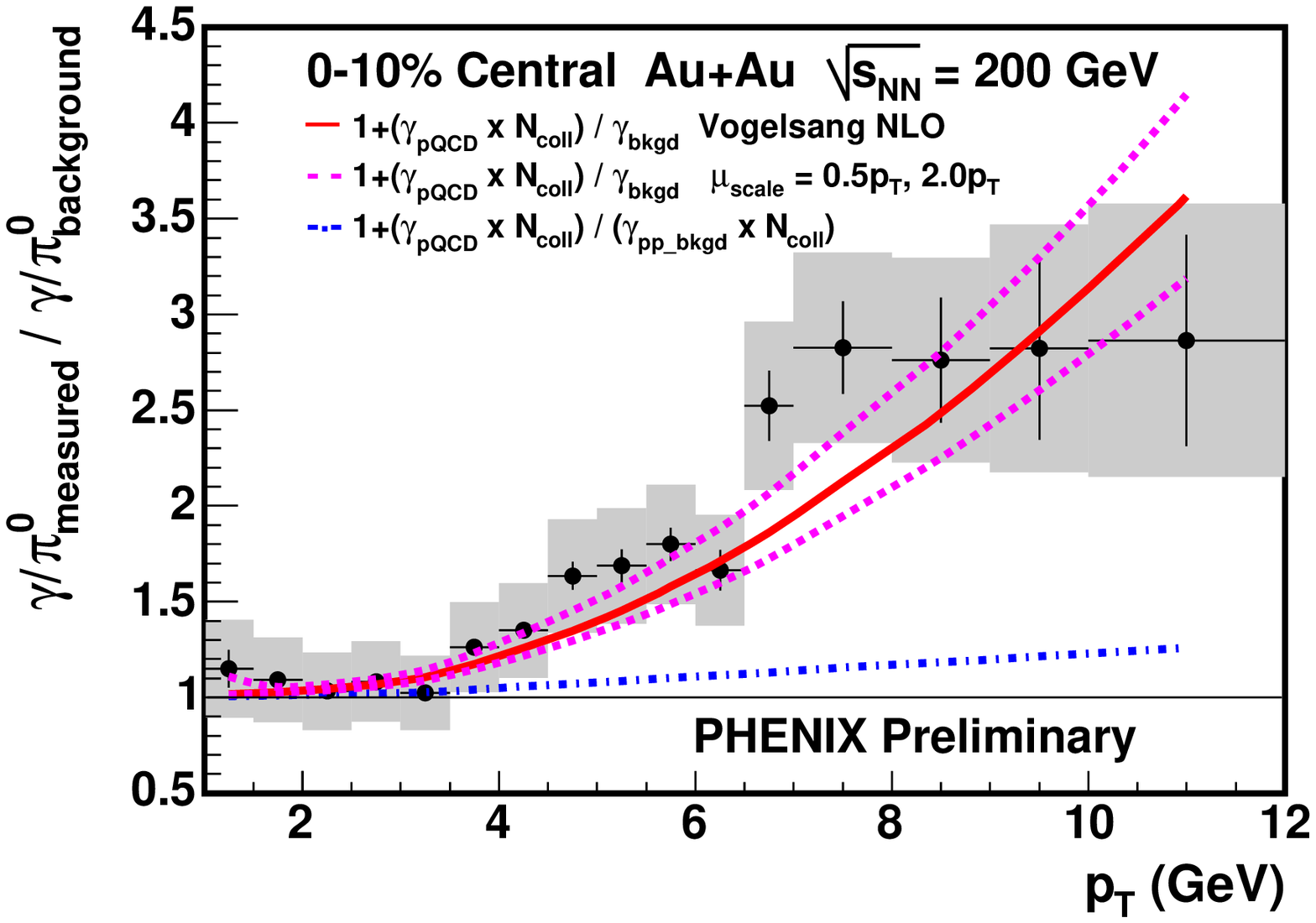}
\caption{
Left: Calculated modification factor of the photon-tagged jet fragmentation
function in central $Au+Au$ collisions at $\protect\sqrt{s}=200$ GeV
for fixed $dE_q/dx=1$ GeV/fm \protect\cite{Wang:1997pe}. Right: Ratio of total to decay photon yield
in central $Au+Au$ collisions at 200 GeV, from PHENIX
\protect\cite{phenix:directgamma}. The lines correspond to direct photon yields from NLO pQCD and decay
photons from the measured $\pi^0$ spectra. Upper curves incorporate
$Au+Au$ jet quenching in the background estimate, lower curves are for binary-scaled background.}
\label{fig:DirectGamma}
\end{figure}

The measurement of direct photon-triggered fragmentation is extremely
challenging because of low cross sections and the difficulty of
isolating direct photons in the heavy-ion collision
environment. However, statistical measurements of the inclusive direct
photon yield are now becoming available. Fig.~\ref{fig:DirectGamma},
right panel, shows the preliminary measured ratio of the yield of total (direct +
decay) photons to expected decay photons, from PHENIX
\cite{phenix:directgamma}. The measured ratio in $p+p$ collisions is consistent with
an NLO calculation of the direct photon yield
\cite{phenix:directgamma}. The figure compares the ratio in central $Au+Au$ 
collisions to NLO calculations of the direct photon yield normalized
alternatively by a background derived from the binary-scaled \pizero\
yield from $p+p$ collisions or by the measured, highly suppressed
\pizero\ yield from central $Au+Au$ collisions. The latter
normalization is clearly favored, demonstrating unambiguously that the
photon excess is not suppressed. This is consistent with the partonic
energy loss picture since the photon does not carry color
charge. Further isolation of direct photons from the QCD fragmentation
background is difficult in heavy ion collisions because standard
techniques, in particular isolation cuts, cannot be directly
applied. Strong jet suppression may help in this regard, however. This
result in any case shows great promise and measurement of the modified
fragmentation function of a jet recoiling from a hard photon is on the
horizon.

\subsubsection*{Heavy Quark Energy Loss}

The current experimental studies of jet quenching have focused on
light quark and gluon jets. The energy loss of gluons and light quarks
differ by a factor of $C_A/C_F=9/4$, but it is difficult to
differentiate quark and gluon jets in practice in $p+p$ and $Au+Au$
collisions. Identified hadron ratios such as $K^-/K^+$ or $\bar{p}/p$
may be used to tag contributions from different flavor jets
\cite{Wang:1998bh,Zhang:2001ce} but this does not provide sharp 
discrimination, due to flavor mixing in the fragmentation of light
quark jets. The study of heavy quark jets offers cleaner flavor
tagging. Recent studies have revealed observable features of heavy
quark energy loss in medium that may provide significant new insight
into the partonic energy loss mechanism
\cite{Dokshitzer:2001zm,Djordjevic:2003qk,Zhang:2003wk,Armesto:2003jh}.

The formation time of gluon radiation off a heavy quark, measured with
respect to the propagation of the heavy quark inside the medium, is
reduced relative to a light quark because of the large quark
mass. LPM interference should therefore be reduced significantly for
heavy quarks at intermediate energy. In addition, the heavy quark mass
suppresses the gluon radiation amplitude at angles smaller than the
ratio of the quark mass to its energy \cite{Dokshitzer:2001zm}. Both
mass effects will lead to different energy loss for a heavy relative to
a light quark in a dense medium.

The mass dependence of the gluon radiation amplitude comes from the
heavy quark propagators. It suppresses the induced gluon spectrum for
small angle radiation for a heavy quark relative to that of a light
quark by a factor
\cite{Zhang:2003wk}
\begin{equation}
f_{Q/q}=\left[\frac{\ell_T^2}{\ell_T^2+z^2
M^2}\right]^4=\left[1+\frac{\theta_0^2}{\theta^2}\right]^{-4}.
\label{f}
\end{equation}
\noindent
Here $\theta=\ell_T/q^-z$ and $\theta_0=M/q^-$  represent
the radiation angle and the ``dead-cone'' within which
the gluon radiation is suppressed. This suppression leads 
to a reduced radiative energy loss of a heavy quark.
To illustrate the mass suppression of radiative energy loss
imposed by the ``dead-cone'', the ratio 
$\langle\Delta z^Q_g\rangle(x_B,Q^2)/\langle\Delta z^q_g\rangle(x_B,Q^2)$
of charm quark and light quark energy loss in DIS off a nucleus
is plotted in Fig.~\ref{fig-hvyloss}, left panel, as functions of $x_B$.

Within the same framework of DIS off nuclei, the gluon formation time
for radiation from a heavy quark
\begin{equation}
\tau_f\equiv\frac{1}{p^+\widetilde{x}_L}
=\frac{2z(1-z)q^-}{\ell_T^2+(1-z)^2M^2},
\end{equation}
is shorter than that for gluon radiation from a light quark, with
significant consequences for heavy quark energy loss.  Because of the
quark mass dependence of the formation time relative to the nuclear
size in the given frame, $m_NR_A/\tau_f p^+ \sim x_BM^2/x_AQ^2$, there
are two distinct limiting behaviors of the energy loss for different
values of $x_B/Q^2$ relative to $x_A/M^2$. When $x_B/Q^2\gg x_A/M^2$
for small quark energy (large $x_B$) or small $Q^2$, the formation
time of gluon radiation off a heavy quark is smaller than the
nuclear size.  In this case, there is no destructive LPM
interference. The heavy quark energy loss
\begin{equation}
\langle\Delta z_g^Q\rangle \sim C_A\frac{\widetilde{C}
\alpha_s^2}{N_c} \frac{x_B}{x_A Q^2}
\end{equation}
is linear in nuclear size $R_A$. In the opposite limit,
$x_B/Q^2\ll x_A/M^2$, for large quark energy (small $x_B$) or
large $Q^2$, the quark mass becomes negligible. The gluon formation
time could still be much larger than the nuclear size. The LPM
interference will limit the available phase space for gluon radiation 
and the heavy quark energy loss
 \begin{equation}
\langle\Delta z_g^Q\rangle \sim C_A \frac{ \widetilde{C}
\alpha_s^2}{N_c} \frac{x_B}{x_A^2 Q^2}
\end{equation}
now has a quadratic dependence on the nuclear size, similar to the
light quark energy loss. Fig.~\ref{fig-hvyloss}, right panel, shows numerical
results for the $R_A$ dependence of charm quark energy loss rescaled
by $\widetilde{C}(Q^2)C_A\alpha_s^2(Q^2)/N_C$, for different values of
$x_B$ and $Q^2$. The $R_A$ dependence is quadratic for large values of
$Q^2$ or small $x_B$, but becomes almost linear for small $Q^2$
or large $x_B$. The charm quark mass is set at $M=1.5$ GeV in the
numerical calculation.

\begin{figure}
\centering
\includegraphics[width=.48\textwidth]{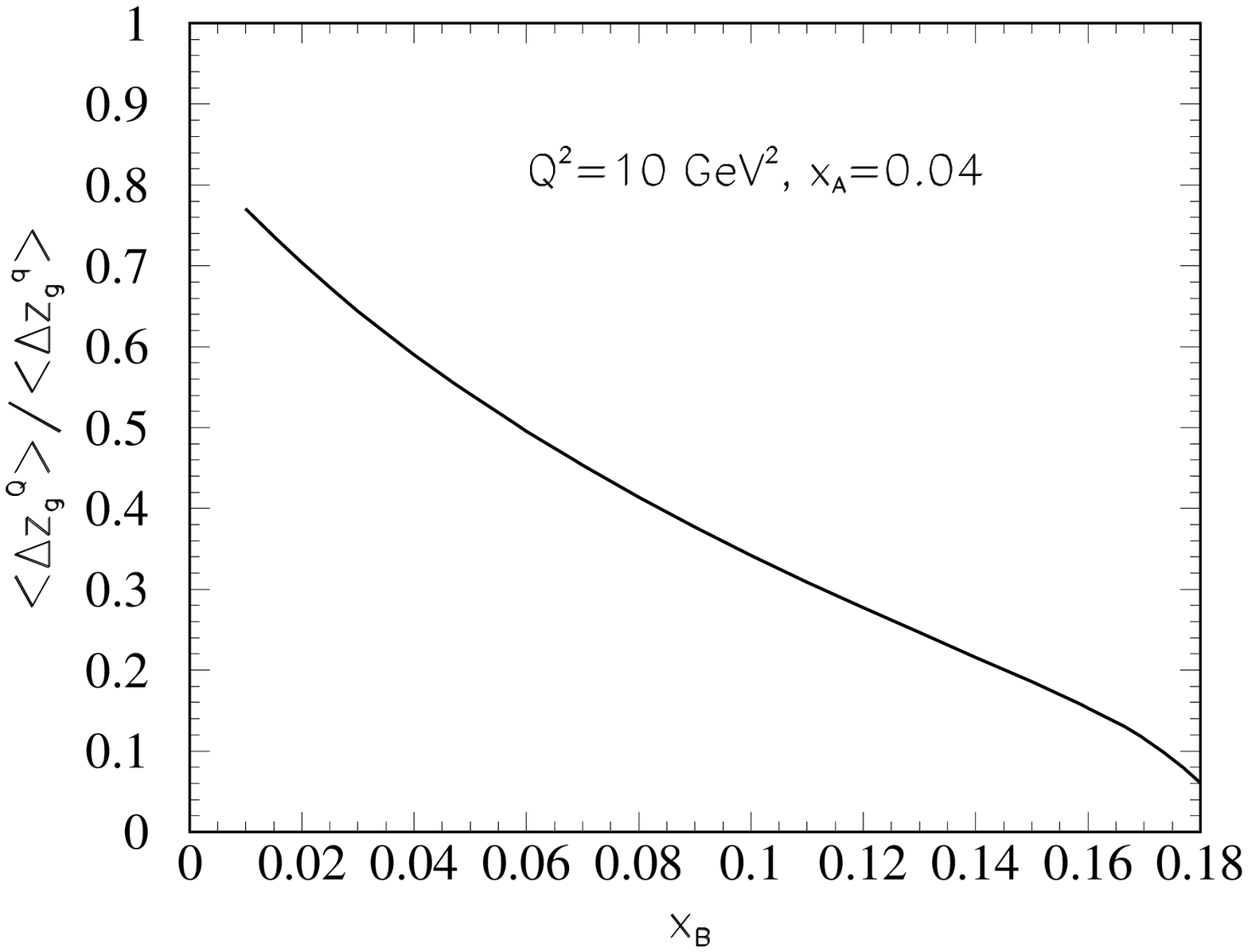}
\includegraphics[width=.48\textwidth]{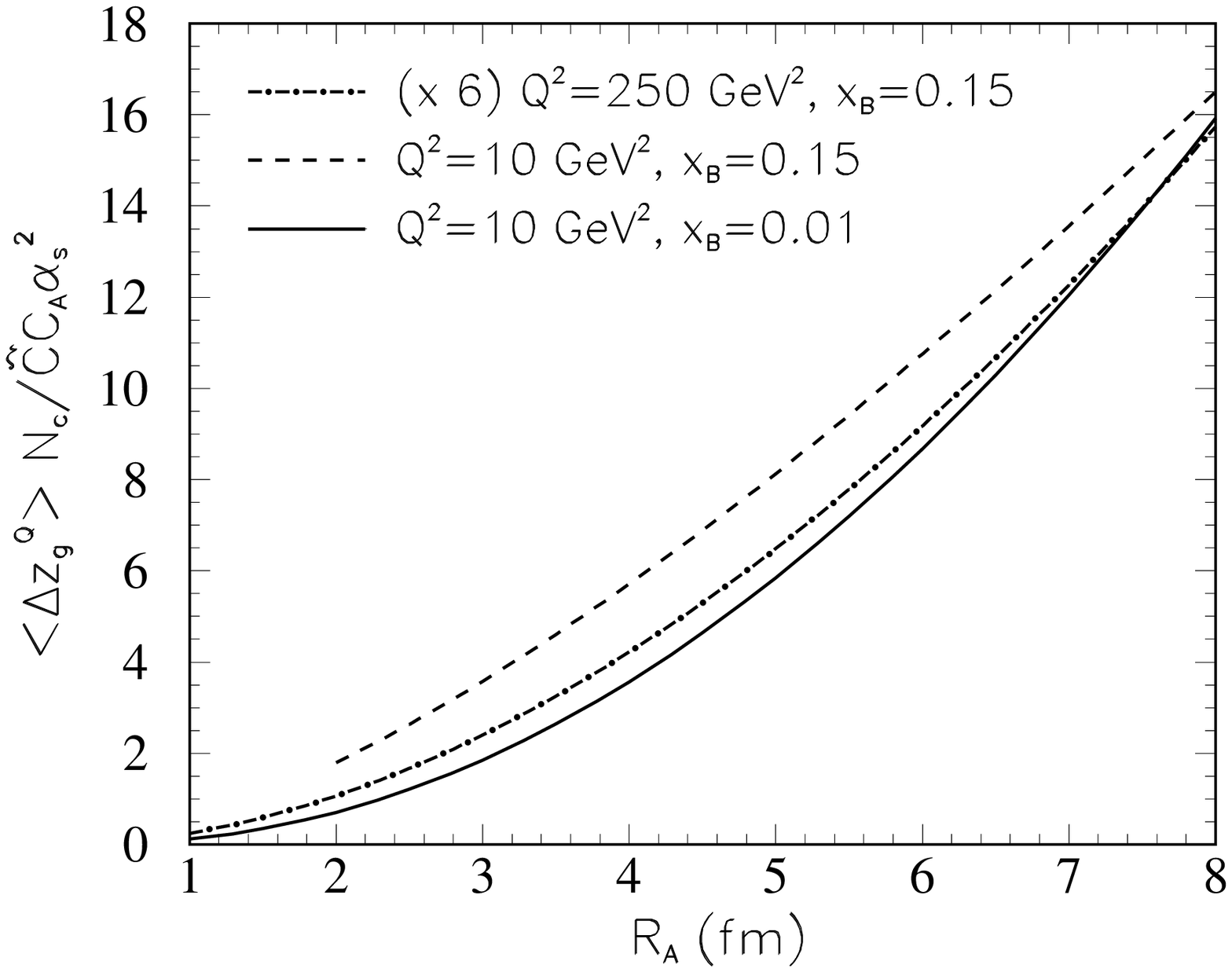}
\caption{
Left: The $x_B$ dependence of the ratio between charm quark and light
quark energy loss in DIS off a heavy nucleus. Right: Dependence of
charm quark energy loss on nuclear size $R_A$, for different values of
$Q^2$ and $x_B$.}
\label{fig-hvyloss}
\end{figure}

Apparently, energy loss induced by gluon radiation is significantly
suppressed for heavy relative to light quarks when the momentum scale
$Q$ or the quark initial energy $q^-$ is not large compared to the
quark mass. Only in the limit $M \ll Q, \; q^-$, is the mass effect
negligible and the heavy quark energy loss approaches that of a light
quark. In the kinematic regime accessible at RHIC,
the heavy quark energy loss will be much reduced relative to the light
quarks and gluons. In addition, the mass effect is likely to reduce
the effect of thermal absorption of gluons, whose average energy is
much smaller than the heavy quark mass.

Heavy quark physics at top RHIC energy and design luminosity is
studied primarily through charm production, due to the relatively low
rate of beauty quark production. Several techniques are available to
study charm production directly and indirectly. Fig. \ref{fig:Charm},
left panel, shows the spectrum of electrons from the semi-leptonic
decay of $D$ mesons in 130 GeV $Au+Au$ collisions from PHENIX
\cite{phenix:electron130}, which provides an indirect measurement of
the open charm meson spectrum. Preliminary non-photonic electron
spectra from 200 GeV $p+p$, $d+Au$ and $Au+Au$ collisions have also
been reported \cite{Kelly:2004qw}. Within the current uncertainties
the charm yield in all three systems appears to scale as the number of
binary collisions, though the data are also consistent with small
suppression at large $p_T$ in $Au+Au$.
 
Single electron spectra resulting from semi-leptonic decays are
however rather insensitive to changes in the $D$-meson spectrum
\cite{Batsouli:2002qf} and more precise measurements of charm quark energy 
loss require direct measurement of fully reconstructed $D$ decays into
hadrons. STAR recently reported direct reconstruction of $D$-mesons in
$d+Au$ collisions. Fig.~\ref{fig:Charm}, right panel, shows a preliminary $D$ meson
spectrum constructed from various hadronic decay channels
\cite{Tai:2004bf}. The spectrum is consistent with the single electron
spectrum measured by the same experiment. Such measurements are
considerably more demanding in $Au+Au$ collisions but will provide
direct study of the open charm production and the effect of heavy
quark energy loss.

\begin{figure}
\includegraphics[width=.48\textwidth]{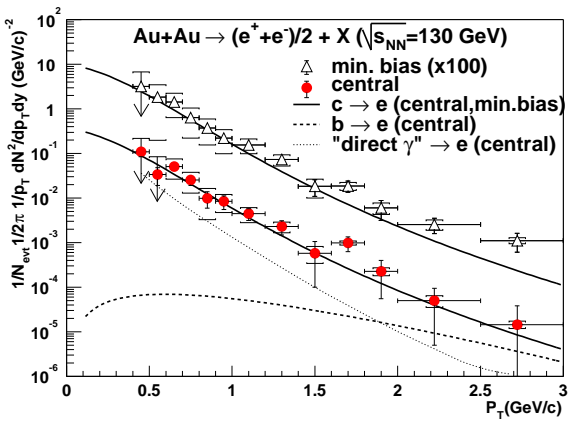}
\includegraphics[width=.48\textwidth]{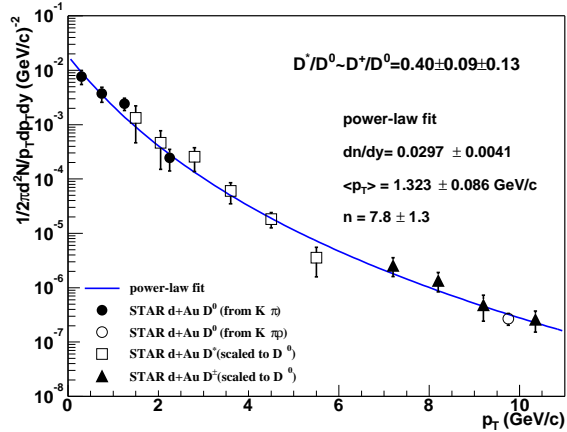}
\caption{Left: Non-photonic single electron spectra  in $Au+Au$ collisions at
130 GeV, from PHENIX \protect\cite{phenix:electron130}.
Right: Spectra of fully reconstructed $D$ meson spectra 
in $d+Au$ collisions at 200 GeV
from STAR \protect\cite{star:QM04summary}.
} 
\label{fig:Charm}
\end{figure}

Parton recombination may also play a role in charm production. Such
effects have already been shown to be significant in the forward
direction for $h+p$ collisions \cite{Braaten:2002yt}. Since
fragmentation functions are harder for heavy than light quarks, heavy
quarks carry a large fraction of the heavy meson momentum in the
wave-function, making it easier for a fast heavy quark (anti-quark) to
pick up a slow light anti-quark (quark) to form a heavy meson. The
recombination effect is therefore expected to be more significant for
$D$ mesons than light quark hadrons. As in the intermediate $p_T$
region of light hadron spectra, discrimination of the effects of
partonic energy loss and parton recombination may be accomplished most
clearly via azimuthal anisotropy of the $D$ meson spectra. Some
recombination models have predicted small but finite azimuthal
anisotropy of the $D$ meson spectra due to recombination with thermal
light quarks \cite{Lin:2003jy}. This measurement is however very
demanding and most likely requires upgrades to the present RHIC
detectors.

\subsubsection*{Dihadron Fragmentation Functions and Angular Correlations}

The suppression of single inclusive hadron spectra or the modification
of the single hadron fragmentation function provide direct measurement
of the gluon density of the dense medium. Broadening of the jet cone
in principle could probe the average transverse momentum transfer to
the jet parton, which can be used to directly estimate the energy
density rather than the gluon density of the medium
\cite{Baier:1999ds,Salgado:2003rv}. Such a study may be possible using
dihadron correlations. Dihadrons can result from the hadronization of
a single leading parton or from two independent partons, one leading
and the other a radiated gluon. The dihadron fragmentation or
correlation function can therefore probe multiple scattering and
induced gluon radiation in the medium. The first mechanism dominates
when both hadrons carry a large fraction of the initial parton
energy. This is likely the case for the same-side correlations
measured by STAR (Fig.~\ref{fig:CorrelationsPR}) and may explain why
the correlation does not change despite the energy loss expected to be
suffered by the leading parton. If the associated (secondary) hadron
carries smaller fractional momentum it could come from radiated
gluons. In this case the dihadron fragmentation function should be
enhanced, as seen by a recent study of its evolution
\cite{Majumder:2004wh}.  Due to the transverse momentum transfer from
multiple scattering, the angular correlation between the hadrons
should be also broadened. This is best seen by looking at the soft
hadrons in a jet, though precision measurements of such observables
are extremely challenging.

\begin{figure}
\includegraphics[width=.45\textwidth]{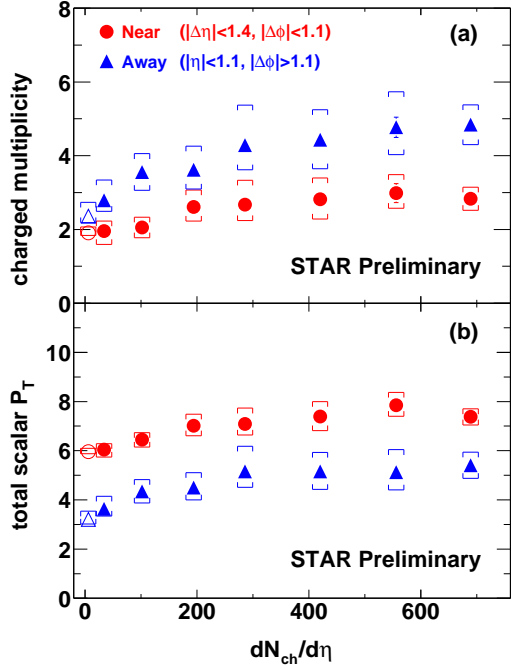}
\includegraphics[width=.45\textwidth]{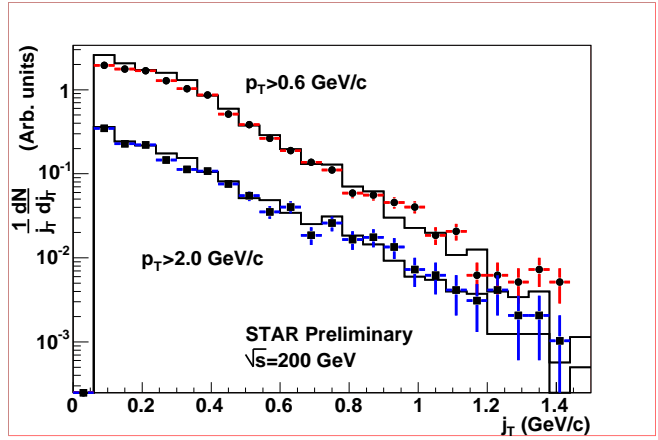}
\caption{
Left: charged hadron multiplicity (upper) and total scalar $p_T$
(lower) of the near and away-side dihadron distributions relative to a
trigger hadron, integrated over broad $(\Delta\eta,\Delta\phi)$
intervals for $4<p_T^{trig}<6$ GeV/$c$ and $0.15<p_T<4$ GeV/$c$ in
$p+p$ (open symbols) and $Au+Au$ collisions, from STAR
\protect\cite{star:mombalance2}. See text for details.  Right: Charged
hadron \jT\ distributions [Eq. (\ref{eq:jT})] from fully reconstructed
jets in $p+p$ collisions at \sqrts=200 GeV from STAR
\protect\cite{Henry:2004kc}, compared to simulations based on 
PYTHIA \protect\cite{Sjostrand:1987su,Sjostrand:2001yu}. 
The two distributions have different lower bounds on the hadron \pT\
relative to the beam axis.}
\label{fig:STARjT}
\end{figure}

A first attempt to measure the radiated energy from the associated
hadrons (Fig. \ref{fig:avet-jet}) is seen in Fig. \ref{fig:STARjT},
left panel, which shows the integrated multiplicity and summed
``scalar \pT'' of hadrons within $0.15<p_T<4$ GeV/$c$ correlated with
a trigger with $\pT^{\rm trig} \gt 4$ GeV/c \cite{star:mombalance2}. The
integration is performed over phase space broader than the normal jet
size in vaccum, with ``near-side'' corresponding to
$|\Delta\phi|\lt1.1$ and $|\Delta\eta|\lt1.4$ and ``away-side''
corresponding to $|\Delta\phi|\gt1.1$ and $|\eta|\lt1.1$. Background
yield is subtracted assuming that the correlated yield is negligible
in the region $0.9\lt|\Delta\phi|\lt1.3$. The associated multiplicity
and summed
\pT\ above background show increases from $p+p$ to $Au+Au$ collisions
qualitatively similar to the theoretical expectations from partonic
energy loss, though the moderate trigger \pT\ makes direct comparison
with such calculations difficult. Future analyses will increase the
trigger \pT\ and reduce the systematic uncertainties associated with
the background definition. Effects of parton recombination in the jet
structure must also be addressed in order to extract net effect of
partonic energy loss.

\subsubsection*{Full jet reconstruction}

As we have discussed extensively, the study of correlations among jet
fragments on a statistical basis provides key observables for studying
hard probes in collisions of heavy nuclei. Event-wise full jet
reconstruction with good energy resolution is exceedingly difficult in
all but perhaps the most peripheral collisions of heavy nuclei, due to the
large multiplicity and the complexity of the underlying
event. Standard jet reconstruction techniques can however be applied
to collisions of simpler systems such as $p+p$ and
$d+Au$. Fig. \ref{fig:STARjT}, right panel, shows the preliminary results on the
\jT\ distribution in fully reconstructed jets in 200 GeV $p+p$
collisions \cite{Henry:2004kc}, where \jT\ is the component of hadron
momentum perpendicular to the jet thrust axis:
\begin{equation}
\jT=\sqrt{p_h^2-\left(\frac{p_h\cdot p_{jet}}{p_{jet}}\right)^2}.
\label{eq:jT}
\end{equation}
$p_h$ is the hadron momentum and $p_{jet}$ is the jet
momentum. The jets are at mid-rapidity and have measured transverse
energy $\langle\ET\rangle\sim11$ GeV. The figure shows comparison to
PYTHIA-based simulations \cite{Sjostrand:1987su} 
which describe the data well, including
variations in the \jT\ distribution due to variations of the lower limit of the hadron
\pT\ relative to the beam axis (the kinematic ``seagull effect''). Aside from the importance of
jets for studying QCD processes at RHIC energies, jet measurements in
simpler systems at RHIC (potentially including the collisions of light
nuclei) provide essential data-based calibrations of the more limited
multi-hadron correlation observables that are accessible in the
collisions of heavy nuclei \cite{Henry:2004kc}.

\section{Summary}

The Relativistic Heavy Ion Collider has initiated a new era in the
study of QCD matter under extreme conditions. The RHIC accelerator has
unprecedented flexibility for a collider, having generated significant
integrated luminosity for $Au+Au$ collisions at several energies as
well as for polarized protons and the very important $d+Au$ control
experiment. The four RHIC experiments have produced a large body of
high quality data. There is considerable overlap in their physics
coverage and it is notable that the results are in agreement on all
major physics points.

The data collected and analysed in the first three years of RHIC
operations indicate that a dense, equilibrated system is generated
briefly in the most violent, head-on collisions of heavy nuclei at top
RHIC energies. The most economical explanation of the observed
phenomena is that a state of matter dominated by colored (partonic)
degrees of freedom has been produced, though direct observation of the
deconfinement transition awaits further experimental and theoretical
developments. For long wavelength excitations (low $Q^2$ or soft
probes) the matter evidently responds as a near-ideal, strongly
coupled fluid, while for short wavelength (high $Q^2$) probes it is
highly dissipative. The medium is markedly different from the ideal
non-interacting gas expected from QCD at asymptotically high
temperature.

In this review we have presented only a limited subset of the
available results, concentrating on the most mature measurements
that directly address new phenomena. The evidence that a novel state of
matter has been created is based on several lines of argument (see
also \cite{Gyulassy:2004vg}):

\begin{itemize}
\item Collective behaviour: the broad success of hydrodynamic calculations
in describing the inclusive spectra and azimuthal anisotropies of soft
hadrons indicates that local thermal equilibrium is established at
time $\tau\lt1$ fm/c after the collision and that the matter expands
as a near-ideal fluid.
\item Partonic energy loss: the strong suppression of inclusive yields and 
correlations of high \pT\ hadrons shows that high energy partons
dissipate significant energy in the medium, providing direct evidence
of strong interaction among partons that is essential to establish
thermalization. The absence of high \pT\ suppression in the $d+Au$
control experiment confirms that the suppression is due to final state
interactions in dense matter. Hadronic absorption calculations cannot
account qualitatively for the systematic behavior of the phenomena,
and the suppression agrees with a picture of partonic interactions
with a medium having high color charge density.
\item Energy density: the independent analyses of transverse energy 
production, hydrodynamic flow, and partonic energy loss result in a consistent
estimate of the energy density early in the collision of 5-10
GeV/fm$^3$, well beyond the critical value for transition to a 
deconfined phase expected from Lattice QCD.
\end{itemize}

Nevertheless, there remain important open questions. The mechanisms
underlying the apparent very rapid thermalization are at present
unclear. The role of finite viscosity, especially at the later
hadronic stage of the expansion, remains to be fully understood.  Most
importantly, sensitivity to the Equation of State has not yet been
systematically explored. New data are critical for addressing these
issues, in particular more precise measurements of the low \pT\
spectra and asymmetries of multistrange baryons and charmed
mesons. New theoretical developments and systematic studies that are
constrained by the broad range of data now available are likewise
needed.

In the area of hard probes, striking effects have been observed but
precision data in a larger $p_T$ range are needed to explore the
detailed properties of jet quenching and their connection to other
properties of the dense matter. The region $2\lt\pT\lt6$ GeV/c has
significant contributions from non-perturbative processes, perhaps
revealing novel hadronization mechanisms. However, all studies to date
of azimuthal anisotropies and correlations of ``jets'' have by
necessity been constrained to this region, with only the inclusive
spectra extending well beyond the range where non-perturbative
processes are seen to play a role. High statistics data sets for much
higher \pT\ hadrons are needed to fully exploit azimuthal asymmetries
and correlations as measurements of partonic energy loss. Heavy quark
suppression is theoretically well controlled, and measurement of it
will provide a critical check on our understanding of partonic energy
loss. The {\it differential} measurement of energy loss through
measurement of the emerging away-side jet and the recovery of the
energy radiated in soft hadrons is still in its initial phase of
study. A complete mapping of the modified fragmentation with larger
initial jet energy and with a direct photon trigger will cross check
the energy dependence of energy loss extracted from single inclusive
hadron suppression.  Experiments at different colliding energies are
also essential to reveal the onset of critical phenomena, or at the
minimum to map the variation of jet quenching with initial energy
density and the lifetime of the dense system.

Qualitatively new observables are also on the horizon. With
accumulated luminosity increasing each year, measurements of $J/\psi$
suppression in $Au+Au$ collisions will become more quantitative and
differential. Together with $d+Au$ results and the open charm
measurements it may be possible to disentangle final state
suppression, initial state nuclear absorption, and possibly the
contribution to $J/\psi$ production from charm quark recombination in
the dense medium
\cite{Thews:2000rj,Grandchamp:2002wp}.  These measurements will
provide an independent measurement of the initial conditions of the
dense medium.  Direct thermal photon emission from the plasma at
intermediate $p_T$ could in principle provide direct measurement of
the effective temperature or average parton energy of the interacting
dense matter. Emission from the hadronic phase and photon
production from perturbative hard processes form large backgrounds,
however, making the extraction of the thermal photon yield from the
QGP phase extremely difficult
\cite{Rapp:1999ej,Gale:2003iz}. 
Finally, experimental study of dilepton 
spectra in the low and intermediate mass region will provide
vital information on the medium modification of vector mesons
due to chiral symmetry restoration, as well as the thermal
dilepton emission from the plasma phase.

There has been considerable recent interest in universal properties of
QCD at very low Bjorken \xBj\ \cite{Iancu:2003xm}. Due to the
non-Abelian nature of QCD, the gluon density in hadrons grows rapidly
with decreasing \xBj. However, the gluon self-interaction must
eventually lead the gluon density to saturate. In the saturation
region the system is dense but weakly coupled, controlled by the large
saturation scale. This is a unique QCD regime, called the Color Glass
Condensate (CGC). While arguments supporting the CGC are generic, the
momentum scale at which it occurs must be fixed by measurements. CGC
effects are expected to be amplified in heavy nuclei and evidence for
them has been sought in RHIC data. The question at present is still
open, but it is possible that the CGC is the underlying initial
coherent state of partons in the nucleus from which the incoherent,
thermalized QGP emerges following the collision. It has been suggested
\cite{Kharzeev:2003wz} that measurements in the forward ($d$ direction) region of
$d+Au$ collisions could provide a crucial test of saturation phenomena,
in particular its spill-over to the large $p_T$ region, though that
region is known to be susceptible to other non-perturbative
effects. Measurements in the central rapidity region at LHC energies
may provide clearer evidence of saturation phenomena.

\section{Acknowledgements}

We thank many colleagues for stimulating discussions and for
providing many of the plots used in this review.
We thank
C. Gagliardi,
R. Hwa,
D. Kharzeev,
C.-M. Ko,
D. Magestro,
A. Poskazner,
I. Tserruya,
R. Venugopalan,
I. Vitev,
and F.-Q. Wang
for comments on the manuscript.
This work was supported by the Director, Office of Energy
Research, Office of High Energy and Nuclear Physics, Divisions of 
Nuclear Physics, of the U.S. Department of Energy under Contract No.\
DE-AC03-76SF00098 and DE-FG03-93ER40792.




\bibliographystyle{h-physrev}


\bibliography{BRAHMS,PHENIX,PHOBOS,STAR,MiscExperiment,Theory}

\end{document}